\documentclass[acmtog, nonacm]{acmart}

\usepackage{multirow} %
\usepackage{pifont} %
\usepackage{soul}
\usepackage[ruled]{algorithm2e} %

\newcommand{\refTab}[1]{Table~\ref{#1}}
\newcommand{\Expect}[2]{\underset{#1}{\mathbb{E}}\left[#2\right]}

\newcommand{\bfx}{{\bf x}}

\definecolor{mus}{HTML}{0000FF}
\definecolor{im}{HTML}{CC0066}
\definecolor{rn}{HTML}{663399}
\definecolor{rev}{HTML}{0066CC}
\definecolor{lightgray}{gray}{0.90}
\definecolor{DarkGreen}{rgb}{0.0,0.6,0.0}

\makeatletter
\newcommand{\ie}{\emph{i.e.}, \@ifnextchar.{\!\@gobble}{}}
\newcommand{\eg}{\emph{e.g.}, \@ifnextchar.{\!\@gobble}{}}
\newcommand{\etc}{etc\@ifnextchar.{}{.\@}}
\makeatother

\DeclareRobustCommand{\captiontitle}[1]{{\textit{~#1}}}

\DeclareGraphicsExtensions{.png,.jpg,.pdf,.ai,.psd}
\SetAlFnt{\small}
\SetAlCapFnt{\small}
\SetAlCapNameFnt{\small}
\SetAlCapHSkip{0pt}
\newcommand{\rot}[1]{\rotatebox[origin=c]{90}{#1}}
\newcommand{\centeredtab}[1]{\setlength{\tabcolsep}{0pt}\begin{tabular}{c} #1 \end{tabular}}
\definecolor{mycyan}{HTML}{00bcdb}

\graphicspath{{figures/}}

\author{Mustafa B. Yaldiz}
\orcid{0000-0001-7795-2454}
\affiliation{%
  \institution{University of California San Diego}
  \city{La Jolla}
  \state{CA}
  \country{USA}}
\email{myaldiz@ucsd.edu}

\author{Ishit Mehta}
\orcid{0009-0003-6360-168X}
\affiliation{%
  \institution{University of California San Diego}
  \city{La Jolla}
  \state{CA}
  \country{USA}}
\email{ibmehta@ucsd.edu}

\author{Nithin Raghavan}
\orcid{0000-0003-3280-0750}
\affiliation{%
  \institution{University of California San Diego}
  \city{La Jolla}
  \state{CA}
  \country{USA}}
\email{n2raghavan@ucsd.edu}

\author{Andreas Meuleman}
\orcid{0000-0002-9899-6365}
\affiliation{%
  \institution{Inria, Universit\'e C\^ote d'Azur}
  \city{Sophia Antipolis}
  \country{France}}
\email{andreas.meuleman@gmail.com}

\author{Tzu-Mao Li}
\orcid{0000-0001-5443-470X}
\affiliation{%
  \institution{University of California San Diego}
  \city{La Jolla}
  \state{CA}
  \country{USA}}
\email{tzli@ucsd.edu} 

\author{Ravi Ramamoorthi}
\orcid{0000-0003-3993-5789}
\affiliation{%
  \institution{University of California San Diego}
  \city{La Jolla}
  \state{CA}
  \country{USA}}
\email{ravir@ucsd.edu}

\begin{document}

\title{Spectral Prefiltering of Neural Fields}
\begin{abstract}
Neural fields excel at representing continuous visual signals but typically operate at a single, fixed resolution.
We present a simple yet powerful method to optimize neural fields that can be prefiltered in a single forward pass.
Key innovations and features include: 
(1) We perform convolutional filtering in the input domain by analytically scaling
Fourier feature embeddings with the filter’s frequency response.
(2) This closed-form modulation generalizes beyond Gaussian filtering and
supports other parametric filters (Box and Lanczos) that are unseen at training time.
(3) We train the neural field using single-sample Monte Carlo estimates of the
filtered signal.
Our method is fast during both training and inference, 
and imposes no additional constraints on the network architecture. We show
quantitative and qualitative improvements over existing methods for neural-field filtering.
\end{abstract}

\begin{teaserfigure}
  \centering
  \includegraphics[width=\linewidth]{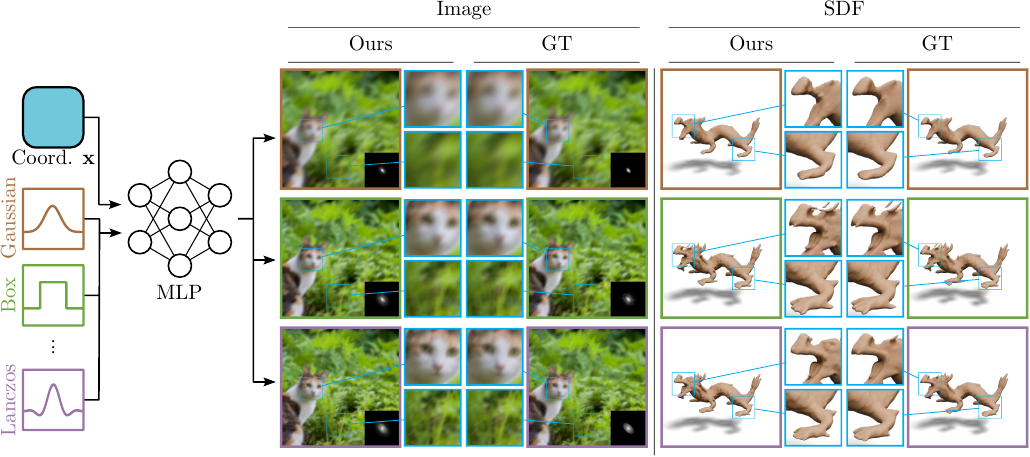}
  \Description{Overview teaser figure showing neural fields trained with Gaussian filters and tested on Gaussian, Box, and Lanczos; includes image and SDF reconstructions.}
  \caption{
  \label{fig:teaser}
  We present a training method for neural fields that enables linear
  prefiltering with multiple reconstruction filters.
  At training time, the neural field sees parameters of a single symmetric filter.
  At test time, we support prefiltering a variety of unseen filters (\textit{e.g.,} Box or Lanczos).
  Here, we show neural fields trained on an image (with bottom-right insets of frequency spectrum) and signed distance function using Gaussian filters, with generalization on Box and Lanczos filters.
  Images from Adobe FiveK; \textcopyright{} original photographers/Adobe.
  Mesh models from the Stanford 3D Scanning Repository; \textcopyright{} Stanford Computer Graphics Laboratory.
  (Project page: \url{https://myaldiz.info/assets/spnf/})
  }
\end{teaserfigure}

\maketitle

\section{Introduction}
\label{sec:intro}
Neural fields are now widely adopted in visual
computing~\cite{Xie:2022:NFV}.
They are used as continuous functions that map coordinates from an input domain
(\eg pixel locations) to the corresponding signal values (\eg radiance).
Generally, they provide point-wise estimates of the signal. As such, naively upsampling or downsampling the neural field produces sampling artifacts, preventing applications such as mipmapping.
Previous works~\cite{Fathony:2020:MFN,Lindell:2022:BBC,mujkanovic2024neural} aim to address the demand for \emph{resolution-aware} neural fields.
These methods impose significant restrictions on their network architectures, and often are restricted to a specific type of filter like Gaussian. 
In contrast, many graphics applications, require alternative filters that trade between sharpness and ringing.
We propose a method for fitting neural fields that
enables accurate filtering with a variety of low-pass symmetric reconstruction kernels,
while imposing few constraints on the neural network architecture.
Given the parameters of a filter and a spatial coordinate, our network predicts filtered signal values in a single evaluation (Fig.~\ref{fig:teaser}).

Our key idea is to integrate the Fourier feature encoding~\cite{tancik2020fourier} and derive an analytical formula to modify the coefficients. 
We show that by supervising the network with one type of low-pass
filter (\eg~Gaussian), it naturally generalizes to different types of low-pass filters (\eg~Box or Lanczos).
We estimate the filtered signal and train the network
with a Monte Carlo estimator of the convolution.

We demonstrate our method for prefiltering 2D images and 3D signed distance functions.
Since we do not impose any restriction on the network architecture apart from the Fourier feature encoding, we achieve significantly higher quality results than prior work~\cite{mujkanovic2024neural}.
Furthermore, our Monte Carlo estimator only requires a single sample from the convolution filter; hence it induces little performance cost during training even when the signal is expensive to evaluate.

In summary, our contributions are:
\begin{enumerate}
    \item An analytical prefiltering approach for neural fields using Fourier feature encoding,
    \item A training regime for prefiltered neural fields that generalizes to a variety of linear and symmetric convolutional filters.
\end{enumerate}

\section{Related Work}
\label{sec:relatedwork}

\begin{figure}[t]
    \centering
    \includegraphics[width=\linewidth,page=1]{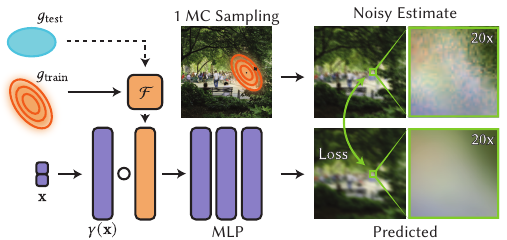}
    \caption{
    \captiontitle{Method Overview.}
    Our method rests on two key ideas: (i) using the analytic Fourier transform
    ($\mathcal{F}$) of a symmetric linear filtering kernel to modulate
    Fourier-feature embeddings, and (ii) using single-sample Monte Carlo (1 MC)
    estimates of the filtered signal for supervision.  
    A neural field trained with one filter type $g_\text{train}$, generalizes to  
    unseen filters $g_\text{test}$.
    Images from Adobe FiveK; \textcopyright{} original photographers/Adobe.
    }
    \label{fig:overview}
\end{figure}

\paragraph{Multi-scale representations}
Computer graphics and vision methods often rely on data structures (\eg~mipmapping) that represent a signal at multiple scales~\cite{williams1983pyramidal, witkin1987scale} to avoid expensive postfiltering. 
The multi-scale representations can be useful for texture filtering~\cite{williams1983pyramidal,Greene:1986:CRO,Heckbert:1989:FTM}, image processing~\cite{Adelson:1984:PMI,lowe2004sift}, level of detail~\cite{Hoppe:1996:PM}, and multi-resolution editing of geometry~\cite{Zorin:1997:IMM}.
Our work shares the same motivation as early texture filtering works. 
Post-filtering neural fields requires approximating the filtering integral through either a) cubature discretization; which is both memory and compute expensive or b) Monte Carlo sampling, which results in excessive noise (see Fig.~\ref{fig:overview})
We focus on building a multi-scale representation of a single-pass coordinate neural network~\cite{Song:2015:VRF} for a given filtering kernel.

\paragraph{Neural fields}
Neural fields compactly represent continuous signals using multilayer perceptrons conditioned on spatial coordinates~\cite{Xie:2022:NFV}. Their versatility makes them appealing for representing images~\cite{Song:2015:VRF,Belhe:2023:D2N}, geometry~\cite{park2019deepsdf,Sivaram:2024:NGF}, light fields~\cite{sitzmann2021light}, radiance fields~\cite{mildenhall2021nerf}, and spatially varying reflectance~\cite{Rainer:2019:NBC,bi2020neural}.

\paragraph{Input embeddings}
Directly mapping coordinates to the output using a neural network can fail to represent high-frequency details due to the spectral bias of multilayer perceptrons~\cite{rahaman2019spectral}.
To alleviate this, modern approaches often map spatial coordinates through diverse activations as embeddings (\eg~Gaussian, sine)~\cite{stanley2007compositional,rahimi2007random,vaswani2017attention}.
In this work, we focus on the Fourier feature mapping introduced by Tancik et al.~\shortcite{tancik2020fourier}, which encodes input coordinates using sines and cosines at multiple frequencies.
We show that this representation enables the derivation of closed-form expressions for modifying feature weightings in accordance with a given filter kernel, leading to significant improvements in filtering accuracy.

\paragraph{Scale-aware neural rendering}
Mip-NeRF~\cite{barron2021mip} introduced conical ray integration to reduce aliasing via scale-aware positional encodings. 
Zip-NeRF~\cite{barron2023zip}, Tri-MipRF~\cite{hu2023tri}, and Rip-NeRF~\cite{liu2024rip} extend this idea using hierarchical grids, mipmaps, and directional ripmaps.
These methods optimize a photometric loss on down-sampled training images.
They focus on anti-aliasing in radiance fields, but lack extensive analysis on whether internal signals such as volumetric density are correctly filtered, nor can they change the filtering at test-time.

\paragraph{Learned multi-scale representations}
\citet{Song:2015:VRF} represent a multilayer-perceptron-based mipmap using coordinates $(x, y, l)$ where $l$ is a continuous mipmapping level.
Recently, several neural multi-scale representations are proposed~\cite{Fathony:2020:MFN,Lindell:2022:BBC,Shekarforoush:2022:RMF,Yang:2022:PNF,Saragadam:2022:MMI}.
However, they typically put significant constraints on the network architectures, and can only handle limited anisotropy.
Beyond multi-scale architectures, \citet{xu2022signal} directly learn operators that act on the neural fields.
\citet{nsampi2023neural} derive continuous convolutions for neural fields by convolving repeated derivatives of the kernel with repeated antiderivatives of the signal, which is exact for piecewise-polynomial kernels.
\citet{Lindell:2022:BBC} directly modify the network architecture to handle box-filtering and Gaussian-filtering but their method is not continuous in filter space.
Closest to our work, NGSSF~\cite{mujkanovic2024neural} learns an anisotropic Gaussian scale-space via a scale-conditioned MLP, but remains restricted to Gaussian filters and relies on Lipschitz regularization plus a calibration stage for continuous smoothing. 
In contrast, our approach uses frequency-aware Fourier features and a single-sample Monte Carlo estimator to achieve filter-agnostic and anisotropic prefiltering without further architectural constraints, resulting in both greater flexibility and superior reconstruction quality.

\section{Background}
\label{sec:background}

Let $\mathbf{x} \in \mathbb{R}^{d_i}$ be a \(d_i\)-dimensional \emph{input coordinate}, and consider a continuous signal
$f : \mathbb{R}^{d_i} \to \mathbb{R}^{d_o}$,
where \(d_i\) and \(d_o\) are typically small (\eg~\(d_i=2,\,d_o=3\) for RGB images, and \(d_i=3,\,d_o=1\) for a signed-distance field). 
A \emph{neural field} approximates this signal:
\begin{equation}
F_\theta( \bfx ) \;\approx\; f( \bfx ),
\end{equation}
where \(\theta\) are the learned parameters. 
To incorporate a filter $K$ parameterized by a parameter \(\Sigma\) (\eg~a Gaussian kernel) the {\it filtered signal} is defined as the following convolution:
\begin{equation}
f_{K,\Sigma} ( \bfx ) \; \dot= \; (K_{\Sigma} * f)(\mathbf{x})
= \int K_{\Sigma}\bigl(\mathbf{x} - \mathbf{x}'\bigr)\,f(\mathbf{x}')\,\mathrm{d}\mathbf{x}'.
\end{equation}
Throughout this paper, we assume the spatial kernel \(K_\Sigma\) is \emph{symmetric}, \ie~ \(K_\Sigma(\mathbf{x}) = K_\Sigma(-\mathbf{x})\).
After Fourier transform, its frequency response \(\mathcal{F}\{K_\Sigma\}(\boldsymbol\omega)\) is therefore real and even, yielding zero phase shift which simplifies our frequency representation (see Sec.~\ref{sec:prefilter_fourier}).
In the anisotropic filter case, symmetry still holds, but magnitude varies by direction (\eg~elliptical Gaussian).

Conventionally, filtering a signal can be done through Monte Carlo estimation of the convolution integral \cite{hermosilla2018monte}.
Naive Monte Carlo estimation for filtering neural fields suffers from two standard drawbacks, (i) high variance and (ii) reliance on full network evaluation for each Monte Carlo sample, making it expensive.
To address these limitations, we seek to prefilter neural fields with a single
compact model that approximates the filtered signal:
\begin{equation}
F_{K,\Sigma}( \bfx; \theta)
\approx
f_{K,\Sigma}(\bfx).
\end{equation}
Our goal is to learn the neural field such that it produces filtered output in a
single forward pass and is continuous in the parameter space $\Sigma$.

A common way to avoid per-query integration at inference is to relocate the convolution to the input features, 
\ie~convolve the embedding once and then feed it to the network.
Prior works~\cite{barron2021mip, liu2024rip, wu2024neural} learn integrated input feature embeddings 
(\eg~positional encoding~\cite{vaswani2017attention}, hashgrids, \etc) to filter neural fields:
\begin{equation}
    \gamma_{K, \Sigma}(\mathbf{x})
    =\int K_{\Sigma}\bigl(\mathbf{x}-\mathbf{x}'\bigr)\,\gamma'(\mathbf{x}')\,\mathrm{d}\mathbf{x}'.
\label{eq:feature_conv}
\end{equation}
Here, $\gamma'$ is a feature embedding at original resolution and $\gamma(\cdot,\Sigma)$ is the convolved embedding.
These features are then passed through the network (MLP) to approximate the filtered signal in one pass:
\begin{equation}
    F_{K,\Sigma}( \bfx; \theta)
    =
    \mathrm{MLP}_\theta\bigl(\gamma_{K,\Sigma}(\bfx)\bigr).
\label{eq:mlp_on_convolved_features}
\end{equation}
We build on the feature-space view in Equations~\eqref{eq:feature_conv}–\eqref{eq:mlp_on_convolved_features}: we choose an embedding whose convolution with $K_\Sigma$ is available in closed form, enabling a single-pass filtered output.
We develop this idea further in Section~\ref{sec:prefilter_fourier}, showing how it generalizes beyond a single kernel to other symmetric filter families with continuous control over $\Sigma$.

\section{Method}
\label{sec:method}
\begin{figure}[t]
    \includegraphics[width=\linewidth,page=1]{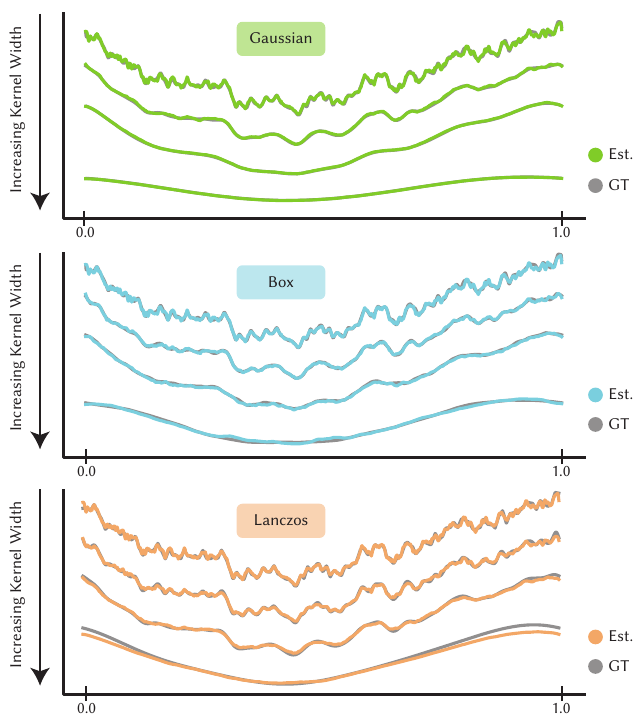}
    \caption{
    \captiontitle{1D Example.}
    Given a discretely sampled 1D signal and a finite set of uniformly sampled
    Gaussian kernels, we train a neural field to encode the continuous scale
    space of the signal (\emph{top}).
    By modulating the Fourier feature embeddings as per
    \S~\ref{sec:prefilter_fourier}, we prefilter the neural field with Box
    (\emph{middle}) and Lanczos (\emph{bottom}) filters with no additional
    supervision.
    }
    \label{fig:1d}
\end{figure}

Given a symmetric kernel $K_\Sigma$, our method:
(i) prefilters Fourier features that are subsequently used by a neural network to predict the filtered signal;
(ii) trains on a single-sample Monte Carlo estimate of $(K_\Sigma * f)$ (Sec.~\ref{sec:monte_carlo}).
Implementation details are provided in Section~\ref{sec:train_impl}.
At test time, the same model supports arbitrary covariance $\Sigma$ (isotropic or anisotropic) and previously unseen filters, such as Box and Lanczos, in a single forward pass (see Fig.~\ref{fig:overview}).

\subsection{Prefiltering Fourier Features}
\label{sec:prefilter_fourier}
Our goal is to use a feature encoding \(\gamma'(\mathbf{x})\) with the following two properties:
\begin{enumerate}
    \item \textbf{Closed-form filtering.}  
    Convolution of $\gamma'$
 with a given filter can be computed analytically, avoiding the runtime overhead and accuracy tradeoffs of Monte Carlo (MC) or quadrature-based methods.
    \item \textbf{Spectral-bias control.}  
    The proposed encoding ensures robustness to MC noise during training and provides a direct mechanism to tune the MLP’s sensitivity to different frequency bands.
\end{enumerate}
Random Fourier feature embeddings meet these requirements:
\begin{equation}
\gamma'(\mathbf{x}) = \bigl[a_i\cos(2\pi\,\mathbf{b}_i^\top\mathbf{x}),\,a_i\sin(2\pi\,\mathbf{b}_i^\top\mathbf{x})\bigr]_{i=1}^m,
\label{eq:fourier}
\end{equation}
where the choice of frequencies \(\{\mathbf{b}_i\}\) and amplitudes \(\{a_i\}\) controls the spectral bias~\cite{tancik2020fourier}.
In this work, we extend this idea by deriving an exact, filter-dependent modulation of \(\{a_i\}\), giving precise control over the network-output's frequency spectrum.
Consistent with Equation~\eqref{eq:feature_conv} in Section~\ref{sec:background}, 
we \emph{instantiate} $\gamma(\mathbf{x},\Sigma)$ by exactly convolving this embedding with $K_\Sigma$.

In Equation~\eqref{eq:fourier}, each entry $g_i$ of $\gamma'(\mathbf{x})$ is a sinusoid at frequency $\mathbf{b}_i$, whose spectrum is a pair of impulses at $\pm\mathbf{b}_i$.
By the convolution theorem,
\begin{equation}
\mathcal{F}\{K_\Sigma * g_i\} \;=\; \mathcal{F}\{K_\Sigma\}\cdot \mathcal{F}\{g_i\}.
\end{equation}
For a \emph{symmetric} kernel, $\mathcal{F}\{K_\Sigma\}$ is real and even, so multiplying the two impulses at $\pm\mathbf{b}_i$ by $\mathcal{F}\{K_\Sigma\}(\mathbf{b}_i)$ only rescales the amplitude (no phase change). Inverting the transform gives:
\begin{equation}
(K_\Sigma * g_i)(\mathbf{x}) \;=\; \mathcal{F}\{K_\Sigma\}(\mathbf{b}_i)\, g_i(\mathbf{x}).
\end{equation}
Thus feature-space convolution amounts to scaling each cosine/sine pair at $\mathbf{b}_i$ by the filter’s magnitude at that frequency. 
We set the Fourier-feature amplitudes to the kernel’s magnitude at each frequency,
\begin{equation}
a_i(K_\Sigma) \;=\; \mathcal{F}\{K_\Sigma\}(\mathbf{b}_i).
\label{eq:downweight_formula}
\end{equation}
Using Equation~\eqref{eq:fourier} with $a_i = a_i(K_\Sigma)$ yields an embedding whose convolution with $K_\Sigma$ is exact. 
When supervised with an estimate of the filtered signal, the network learns to condition on this prefiltered embedding so that a single forward pass outputs $(K_\Sigma * f)(x)$. 
Matching $a_i$ to $\mathcal{F}\{K_\Sigma\}(\mathbf{b}_i)$ provides direct spectral-bias control.

\paragraph{Kernel-specific magnitudes.}
We now list the closed-form $a_i(K_\Sigma)$ used in this paper.
Given that $\Sigma\in\mathbb{R}^{n\times n}$ is a symmetric positive–definite covariance matrix,
we extend 1D kernels (Gaussian, Box, Lanczos) to $n$D by evaluating them at the Mahalanobis distance $\|\mathbf{x}\|_{\Sigma}=\sqrt{\mathbf{x}^\top\Sigma^{-1}\,\mathbf{x}}$, yielding anisotropic $n$D filters.
We provide full derivations and constants in the supplemental material.
For a Gaussian kernel:
\begin{equation}
a_i(K_\Sigma) = \exp\left(-2\pi^2 \mathbf{b}_i^\top \Sigma \mathbf{b}_i\right).
\label{eq:downweight}
\end{equation}
For an $n$-dimensional Box kernel, where $J_{n/2}$ is the Bessel function and $\Gamma$ is the Gamma function:
\begin{equation}
a_i(K_\Sigma) =
\frac{\Gamma\!\bigl(\tfrac{n}{2}+1\bigr)}
     {\pi^{n/2}}\,
\frac{J_{n/2}\!\Bigl(2\pi\sqrt{\mathbf{b}_i^{\!\top}\Sigma\mathbf{b}_i}\Bigr)}
     {\left(\sqrt{\mathbf{b}_i^\top \Sigma \mathbf{b}_i}\right)^{n/2}}.
\end{equation}

For a Lanczos kernel, let $p>0$ denote the Lanczos \emph{order} (we use $p$ to avoid a clash with the Fourier–feature weights $a_i$),
The Fourier transform of the $n$D Lanczos kernel is then:
\begin{equation}
a_i(K_\Sigma)
= \frac{p}{z_n}\,
\max\!\Big(
  \min\!\big(
    \frac{p+1}{2p} - \sqrt{\mathbf{b}_i^{\!\top}\Sigma\mathbf{b}_i},\ \min(1,\frac{1}{p})
  \big),\ 0
\Big).
\end{equation}
The constant $z_n$ is a dimension $n$ dependent normalization factor for the Lanczos kernel.
Finally, following Equation~\eqref{eq:mlp_on_convolved_features} in Section~\ref{sec:background}, 
we pass the convolved features through the network to approximate the filtered signal.

As an illustration, consider the toy example in Fig.~\ref{fig:1d}, where we
encode the scale-space of a 1D signal in a neural field.
We analytically integrate the Fourier features for a given filtering kernel,
which acts as proxy for prefiltering the signal.
We observe that using the proposed feature encoding modulation
(Eq.~\eqref{eq:downweight_formula}), enables prefiltering the neural field with
additional filters without fine-tuning or re-training the MLP network.

\subsection{Monte Carlo–Based Training}
\label{sec:monte_carlo}

The previous section defines a deterministic, filter-conditioned encoding $\gamma_{K,\Sigma}(\mathbf{x})$.
In our training scheme, we optimize only the MLP parameters $\theta$ to map $\gamma_{K,\Sigma}(\mathbf{x})$ to the filtered signal; 
the encoding itself is fixed given $(\mathbf{x},K,\Sigma)$.

Most scale-aware neural-field methods avoid explicit or approximated convolution of the ground truth signal by baking the filter, either directly into the dataset (via precomputation) or by incorporating the filter into the network architecture.
While these methods show good results, they nonetheless come with limitations as discussed in Section~\ref{sec:relatedwork}.

\paragraph{Objective and estimator.}
An unbiased estimate of the exact convolution can be given via Monte Carlo estimation; 
we draw \(N\) samples \(\{\mathbf{x}'_i\}\) from a probability distribution \(p(\cdot  \mid \mathbf{x})\) centered at the point of convolution \( \bfx \),%
\begin{align}
f_{K,\Sigma}( \bfx ) &= \Expect{\bfx' \sim p(\cdot \mid x)}{\frac{K_{\Sigma}(\bfx - \bfx')}{p(\bfx'|\bfx)}f(\bfx')} \\
&\approx \frac{1}{N}\sum_{i=1}^N \frac{K_{\Sigma}(\bfx - \bfx'_i)}{p(\bfx'|\bfx)}f(\bfx'_i),
\quad \bfx'_i\sim p(\cdot  \mid \bfx).
\label{eq:mc_conv}
\end{align}
We choose the density $p$ to be proportional to $|K_\Sigma|$.
However, a low-variance estimate given by dense multi-sampling is prohibitively slow.
We therefore \emph{train with a single Monte Carlo sample} ($N{=}1$) at each iteration.

\paragraph{Sampling procedure.}
During training, we uniformly sample a batch of coordinates $\bfx$ from the valid signal area,
and we sample positive semidefinite covariances $\Sigma$ by drawing principal variances uniformly in log-space with a random rotation.

(i) Given a filter (Gaussian, Box, Lanczos), we first normalize the filter $|K_\Sigma|$, treat it as a distribution and sample $\bfx'$.
Distributions of Gaussian and Box filters yield normal and uniform sampling respectively, which is easy to sample.
For Lanczos filter, we use rejection sampling, and cache those samples. 
During training and evaluation of Lanczos filter, we draw cached samples at random and stretch them according to the covariance.

(ii) Next, we sample the corresponding signal at $f(\bfx')$ as a MC estimate of the signal.
Our single-sample Monte Carlo training is similar to the coordinate perturbation introduced by \citet{ling2025stochastic} (See Eq.~\eqref{eq:objective}).
We also perturb the sampling coordinate, but unlike their method, we modulate the encoding according to the kernel $K_\Sigma$.

\subsection{Implementation Details}
\label{sec:train_impl}
We use standard settings for stable single-sample training.

\paragraph{Optimizer and schedule.}
We use Adam~\cite{kingma2014adam} for its robustness to noise. 
We found that an exponential learning-rate decay is critical: larger early steps drive progress despite target noise, while smaller later steps average out variance. 
In our experiments, adding exponential decay to $1e{-}3$ times of the original learning rate improved PSNR (\eg~+2.19\,dB on \emph{Alien}) and reduced speckle. After sweeping learning rates from $5\times10^{-5}$ to $5\times10^{-1}$, stochastic gradient descent (with/without momentum) either converged to poor minima or diverged, so we adopt Adam with learning rate $5 \times 10^{-4}$ with batch size $100k$ for images and $200k$ for SDF by default.

\paragraph{Loss.}
We minimize the minibatch MSE between the MLP on the prefiltered encoding and a single-sample MC target:
\begin{equation}
\begin{aligned}
\hat{\mathcal{L}}=\frac{1}{|\mathcal{B}|}\sum_{(\mathbf{x},K,\Sigma)\in\mathcal{B}}\Big\|\mathrm{MLP}_\theta\!\big(\gamma_{K,\Sigma}(\mathbf{x})\big)-\frac{K_\Sigma(\mathbf{x}-\mathbf{x}')}{p(\mathbf{x}'\,|\,\mathbf{x},K_\Sigma)}\,f(\mathbf{x'})\Big\|_2^2,\\
\mathbf{x}'\sim p(\cdot\,|\,\mathbf{x},K_\Sigma)\propto |K_\Sigma|.
\end{aligned}
\label{eq:objective}
\end{equation}
Because the MC target is unbiased for $(K_\Sigma * f)(\mathbf{x})$, this MSE is an unbiased objective.
For datasets with heavy-tailed, high-dynamic-range noise, one can instead explore objectives from Noise2Noise that down-weight extreme outliers~\cite{mansour2023zero}.

\paragraph{Architecture details.}
Unless otherwise noted, we follow the configuration from NGSSF~\cite{mujkanovic2024neural}: a 3-layer multi-layer perceptron of width 1024 and ($m{=}512$) 1024 channel Fourier features. 
We use a basis scale of $2000$ for images and $40$ for SDFs. The Fourier basis scale should match the signal’s frequency distribution: if set too small, the model is biased toward low frequencies (underfitting high-frequency detail); if set too large, it over-emphasizes high-frequency noise due to MC sampling.
We found that our Fourier feature prefiltering helps suppress the noise since it biases updates towards low-frequency content.

\section{Results}
\label{sec:results}
\begin{table*}
    \centering
    \caption{
    \captiontitle{Image filtering with isotropic Gaussian kernels.}
    We test our model across \textit{isotropic} kernels by averaging metrics across 100 images on 4($+$original signal) different isotropic kernels.
    Indicated by $\sigma^2$-cont., the NGSSF~\cite{mujkanovic2024neural} and NFC~\cite{nsampi2023neural} baselines are the only other methods that operate continuously in isotropic scale space. 
    Other methods, namely BACON~\cite{Lindell:2022:BBC}, MINER~\cite{Saragadam:2022:MMI}, INSP~\cite{xu2022signal}, illustrate the tradeoff between changing the network architecture and filtering quality.
    Best and second best are \textbf{bold} and \underline{underlined}.
    Our method outperforms these alternatives on filtering with changing scales. While MINER yields marginally higher PSNR at representing the original signal, it lacks continuous control over $\sigma^2$ scale space.
    }
    \renewcommand{\tabcolsep}{2pt} %
    \renewcommand{\arraystretch}{0.9}
    \label{tab:quant_img_iso}
    
    \begin{tabular}{lcrrrrrrrrrrrrrrrr}
        \toprule
        \multirow{2}[2]{*}{Method} & \multirow{2}[2]{*}{$\sigma^2$-cont.} & \multicolumn{3}{c}{$\sigma^2=0$} & \multicolumn{3}{c}{$\sigma^2=10^{-4}$} & \multicolumn{3}{c}{$\sigma^2=10^{-3}$} & \multicolumn{3}{c}{$\sigma^2=10^{-2}$} & \multicolumn{3}{c}{$\sigma^2=10^{-1}$} \\
        \cmidrule(lr){3-5}
        \cmidrule(lr){6-8}
        \cmidrule(lr){9-11}
        \cmidrule(lr){12-14}
        \cmidrule(lr){15-17}
        & & \footnotesize{PSNR$\uparrow$} & \footnotesize{LPIPS$\downarrow$} & \footnotesize{SSIM$\uparrow$} & \footnotesize{PSNR$\uparrow$} & \footnotesize{LPIPS$\downarrow$} & \footnotesize{SSIM$\uparrow$} & \footnotesize{PSNR$\uparrow$} & \footnotesize{LPIPS$\downarrow$} & \footnotesize{SSIM$\uparrow$} & \footnotesize{PSNR$\uparrow$} & \footnotesize{LPIPS$\downarrow$} & \footnotesize{SSIM$\uparrow$} & \footnotesize{PSNR$\uparrow$} & \footnotesize{LPIPS$\downarrow$} & \footnotesize{SSIM$\uparrow$} \\
        \midrule
        BACON [\citeyear{Lindell:2022:BBC}] & \ding{55} & 32.89 & 0.308 & 0.823 & \underline{38.95} & 0.235 & \underline{0.955} & 36.48 & 0.123 & 0.953 & 30.59 & 0.086 & 0.895 & 25.36 & 0.100 & 0.601 \\
        MINER [\citeyear{Saragadam:2022:MMI}] & \ding{55} & \textbf{41.19} & \textbf{0.088} & \textbf{0.963} & 37.38 & 0.259 & 0.945 & \underline{36.99} & 0.097 & \underline{0.959} & 25.89 & 0.205 & 0.815 & 24.38 & 0.156 & 0.567 \\
        INSP [\citeyear{xu2022signal}] & \ding{55} & 30.57 & 0.454 & 0.770 & 30.14 & 0.420 & 0.838 & 23.77 & 0.546 & 0.725 & 20.75 & 0.546 & 0.627 & 23.37 & 0.381 & 0.633 \\
        NFC [\citeyear{nsampi2023neural}] & \checkmark & 20.75 & 0.703 & 0.533 & 26.49 & 0.224 & 0.839 & 36.05 & \underline{0.071} & 0.949 & 39.74 & \underline{0.011} & \underline{0.965} & \underline{41.06} & \underline{0.006} & \underline{0.965} \\
        NGSSF [\citeyear{mujkanovic2024neural}] & \checkmark & 33.85 & 0.305 & 0.854 & 35.05 & \underline{0.207} & 0.942 & 34.74 & 0.077 & 0.954 & \underline{35.06} & 0.023 & 0.949 & 34.99 & 0.020 & 0.878 \\
        Ours & \checkmark & \underline{38.23} & \underline{0.193} & \underline{0.918} & \textbf{43.83} & \textbf{0.192} & \textbf{0.971} & \textbf{48.91} & \textbf{0.064} & \textbf{0.991} & \textbf{53.09} & \textbf{0.009} &\textbf{ 0.997} & \textbf{53.81} & \textbf{0.005} & \textbf{0.997 }\\
        \bottomrule
    \end{tabular}
\end{table*} 

\begin{table}
    \centering
    \caption{
    \captiontitle{Image filtering with anisotropic Gaussian kernels.}
    We test our model across \textit{anisotropic} kernels by averaging metrics across 100 images on 100 different anisotropic kernels with varying scales and orientations.
    Best scores are \textbf{bold}.
    We achieve greater accuracy on all metrics compared to prior work while remaining continuous in filter space.
    }
    \label{tab:quant_img_aniso}
    \begin{tabular}{lcrrr}
        \toprule
        & $\Sigma$-cont. & PSNR$\uparrow$ & LPIPS$\downarrow$ & SSIM$\uparrow$ \\
        \midrule
        PNF [\citeyear{Yang:2022:PNF}] & \ding{55} & 24.15 & 0.571 & 0.704 \\
        NFC [\citeyear{nsampi2023neural}] & \ding{55} & 30.31 & 0.094 & 0.857 \\
        NGSSF [\citeyear{mujkanovic2024neural}] & \checkmark & 34.82 & 0.069 & 0.940 \\
        Ours  & \checkmark & \textbf{49.33} & \textbf{0.054} & \textbf{0.991} \\
        
        \bottomrule
    \end{tabular}
\end{table}

We first compare against prior methods under Gaussian smoothing (Sec.~\ref{sec:results_images_gaussian}). 
We then study generalization to other filter families, Box and Lanczos (Sec.~\ref{sec:generalization_exp}). 

We evaluate on images and signed distance fields (SDFs).
For images, we follow the NGSSF evaluation benchmark~\cite{mujkanovic2024neural} on 100 high-resolution $2048\times2048$ Adobe FiveK images~\cite{bychkovsky2011learning}.
We treat signals as periodic and evaluate pixels whose convolution windows lie within the image boundaries.
We report PSNR (Peak Signal-to-Noise Ratio), SSIM~\cite{wang2004image}, and LPIPS~\cite{zhang2018unreasonable}.

For SDFs, following the NGSSF benchmark~\cite{mujkanovic2024neural} and training coordinate sampling,
we voxelize meshes into SDFs at $1024^3$ during training and $256^3$ at test time.
We extend the ground truth SDF values $1.2$ times beyond the normalized coordinates, calculate filtered fields, and again crop borders.
We use the Lucy, Dragon, Thai Statue, and Armadillo meshes from the Stanford 3D Scanning Repository (models courtesy of the \textcopyright{} Stanford Computer Graphics Laboratory).
We report MSE (Mean Squared Error), Chamfer distance~\cite{fan2017point}, and IoU~\cite{mescheder2019occupancy}.

\subsection{Gaussian Filtering}
\paragraph{Image filtering}
\label{sec:results_images_gaussian}

We report quantitative results for Gaussian smoothing in \refTab{tab:quant_img_iso} (isotropic) and \refTab{tab:quant_img_aniso} (anisotropic).
In addition, the first two rows (Gaussian) and first two columns (NGSSF and Trained w/ Gaussian) of Figure~\ref{fig:image_results} show qualitative comparisons.
We include additional qualitative results in the supplemental material.

In the isotropic setting, against the most relevant continuous-scale baselines—NFC~\cite{nsampi2023neural} and NGSSF~\cite{mujkanovic2024neural}—our method improves PSNR at least by \textbf{+4.9} to \textbf{+13.4}\,dB across scales.
It also outperforms discrete-scale architectures (BACON~\cite{Lindell:2022:BBC}, MINER~\cite{Saragadam:2022:MMI}, INSP~\cite{xu2022signal}) when filtering at varying scales.
MINER gives higher PSNR on the \emph{unfiltered} signal only (\(\sigma^2{=}0\); +2.96\,dB), which reflects differences in representation capacity rather than filtering behavior.

For anisotropic Gaussian smoothing, our method yields large, consistent gains over all baselines (\refTab{tab:quant_img_aniso}; see also the Gaussian rows/columns in Figure~\ref{fig:image_results}). 
Quantitatively, we improve PSNR by \textbf{+14.5}\,dB over NGSSF~\cite{mujkanovic2024neural} (49.33 vs.\ 34.82\,dB), and by \textbf{+19.0}/\textbf{+25.2}\,dB over NFC~\cite{nsampi2023neural}/PNF~\cite{Yang:2022:PNF} respectively, with corresponding LPIPS and SSIM also best (\eg~LPIPS $0.054$ vs.\ $0.069$ for NGSSF; SSIM $0.991$ vs.\ $0.940$). 

Among prior methods, NGSSF is the strongest baseline for Gaussian filtering in both isotropic and anisotropic settings.
However, its calibration scheme does not perfectly match the ground-truth filtering magnitude across scales and underfits frequency content which is visible near edges of the filtered signal.

\paragraph{SDF smoothing}
\label{sec:results_sdf} 
\begin{table*}
    \centering
    \caption{
    \captiontitle{SDF filtering with isotropic Gaussian kernels.}
    Metrics are averaged over \emph{Lucy, Dragon, Thai Statue, Armadillo} SDFs.
    Columns correspond to $\sigma^2\!\in\!\{0,10^{-4},10^{-3},10^{-2}\}$; $\sigma^2{=}0$ denotes unfiltered reconstruction.
    “$\sigma^2$-cont.” indicates continuity w.r.t.\ the scalar variance.
    Best and second best are \textbf{bold} and \underline{underlined}.
    MINER reports the lowest MSE at $\sigma^2\!\in\!\{0,10^{-4},10^{-3}\}$ and NFC at $\sigma^2\!=\!10^{-2}$, 
    whereas our method yields the best \emph{geometry}—lowest Chamfer and highest IoU—at all non-zero blur levels while remaining continuous in $\sigma^2$.
    }
    \label{tab:quant_geom_iso}
    \renewcommand{\tabcolsep}{0.1 cm} %
    \begin{tabular}{lcrrrrrrrrrrrrr}
        \toprule
        \multirow{2}[2]{*}{Method} & \multirow{2}[2]{*}{$\sigma^2$-cont.} & \multicolumn{3}{c}{$\sigma^2=0$} & \multicolumn{3}{c}{$\sigma^2=10^{-4}$} & \multicolumn{3}{c}{$\sigma^2=10^{-3}$} & \multicolumn{3}{c}{$\sigma^2=10^{-2}$} \\
        \cmidrule(lr){3-5}
        \cmidrule(lr){6-8}
        \cmidrule(lr){9-11}
        \cmidrule(lr){12-14}
        & & \footnotesize{MSE$\downarrow$} & \footnotesize{Cham.$\downarrow$} & \footnotesize{IoU$\uparrow$} & \footnotesize{MSE$\downarrow$} & \footnotesize{Cham.$\downarrow$} & \footnotesize{IoU$\uparrow$} & \footnotesize{MSE$\downarrow$} & \footnotesize{Cham.$\downarrow$} & \footnotesize{IoU$\uparrow$} & \footnotesize{MSE$\downarrow$} & \footnotesize{Cham.$\downarrow$} & \footnotesize{IoU$\uparrow$} \\
        \midrule
        BACON [\citeyear{Lindell:2022:BBC}] & \ding{55} & 2.5e-3 & \underline{1.3e-3} & \textbf{0.99} & 4.0e-3 & 2.2e-3 & \underline{0.97} & 8.3e-2 & 1.5e-2 & 0.84 & 2.6e-4 & 4.9e-2 & 0.53 \\
        MINER [\citeyear{Saragadam:2022:MMI}] & \ding{55} & \textbf{1.6e-7} & \textbf{1.1e-3} & \underline{0.98} & \textbf{3.3e-7} & \underline{1.4e-3} & \textbf{0.98} & \textbf{4.1e-6} & 8.0e-3 & 0.92 & 1.8e-4 & 6.1e-2 & 0.52 \\
        INSP [\citeyear{xu2022signal}] & \ding{55} & 1.2e-1 & \underline{1.3e-3} & \textbf{0.99} & 4.3e-2 & 4.4e-3 & 0.95 & 3.6e-2 & 1.1e-2 & 0.88 & 3.1e-2 & 3.7e-2 & 0.64 \\
        NFC [\citeyear{nsampi2023neural}] & \checkmark & 3.7e-3 & 5.7e-3 & 0.89 & 2.5e-5 & 4.8e-3 & 0.92 & 1.4e-5 & \underline{2.2e-3} & \underline{0.97} & \textbf{1.0e-5} & \underline{2.3e-2} & \underline{0.77} \\
        NGSSF [\citeyear{mujkanovic2024neural}] & \checkmark & 8.3e-5 & 3.9e-3 & 0.94 & 6.0e-5 & 5.5e-3 & 0.92 & 6.5e-4 & 1.6e-2 & 0.83 & 1.1e-2 & 1.3e-1 & 0.32 \\
        Ours & \checkmark & \underline{1.6e-5} & 1.7e-3 & \underline{0.98} & \underline{1.1e-5} & \textbf{1.2e-3} & \textbf{0.98} & \underline{9.5e-6} & \textbf{1.8e-3} & \textbf{0.98} & \underline{1.8e-5} & \textbf{3.6e-3} & \textbf{0.95} \\
        \bottomrule
    \end{tabular}
\end{table*}

\begin{table}
    \centering
    \caption{
    \captiontitle{SDF filtering with anisotropic Gaussian kernels.}
    Metrics are averaged over covariances $\Sigma$ on 100 different anisotropic kernels with varying scales and orientations.
    “$\Sigma$-cont.” indicates continuity over the full covariance.
    Best scores are \textbf{bold}.
    Our method achieves greater accuracy on all metrics while remaining continuous in filter space.
    }
    \label{tab:quant_geom_aniso}
    \begin{tabular}{lcrrr}
        \toprule
        & $\Sigma$-cont. & MSE$\downarrow$ & Cham.$\downarrow$ & IoU$\uparrow$ \\
        \midrule
        NFC [\citeyear{nsampi2023neural}] & \ding{55} & 7.1e-2 & 4.6e-1 & 0.08 \\
        NGSSF [\citeyear{mujkanovic2024neural}] & \checkmark & 2.8e-3 & 1.2e-1 & 0.42 \\
        Ours & \checkmark & \textbf{2.2e-5} & \textbf{3.6e-3} & \textbf{0.83} \\
        \bottomrule
    \end{tabular}
\end{table}

We present the quantitative results of isotropic and anisotropic Gaussian filtering of SDFs in Tables \ref{tab:quant_geom_iso} and \ref{tab:quant_geom_aniso} respectively.
Also the first row (Gaussian) of Figure~\ref{fig:sdf_results} shows qualitative SDF results.

Across isotropic kernels (Table~\ref{tab:quant_geom_iso}), our method gives the best overall geometry—lowest Chamfer distance and highest IoU—at all non-zero blur levels. 
MINER reports the lowest MSE at $\sigma^2\!\in\!\{0,10^{-4},10^{-3}\}$ and NFC has the lowest MSE at $\sigma^2\!=\!10^{-2}$, but both either lack continuity in $\sigma^2$ or degrade geometry (higher Chamfer, lower IoU). 
In contrast, our model remains continuous in filter space and preserves geometry across scales. For anisotropic kernels (Table~\ref{tab:quant_geom_aniso}), our method improves over NGSSF and NFC by large margins on all metrics (\eg~MSE $2.2\!\times\!10^{-5}$ vs.\ $2.8\!\times\!10^{-3}$ for NGSSF; Chamfer $3.6\!\times\!10^{-3}$ vs.\ $1.2\!\times\!10^{-1}$; IoU $0.83$ vs.\ $0.42$), removing streaking and floaters while matching the ground-truth shape more closely.

\subsection{Filter Generalization}
\label{sec:generalization_exp}
\begin{table*}
    \caption{
    \label{tab:training_filter_results_1}
    \captiontitle{Filter generalization.}
    We train four versions of our image-regression model—one each with Gaussian, Box, and Lanczos filters, plus a “Multiple” model trained on all three. 
    At test time, we evaluate each model using Gaussian, Box, and Lanczos kernels. 
    The results show that even models trained on single filters generalize well to unseen filters,  nearly matching the quality of training on all filters simultaneously.
    }
    \begin{tabular}{lrrrrrrrrr}
    \toprule
    \multirow{2}{*}{Training Filter} &
      \multicolumn{3}{c}{Gaussian} &
      \multicolumn{3}{c}{Box} &
      \multicolumn{3}{c}{Lanczos} \\
    \cmidrule(lr){2-4}
    \cmidrule(lr){5-7}
    \cmidrule(lr){8-10}
    & PSNR$\uparrow$ & LPIPS$\downarrow$ & SSIM$\uparrow$
    & PSNR$\uparrow$ & LPIPS$\downarrow$ & SSIM$\uparrow$
    & PSNR$\uparrow$ & LPIPS$\downarrow$ & SSIM$\uparrow$ \\
    \midrule
    Multiple & 43.82 & 0.128 & 0.969 & 40.22 & \textbf{0.111 }& 0.963 & 39.21 & \textbf{0.104 }& 0.963 \\
    Gaussian & \textbf{44.71} & 0.127 & \textbf{0.972} & 40.23 & \textbf{0.111} & 0.962 & 37.33 & 0.106 & 0.955 \\
    Box & 43.81 & \textbf{0.125} & \textbf{0.972} & \textbf{41.58} & 0.113 & \textbf{0.966} & 36.09 & 0.108 & 0.948 \\
    Lanczos & 38.03 & 0.128 & 0.952 & 35.24 & 0.120 & 0.938 & \textbf{41.47} & 0.107 & \textbf{0.967} \\
    \bottomrule
    \end{tabular}
\end{table*}

We test generalization across low-pass filter families, Gaussian, Box, and Lanczos, under two training schemes: 
(i) training with a single filter family and evaluating on the others, and 
(ii) training jointly with multiple filter families, Gaussian, Box, and Lanczos, at each iteration. 
(See also the 1D example in Fig.~\ref{fig:1d}).

\begin{figure*}
    \centering
    \setlength{\tabcolsep}{1pt}
    \renewcommand{\arraystretch}{0.6}
    \fboxsep=-0.1pt
    \begin{tabular}{cccccccc}
        & & & \multicolumn{2}{c}{NFC} &  & \multirow{2}{*}{\begin{tabular}{c}\vspace{-3.5pt}\\Ours\\Gaussian Trained\end{tabular}} \\
        \cmidrule(lr){4-5}
        & & Kernel & Fitted Kernel & Prediction & NGSSF & & GT \\
        \centeredtab{\rot{Gaussian}} & &
        \centeredtab{
            \includegraphics[width=0.155\textwidth, height=0.155\textwidth]{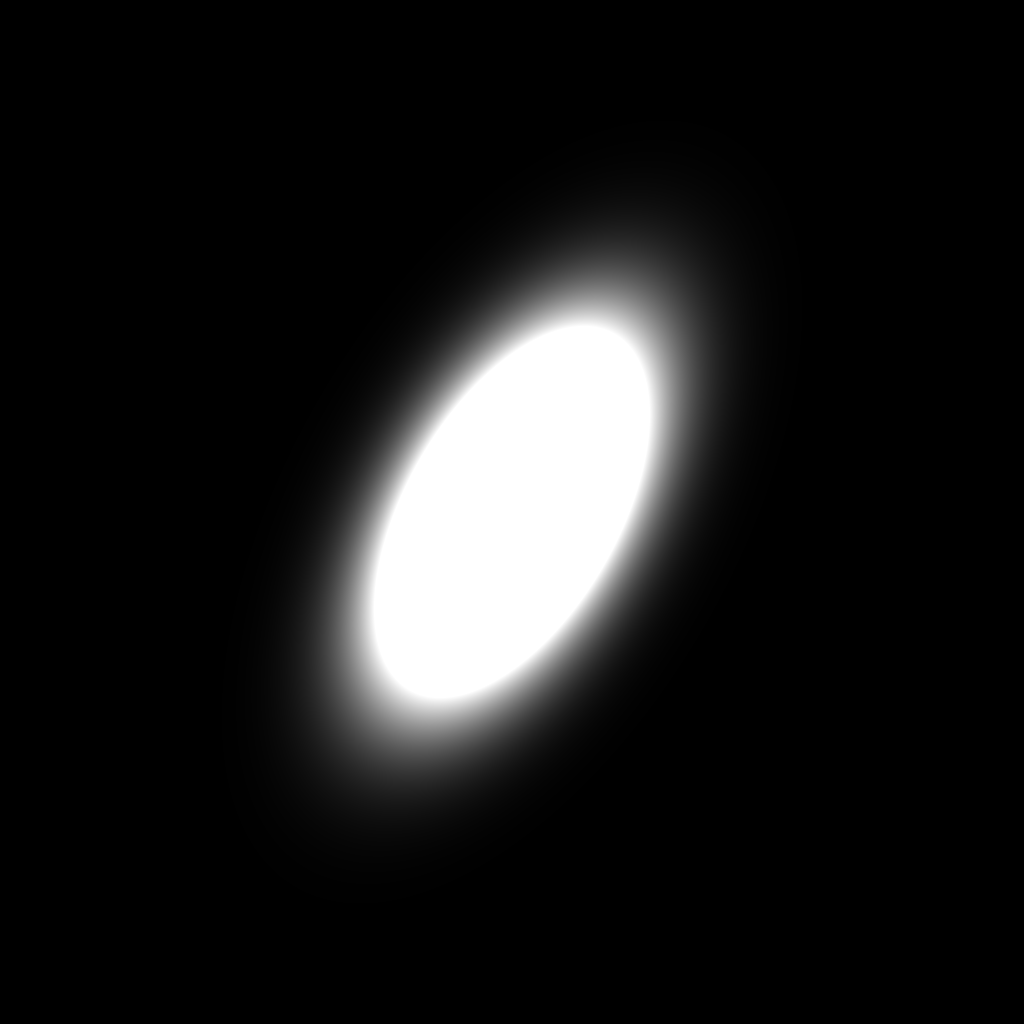} \\
            \includegraphics[width=0.155\textwidth, height=0.155\textwidth]{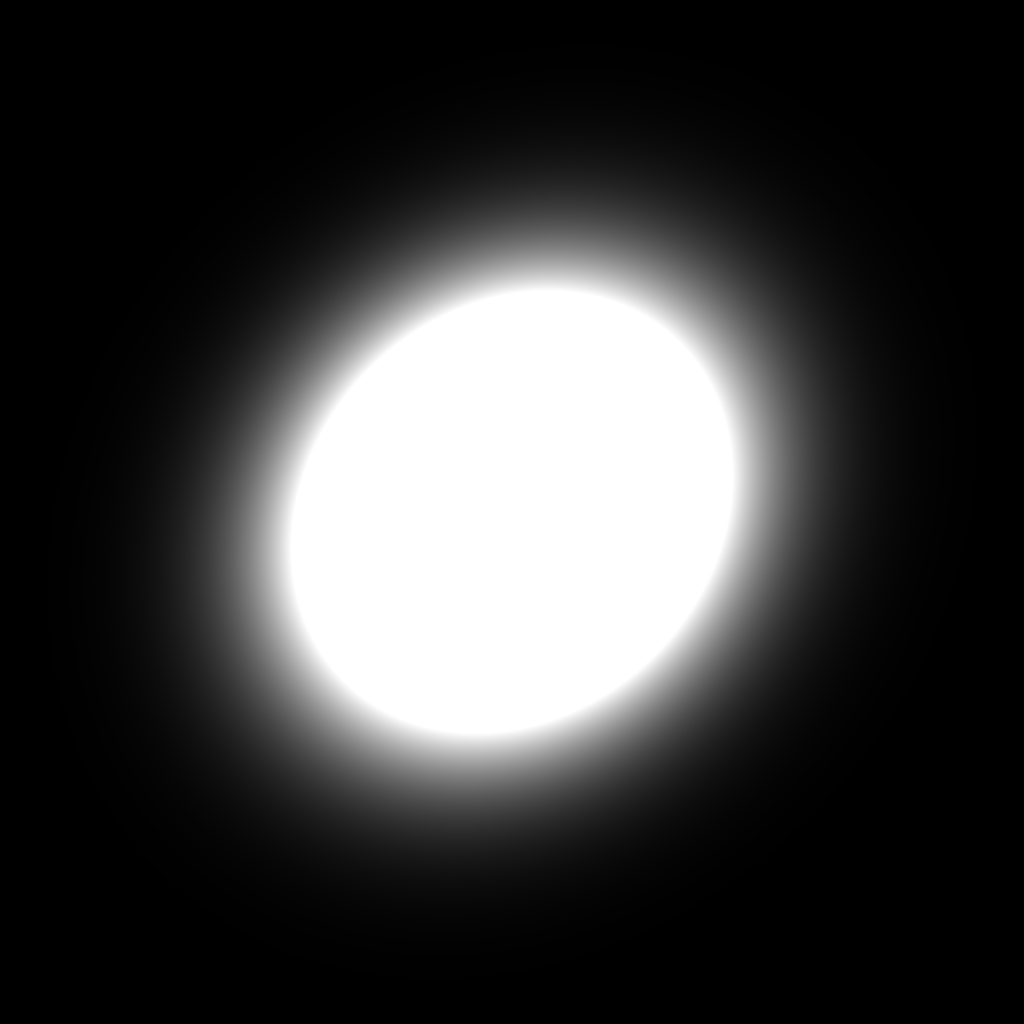}} & 
        \centeredtab{
            \includegraphics[width=0.155\textwidth, height=0.155\textwidth]{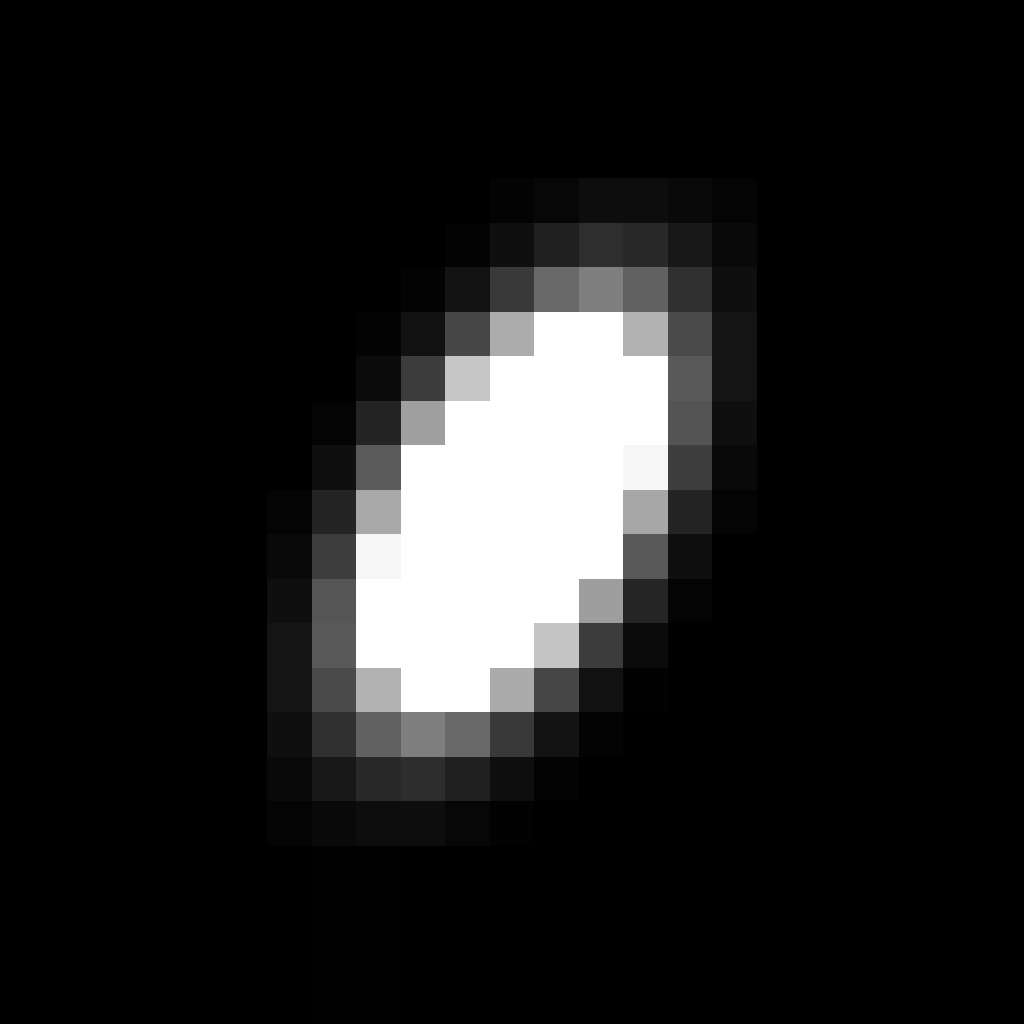} \\
            \includegraphics[width=0.155\textwidth, height=0.155\textwidth]{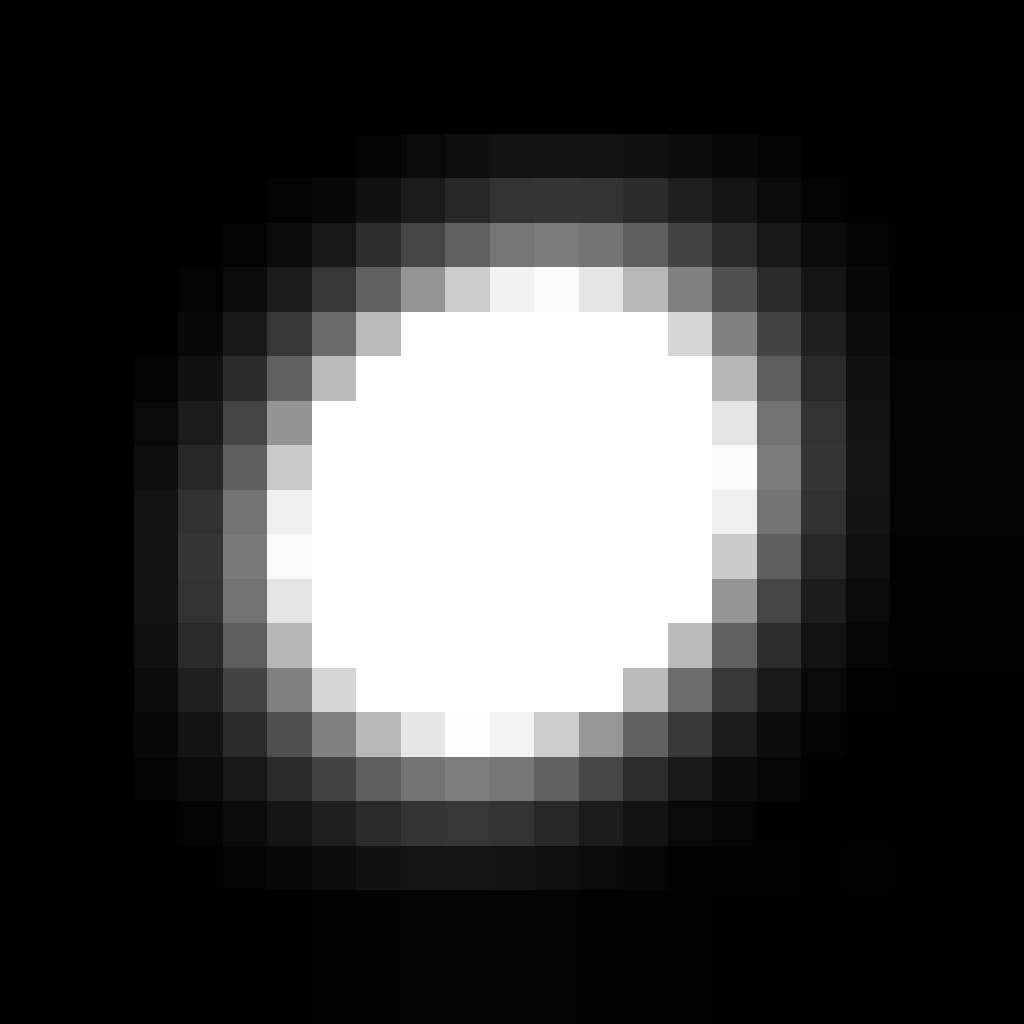}} & 
        \centeredtab{
            \includegraphics[width=0.155\textwidth, height=0.155\textwidth]{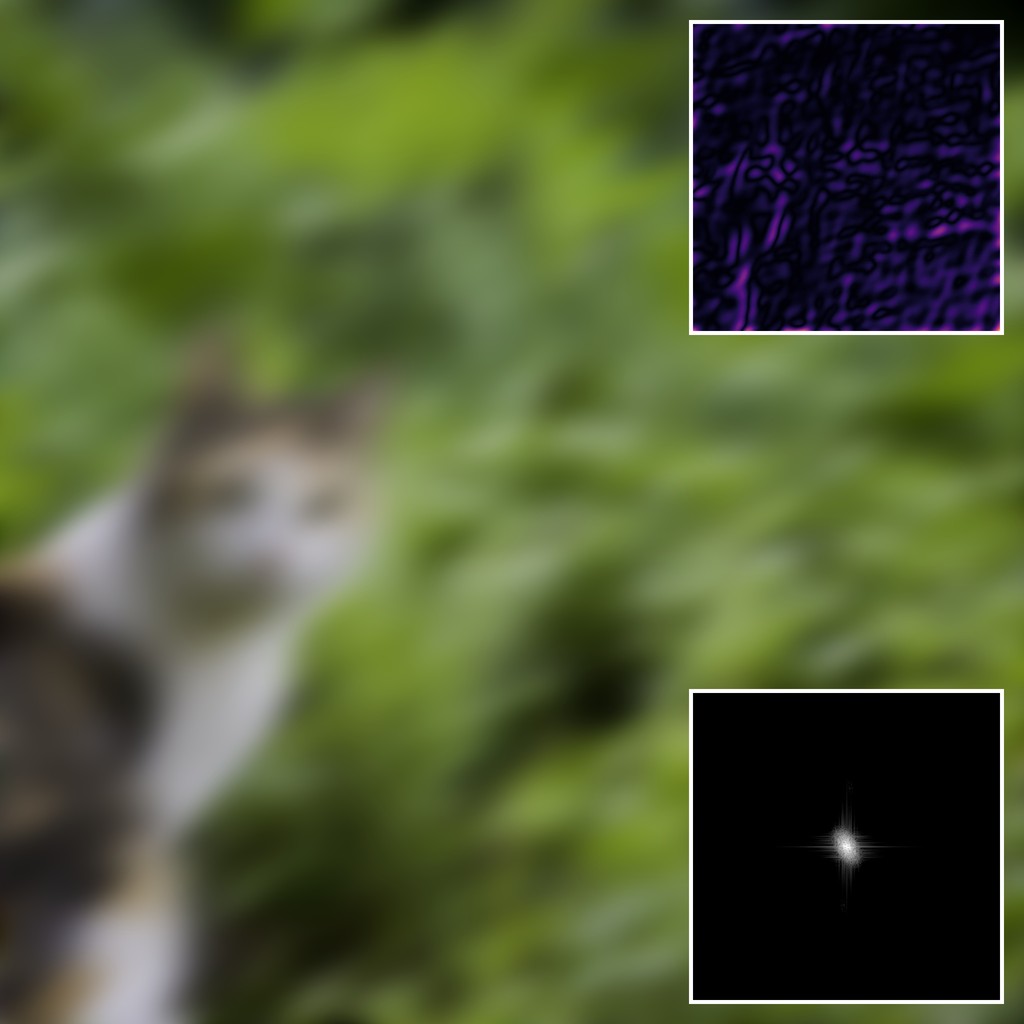} \\
            \includegraphics[width=0.155\textwidth, height=0.155\textwidth]{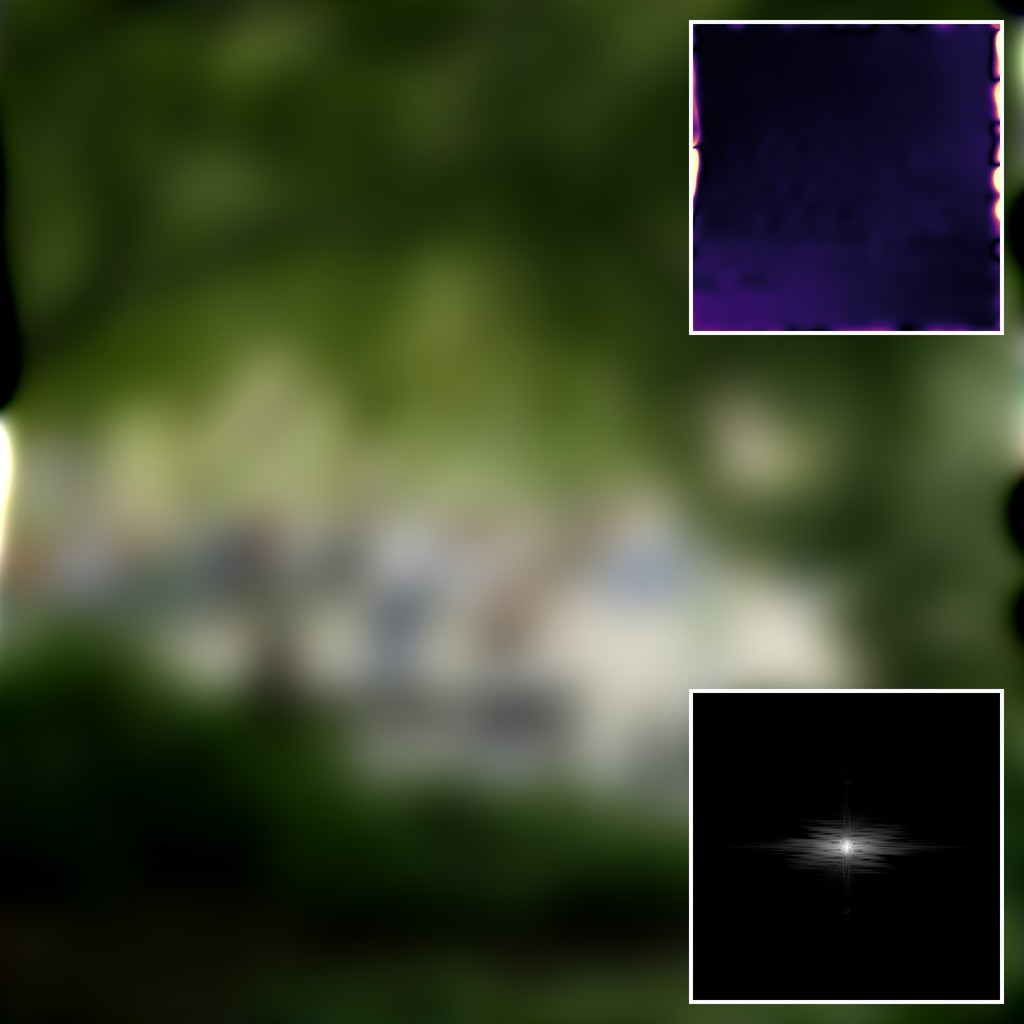}} & 
        \centeredtab{
            \includegraphics[width=0.155\textwidth, height=0.155\textwidth]{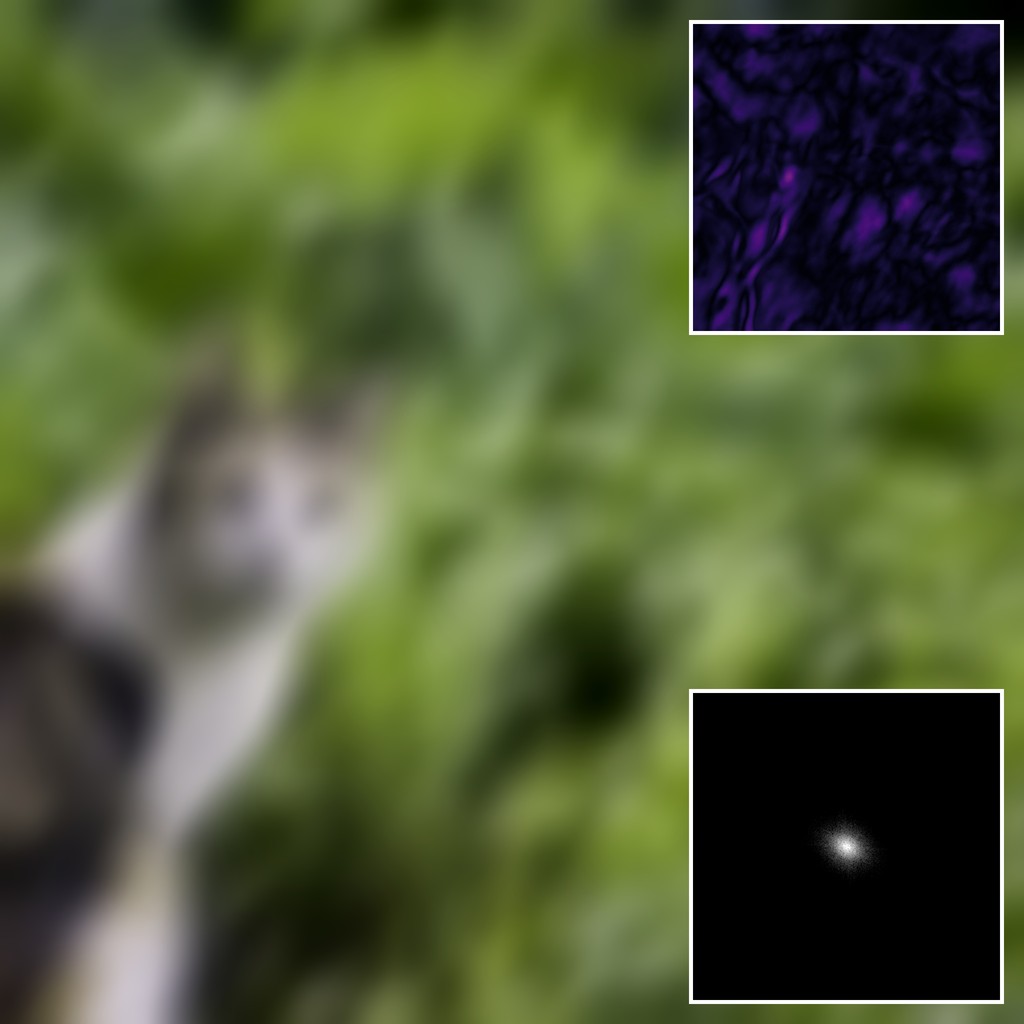} \\
            \includegraphics[width=0.155\textwidth, height=0.155\textwidth]{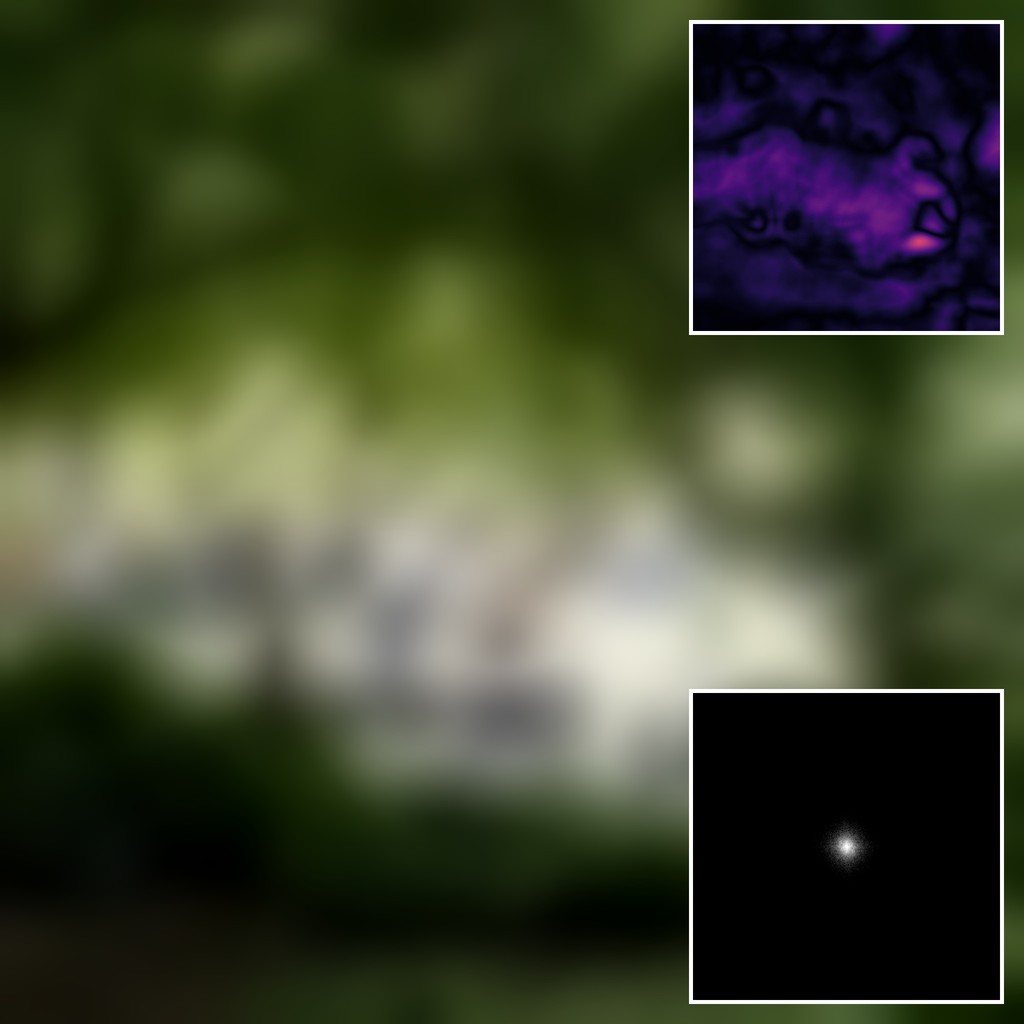}} & 
        \centeredtab{
            \includegraphics[width=0.155\textwidth, height=0.155\textwidth]{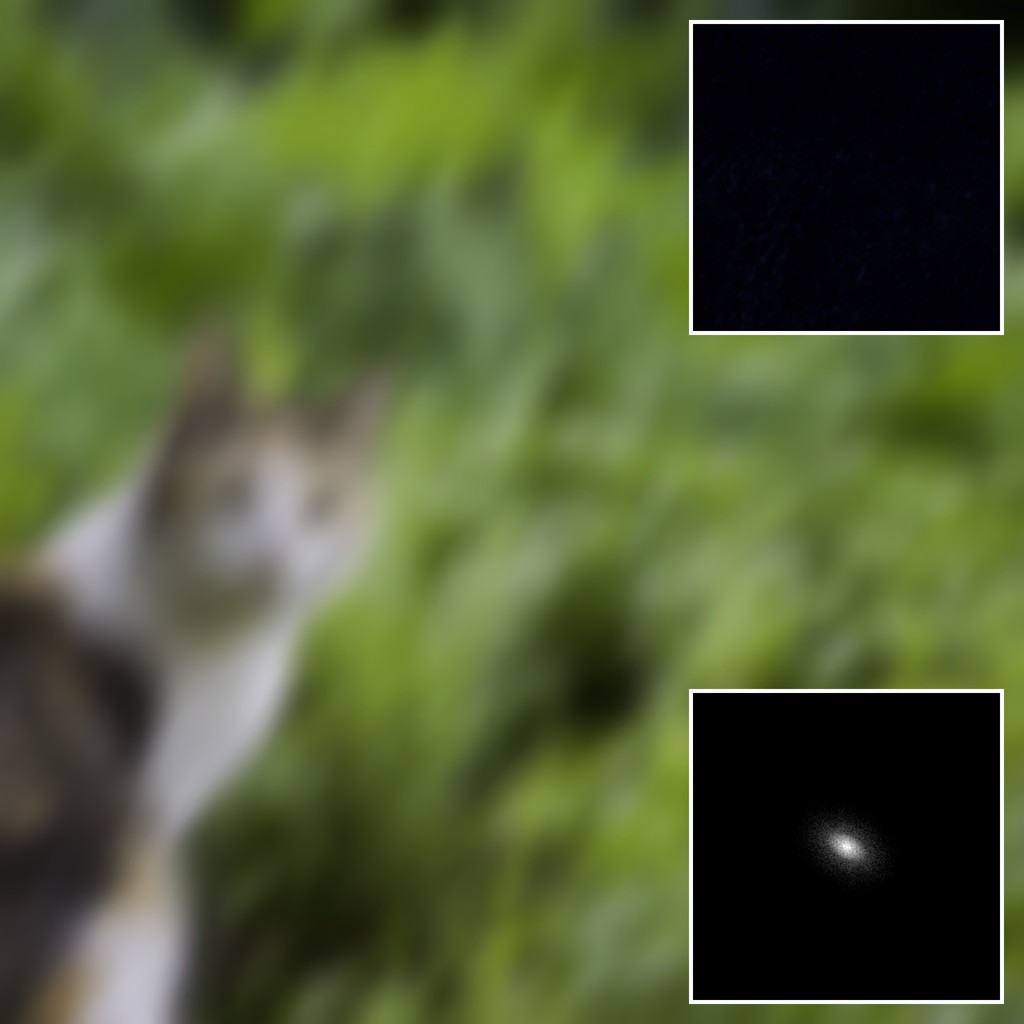} \\
            \includegraphics[width=0.155\textwidth, height=0.155\textwidth]{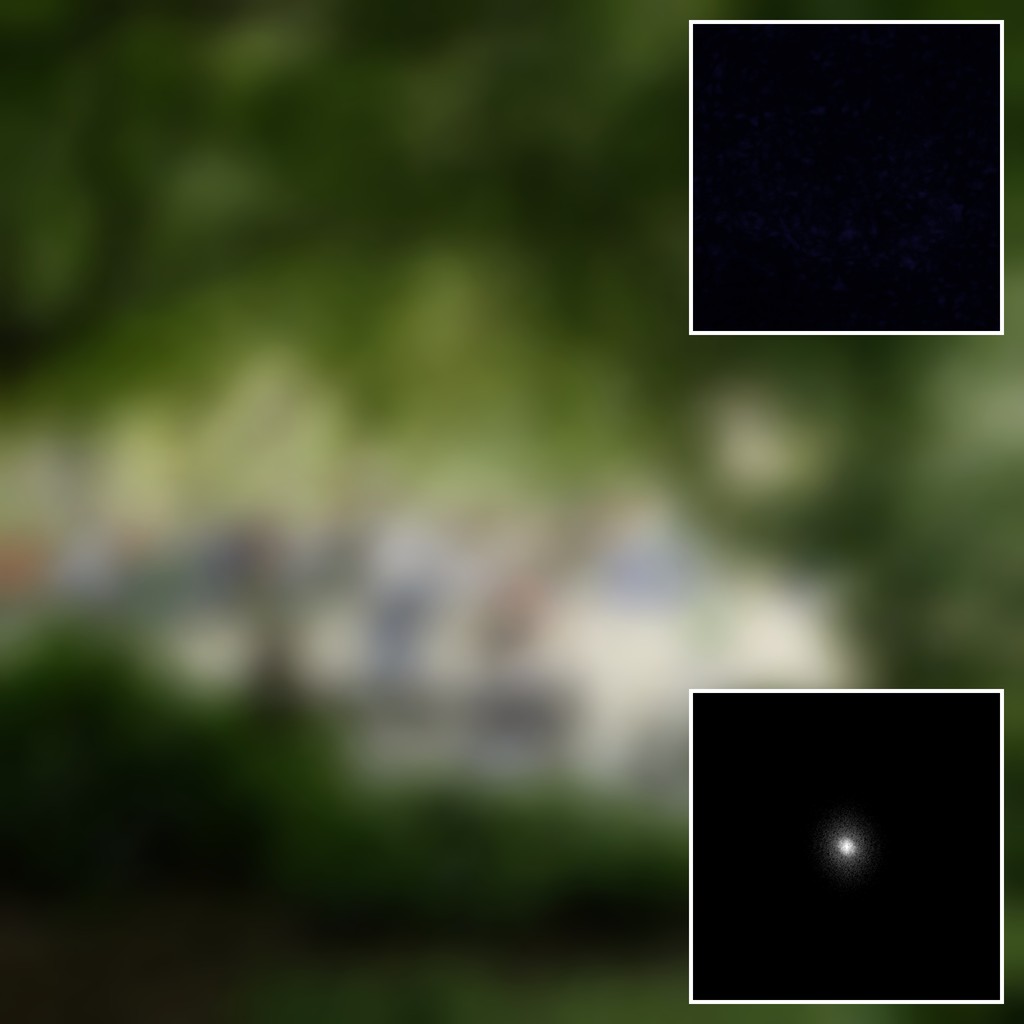}} &
        \centeredtab{
            \includegraphics[width=0.155\textwidth, height=0.155\textwidth]{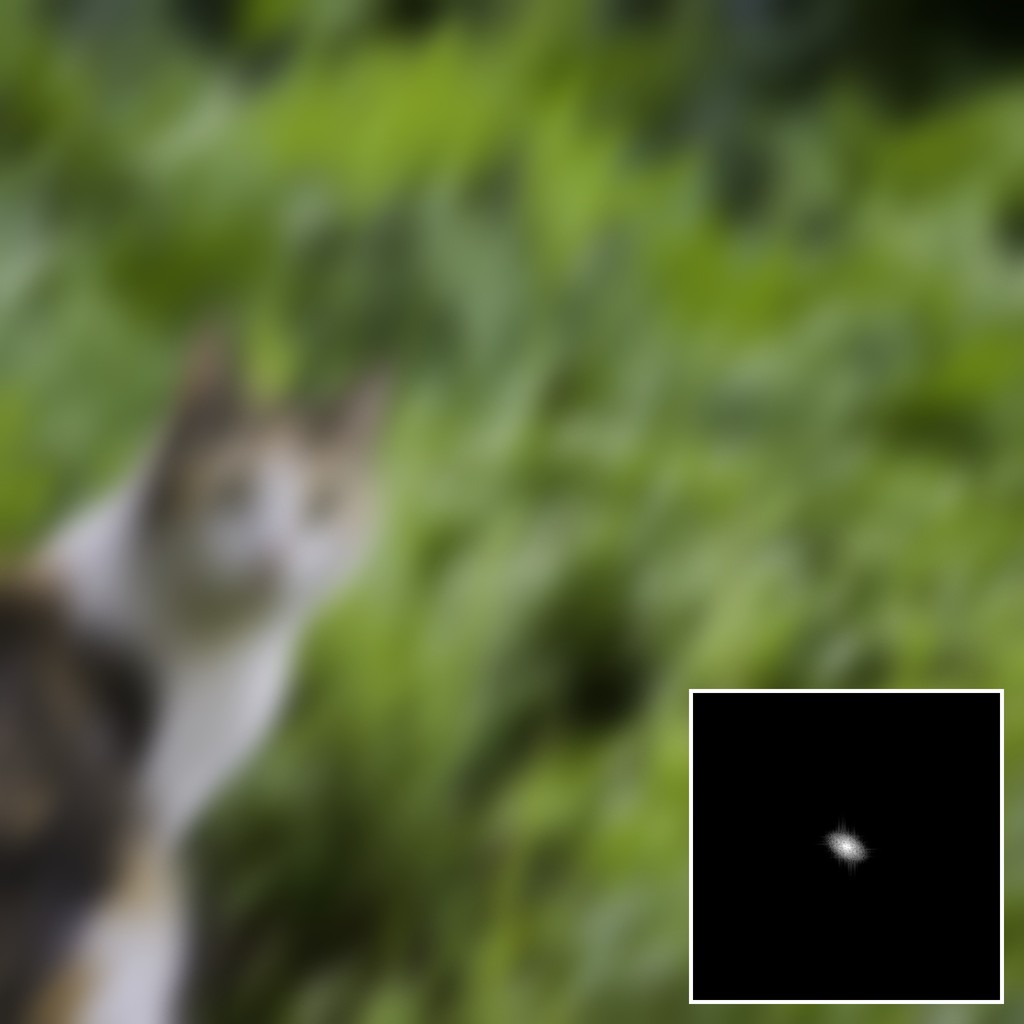} \\
            \includegraphics[width=0.155\textwidth, height=0.155\textwidth]{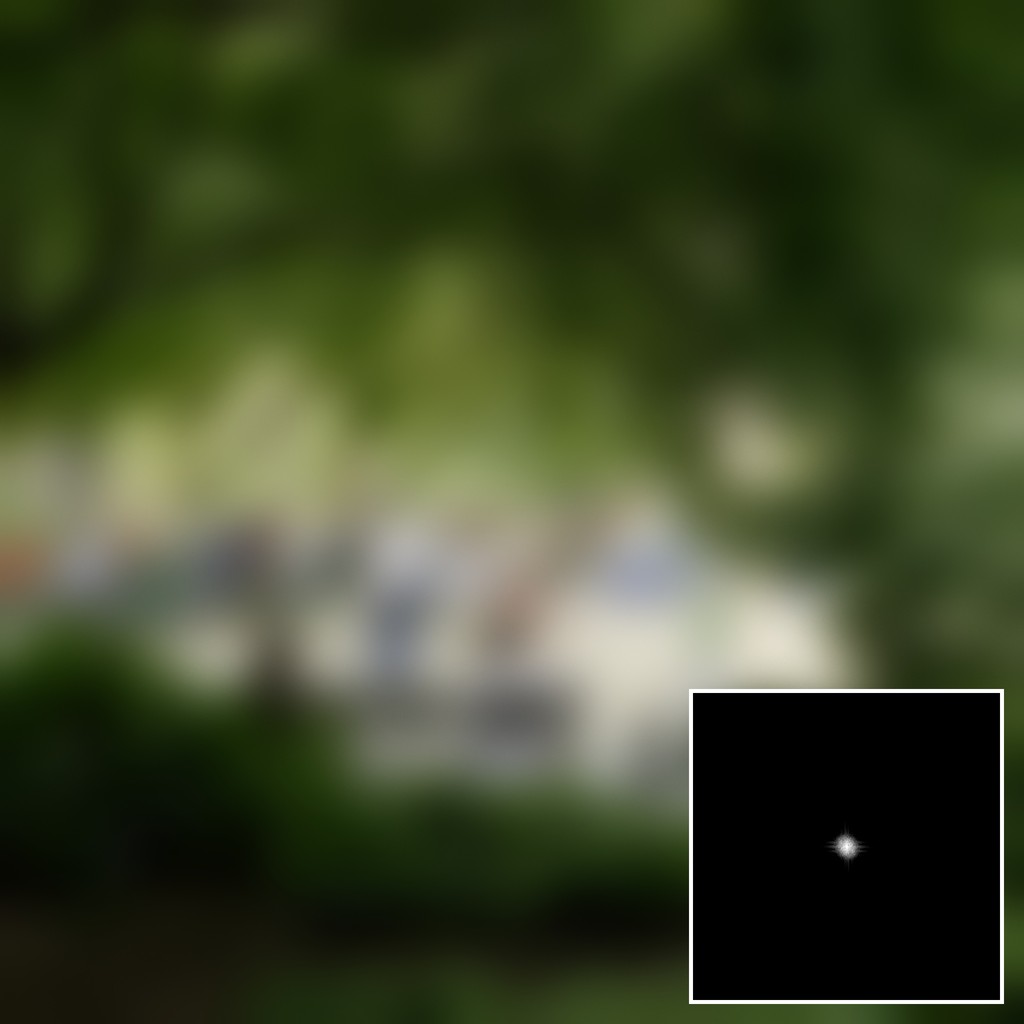}} \\
        \cmidrule(lr){3-8}

        \centeredtab{\rot{Box}} & &
        \centeredtab{
            \includegraphics[width=0.155\textwidth, height=0.155\textwidth]{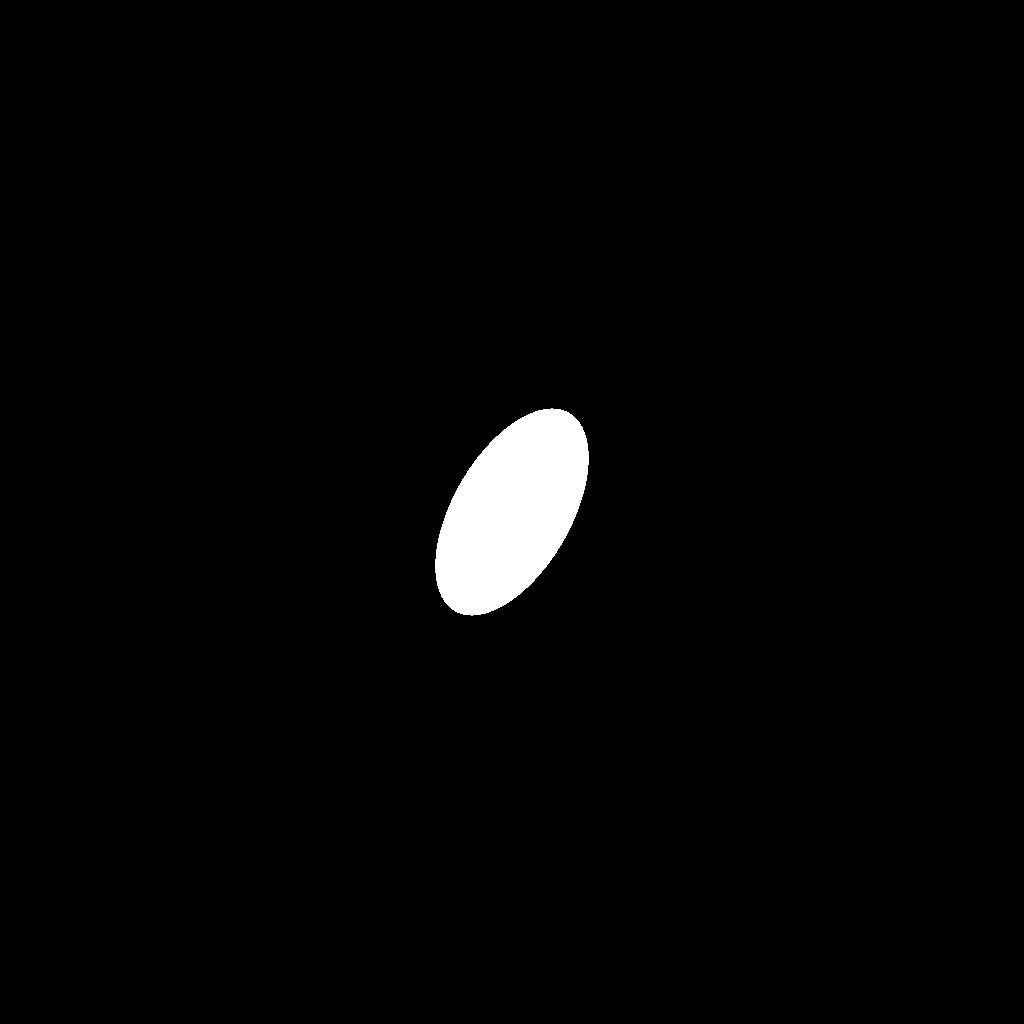} \\
            \includegraphics[width=0.155\textwidth, height=0.155\textwidth]{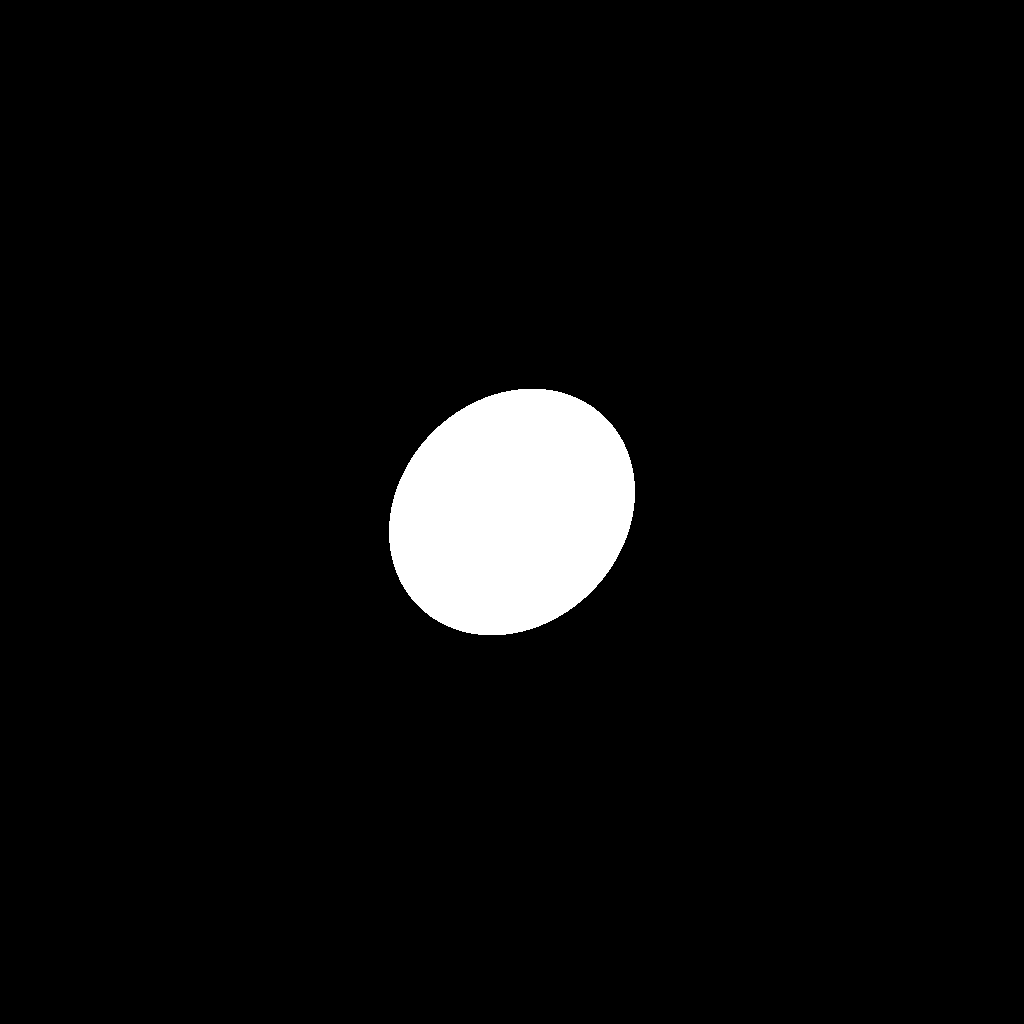}} & 
        \centeredtab{
            \includegraphics[width=0.155\textwidth, height=0.155\textwidth]{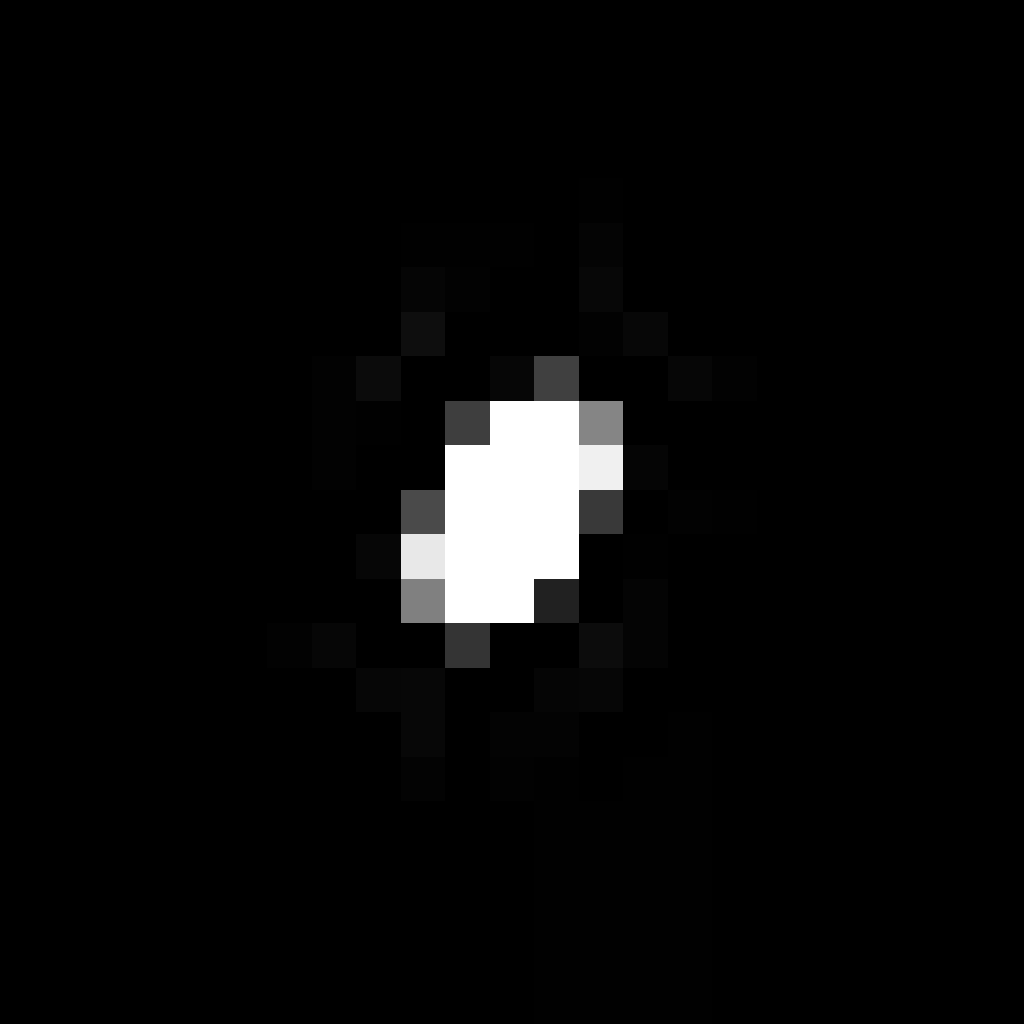} \\
            \includegraphics[width=0.155\textwidth, height=0.155\textwidth]{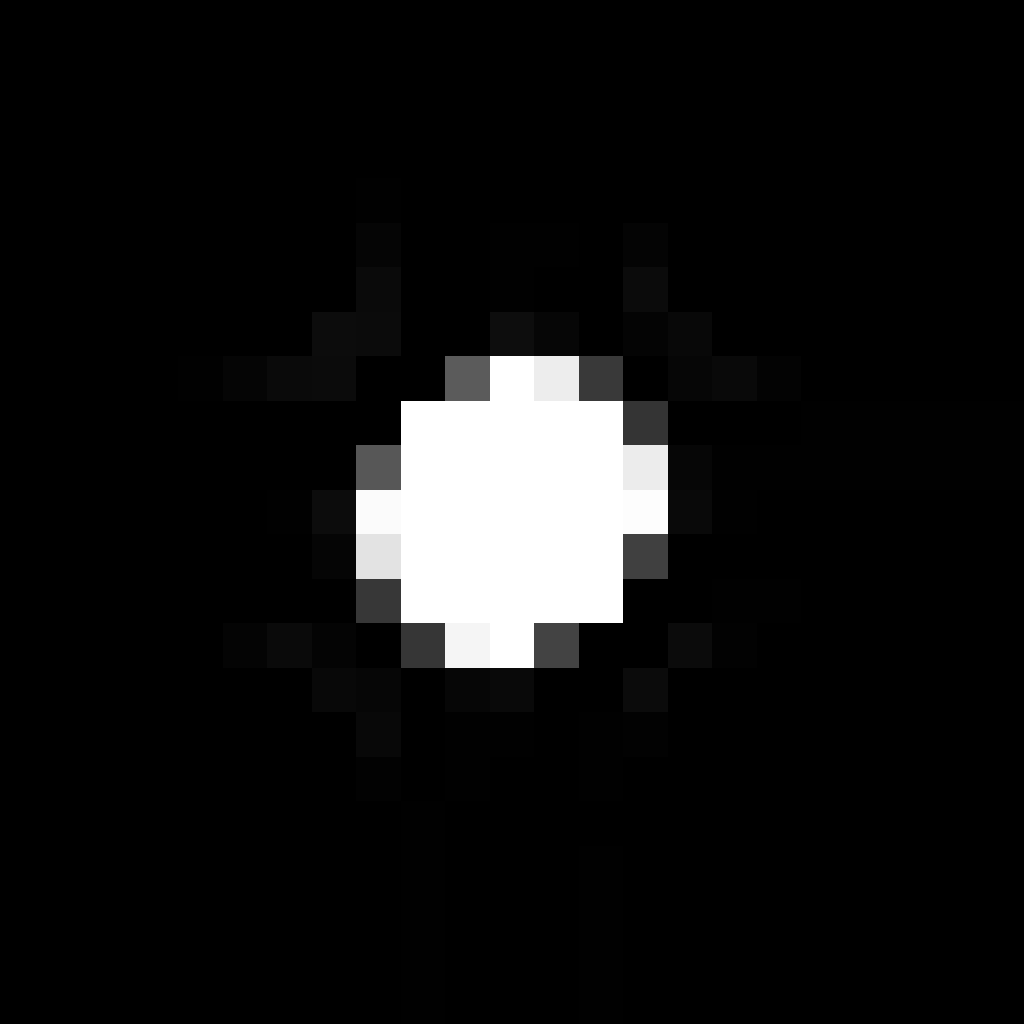}} & 
        \centeredtab{
            \includegraphics[width=0.155\textwidth, height=0.155\textwidth]{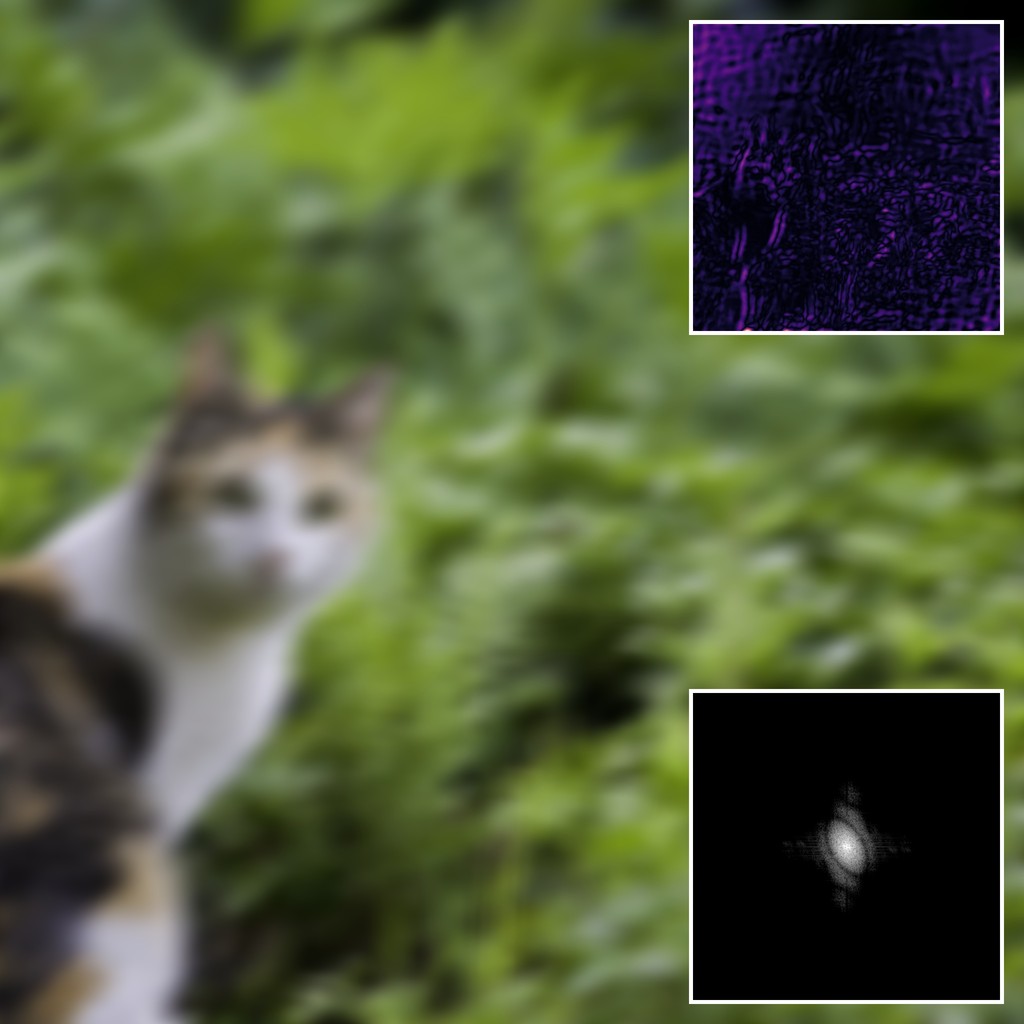} \\
            \includegraphics[width=0.155\textwidth, height=0.155\textwidth]{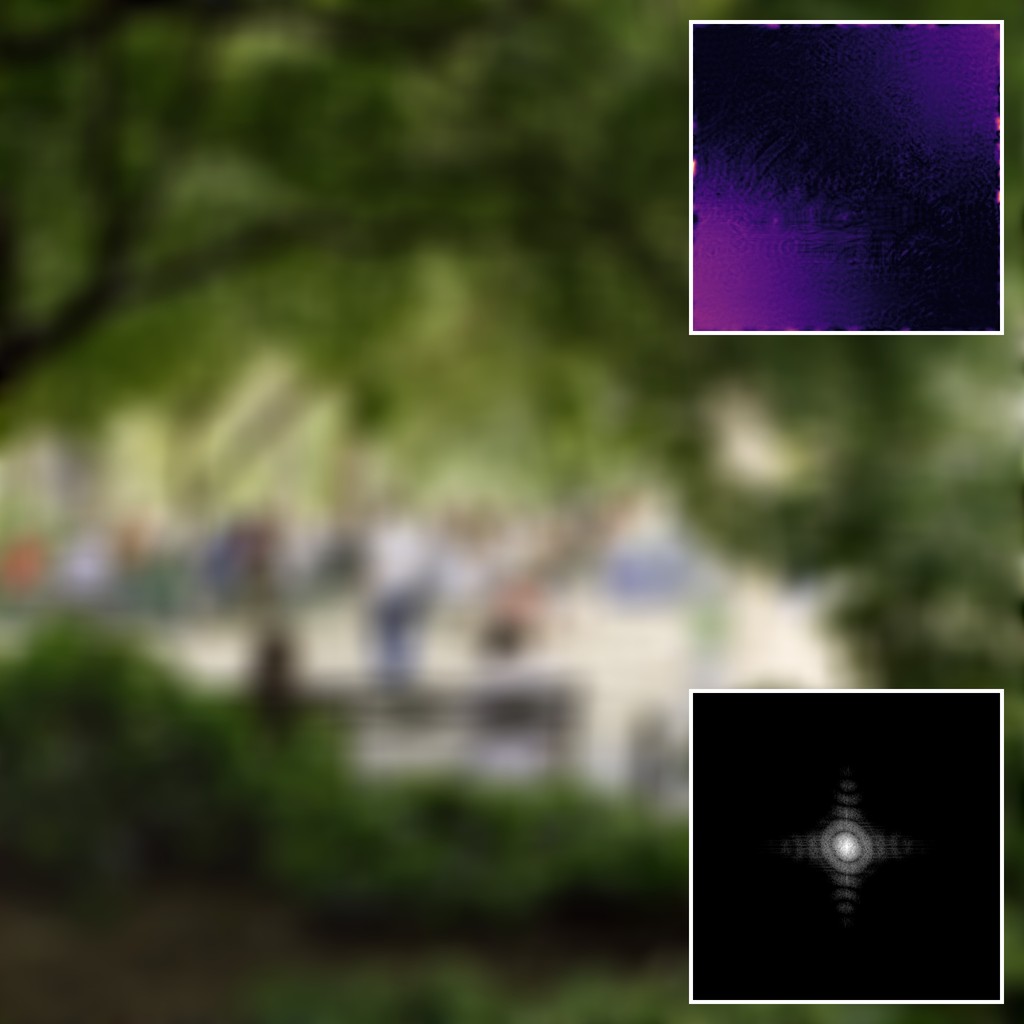}} & 
        \centeredtab{
            \includegraphics[width=0.155\textwidth, height=0.155\textwidth]{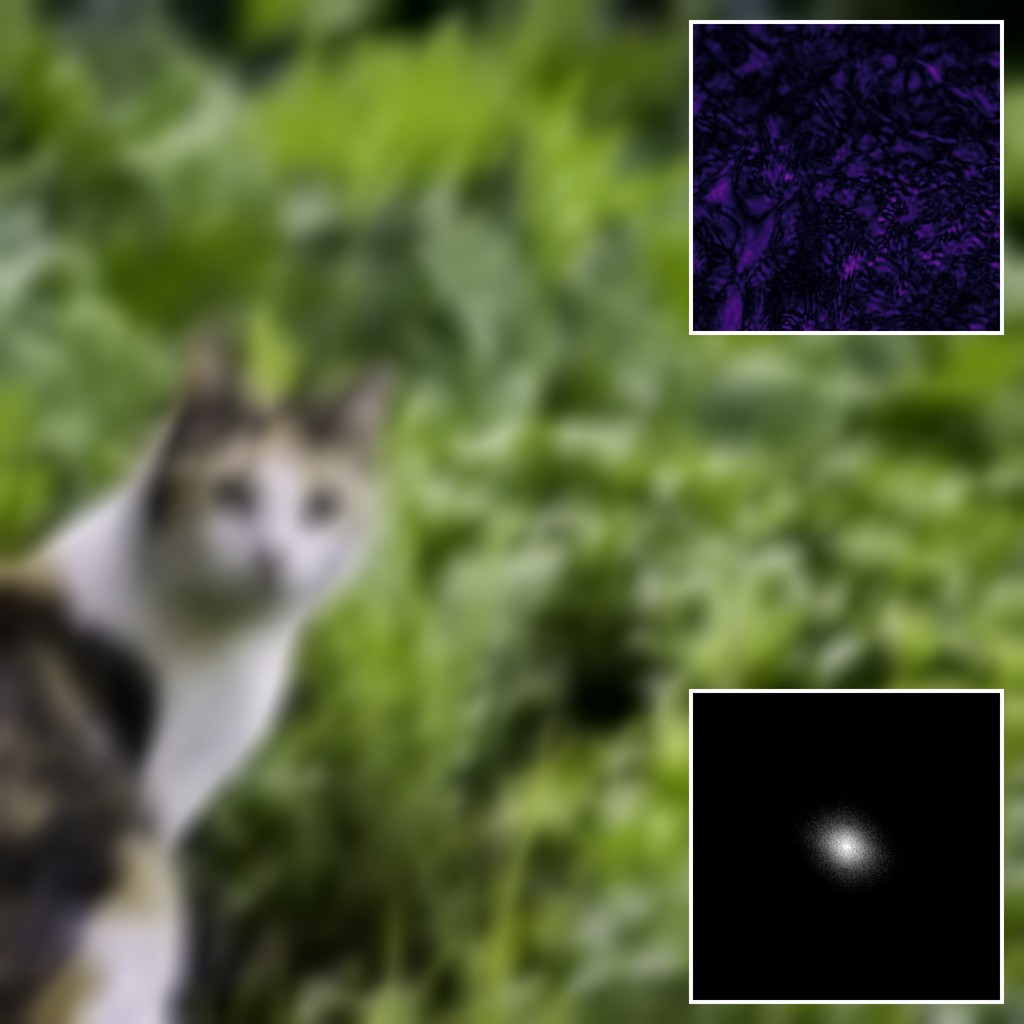} \\
            \includegraphics[width=0.155\textwidth, height=0.155\textwidth]{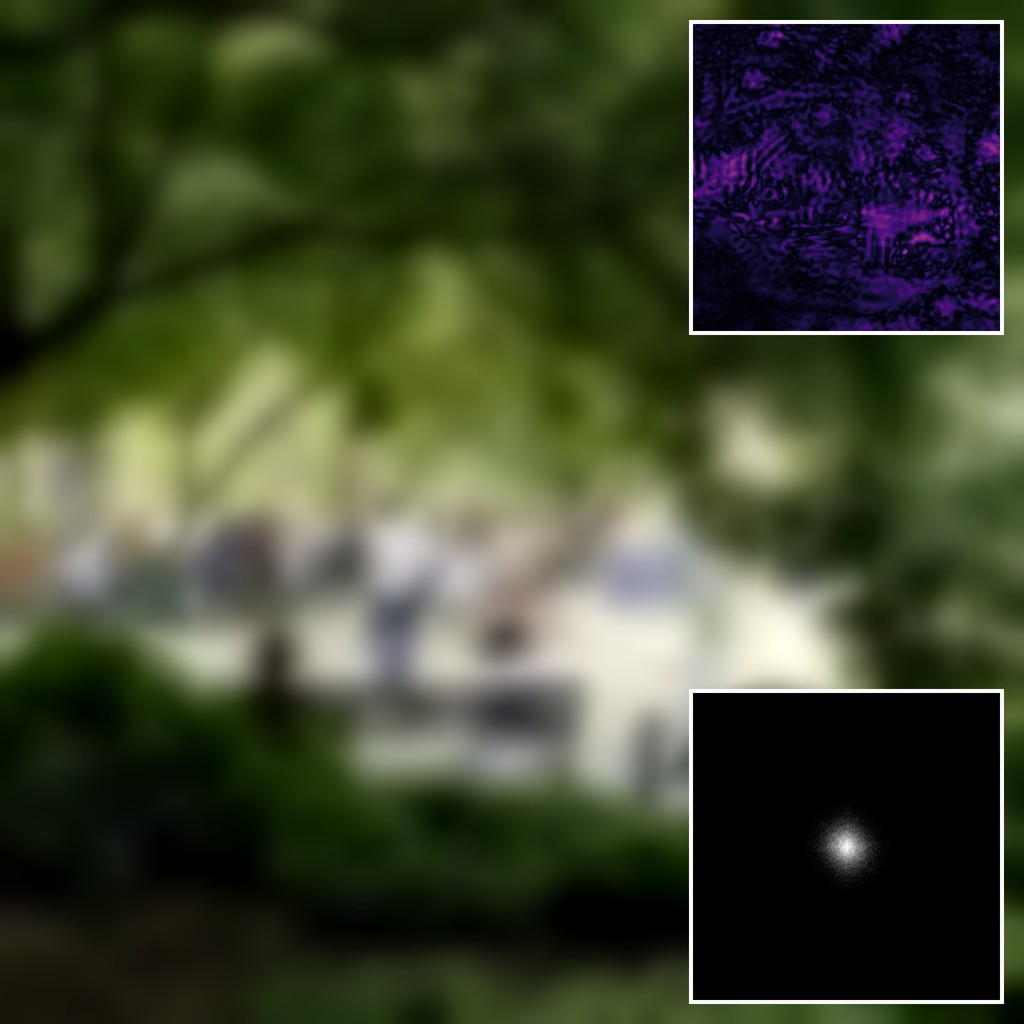}} & 
        \centeredtab{
            \includegraphics[width=0.155\textwidth, height=0.155\textwidth]{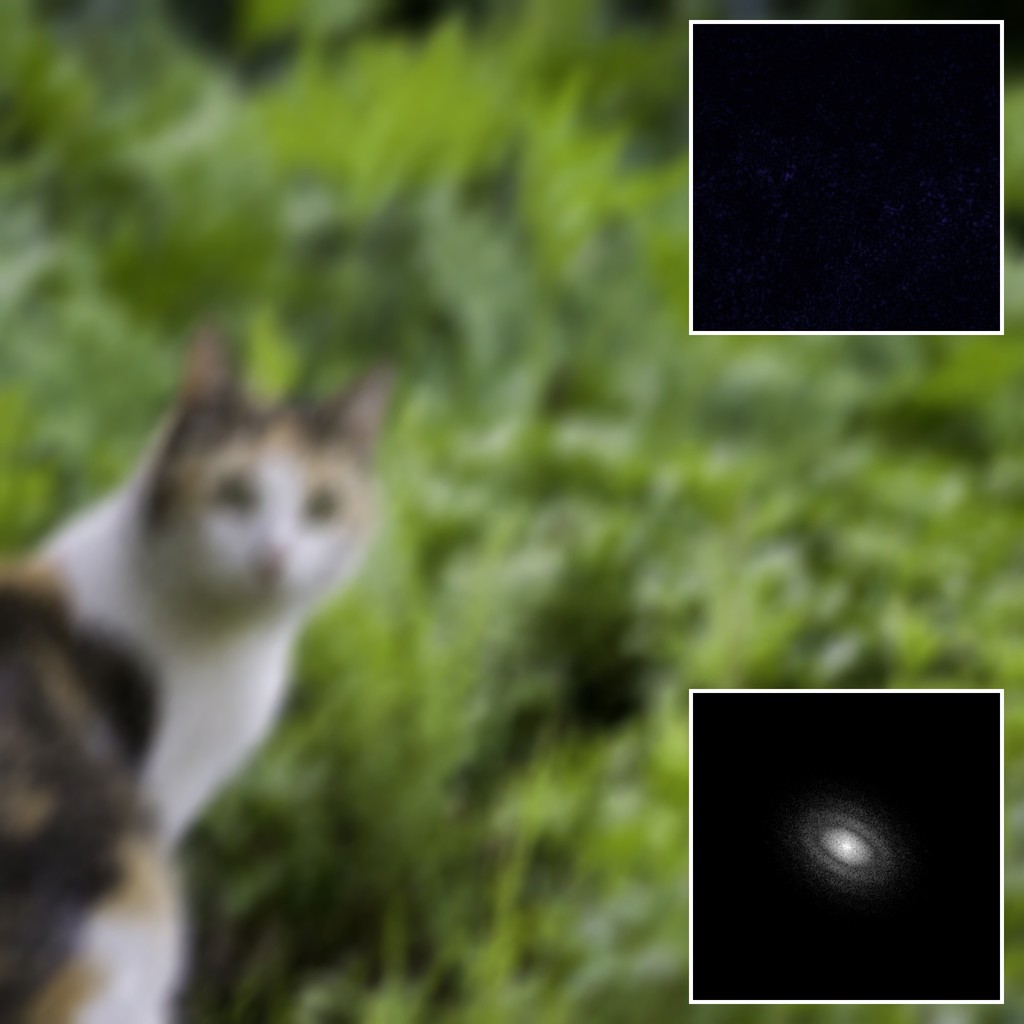} \\
            \includegraphics[width=0.155\textwidth, height=0.155\textwidth]{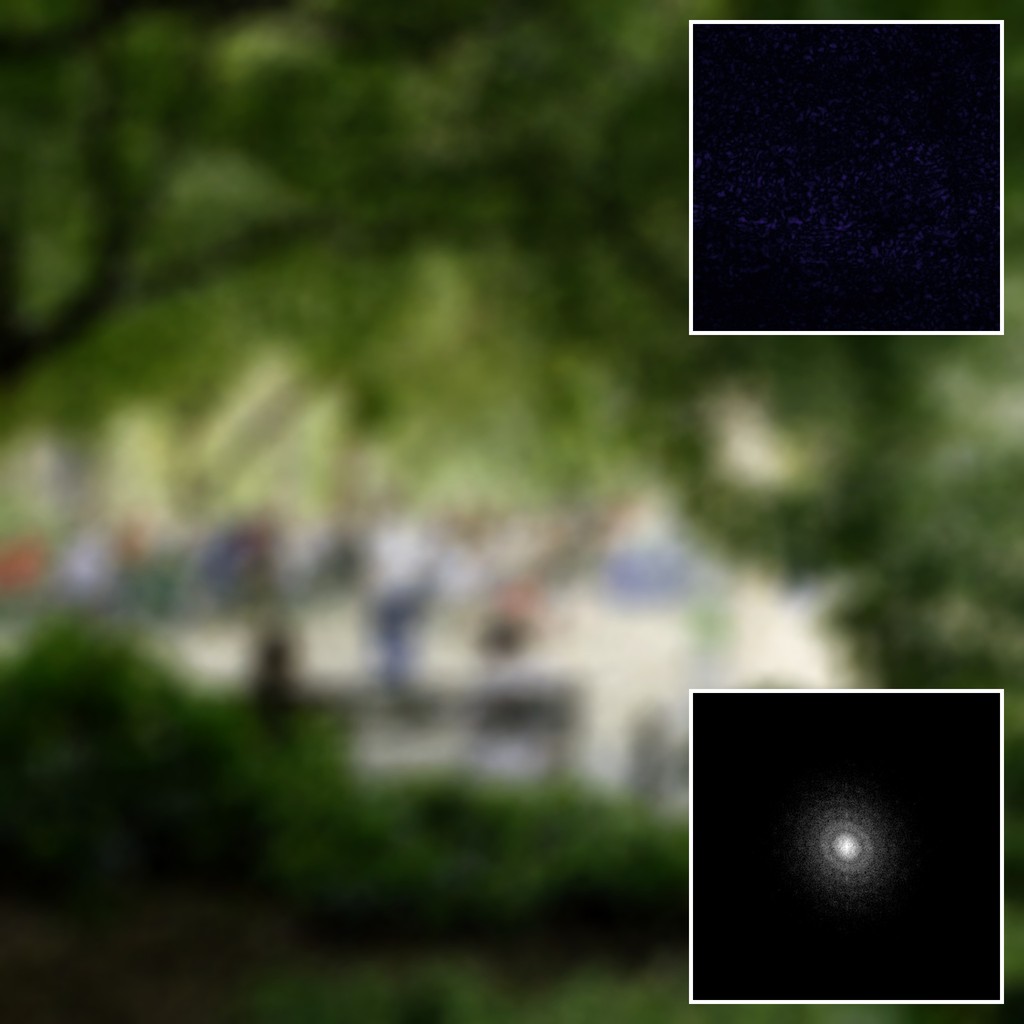}} &
        \centeredtab{
            \includegraphics[width=0.155\textwidth, height=0.155\textwidth]{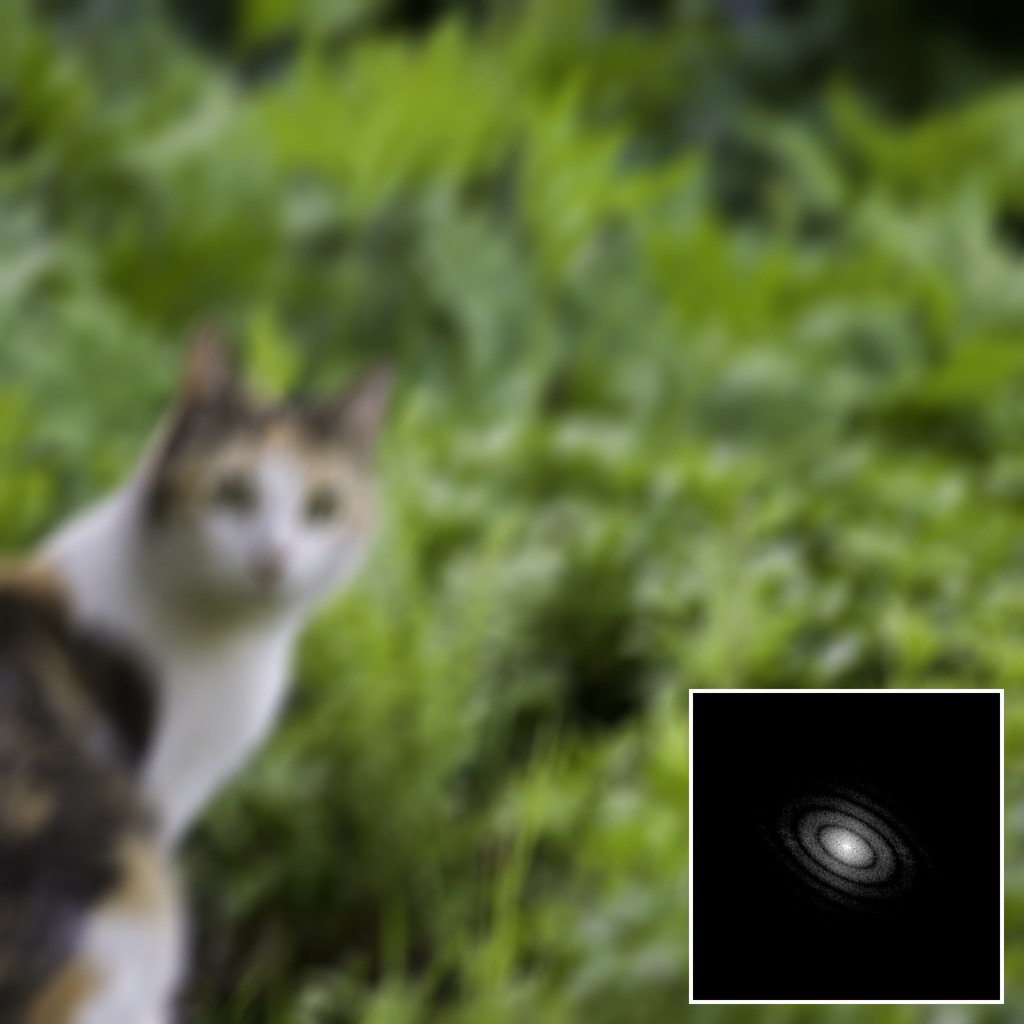} \\
            \includegraphics[width=0.155\textwidth, height=0.155\textwidth]{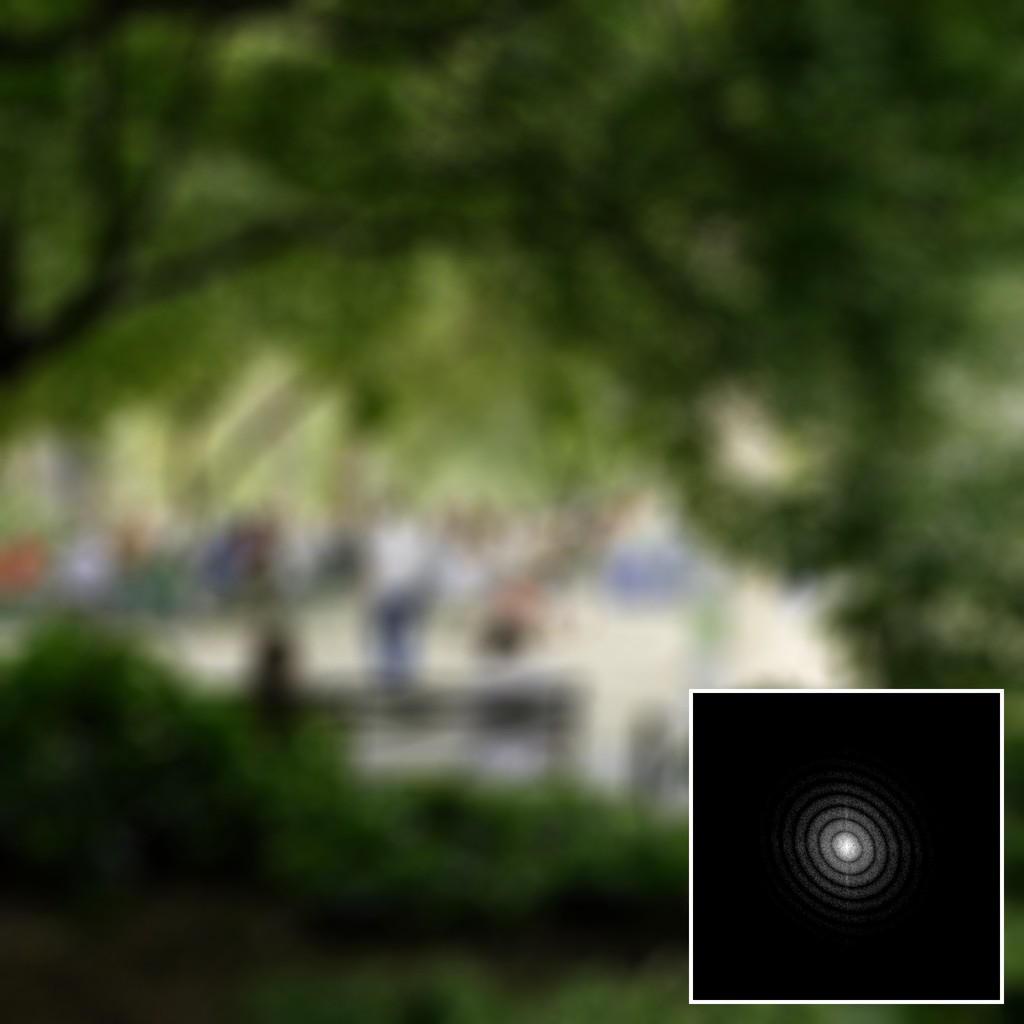}} \\
        \cmidrule(lr){3-8}

        \centeredtab{\rot{Lanczos}} & &
        \centeredtab{
            \includegraphics[width=0.155\textwidth, height=0.155\textwidth]{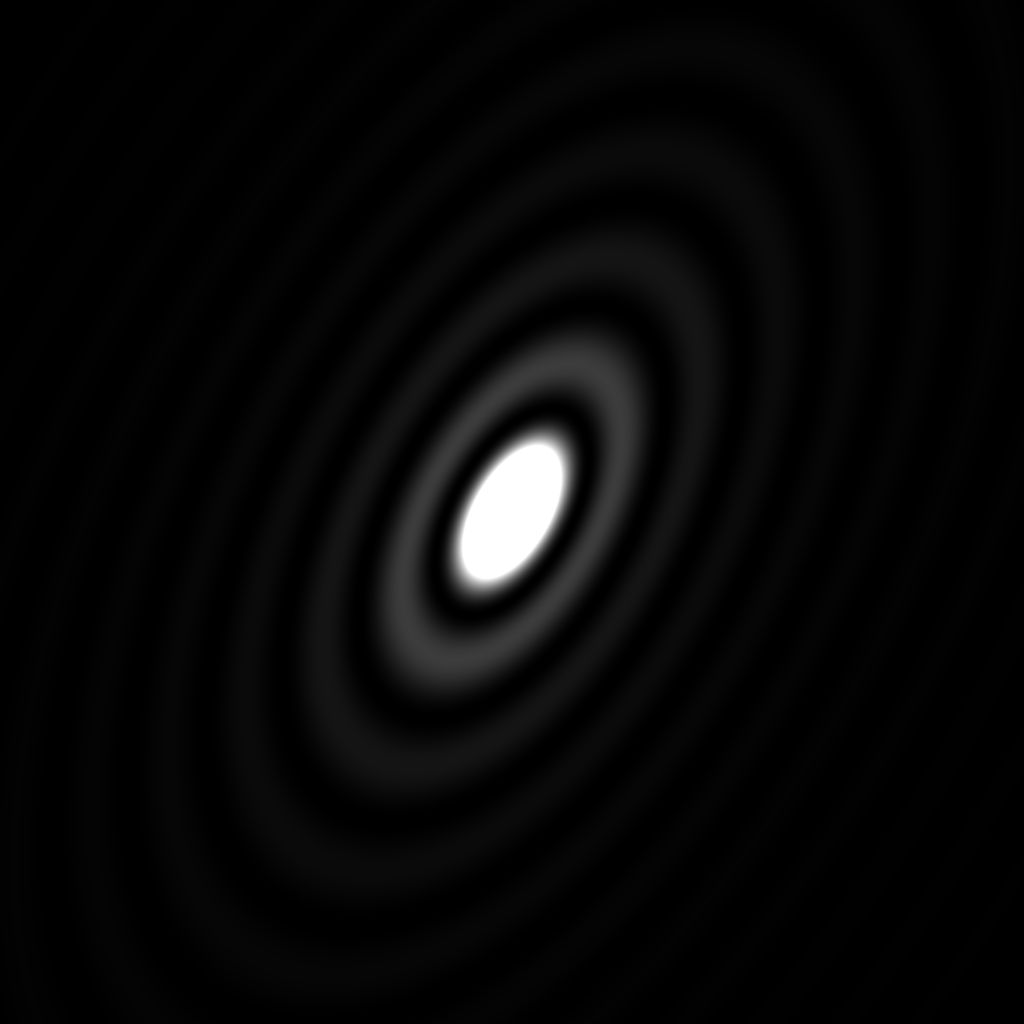} \\
            \includegraphics[width=0.155\textwidth, height=0.155\textwidth]{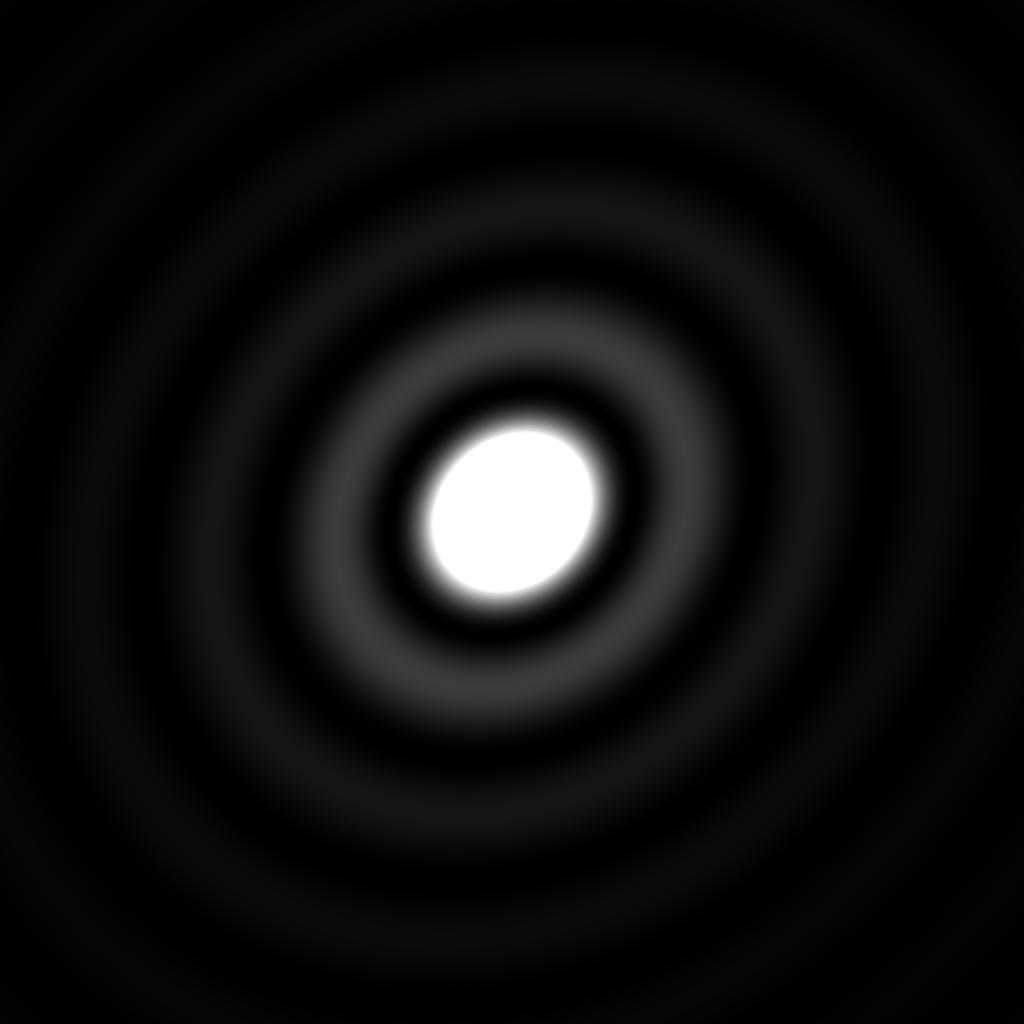}} & 
        \centeredtab{
            \includegraphics[width=0.155\textwidth, height=0.155\textwidth]{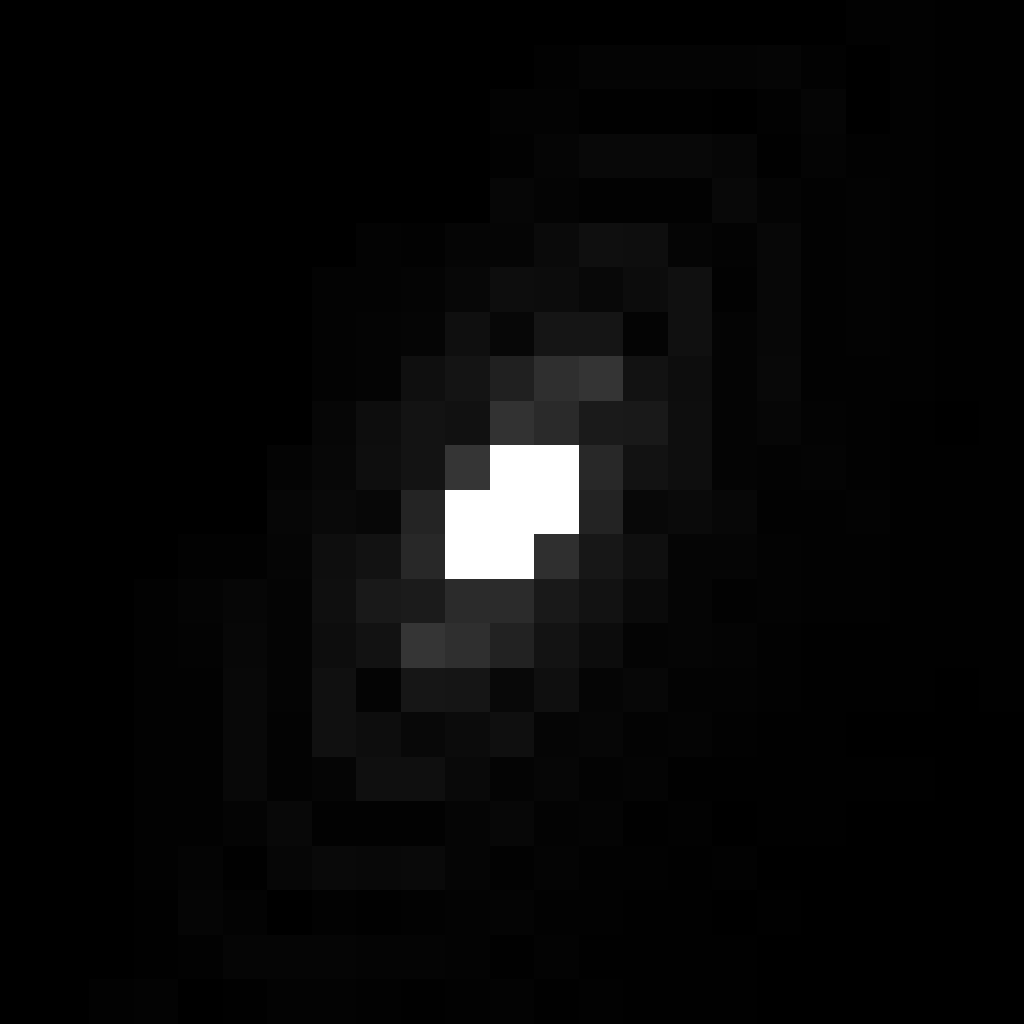} \\
            \includegraphics[width=0.155\textwidth, height=0.155\textwidth]{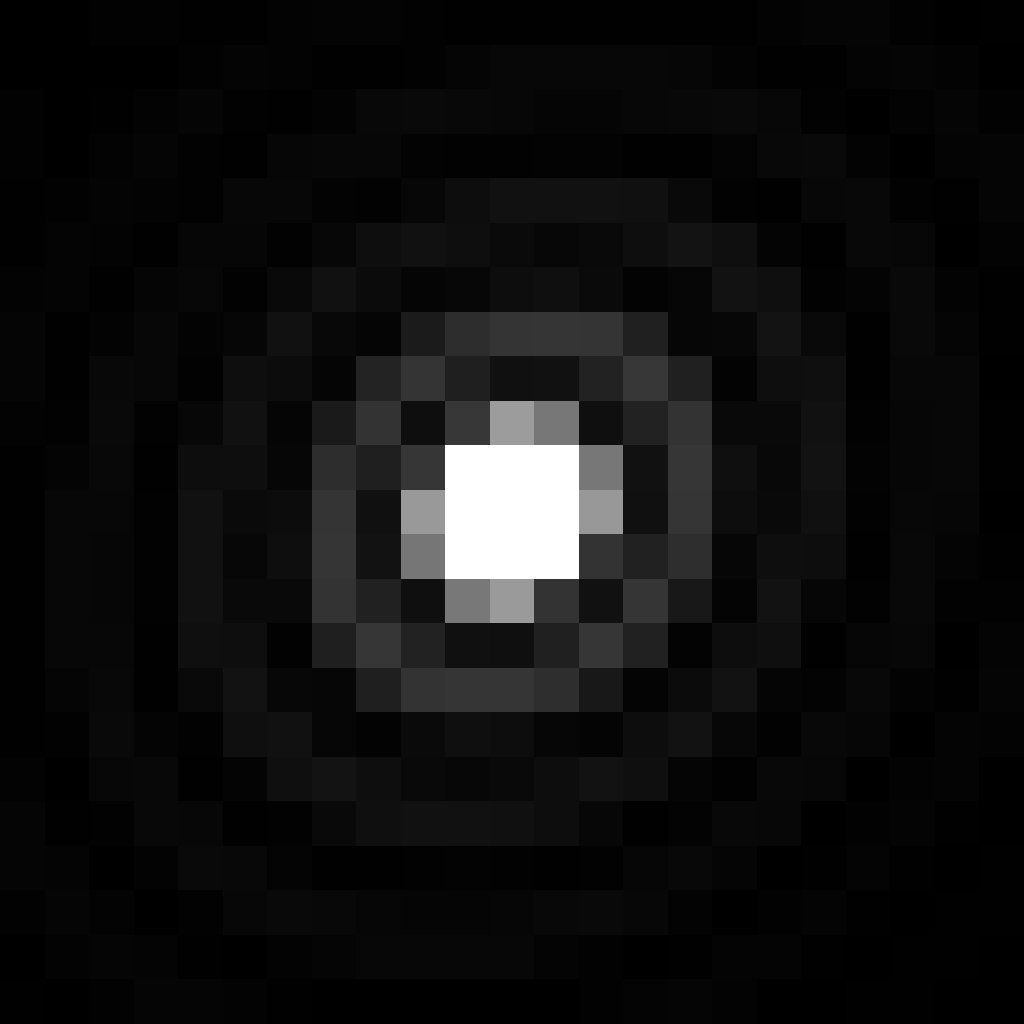}} & 
        \centeredtab{
            \includegraphics[width=0.155\textwidth, height=0.155\textwidth]{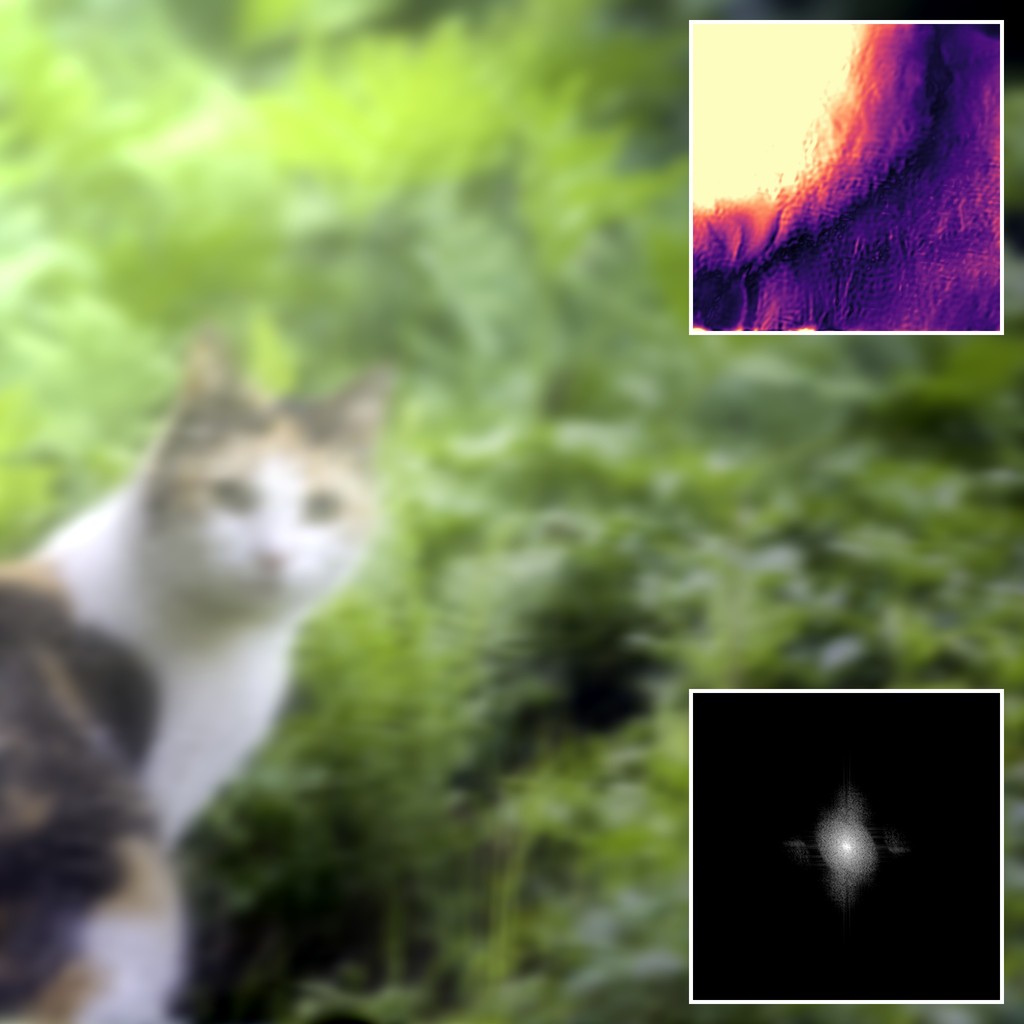} \\
            \includegraphics[width=0.155\textwidth, height=0.155\textwidth]{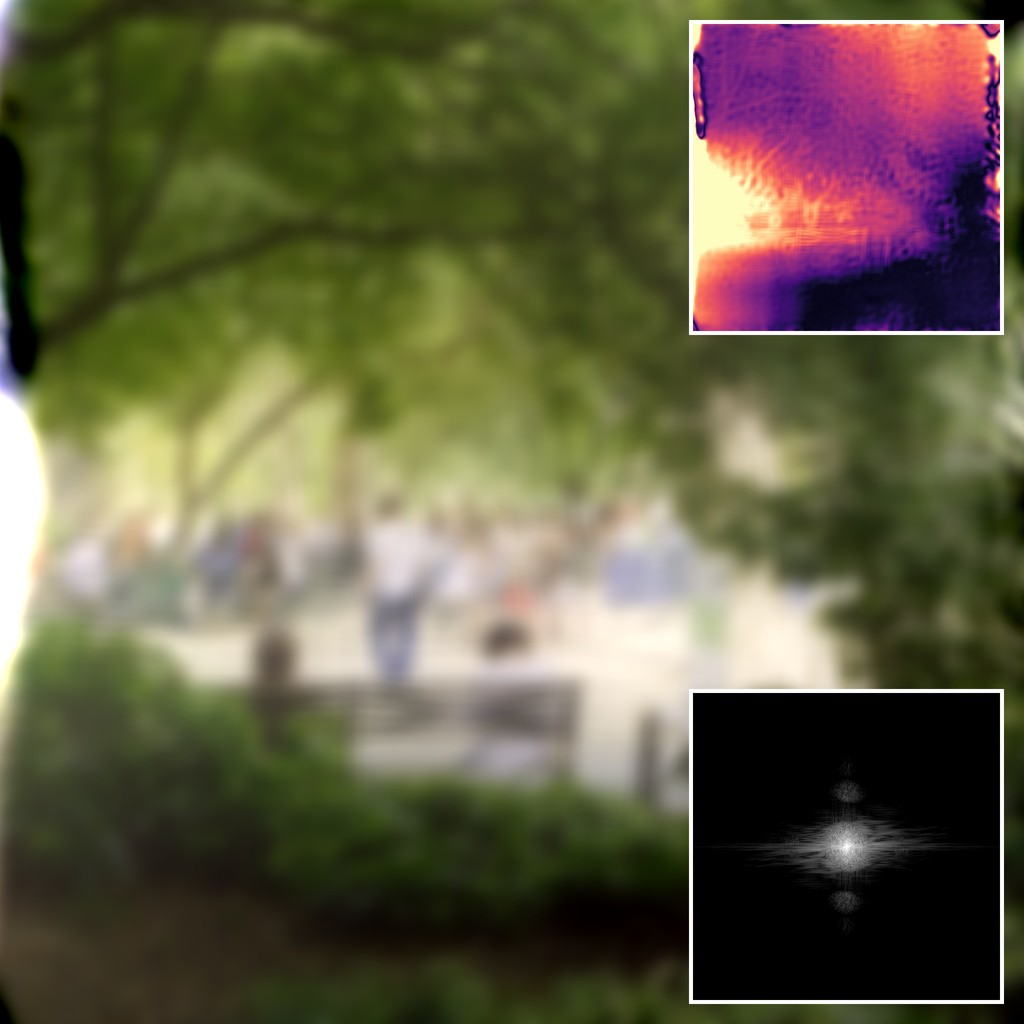}} & 
        \centeredtab{
            \includegraphics[width=0.155\textwidth, height=0.155\textwidth]{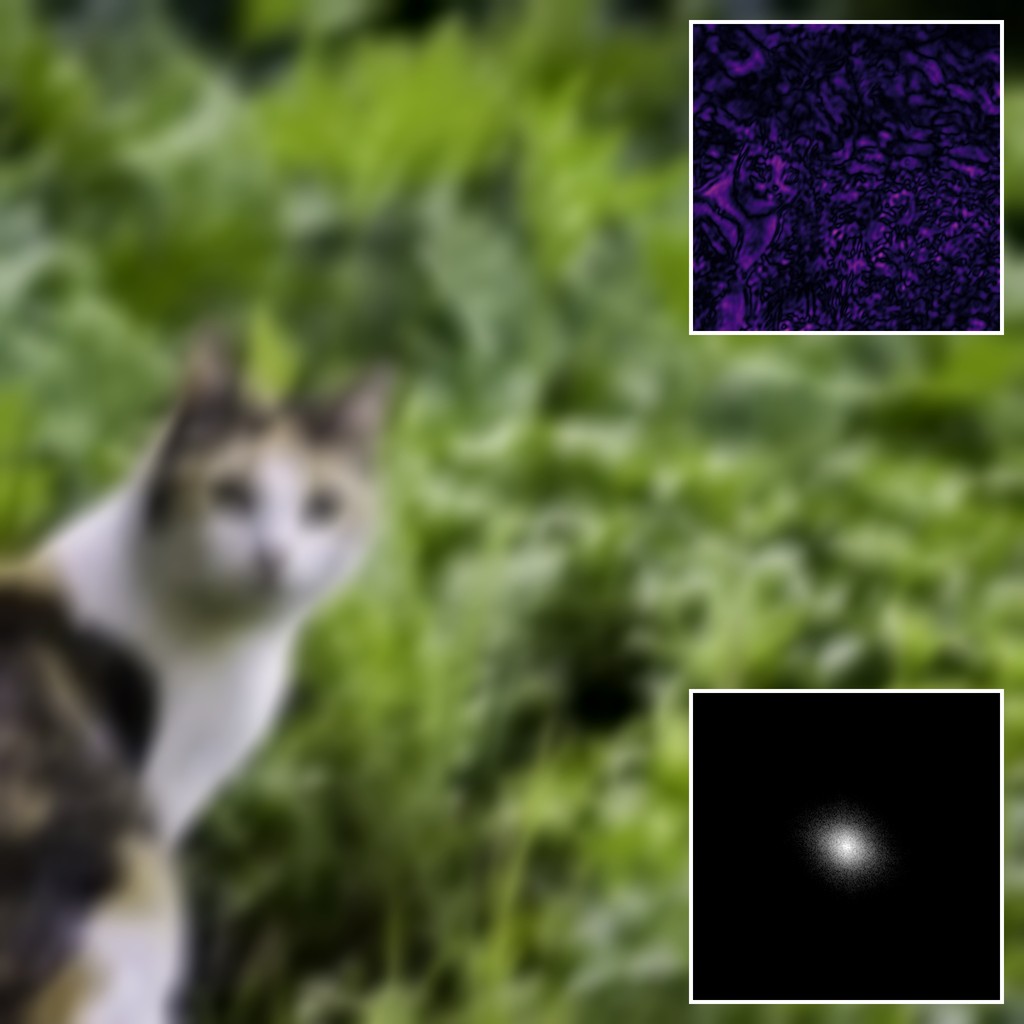} \\
            \includegraphics[width=0.155\textwidth, height=0.155\textwidth]{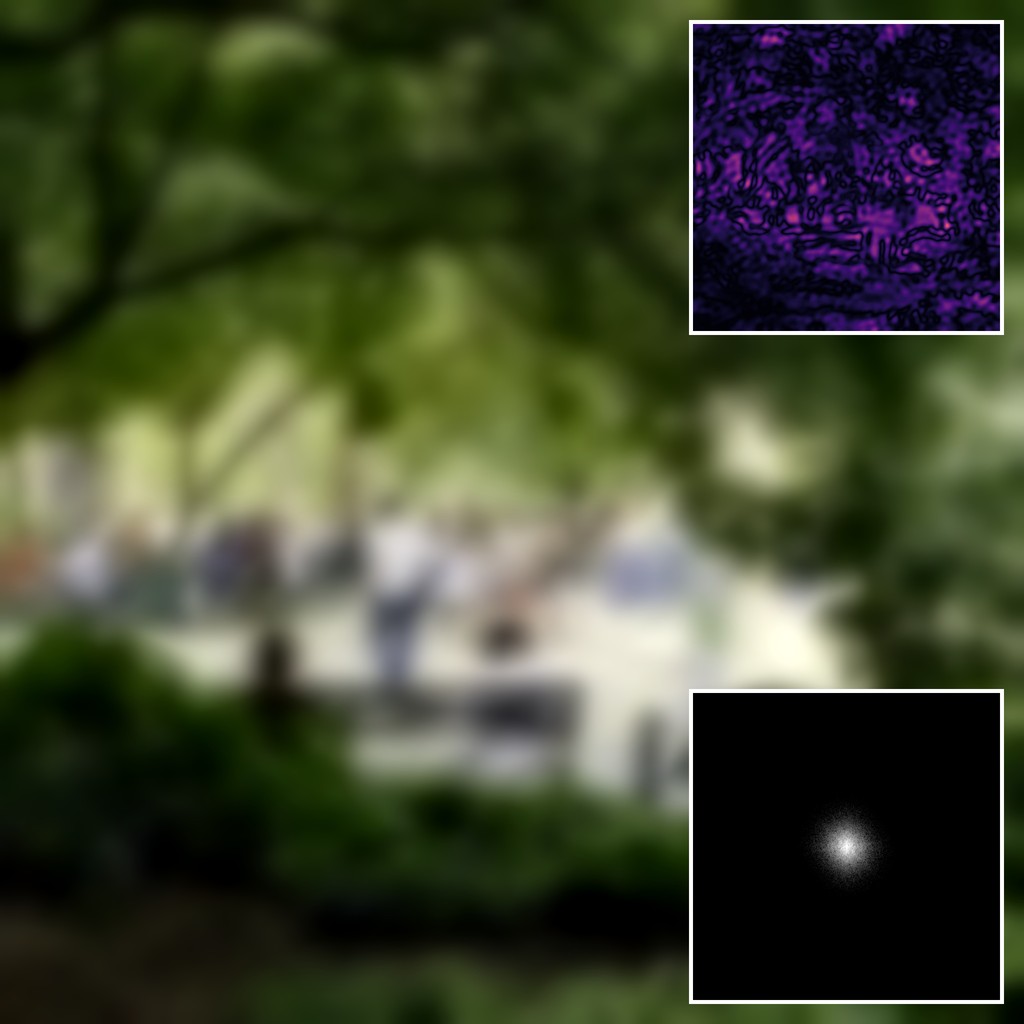}} & 
        \centeredtab{
            \includegraphics[width=0.155\textwidth, height=0.155\textwidth]{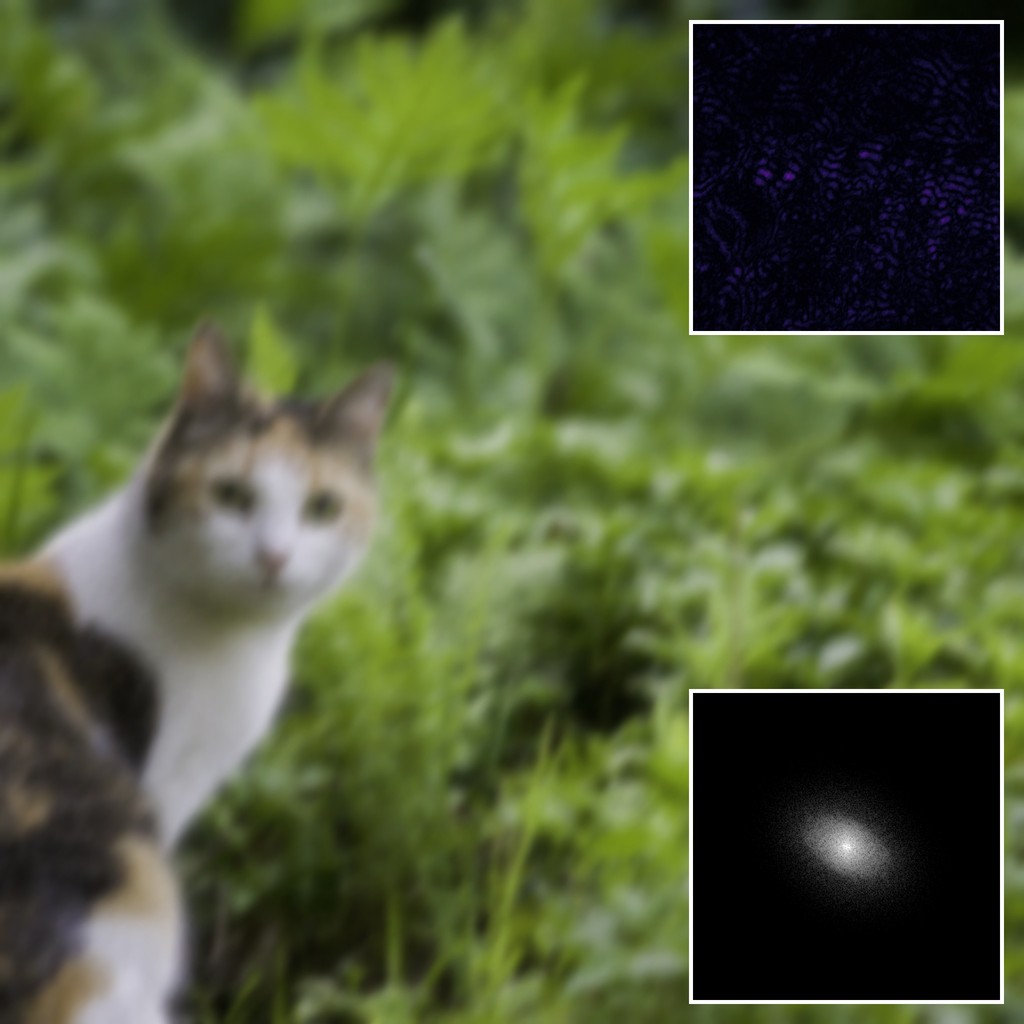} \\
            \includegraphics[width=0.155\textwidth, height=0.155\textwidth]{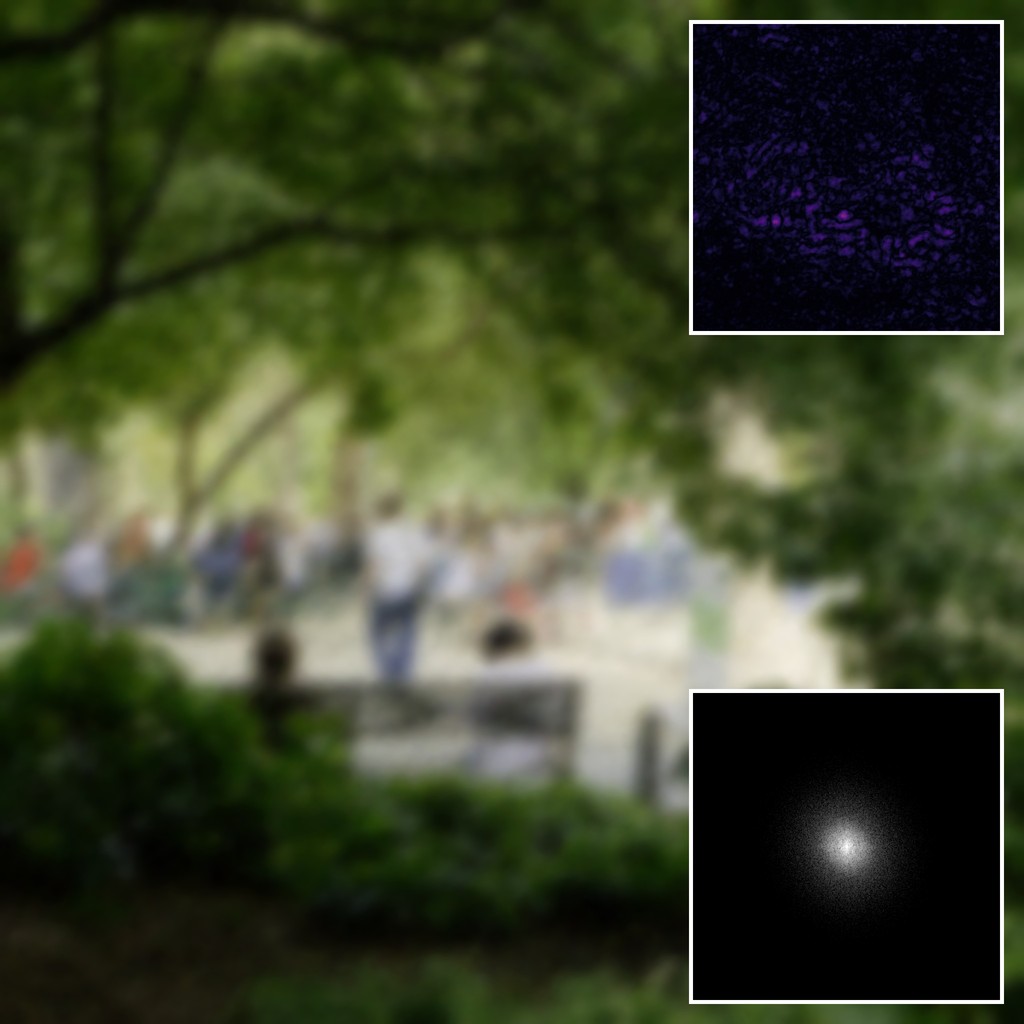}} &
        \centeredtab{
            \includegraphics[width=0.155\textwidth, height=0.155\textwidth]{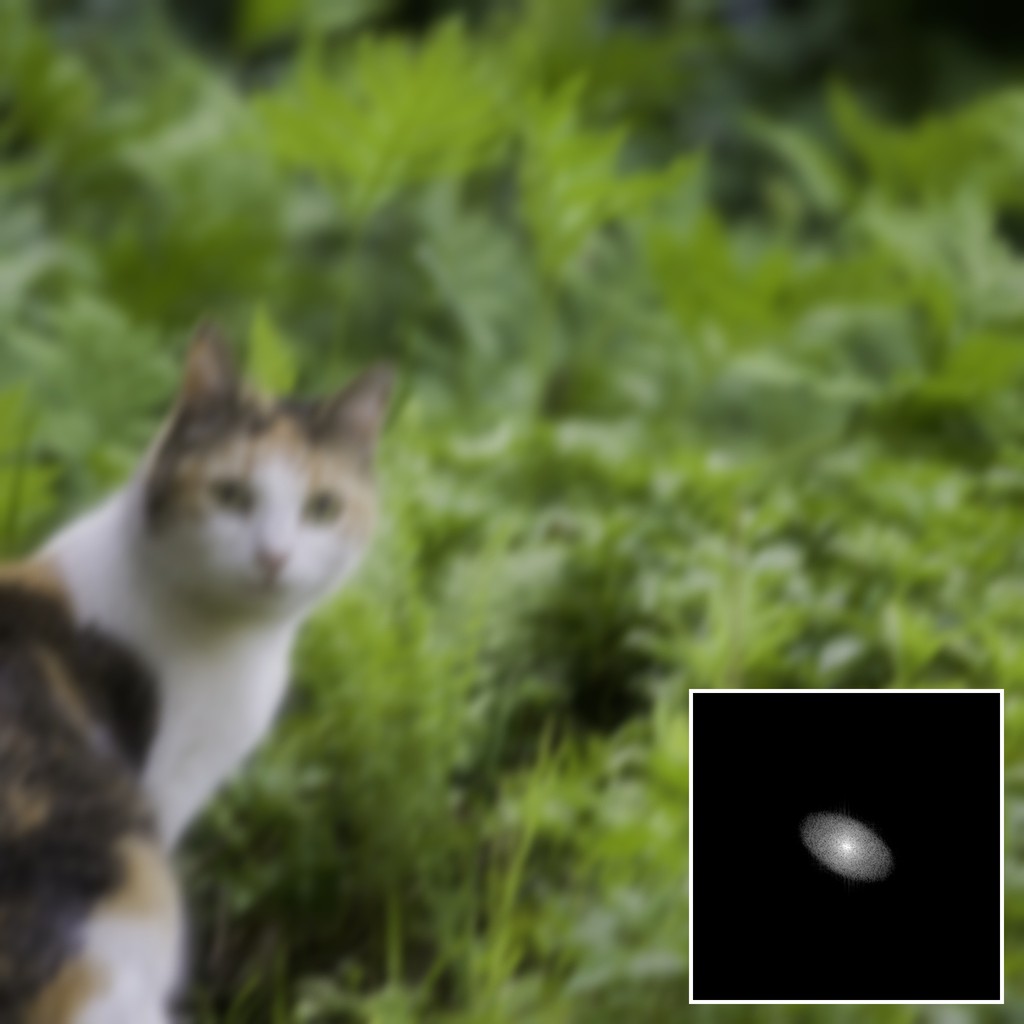} \\
            \includegraphics[width=0.155\textwidth, height=0.155\textwidth]{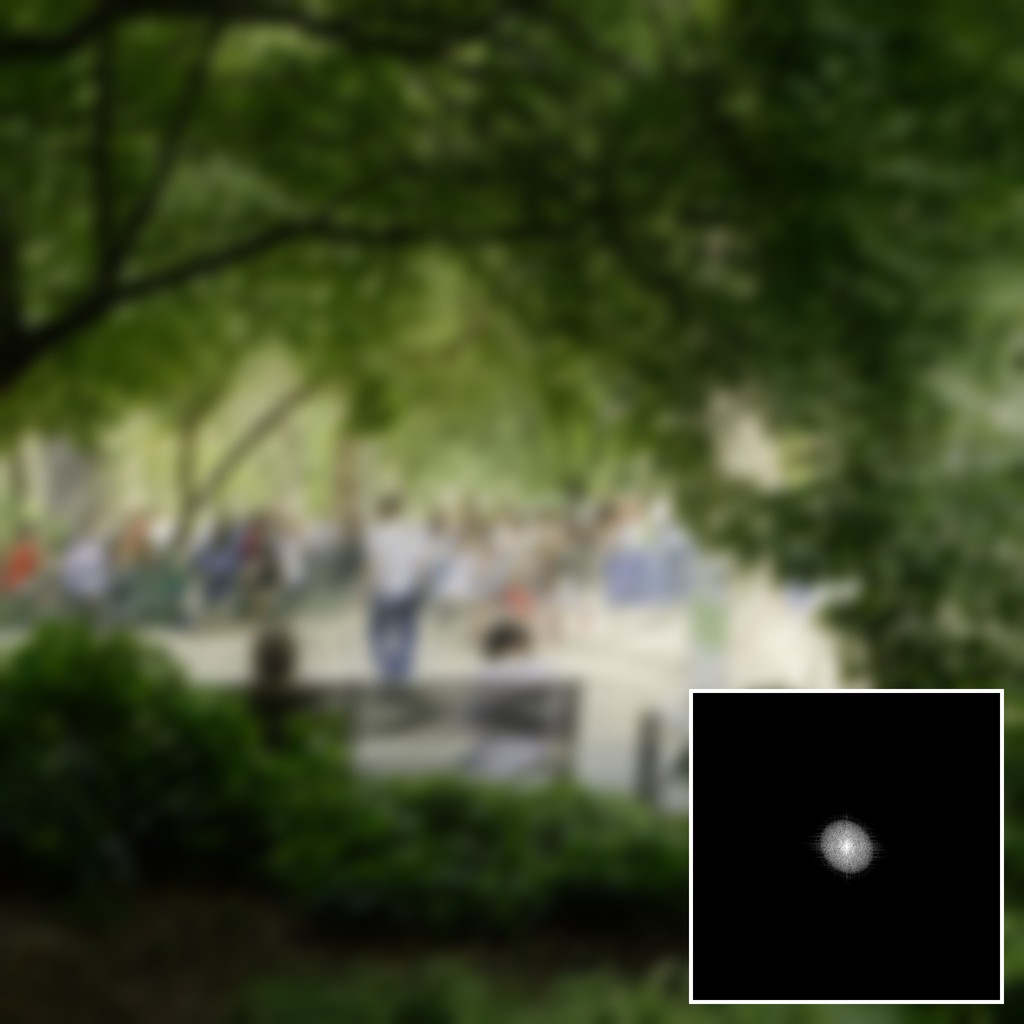}}
        \\
        & &  
        \multicolumn{2}{c}{
            \begin{tabular}{ccc}
                \centeredtab{0.0} & 
                \centeredtab{\fbox{\includegraphics[width=0.155\textwidth]{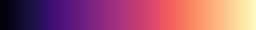}}} & 
                \centeredtab{0.2}
            \end{tabular}} \\
    \end{tabular}
    \caption{
    \label{fig:image_results}
        Comparisons against Neural Field Convolutions (NFC)~\cite{nsampi2023neural} and Neural Gaussian Scale-Space Fields (NGSSF)~\cite{mujkanovic2024neural} for image filtering across Gaussian, Box, and Lanczos kernels. 
        Our model supports controllable smoothing across families and anisotropic covariances in a single forward pass. NFC parameterizes filters with Dirac impulses; it is reliable for isotropic/mild kernels but is capacity-limited for anisotropic and non-polynomial families. 
        NGSSF is tuned for Gaussian smoothing; we recalibrated its encoding for each family, but the filter family cannot be switched explicitly at test time. 
        Bottom-right insets show frequency spectra; top-right insets show mean error. See the supplemental and website for additional results.
        Images from Adobe FiveK; \textcopyright{} original photographers/Adobe.
    }
\end{figure*}

\paragraph{Baselines.}
\emph{Neural Field Convolutions (NFC)}~\cite{nsampi2023neural} models a filter as a sparse set of Dirac impulses, interpolated through a piecewise-polynomial approximation to the target kernel. 
This construction yields continuity in \emph{isotropic} and \emph{axis-aligned} settings, but for general \emph{anisotropic} kernels, it requires re-optimizing Dirac locations and weights for each covariance.
Therefore, the representation is not continuous across anisotropic scale space. 
In practice, following the prescribed guidelines, we use piecewise-linear kernels only for 2D isotropic comparisons; for anisotropic cases, piecewise-constant kernels are required because higher-order models fail to optimize reliably.
Even with substantial per-kernel tuning, representing non-polynomial filter families (\eg Gaussian and Lanczos) and highly anisotropic filters remain limited by the NFC construction (see the optimized Dirac deltas in Fig.~\ref{fig:image_results}).

\emph{Neural Gaussian Scale-Space Fields (NGSSF)}~\cite{mujkanovic2024neural} learns a scale-conditioned MLP that is well suited to general low-pass filtering. 
However, it does not provide an explicit control to switch the \emph{type} of smoothing at test time. 
To compare across families, we therefore recalibrate NGSSF’s encoding with a Monte Carlo estimate of the filtered signal for the target family. 
While this calibration aligns the overall spectral shape, the nature of the smoothing (\eg Gaussian, Box, and Lanczos), it is not directly controllable and residual estimation errors remain (see Fig.~\ref{fig:image_results}).

\paragraph{Results on images.}
We assess generalization for the \emph{image regression} experiments quantitatively in Table~\ref{tab:training_filter_results_1} and qualitatively in Figure~\ref{fig:image_results}.
For this task, frequency characteristics are most visible in the frequency-domain insets (bottom right of each panel; see also Figs.~\ref{fig:teaser} and \ref{fig:image_results}).
Single-family training already generalizes well to the other families; joint training provides only marginal gains.
The Lanczos frequency response closely matches the reference spectrum with or without Lanczos‑specific training (See supplemental material for more examples).
However, errors in the Lanczos filter overall are higher than those in the other filters.

\begin{figure*}
    \centering
    \setlength{\tabcolsep}{1pt}
    \renewcommand{\arraystretch}{0.6}
    \begin{tabular}{cccccccc}
        & & \multicolumn{3}{c}{Light filtering} & \multicolumn{3}{c}{Heavy filtering} \\
        \cmidrule(lr){2-5} \cmidrule(lr){6-8} 
        & & NGSSF & Ours & GT & NGSSF & Ours & GT \\
        \centeredtab{\rot{Gaussian}} & &
        \centeredtab{
            \includegraphics[width=0.155\textwidth, height=0.155\textwidth]{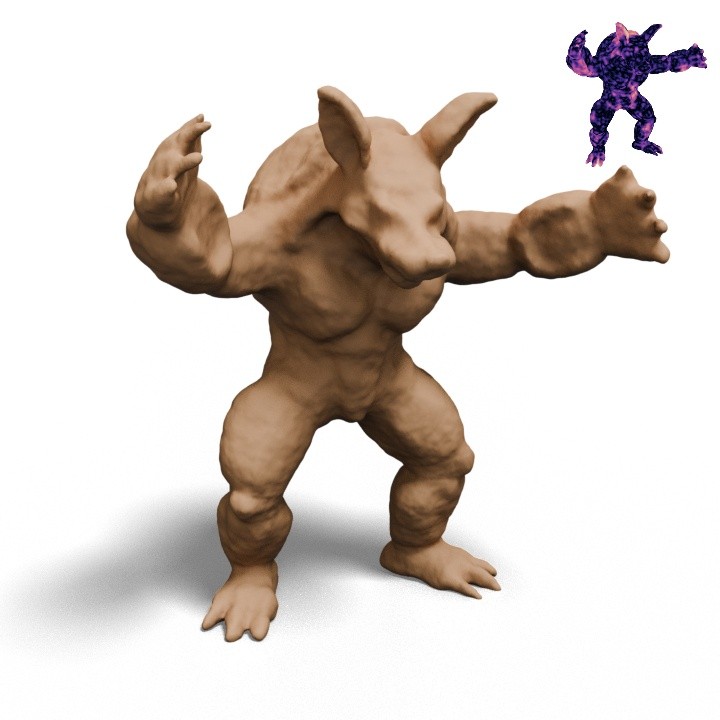} \\
            \includegraphics[width=0.155\textwidth, height=0.155\textwidth]{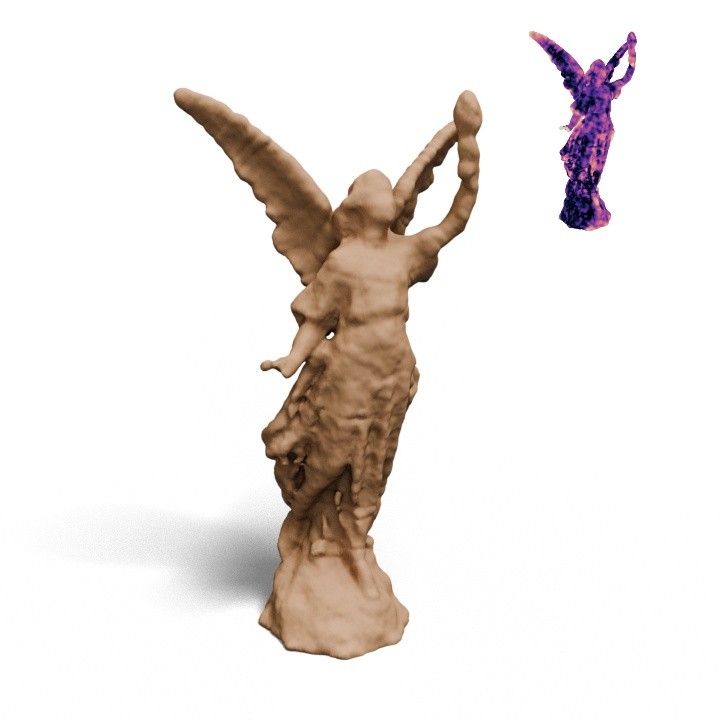}} & 
        \centeredtab{
            \includegraphics[width=0.155\textwidth, height=0.155\textwidth]{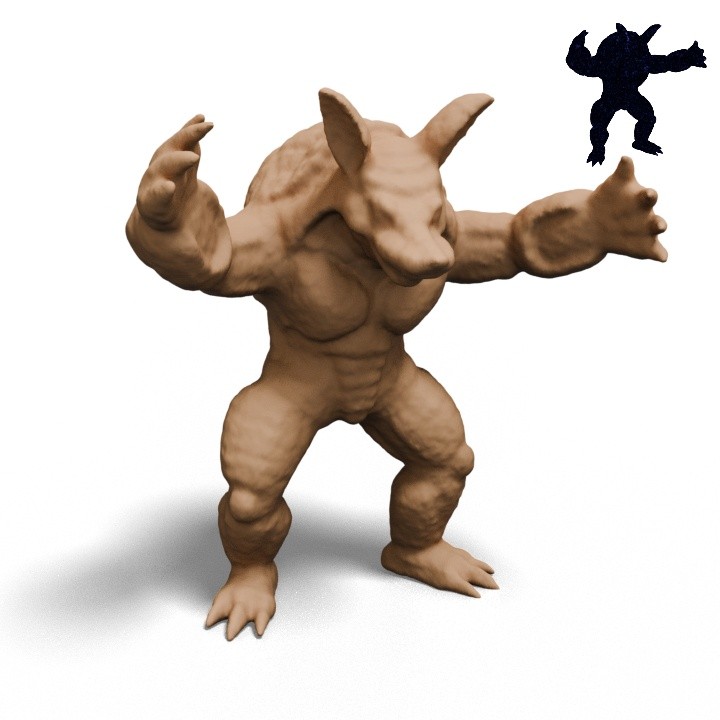} \\
            \includegraphics[width=0.155\textwidth, height=0.155\textwidth]{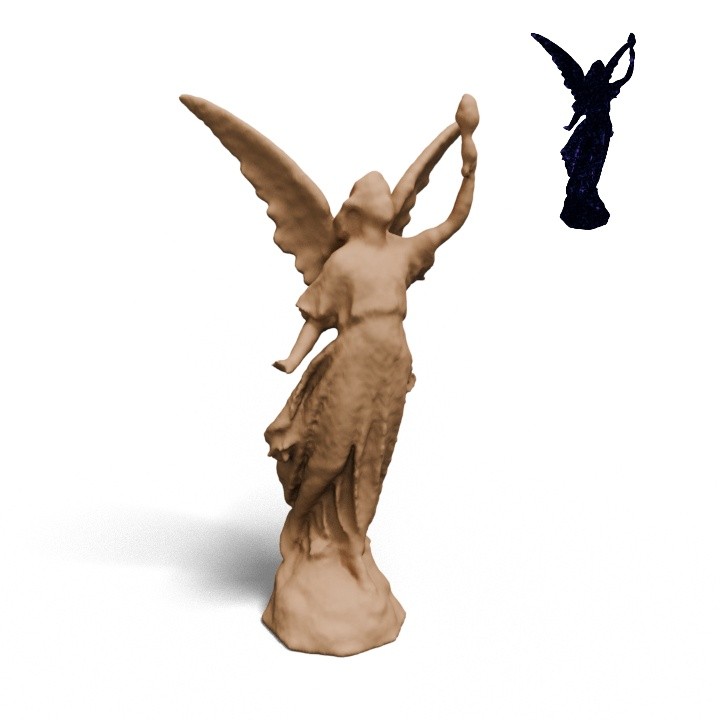}} & 
        \centeredtab{
            \includegraphics[width=0.155\textwidth, height=0.155\textwidth]{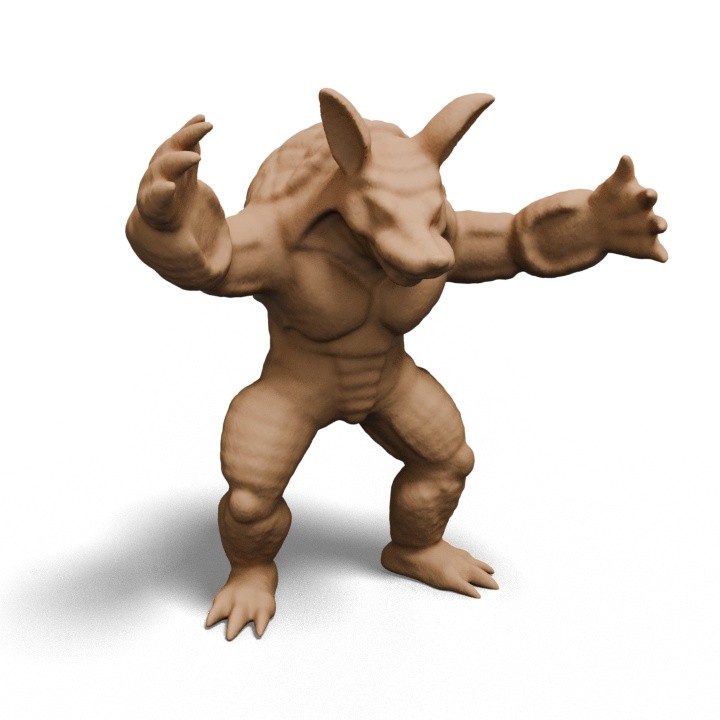} \\
            \includegraphics[width=0.155\textwidth, height=0.155\textwidth]{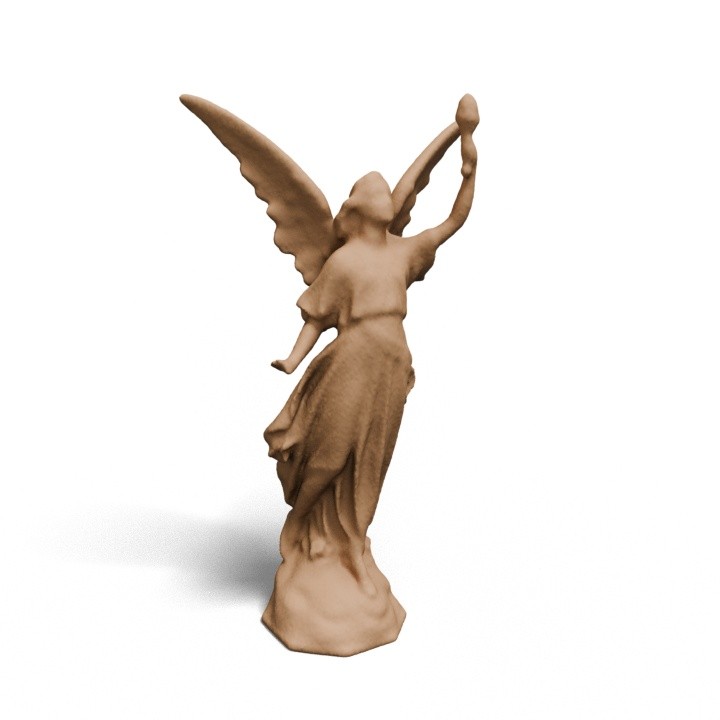}} & 
        \centeredtab{
            \includegraphics[width=0.155\textwidth, height=0.155\textwidth]{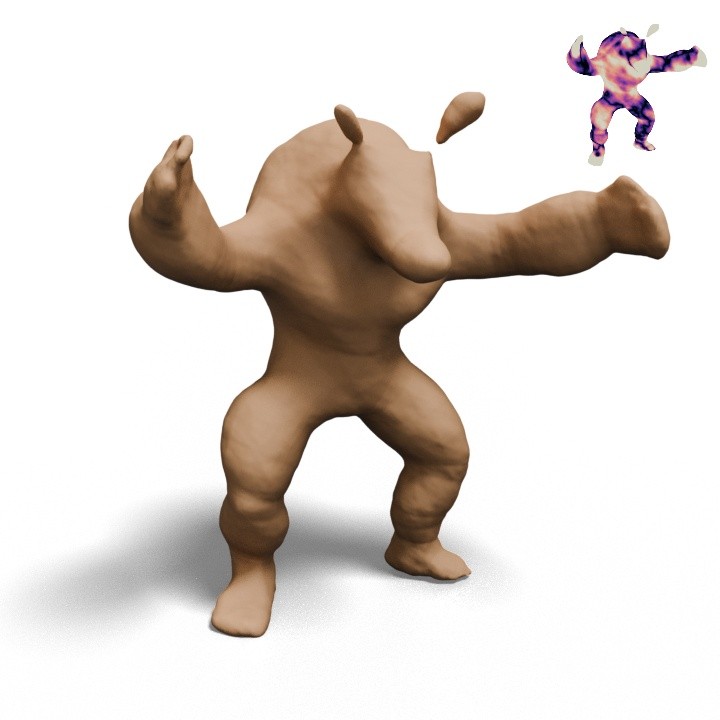} \\
            \includegraphics[width=0.155\textwidth, height=0.155\textwidth]{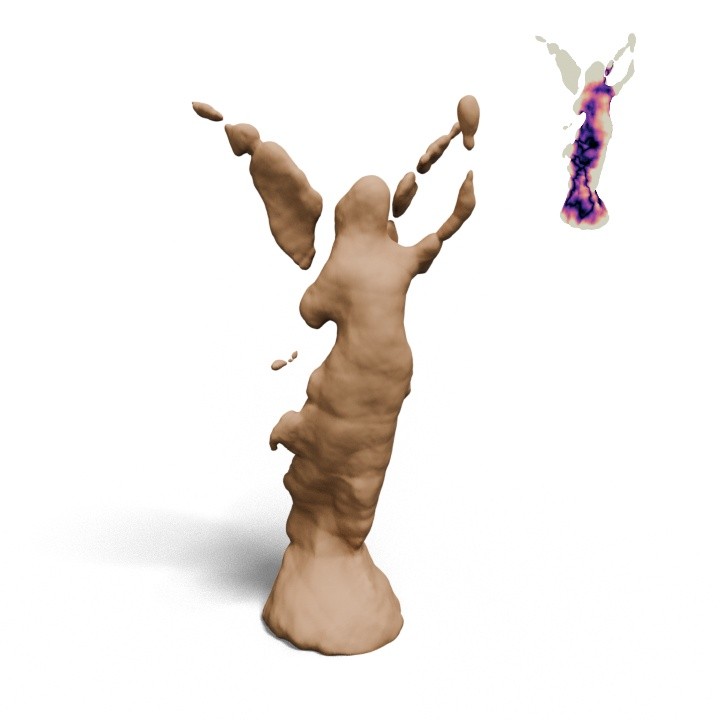}} & 
        \centeredtab{
            \includegraphics[width=0.155\textwidth, height=0.155\textwidth]{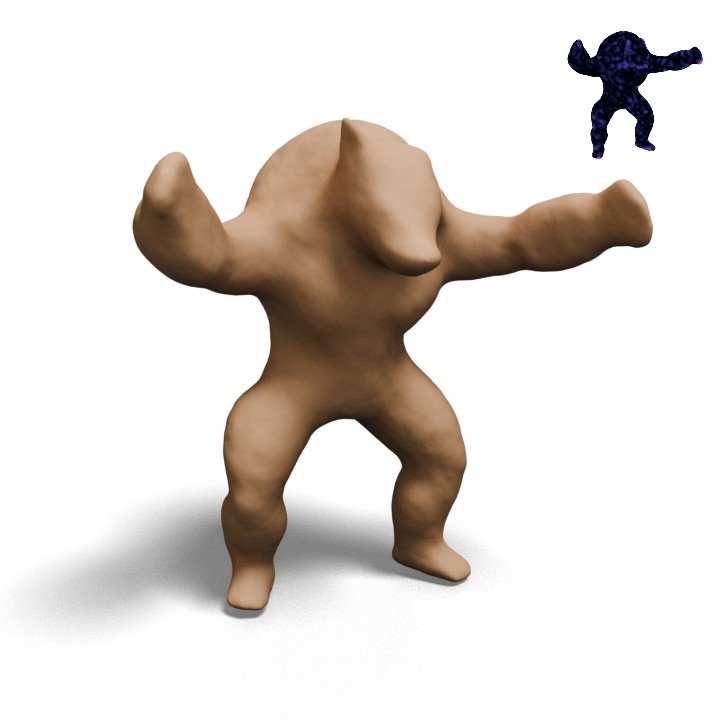} \\
            \includegraphics[width=0.155\textwidth, height=0.155\textwidth]{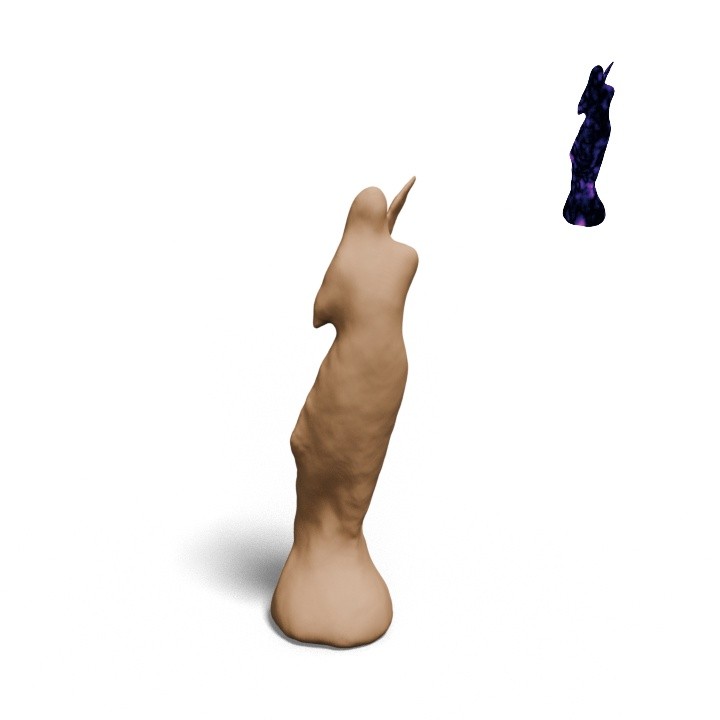}} & 
        \centeredtab{
            \includegraphics[width=0.155\textwidth, height=0.155\textwidth]{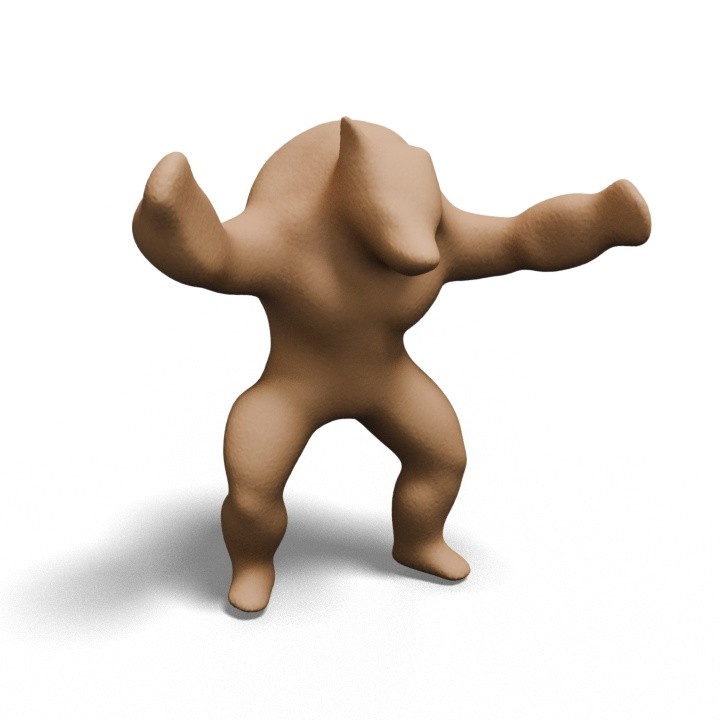} \\
            \includegraphics[width=0.155\textwidth, height=0.155\textwidth]{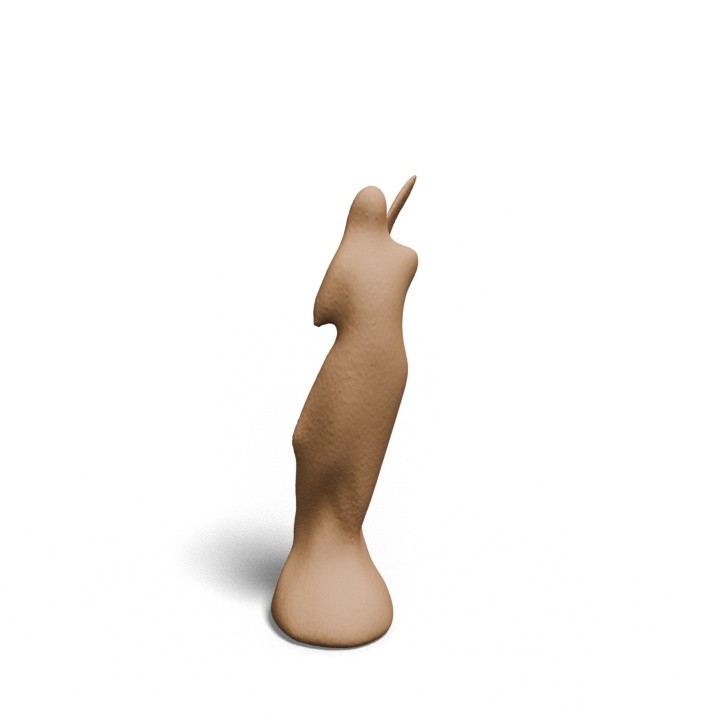}}\\
        \cmidrule(lr){3-8}
        \centeredtab{\rot{Box}} & &
        \centeredtab{
            \includegraphics[width=0.155\textwidth, height=0.155\textwidth]{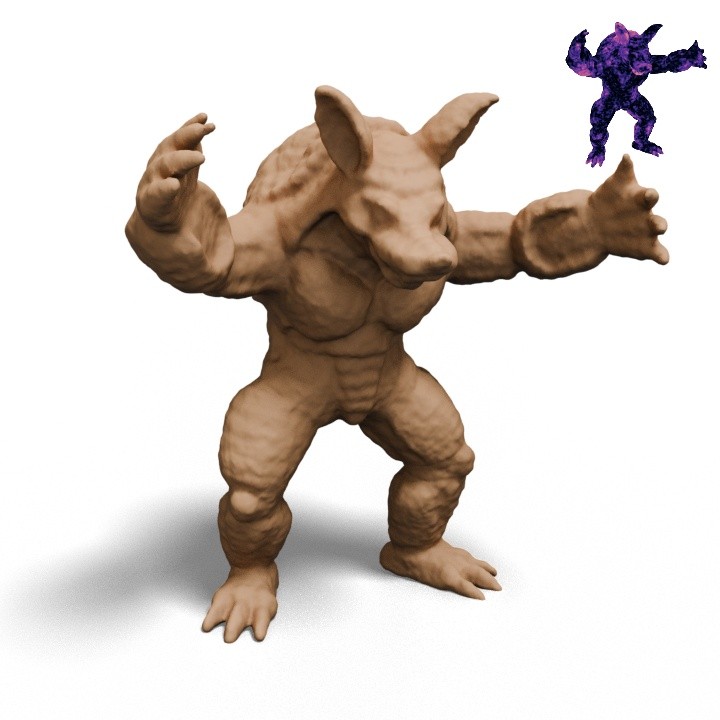} \\
            \includegraphics[width=0.155\textwidth, height=0.155\textwidth]{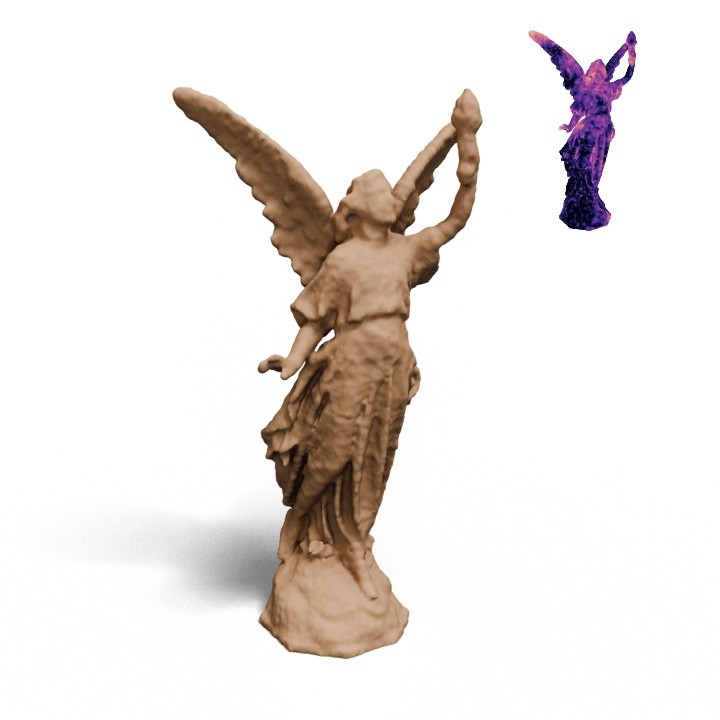}} & 
        \centeredtab{
            \includegraphics[width=0.155\textwidth, height=0.155\textwidth]{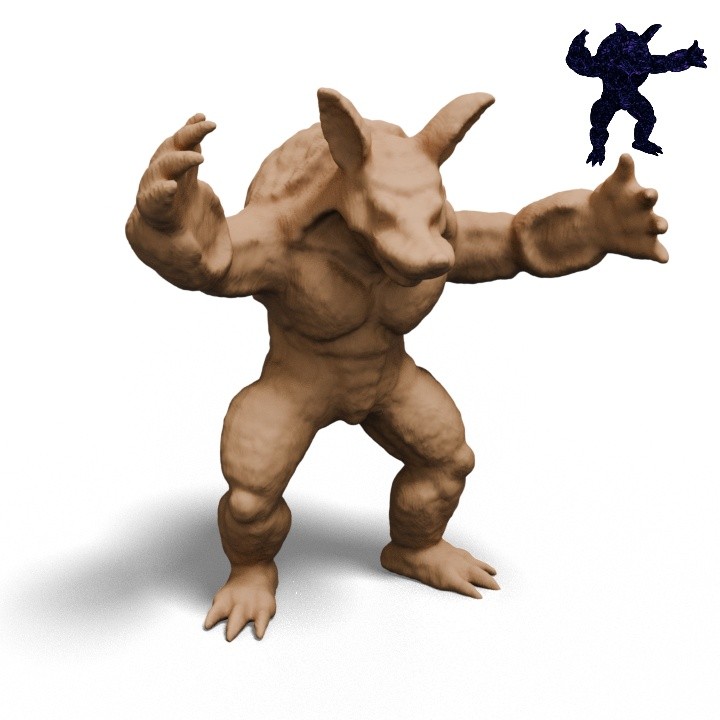} \\
            \includegraphics[width=0.155\textwidth, height=0.155\textwidth]{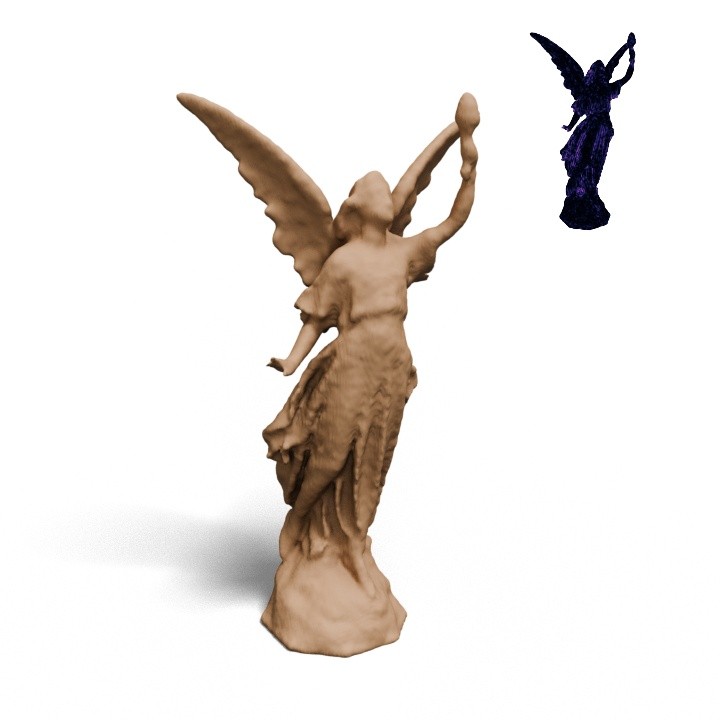}} & 
        \centeredtab{
            \includegraphics[width=0.155\textwidth, height=0.155\textwidth]{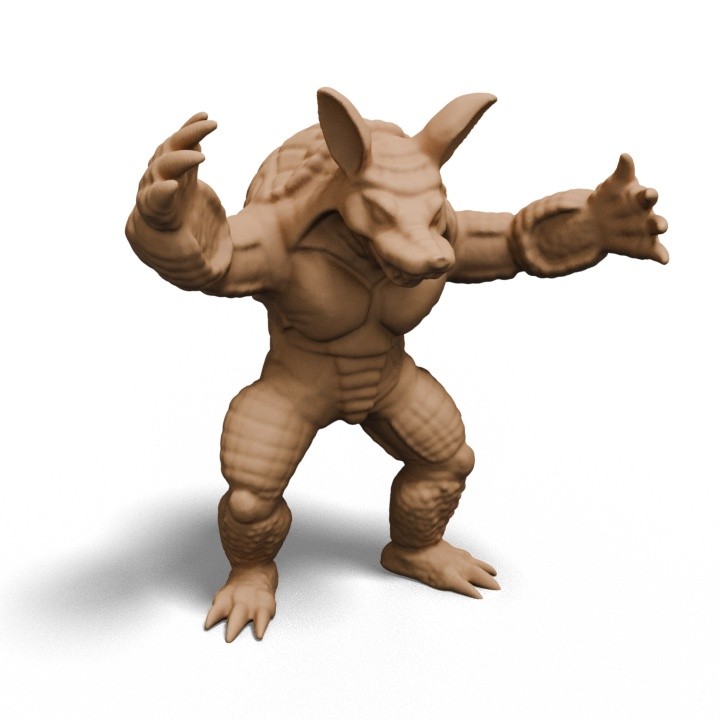} \\
            \includegraphics[width=0.155\textwidth, height=0.155\textwidth]{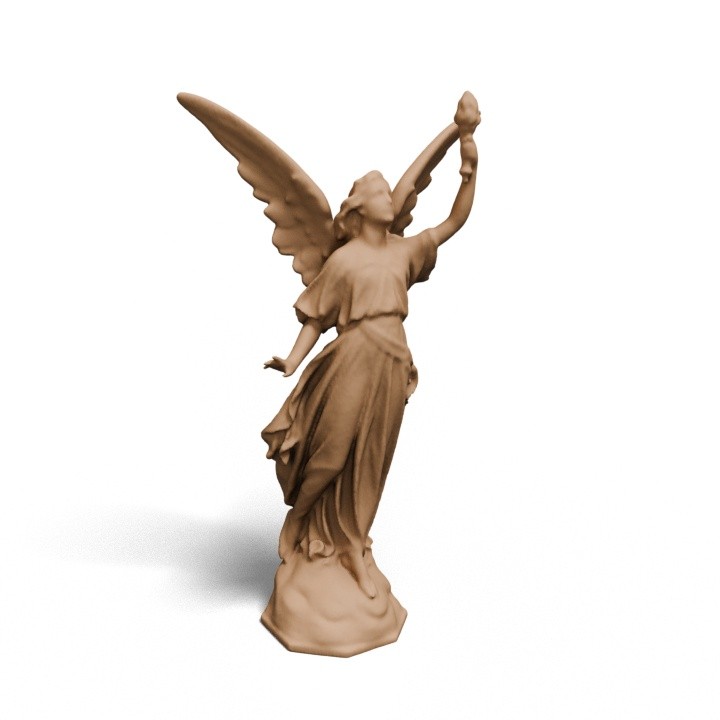}} & 
        \centeredtab{
            \includegraphics[width=0.155\textwidth, height=0.155\textwidth]{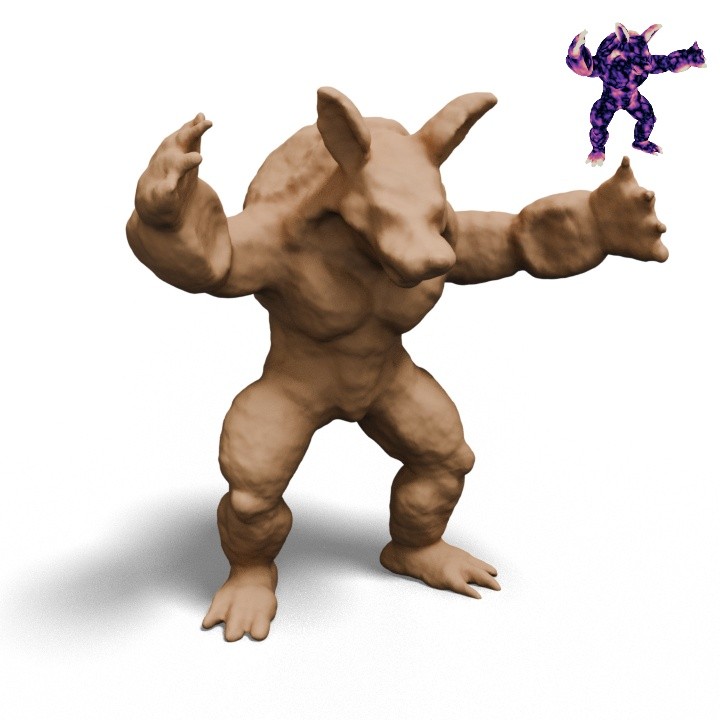} \\
            \includegraphics[width=0.155\textwidth, height=0.155\textwidth]{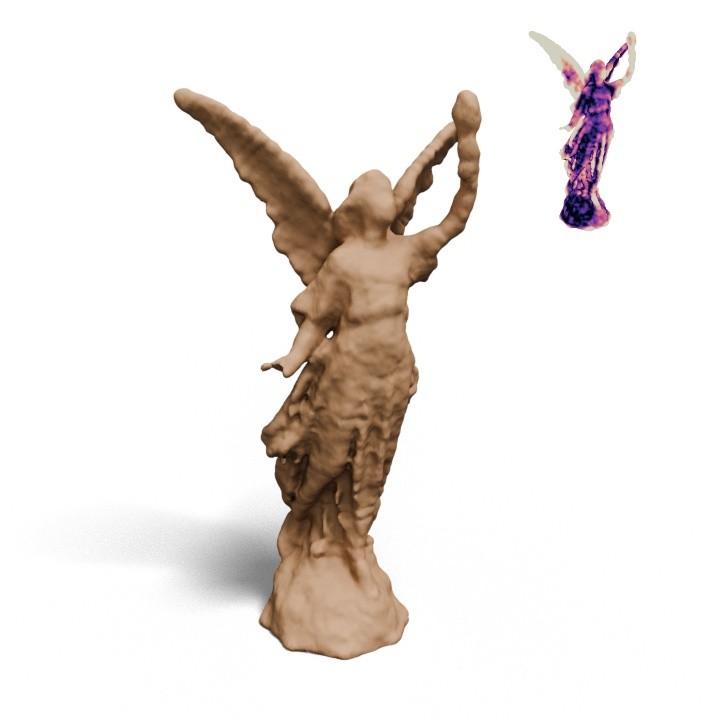}} & 
        \centeredtab{
            \includegraphics[width=0.155\textwidth, height=0.155\textwidth]{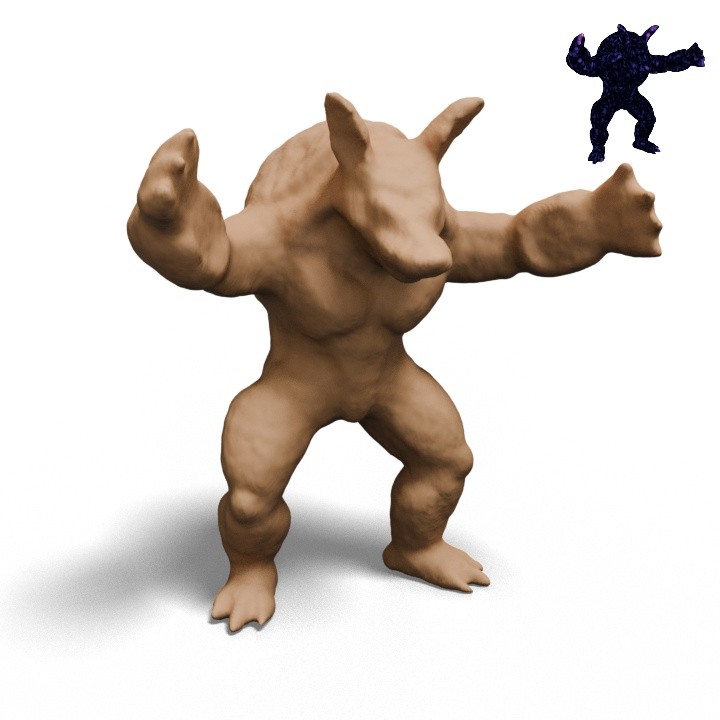} \\
            \includegraphics[width=0.155\textwidth, height=0.155\textwidth]{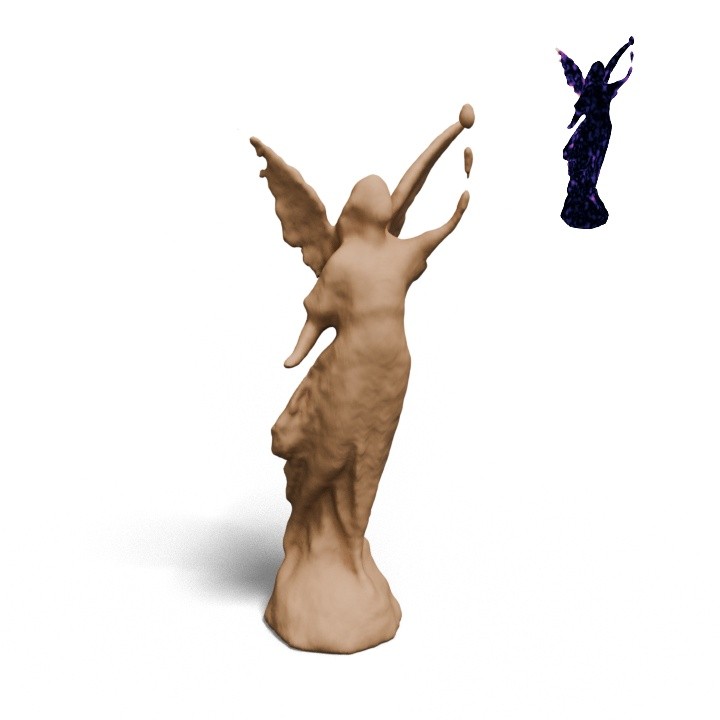}} & 
        \centeredtab{
            \includegraphics[width=0.155\textwidth, height=0.155\textwidth]{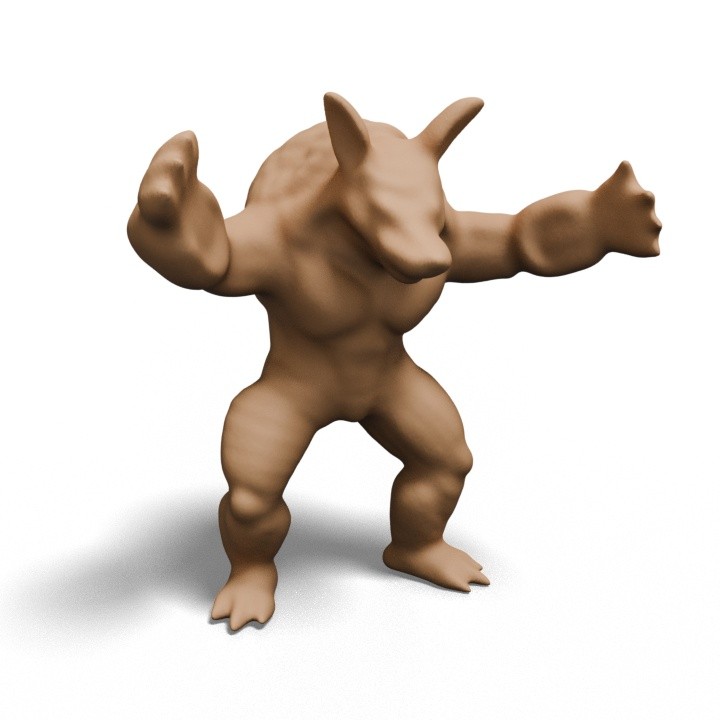} \\
            \includegraphics[width=0.155\textwidth, height=0.155\textwidth]{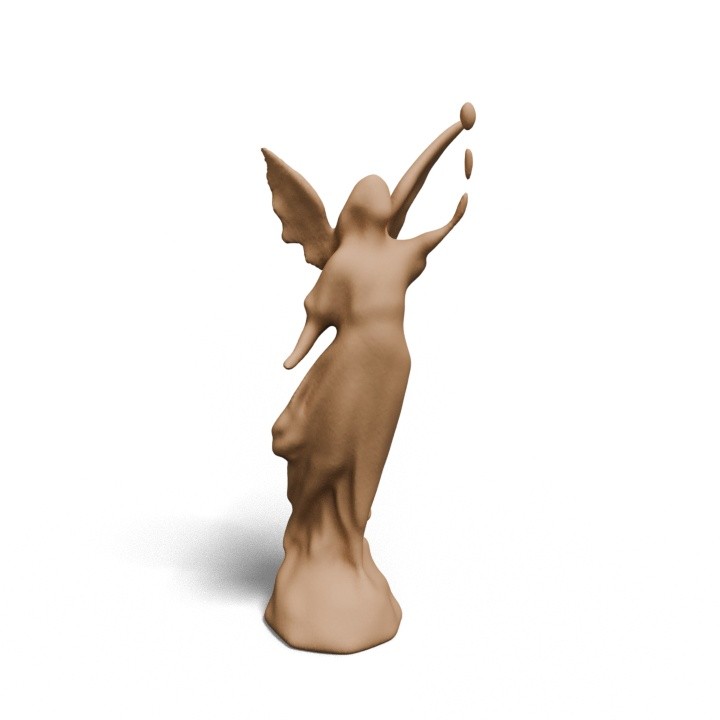}}\\
        \cmidrule(lr){3-8}
        \centeredtab{\rot{Lanczos}} & &
        \centeredtab{
            \includegraphics[width=0.155\textwidth, height=0.155\textwidth]{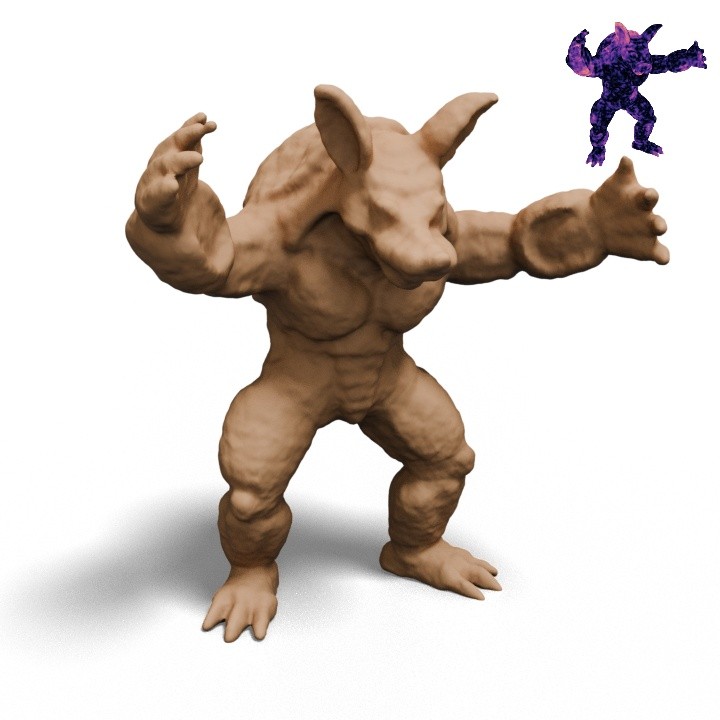} \\
            \includegraphics[width=0.155\textwidth, height=0.155\textwidth]{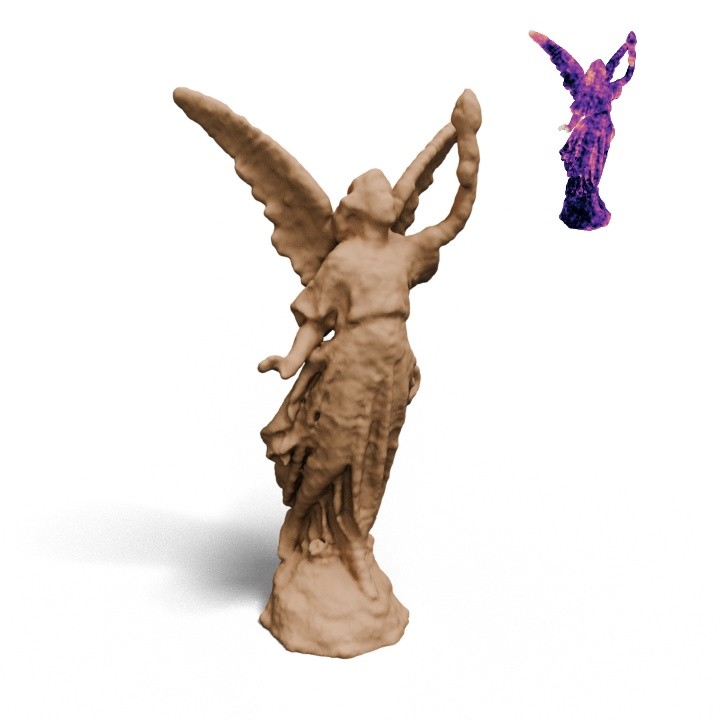}} & 
        \centeredtab{
            \includegraphics[width=0.155\textwidth, height=0.155\textwidth]{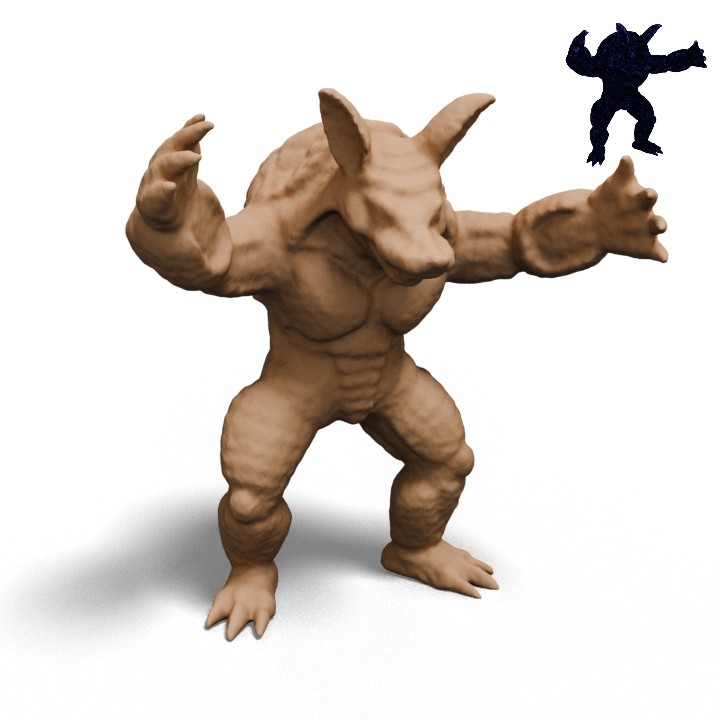} \\
            \includegraphics[width=0.155\textwidth, height=0.155\textwidth]{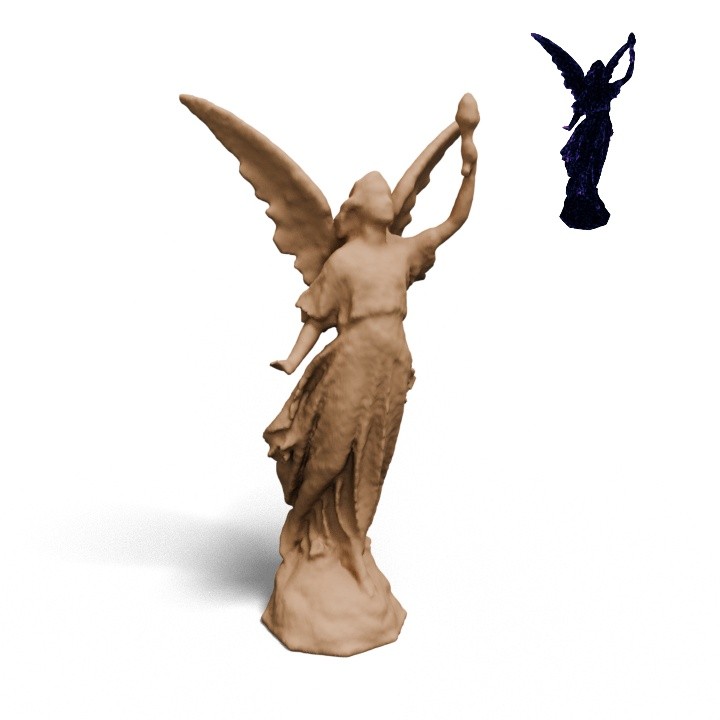}} & 
        \centeredtab{
            \includegraphics[width=0.155\textwidth, height=0.155\textwidth]{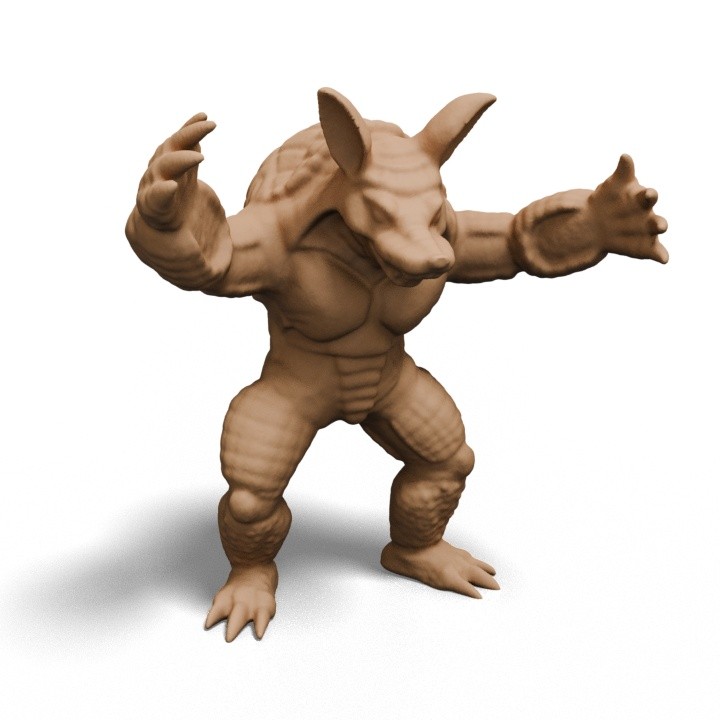} \\
            \includegraphics[width=0.155\textwidth, height=0.155\textwidth]{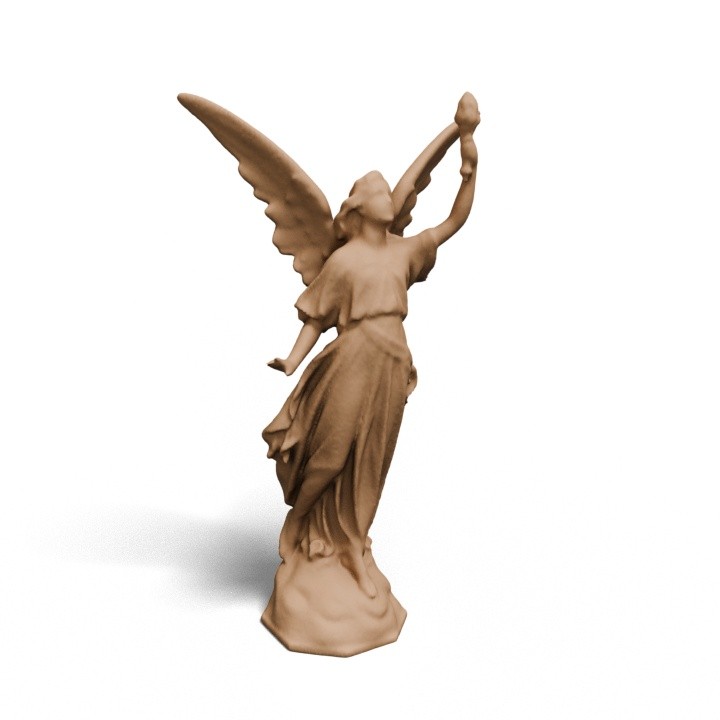}} & 
        \centeredtab{
            \includegraphics[width=0.155\textwidth, height=0.155\textwidth]{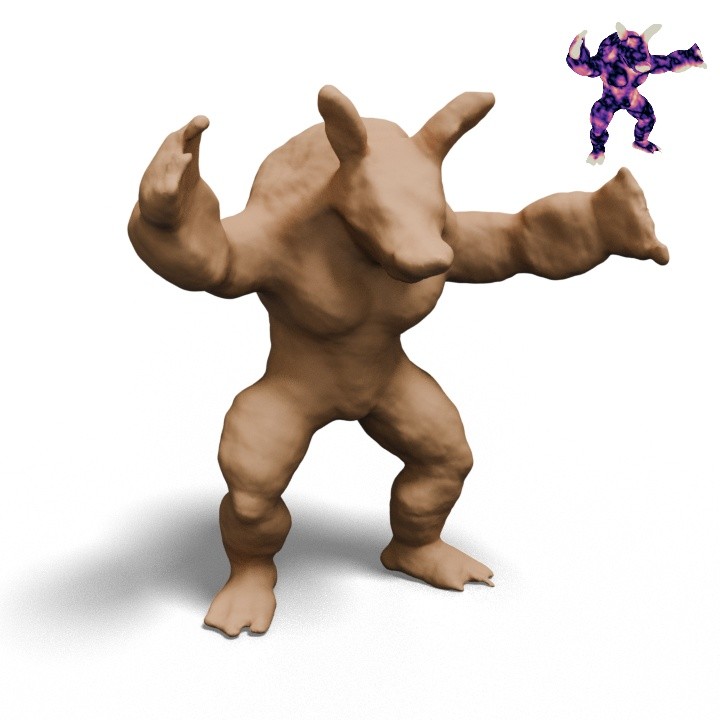} \\
            \includegraphics[width=0.155\textwidth, height=0.155\textwidth]{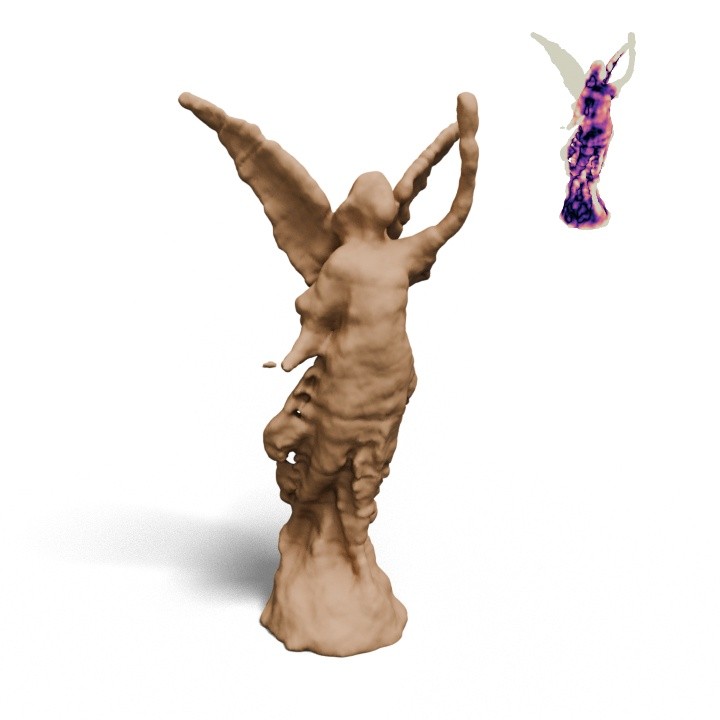}} & 
        \centeredtab{
            \includegraphics[width=0.155\textwidth, height=0.155\textwidth]{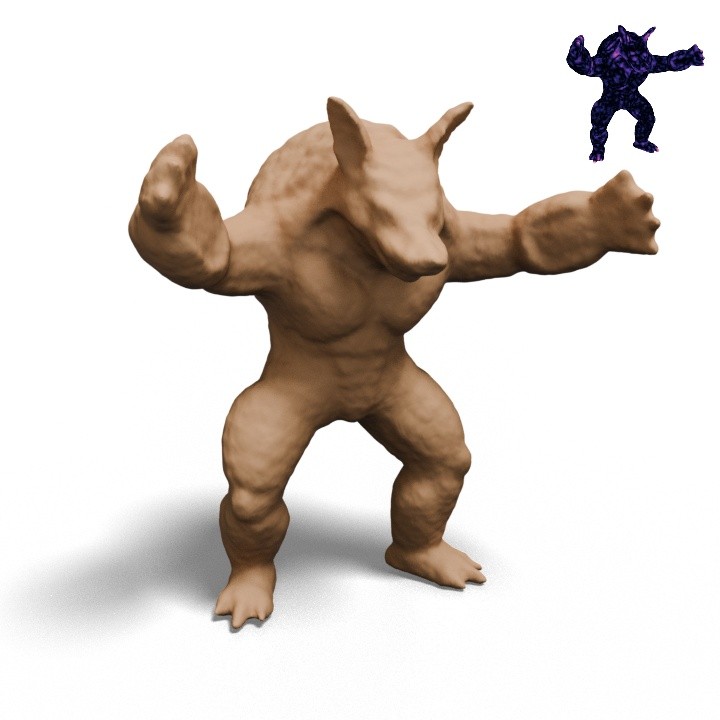} \\
            \includegraphics[width=0.155\textwidth, height=0.155\textwidth]{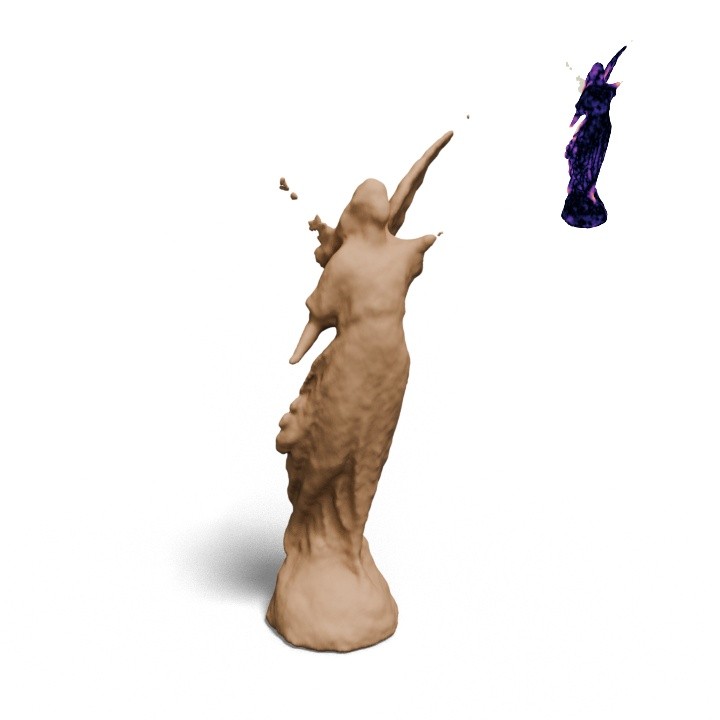}} & 
        \centeredtab{
            \includegraphics[width=0.155\textwidth, height=0.155\textwidth]{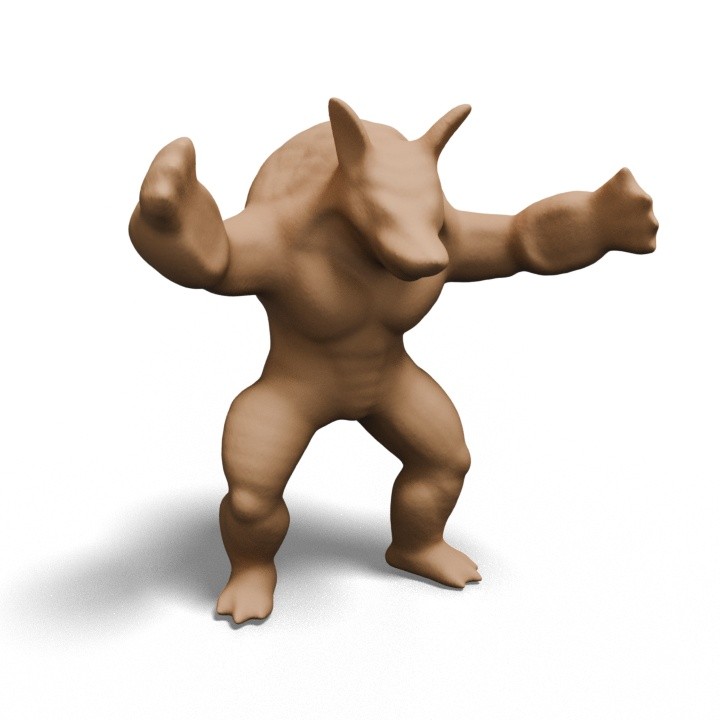} \\
            \includegraphics[width=0.155\textwidth, height=0.155\textwidth]{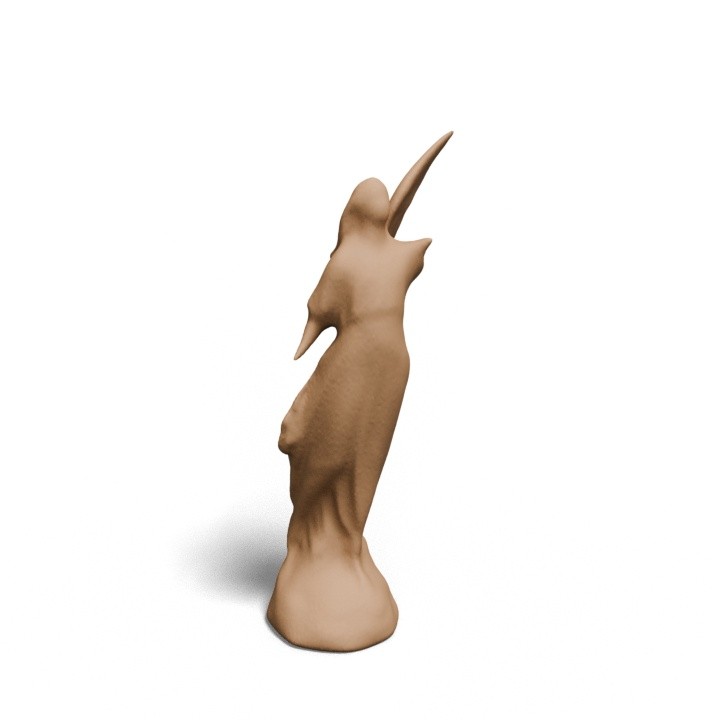}}\\
    \end{tabular}
    \caption{\label{fig:sdf_results}
        \captiontitle{Filter generalization on SDFs.}
        Comparisons against recalibrated Neural Gaussian Scale-Space Fields~\cite{mujkanovic2024neural} (NGSSF) for SDF filtering with different kernels. 
        Insets at the top-right represent the distance to the ground truth mesh, darker is better.
        Our model shows closer geometry to the ground-truth compared to NGSSF.
        Mesh models from the Stanford 3D Scanning Repository; \textcopyright{} Stanford Computer Graphics Laboratory.
    } 
\end{figure*}

\paragraph{Results on SDFs.}
We present generalization capabilities of our model for SDFs in Figure \ref{fig:sdf_results}.
Our model matches overall smoothing shape and characteristics of each filter, and achieves perceptually convincing results.
Fine-scale artifacts persist in some cases, which we attribute to training with higher MC variance on SDFs.
Importance sampling high-variance regions such as sharp surface changes and better tuning Fourier basis scale to match signal frequency characteristics (see Sec.~\ref{sec:train_impl}) help mitigate these artifacts.

\paragraph{Summary.}
Across both images and SDFs, our model trained with frequency-modulated features supports Gaussian, Box, and Lanczos filtering with anisotropic covariances in single forward pass, without architectural changes. 
Compared to NFC’s Dirac-impulse parameterization (capacity-limited and not continuous in anisotropic scale space) and NGSSF’s Gaussian-specific design (recalibration needed and no explicit kernel family control), our approach exposes the frequency response directly in the input encoding, which enables cross-family generalization and controllable smoothing at test time.

\subsection{Ablations}
\label{sec:ablation}
\begin{figure}[htbp]
  \centering
  \def\svgwidth{\linewidth}
  \footnotesize
  \includegraphics[width=\linewidth]{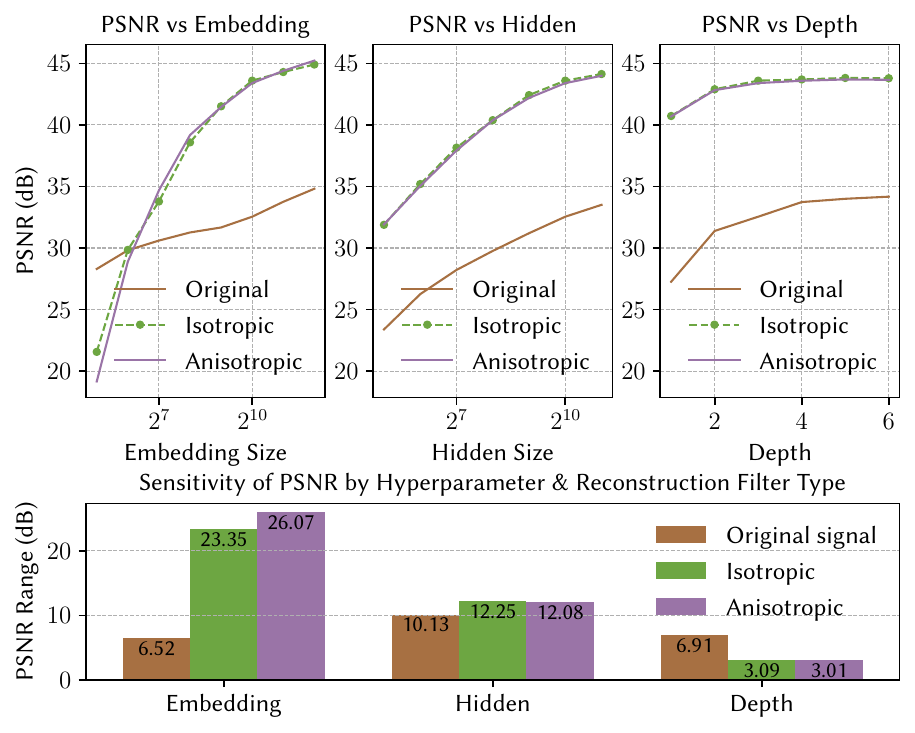}
  \caption{
  \label{fig:psnr_vs_size}
  \captiontitle{Sensitivity Analysis.}
  We evaluate how Fourier-feature embedding size and network architecture affect (1) reconstruction of the original signal, (2) isotropic smoothing, and (3) anisotropic smoothing.
  \emph{Embedding Size.} While embedding size has relatively smaller effect on reconstructing the original signal, the model benefits substantially from larger embeddings in both isotropic and anisotropic smoothing tasks.
  \emph{Hidden Dimension.}
  Increasing the hidden-layer width consistently yields higher PSNR across all tests.
  \emph{Network Depth.} Adding more than three layers provides only marginal PSNR gains, indicating diminishing returns beyond a depth of three.
  The bottom row shows the PSNR range (highest minus lowest) observed over each hyperparameter sweep, illustrating the relative sensitivities.
  }
\end{figure}

In this section, we investigate design choices of our method and ablate them to observe their effectiveness.
For each of these ablations, the model and the experiment parameters were the same as before. 

\paragraph{Model parameters}
To assess which components matter most for (i) reconstructing the original signal, (ii) isotropic filtering, 
and (iii) anisotropic filtering, we run ablations in Figure~\ref{fig:psnr_vs_size} by varying the Fourier-feature (embedding) size, the MLP hidden width, and the network depth. 
Although the model can reconstruct the original signal with relatively small embeddings (\eg~NeRF uses $m{=}10$~\cite{mildenhall2021nerf}), the embedding size is the most sensitive factor for filtering performance. 
Increasing hidden width also helps across all tests, while increasing depth beyond three layers yields only marginal gains.
We perform a depth ablation with matched number of trainable parameters (see supplemental Fig.), varying network depth, and find that a three or four layer MLP yields the best reconstruction quality.

\paragraph{Model components}
\begin{table*}
\caption{
\captiontitle{Quantitative Ablation Study on Alien.}
We analyze the impact of having one Monte Carlo sample, removing the Lipschitz constraint, and using our exact Fourier modulation.
We observe that, while the Lipschitz-constrained MLP achieves better filtering without Monte Carlo, lifting it performs better once single Monte Carlo sampling is enabled. 
The exact Fourier modulation improves results in all cases.
When the three components are enabled, the method achieves the best results in both original signal fitting and filtering. 
}
\label{tab:alien_exact_downweight}
\begin{tabular}{cccrrrrrrrrr}
\toprule
\multirow{2}[2]{*}{\centeredtab{Monte \\ Carlo}} & 
\multirow{2}[2]{*}{\centeredtab{Lipschitz \\ Lift}} & 
\multirow{2}[2]{*}{\centeredtab{Exact \\ Fourier}} & 
\multicolumn{3}{c}{Original} & \multicolumn{3}{c}{Anisotropic} & \multicolumn{3}{c}{Isotropic} \\
\cmidrule(lr){4-6}
\cmidrule(lr){7-9}
\cmidrule(lr){10-12}
& & & \footnotesize{PSNR$\uparrow$} & \footnotesize{LPIPS$\downarrow$} & \footnotesize{SSIM$\uparrow$} &
  \footnotesize{PSNR$\uparrow$} & \footnotesize{LPIPS$\downarrow$} & \footnotesize{SSIM$\uparrow$} &
  \footnotesize{PSNR$\uparrow$} & \footnotesize{LPIPS$\downarrow$} & \footnotesize{SSIM$\uparrow$} \\
\midrule
\ding{55} & \ding{55} & \ding{55} &
26.81 & 0.157 & 0.899 & 32.01 & 0.205 & 0.957 & 31.35 & 0.183 & 0.945 \\
\ding{55} & \checkmark & \ding{55} &
29.30 & 0.046 & 0.946 & 22.58 & 0.374 & 0.883 & 21.34 & 0.331 & 0.839 \\
\ding{55} & \ding{55} & \checkmark &
26.06 & 0.169 & 0.894 & 31.24 & 0.197 & 0.953 & 30.56 & 0.172 & 0.941 \\
\checkmark & \ding{55} & \ding{55} &
25.98 & 0.166 & 0.909 & 32.38 & 0.405 & 0.928 & 33.16 & 0.296 & 0.921 \\
\checkmark & \checkmark & \ding{55} &
31.89 & 0.032 & 0.972 & 36.01 & 0.350 & 0.933 & 37.14 & 0.183 & 0.934 \\
\checkmark & \ding{55} & \checkmark &
27.01 & 0.127 & 0.928 & 42.00 & 0.185 & 0.973 & 41.80 & 0.159 & 0.967 \\
\checkmark & \checkmark & \checkmark &
32.54 & 0.025 & 0.977 & 43.40 & 0.180 & 0.974 & 43.60 & 0.150 & 0.970 \\
\bottomrule
\end{tabular}
\end{table*}

\begin{figure}
  \centering
  \setlength{\tabcolsep}{1pt}
  \renewcommand{\arraystretch}{0.6}
  \begin{tabular}{ccc}
      & Filtered & Original \\
      \centeredtab{\rot{GT}} &
      \centeredtab{\includegraphics[width=0.22\textwidth, height=0.22\textwidth]{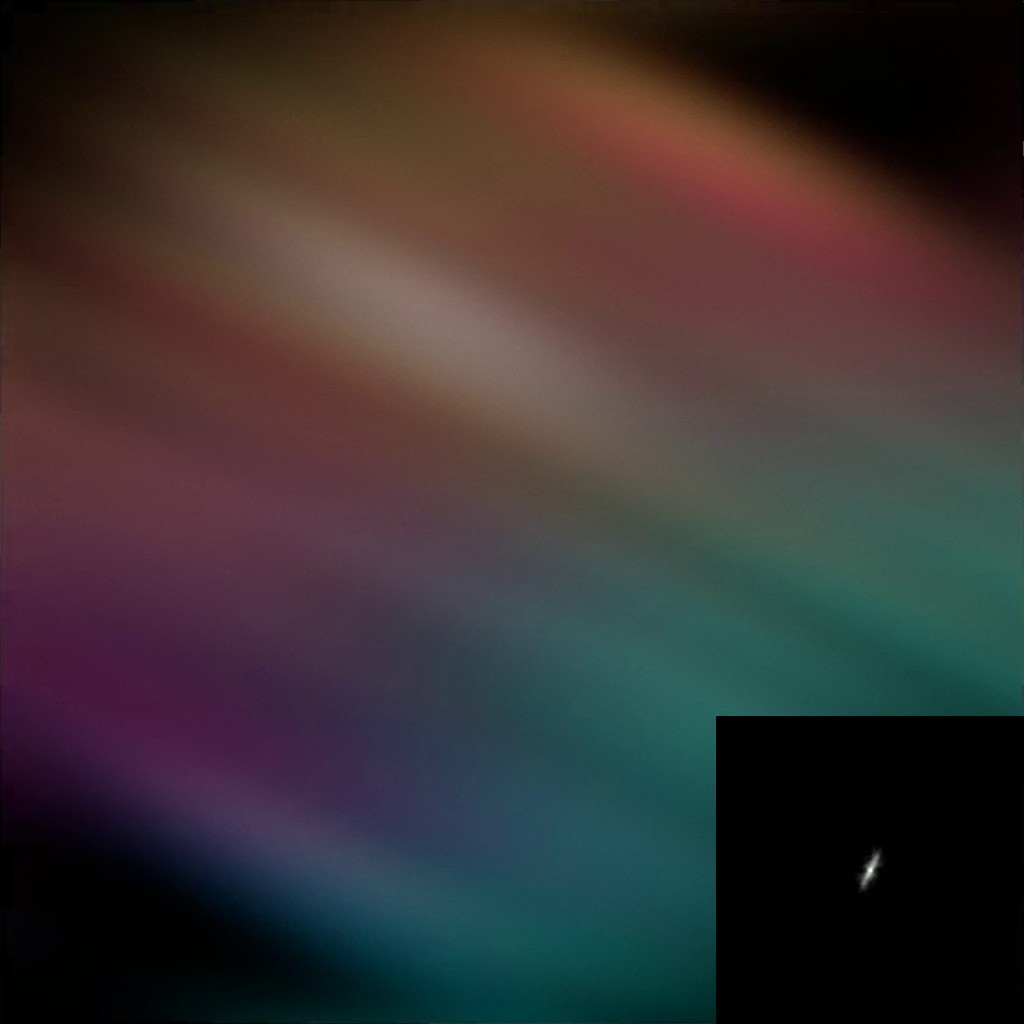}} &
      \centeredtab{\includegraphics[width=0.22\textwidth, height=0.22\textwidth]{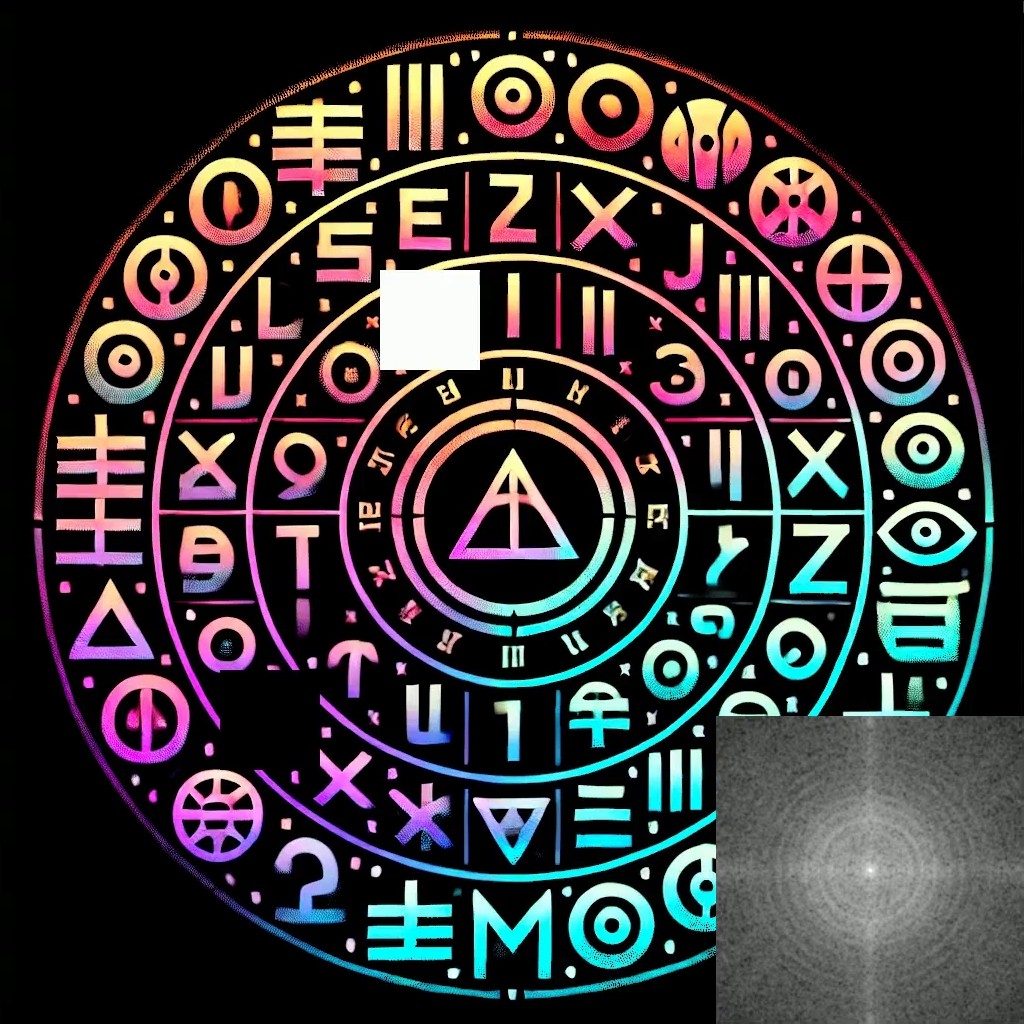}} \\
      \cmidrule(lr){2-3}
      & Prediction & Error ($\times 4$) \\
      \centeredtab{\rot{Without Exact Fourier}} &
      \centeredtab{\includegraphics[width=0.22\textwidth, height=0.22\textwidth]{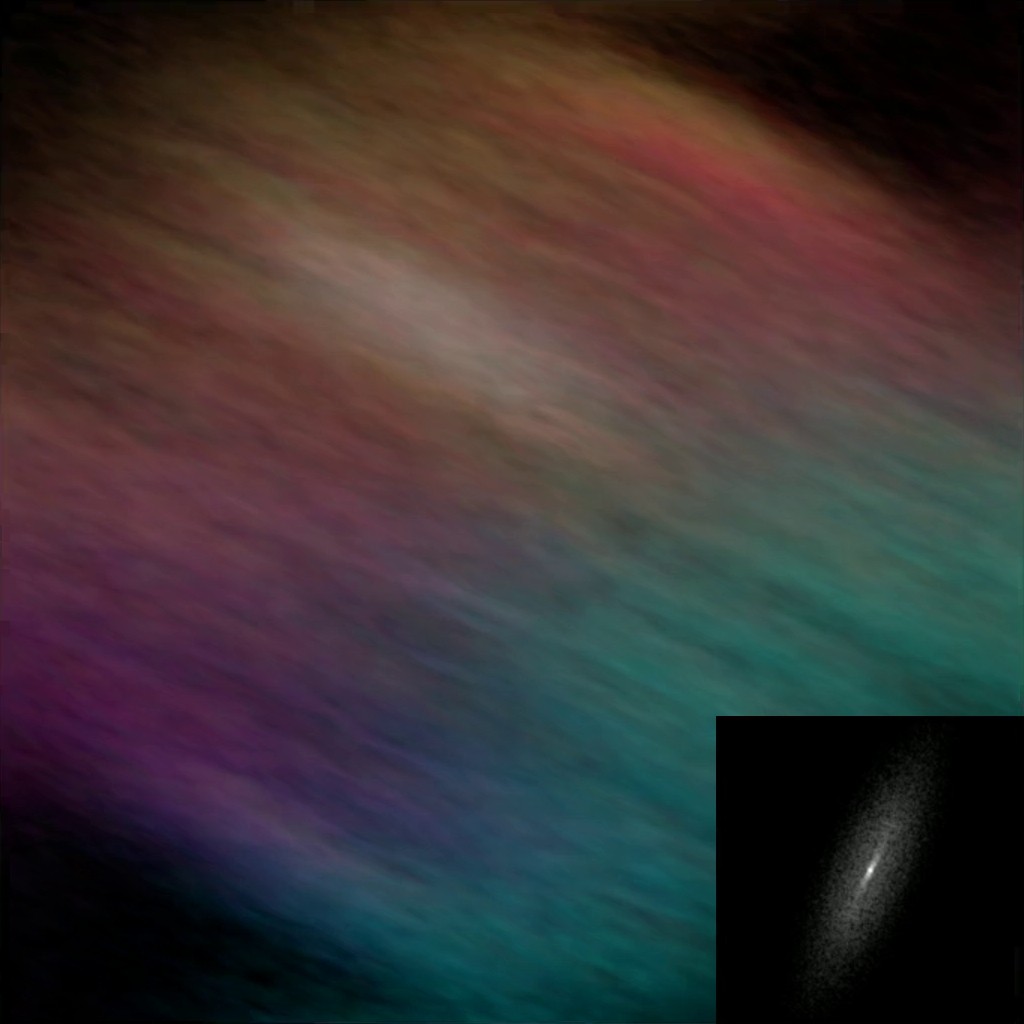}}  & 
      \centeredtab{\includegraphics[width=0.22\textwidth, height=0.22\textwidth]{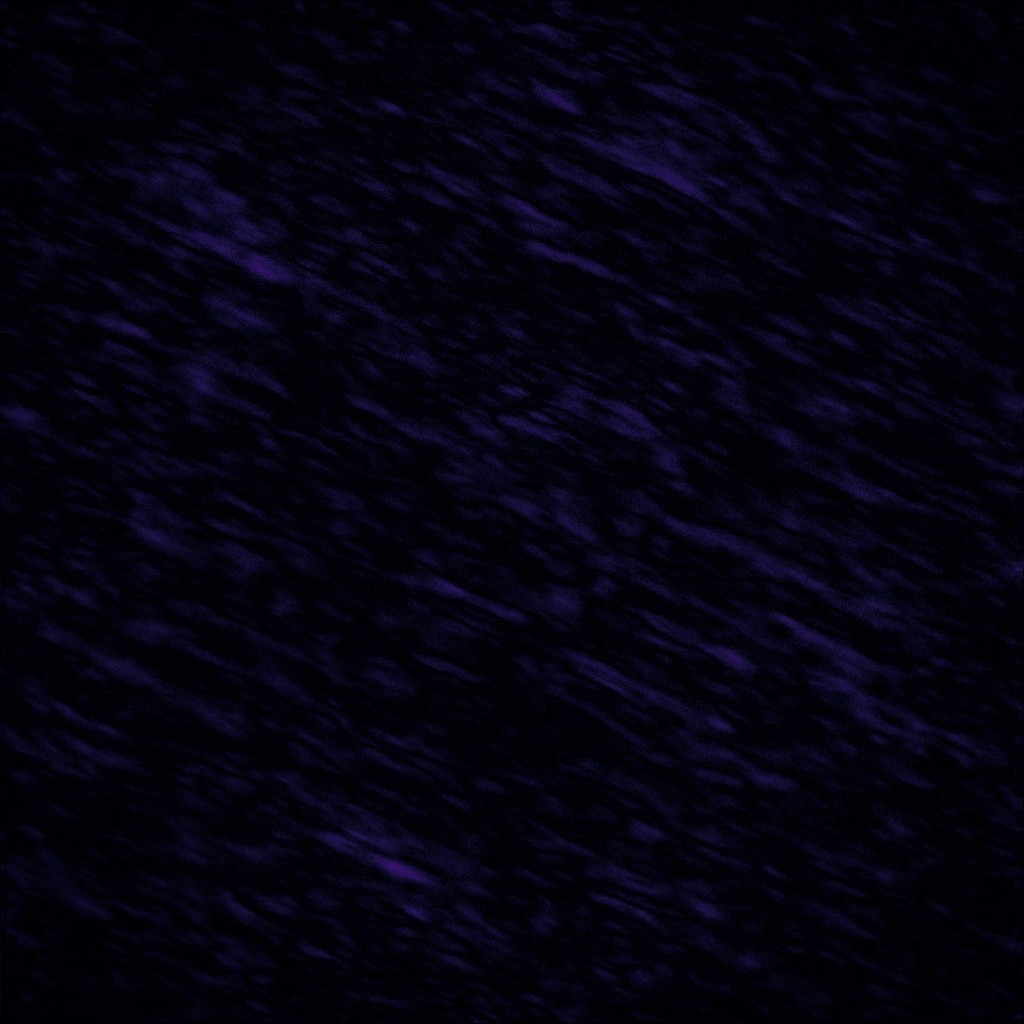}} \\
      \centeredtab{\rot{With Exact Fourier}} &
      \centeredtab{\includegraphics[width=0.22\textwidth, height=0.22\textwidth]{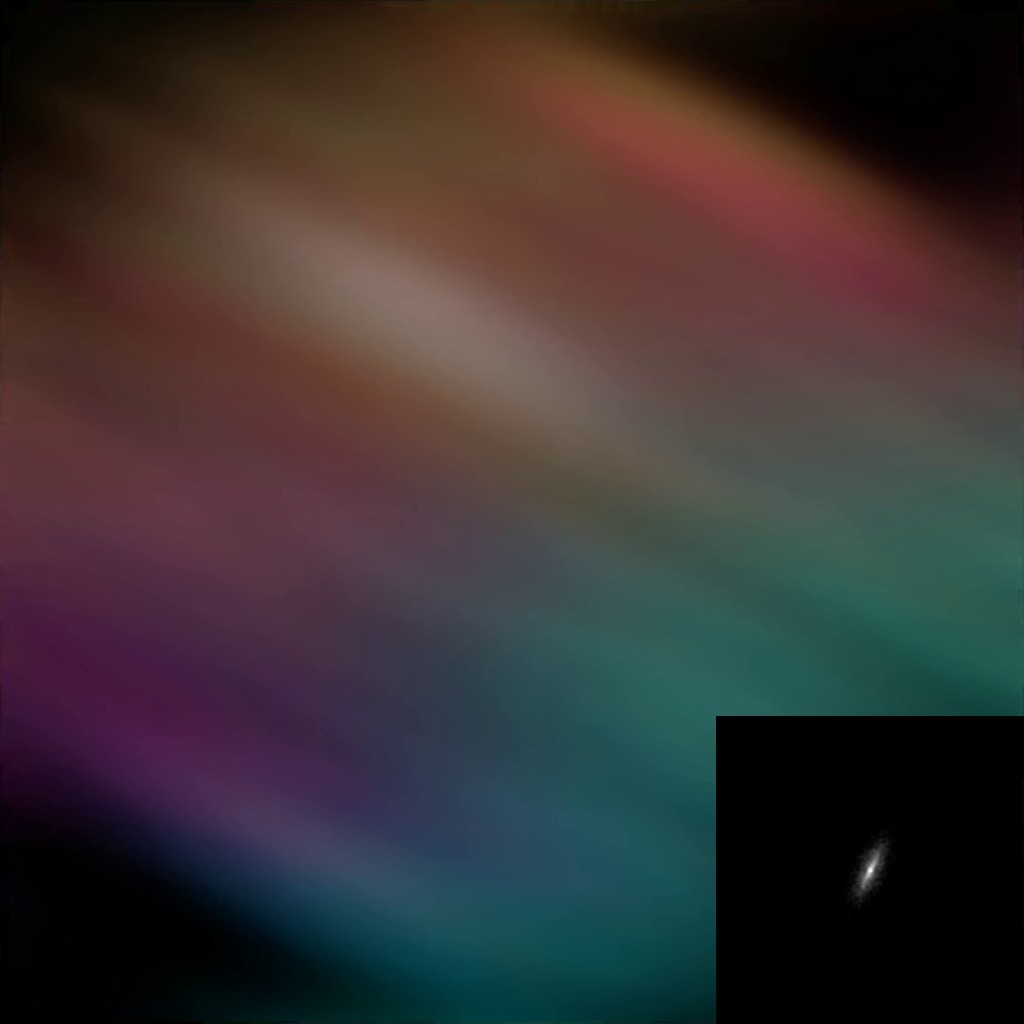}}  & 
      \centeredtab{\includegraphics[width=0.22\textwidth, height=0.22\textwidth]{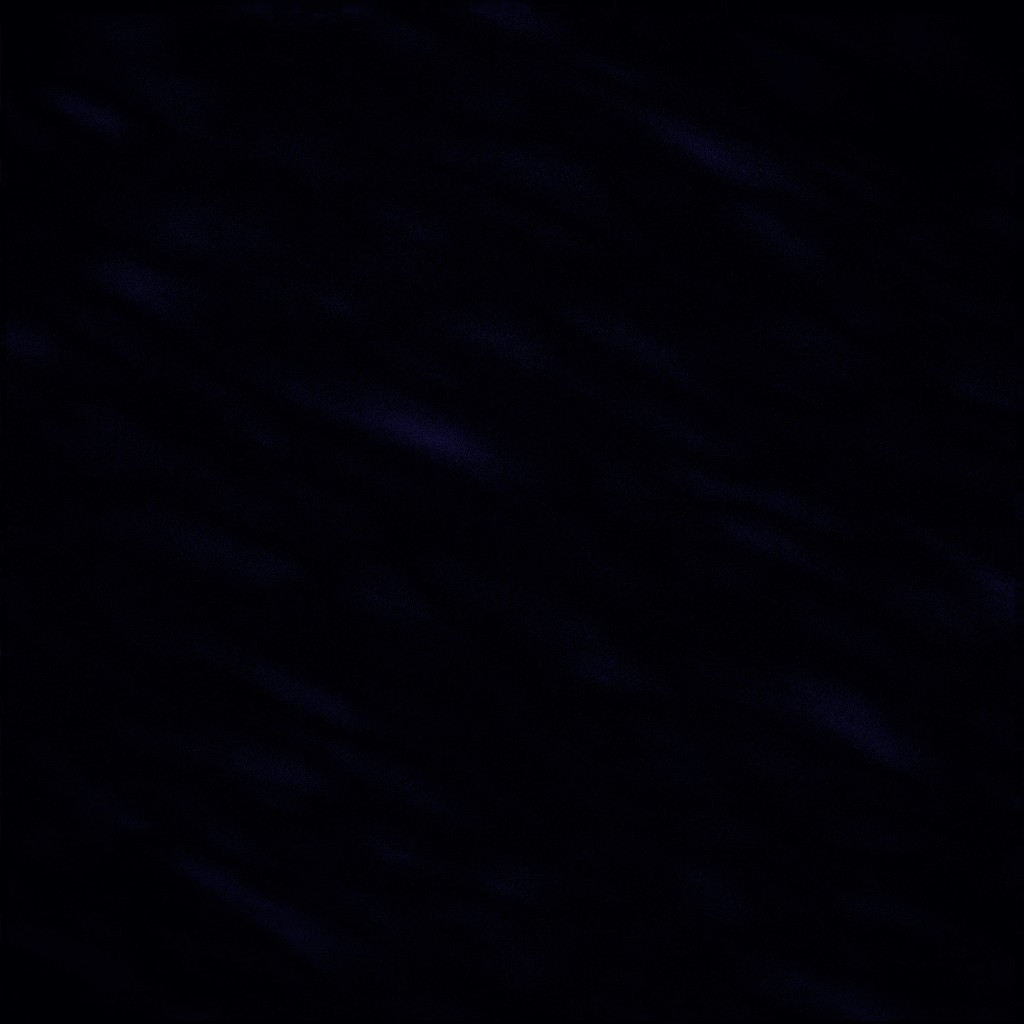}} 
  \end{tabular}
  \caption{\label{fig:ablation}
      \captiontitle{Effect of exact Fourier modulation.}
      We show the impact of exact Fourier modulation on the Alien image.
      The inset on the right bottom represents the frequency spectrum of the image.
      Our model produces much smoother filtering results when encoding is exactly modulated to match the filter's Fourier transform.
}
\end{figure}

We ablate three components on the \emph{Alien} image (Figure~\ref{fig:ablation} and Table~\ref{tab:alien_exact_downweight}): 
(i) single-sample Monte Carlo supervision, 
(ii) exact Fourier down-weighting of the input features via $\mathcal{F}\{K_\Sigma\}$ (Sec.~\ref{sec:prefilter_fourier}), and 
(iii) lifting the Lipschitz constraint used by NGSSF. 
MC alone yields only small gains in filtering quality (\eg~isotropic PSNR $31.35\!\to\!33.16$). 
Exact Fourier down-weighting is effective only when we train with MC: without MC it changes little, while with MC it boosts filtering PSNR to $41.8$–$42.0$\,dB. 
Lifting the Lipschitz constraint further improves results under MC (cf. rows 4$\to$5 and 6$\to$7). 
With all three enabled, the model achieves the best scores on the original and filtered signals (32.54/43.40/43.60\,dB PSNR with the lowest LPIPS and highest SSIM).

\begin{table}[H]
    \centering
    \caption{
    \captiontitle{Effect of the number of MC samples.}
    We conduct an image regression experiment using lower-variance estimates of
    the filtered field, averaged over 100 anisotropic test-time Gaussian kernels
    and evaluated after 100,000 training iterations. 
    We observe that, although increasing the number of Monte Carlo samples
    substantially reduces the variance of the signal estimate, this has a
    minimal impact on the overall quality of the reconstructed result at
    convergence.}
    \begin{tabular}{lcccccc}
        \hline
         & 1 MC & 4 MC & 16 MC & 64 MC & 256 MC & 1024 MC \\
        \hline
        PSNR $\uparrow$
          & 43.64 & 44.32 & 44.57 & 44.70 & 44.68 & 44.73 \\
        \hline
    \end{tabular}
    \label{tab:psnr_vs_mc}
\end{table}

\paragraph{Activation function}
We integrate our exact Fourier encoding with a SIREN‑style sine activation and achieve $43.43$\,dB PSNR on the Alien image under anisotropic smoothing, nearly identical to the $43.40$\,dB of our ReLU model. 

\paragraph{Using a single Monte Carlo sample}
Table~\ref{tab:psnr_vs_mc} shows that presenting the network with a lower-variance estimate of the signal does not significantly improve the results, 
which validates our choice of using a single Monte Carlo sample.

\section{Limitations}

\paragraph{Training/Inference speed.}  Like all multilayer-perceptron-based neural fields, our method incurs higher per-sample evaluation cost compared to grid-based or other discrete representations (\eg~hash grids or voxel grids). 
Compared to NGSSF, removing Lipschitz bounds reduces the training computation by $38.5\%$ (130ms to 80ms forward + backward + optimization step). 
For inference time, our method is as fast or faster than NGSSF (1.8s), BACON (1.6s), PNF (5.7s), INSP (189.5s), NFC (63.9s). 
MINER (0.1s) remains faster at inference due to its multiple sparse tiny MLPs (0.1s vs 1.8s).
This performance gap can limit deployment in real-time or latency-sensitive applications.

\paragraph{Only Symmetric filters.}  Our method relies on the filter’s Fourier
transform being real and even. Asymmetric filters introduce phase shifts that
are not captured by our current encoding, so extending to arbitrary
non-symmetric kernels would require modeling and compensating for phase
information.

\paragraph{High-pass filter support.} 
Our results only indicate performance on low-pass filters. 
A high-pass filter such as difference-of-Gaussians is readily feasible with our method as the filtered response can be achieved as a linear combination of Gaussians with two different scales. 
Since our method operates continuously in scale space, any such parametric combination is supported.
We leave the investigation of supporting arbitrary and asymmetric filters as future work.

\paragraph{Scope and applicability.}
In this work we evaluate only MLP-based neural fields with ReLU and SIREN-like sinusoidal activations. Our approach relies on continuous inputs and analytic frequency responses, and is therefore not directly applicable to grid-parameterized methods (\eg~multi-resolution or voxel/tensor grids) without additional changes. Extending the method to such grids is outside the scope of this paper.

\paragraph{Scope of analysis.} We focus on the empirical
evaluation of our training scheme and its practical generalization to unseen
filters. 
A theoretical investigation into why Fourier-feature modulation so effectively represents multi-scale signals, and establishing the conditions under which it provably
converges represent a promising future direction.

\section{Conclusion}
\label{sec:conclusion}
We have presented a simple yet powerful framework for prefiltering neural fields in the frequency domain by analytically modulating Fourier–feature embeddings with a family of symmetric filter kernels. 
By integrating a closed-form expression for the filter’s frequency response into the first layer of a multi-layer perceptron and supervising with single-sample Monte Carlo estimates, our method supports continuous Gaussian, Box, and Lanczos filters, even those unseen at training time, without imposing architectural constraints or relying on precomputed multiscale datasets. 
Extensive experiments on 2D images and 3D signed-distance fields demonstrate
that our approach delivers higher fidelity and generality than prior scale-aware
neural-field methods, while remaining memory-efficient and straightforward to
implement.

\begin{acks}
We thank Jaiden Ekgasit for help with figure preparation; Baha Eren Yaldiz for running experiments; and Saeyoung Rho, Yash Belhe, Sina Nabizadeh, Zilu Li, Kaiwen Jiang, Wesley Chang, Alex Trevithick, and Sumanth Varambally for comments and discussions.  We also thank the anonymous reviewers for their extensive comments that greatly improved the clarity and analysis in the final paper.

This work was supported in part by ONR grant N00014-23-1-2526; NSF grants (2110409, 2212085, 2042583, 2238839); the European Research Council (ERC) Advanced Grant NERPHYS (101141721, \url{https://project.inria.fr/nerphys/}); gifts from Adobe, Google, and Qualcomm; the Ronald L. Graham Chair; and the UC San Diego Center for Visual Computing. 
We also acknowledge NSF grants (2100237, 2120019) for the NRP Nautilus cluster.
\end{acks}

\bibliographystyle{ACM-Reference-Format}
\bibliography{main}

\clearpage

\setcounter{section}{0} %
\renewcommand{\thesection}{\Alph{section}}   %

\clearpage
\section{Convolution via Fourier Feature Encoding}

\subsection{Fourier transform of a Gaussian}
\label{subsec:gaussian_fourrier}

Here we calculate the Fourier transform of the multivariate Gaussian with covariance matrix $\Sigma^{-1}$ and Fourier basis vector $\mathbf{b}^\top$:
\begin{equation}
\mathcal{F}\{p_X\}(\mathbf{b}) = \int_{\mathbb{R}^d} \frac{1}{(2\pi)^{\frac{d}{2}} |\Sigma|^{\frac{1}{2}}} \exp\left(-\tfrac{1}{2} \mathbf{x}^\top \Sigma^{-1} \mathbf{x}\right) e^{-2\pi i \mathbf{b}^\top \mathbf{x}} \, d\mathbf{x}
\end{equation}

Once we combine the exponentials
\begin{equation}
= \frac{1}{(2\pi)^{\frac{d}{2}} |\Sigma|^{\frac{1}{2}}} \int_{\mathbb{R}^d} \exp\left(-\tfrac{1}{2} \mathbf{x}^\top \Sigma^{-1} \mathbf{x} - 2\pi i \mathbf{b}^\top \mathbf{x}\right) \, d\mathbf{x}.
\end{equation}

We can define the quadratic form in the exponent:
\begin{equation}
-\tfrac{1}{2} \mathbf{x}^\top \Sigma^{-1} \mathbf{x} - 2\pi i \mathbf{b}^\top \mathbf{x} = -\tfrac{1}{2} \left(\mathbf{x} + i 2\pi \Sigma \mathbf{b}\right)^\top \Sigma^{-1} \left(\mathbf{x} + i 2\pi \Sigma \mathbf{b}\right) - 2\pi^2 \mathbf{b}^\top \Sigma \mathbf{b}.
\end{equation}

The integral becomes:
\begin{equation}
\exp\left(-2\pi^2 \mathbf{b}^\top \Sigma \mathbf{b}\right) \int_{\mathbb{R}^d} \exp\left(-\tfrac{1}{2} \left(\mathbf{x} + i 2\pi \Sigma \mathbf{b}\right)^\top \Sigma^{-1} \left(\mathbf{x} + i 2\pi \Sigma \mathbf{b}\right)\right) d\mathbf{x}.
\end{equation}

The resulting is another shifted Gaussian integral, which evaluates to the same normalization factor \((2\pi)^{\frac{d}{2}} |\Sigma|^{\frac{1}{2}}\), yielding:
\begin{equation}
\boxed{
\mathcal{F}\{p_X\}(\mathbf{b}) = \exp\left(-2\pi^2 \mathbf{b}^\top \Sigma \mathbf{b}\right).
\label{eq:gaussian_fourier}
}
\end{equation}

\subsection{Fourier transform of normalized ellipsoid-shaped box filter}

Consider an ellipsoid \( E = \{\mathbf{x} \in \mathbb{R}^n : \mathbf{x}^\top \Sigma^{-1} \mathbf{x} \le 1\} \), where \( \Sigma \) is a symmetric positive-definite matrix. The Fourier transform of its characteristic function \( \chi_E(\mathbf{x}) \) is:
\begin{equation}
\mathcal{F}\{\chi_E\}(\mathbf{b}) = \int_{E} e^{-2\pi i \mathbf{b}^\top \mathbf{x}} d\mathbf{x}.
\end{equation}
  
Diagonalize \( \Sigma \) as \( \Sigma = Q \Lambda Q^\top \), where \( Q \) is orthogonal and \( \Lambda = \text{diag}(\lambda_1, \dots, \lambda_n) \). Let \( y = Q^\top \mathbf{x} \), transforming the ellipsoid into a scaled unit ball:
\begin{equation}
y^\top \Lambda^{-1} y \le 1 \quad \Rightarrow \quad \sum_{i=1}^n \frac{y_i^2}{\lambda_i} \le 1.
\end{equation}

We apply change of variables \( z_i = \frac{y_i}{\sqrt{\lambda_i}} \), so \( \sum_{i=1}^n z_i^2 \le 1 \) (unit ball). The Jacobian determinant, which is product of square root of eigenvalues, is \( |\Sigma|^{1/2} \), giving:
\begin{equation}
d\mathbf{x} = dy, \quad d\mathbf{x} = |\Sigma|^{1/2} dz . 
\end{equation}

Substitute \( x = Qy = Q \sqrt{\Lambda} z \), and let \( \eta = Q^\top \mathbf{b} \):
\begin{equation}
\mathcal{F}\{\chi_E\}(\mathbf{b}) = |\Sigma|^{1/2} \int_{B_1} e^{-2\pi i (\sqrt{\Lambda} z)^\top \eta} dz = |\Sigma|^{1/2} \int_{B_1} e^{-2\pi i z^\top \sqrt{\Lambda} \eta} dz.
\end{equation}
 
The Fourier transform of the unit ball \( B_1 \) in \( \mathbb{R}^n \) is known:
\begin{equation}
\int_{B_1} e^{-2\pi i z^\top \omega} dz = \frac{J_{n/2}(2\pi \|\omega\|)}{\|\omega\|^{n/2}},
\end{equation}
where \( J_{n/2} \) is the Bessel function. 

Let \( \omega = \sqrt{\Lambda} \eta \), so \( \|\omega\| = \sqrt{\eta^\top \Lambda \eta} = \sqrt{\mathbf{b}^\top \Sigma \mathbf{b}} \).
We substitute variables and simplify further:
\begin{equation}
\mathcal{F}\{\chi_E\}(\mathbf{b}) = |\Sigma|^{1/2} \cdot \frac{J_{n/2}\left(2\pi \sqrt{\mathbf{b}^\top \Sigma \mathbf{b}}\right)}{\left(\sqrt{\mathbf{b}^\top \Sigma \mathbf{b}}\right)^{n/2}} .
\end{equation}

In addition, let us calculate the \textbf{volume-normalized ellipsoid} \( \tilde{\chi}_E(x) = \frac{1}{\text{Vol}(E)} \chi_E(x) \), where \( E = \{x \in \mathbb{R}^n : x^\top \Sigma^{-1} x \le 1\} \).

The volume of \( E \) is:
\begin{equation}
\text{Vol}(E) = V_n |\Sigma|^{1/2}, \quad \text{where } V_n = \frac{\pi^{n/2}}{\Gamma\left(\frac{n}{2} + 1\right)},
\end{equation}
and \( V_n \) is the volume of the unit \( n \)-ball.

Substituting and simplifying the constants:
\begin{equation}
\boxed{
\mathcal{F}\{\tilde{\chi}_E\}(\mathbf{b})
     = \frac{\Gamma\!\bigl(\tfrac{n}{2}+1\bigr)}
            {\pi^{n/2}}\,
       \frac{J_{n/2}\!\bigl(2\pi\sqrt{\mathbf{b}^{\!\top}\Sigma\mathbf{b}}\bigr)}
            {\left(\sqrt{\mathbf{b}^\top \Sigma \mathbf{b}}\right)^{n/2}}
\label{eq:box_fourier}
}
\end{equation}

The Fourier transform of the ellipsoid’s characteristic function involves Bessel functions, reflecting its sharp boundary. Unlike Gaussians, which remain Gaussian under Fourier transforms, the ellipsoid’s oscillatory frequency response encodes its geometric shape through \( \Sigma \). The eigenvalues of \( \Sigma \) scale the frequencies, while the Bessel function \( J_{n/2} \) captures radial symmetry through \( \mathbf{b}^\top \Sigma \mathbf{b} \).

Let's substitute and derive normalized Fourier transform expression of an ellipsoid into 2D and 3D cases.

\textbf{Case 1: 2D (n=2)} \( \Gamma\left(\frac{2}{2} + 1\right) = \Gamma(2) = 1 \):
\begin{equation}
\boxed{
\mathcal{F}\{\tilde{\chi}_E\}(\mathbf{b}) = \frac{1}{\pi \sqrt{\mathbf{b}^\top \Sigma \mathbf{b}}} \, J_1\!\left(2\pi \sqrt{\mathbf{b}^\top \Sigma \mathbf{b}}\right)
}
\end{equation}

\textbf{Case 2: 3D (n=3)} 
\(
\Gamma\left(\frac{3}{2} + 1\right) = \Gamma\left(\frac{5}{2}\right) = \frac{3}{4}\sqrt{\pi},
\)  
and the Bessel function \( J_{3/2}(x) \) can be expressed as  
\(
J_{3/2}(x) = \sqrt{\frac{2}{\pi x}} \left(\frac{\sin x}{x} - \cos x\right).
\) 
\begin{equation}
\boxed{
\mathcal{F}\{\tilde{\chi}_E\}(\mathbf{b}) = \frac{3}{4\pi\,\left(\sqrt{\mathbf{b}^\top \Sigma \mathbf{b}}\right)^{3/2}}\,
         J_{3/2}\!\Bigl(2\pi\sqrt{\mathbf{b}^{\!\top}\Sigma\mathbf{b}}\Bigr)
}
\end{equation}

\subsection{Fourier transform of the Lanczos kernel}
\label{subsec:gaussian_fourier}
For a single-dimensional variable $x$, we write the (untruncated) Lanczos kernel of order $a$, $a > 0$ with normalized sinc functions as:
\begin{equation}
    L(x) = {\rm sinc}(x) \,{sinc}(x/a) = \frac{a\sin{\pi x} \sin {\pi x/a}}{\pi^2 x^2}
\end{equation}

Note that this is the pointwise multiplication of two $\rm{sinc}$ functions. It is known that the Fourier transform of this function is a scaled rectangle function, which for a positive constant $c$ is as follows:
\begin{equation}
    {\rm sinc}(cx) = \frac{1}{c} {\rm rect}\left(\frac{b}{c}\right)
\end{equation}
where 
\begin{equation}
    {\rm rect}(x) = 
    \begin{cases} 
      1 & :\;\; |x|\leq \frac{1}{2} \\
      0 & :\;\;{\rm else}
   \end{cases}
\end{equation}

The Lanczos kernel is the primal-domain multiplication of two sinc functions; it is therefore equivalently the frequency-domain convolution of two rectangle functions
\begin{equation}
    a\int_{-\infty}^{\infty} {\rm rect}(b-b'/\pi) \, {\rm rect}(ab'/\pi)\,\,db'
\end{equation}
which is a trapezoid (or a triangle wave if both rectangles coincide exactly).

Since a piecewise definition of this trapezoid is unwieldy, we refer the reader to the description of the trapezoid given in \cite{sarwate2023_convolution} for more detail. 
In particular, for rectangle functions of widths $A$ and $B$, this trapezoidal function begins from $0$ at $b=-\frac{1}{2}(A+B)$, rising to a maximum of ${\rm min}(A,B)$ with a slope of $1$ at $-\frac{1}{2}|A-B|$.
This is even in $b$, so the construction of the positive-domain part of this function is analogous. We identified the following function which matches the above description:
\begin{equation}
 \mathcal{F}(L)(b)=a\max\left( \min\left(\frac{\left(a+1\right)}{2a}\right. \right. \left. \left.- |b|,\min\left(1,\frac{1}{a}\right)\right),0\right)
\label{eq:lanczos_fourier_1d}
\end{equation}

For higher-dimensional cases, we consider the Lanczos kernel as a radial kernel in $\Vert {\bf x}\Vert = \sqrt{{\bf x}^\top {\bf x}}$, which is then shifted by a covariance $\Sigma$ with determinant $|\Sigma|$. 
To find the Fourier transform, we make the change-of-variable ${\bf x} \rightarrow \Sigma^{1/2} {\bf u}$ as before so that $d{\bf x} = |\Sigma|^{1/2} d{\bf u}$ and use the property that the Fourier transform of a function radial in $\bf x$ is also radial in $\bf b$:
\begin{align}
    \mathcal{F}(L(\Vert \Sigma^{-1/2} {\bf x}\Vert)) &= \int_{\mathbb{R}^2}L(\Vert \Sigma^{-1/2} {\bf x}\Vert) \exp(-2\pi i {\bf b}^\top {\bf x})\,\,d{\bf x} \\
    &= |\Sigma|^{1/2}\int_{\mathbb{R}^2}L(\Vert {\bf u}\Vert) \exp(-2\pi i (\Sigma^{1/2}{\bf b})^\top {\bf u}) \,\, d{\bf u} \\
    &= |\Sigma|^{1/2}\mathcal{F}(L)(\Vert \Sigma^{1/2}{\bf b} \Vert)
\end{align}

We combine this with the previous result to obtain the Fourier transform of our multidimensional Lanczos kernel:
\begin{equation}
\mathcal{F}(L)({\bf b})=a|\Sigma|^{1/2}\max\left( \min\left(\frac{\left(a+1\right)}{2a}\right. \right. \left. \left.-\Vert \Sigma^{1/2}{\bf b} \Vert,\min\left(1,\frac{1}{a}\right)\right),0\right)
\label{eq:lanczos_fourier}
\end{equation}

During rejection sampling, we draw samples from the distribution proportional to the \emph{normalized absolute value} of the Lanczos kernel, i.e., $p(x)\propto |L(x)|$.
Therefore, we need the total mass of normalization in our encoding: $\int_{\mathbb{R}} |L(x)|\,dx$.
To balance accuracy and cost, we restrict sampling to $x\in[-2\sqrt{a},\,2\sqrt{a}]$, which captures most of the kernel’s mass. 
For our Fourier encoding, however, we still assume infinite support: truncating in space would convolve the ideal trapezoidal spectrum with a sinc induced by the window, an effect that is negligible at this truncation radius.
Given area can be defined as:
\begin{equation}
     \int_{\mathbb{R}^n} \left|{\rm sinc}(\pi \Vert \Sigma^{-1/2} {\bf x} \Vert)\,{\rm sinc}(\pi \Vert \Sigma^{-1/2} {\bf x} \Vert/a) \right| \,\,d{\bf x}
\end{equation}

\newcommand{\sinc}{\operatorname{sinc}} %

\paragraph{Radial reduction (general \(n\)).}
With the change of variables \(\mathbf{x}=\Sigma^{1/2}\mathbf{u}\) so that \(d\mathbf{x}=|\Sigma|^{1/2}\,d\mathbf{u}\) and \(\|\Sigma^{-1/2}\mathbf{x}\|=\|\mathbf{u}\|\), 
and writing \(r=\|\mathbf{u}\|\), we switch to \(n\)-D spherical coordinates:
\begin{align}
    & = \int_{\mathbb{R}^n} \Bigl|\sinc\bigl(\|\Sigma^{-1/2}\mathbf{x}\|\bigr)\,\sinc\!\bigl(\|\Sigma^{-1/2}\mathbf{x}\|/a\bigr)\Bigr|\,d\mathbf{x} \\
    &= |\Sigma|^{1/2}\,S_{n-1}\!\int_{0}^{2\sqrt{a}} \Bigl|\sinc(r)\,\sinc\!\bigl(r/a\bigr)\Bigr|\,r^{\,n-1}\,dr,
\end{align}
where \(S_{n-1}=\frac{2\pi^{n/2}}{\Gamma(n/2)}\) is the surface area of the unit \((n{-}1)\)-sphere.
Finally we define the Fourier transform of our Lanczos kernel as:
\begin{equation}
\boxed{
\mathcal{F}\{L\}({\bf b}_i)
= \frac{a}{z_n}\,
\max\!\Big(
  \min\!\left(
    \frac{a+1}{2a} - \sqrt{\mathbf{b}_i^{\!\top}\Sigma\mathbf{b}_i},\ \min\left(1,\frac{1}{a}\right)
  \right),\ 0
\Big)
}
\end{equation}
We numerically estimate $z_1=0.9499393398$, $z_2=0.9913304793$, $z_3=1.2732395447$ following: $z_n = S_{n-1}\!\int_{0}^{2\sqrt{a}} \Bigl|\sinc(r)\,\sinc\!\bigl(r/a\bigr)\Bigr|\,r^{\,n-1}\,dr.$

\begin{figure}[htbp]
  \centering
  \includegraphics[width=\linewidth]{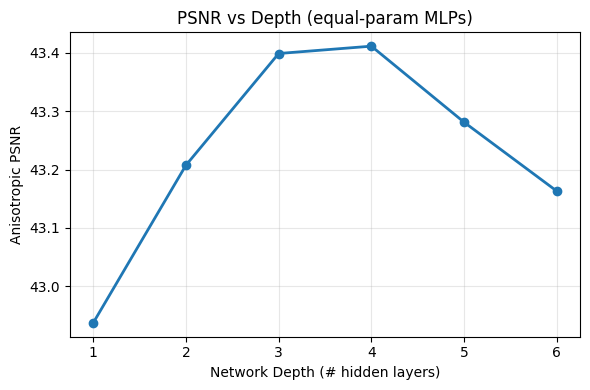}
  \label{fig:psnr_vs_depth}
  \caption{\captiontitle{Depth ablation at matched parameter counts.}
    We vary MLP depth while holding total parameters constant (adjusting width accordingly); a three-layer MLP achieves the optimal reconstruction quality.
    }
\end{figure}

\begin{figure*}
    \centering
    \setlength{\tabcolsep}{1pt}
    \renewcommand{\arraystretch}{0.6}
    \begin{tabular}{cccccccc}
        & & & \multicolumn{4}{c}{Ours} \\
        \cmidrule(lr){4-7} 
        & & NGSSF & Trained with Gaussian & Trained with Box & Trained with Lanczos & Trained with all & GT \\
        \centeredtab{\rot{Gaussian}} & &
        \centeredtab{
            \includegraphics[width=0.155\textwidth, height=0.155\textwidth]{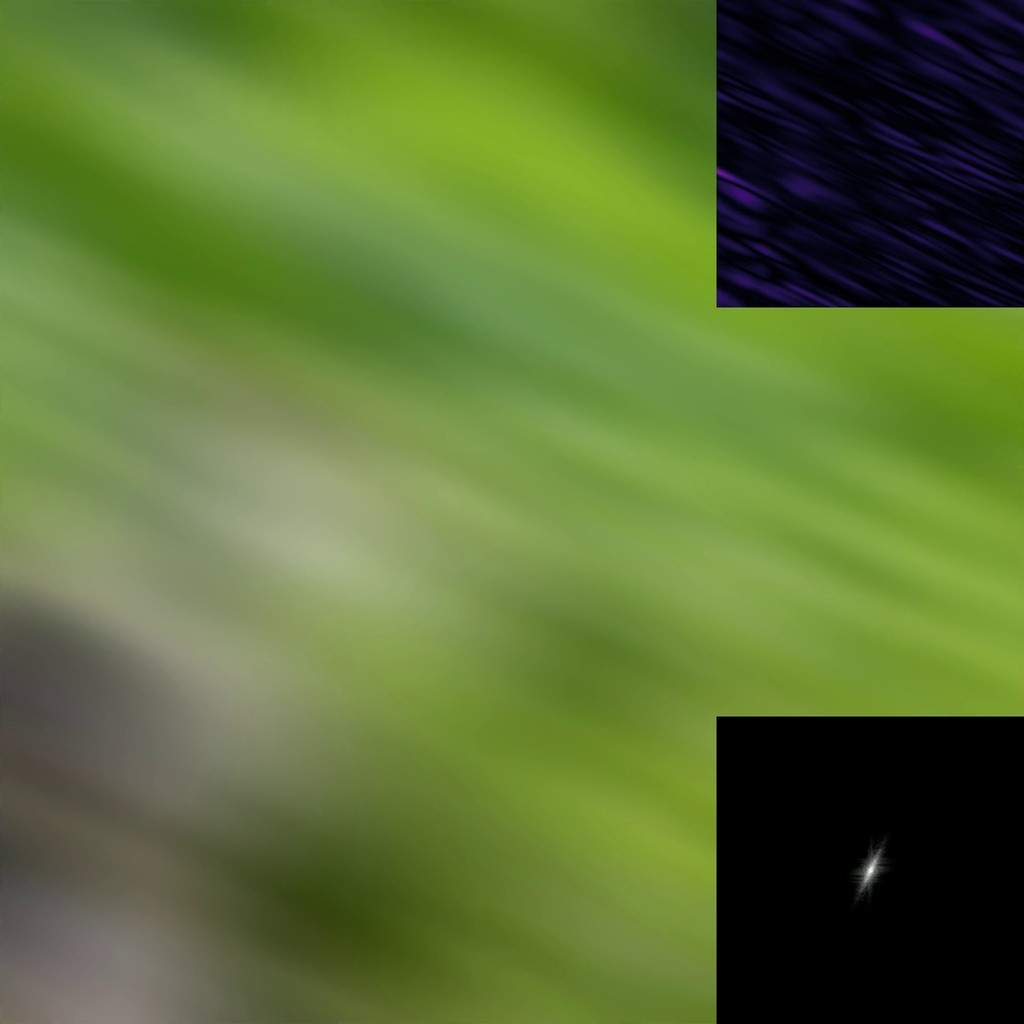} \\
            \includegraphics[width=0.155\textwidth, height=0.155\textwidth]{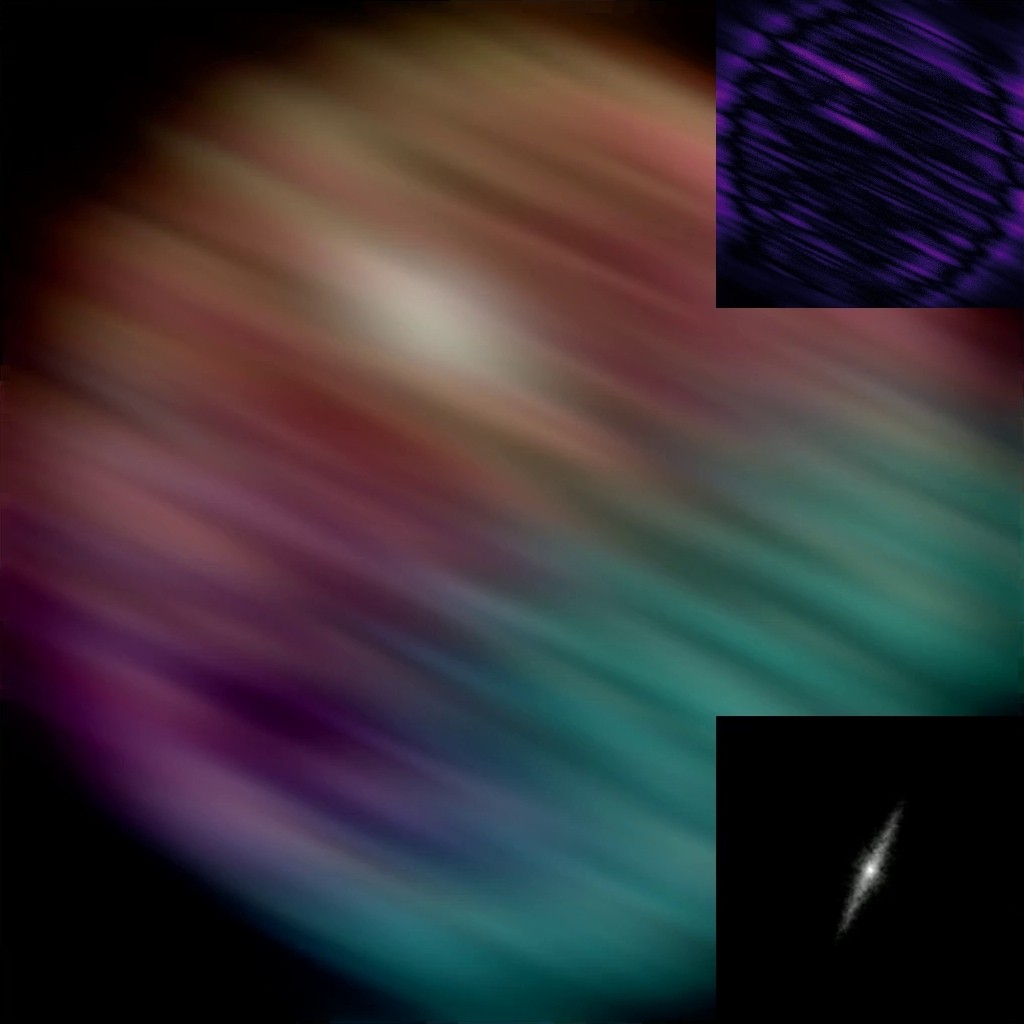}} & 
        \centeredtab{
            \includegraphics[width=0.155\textwidth, height=0.155\textwidth]{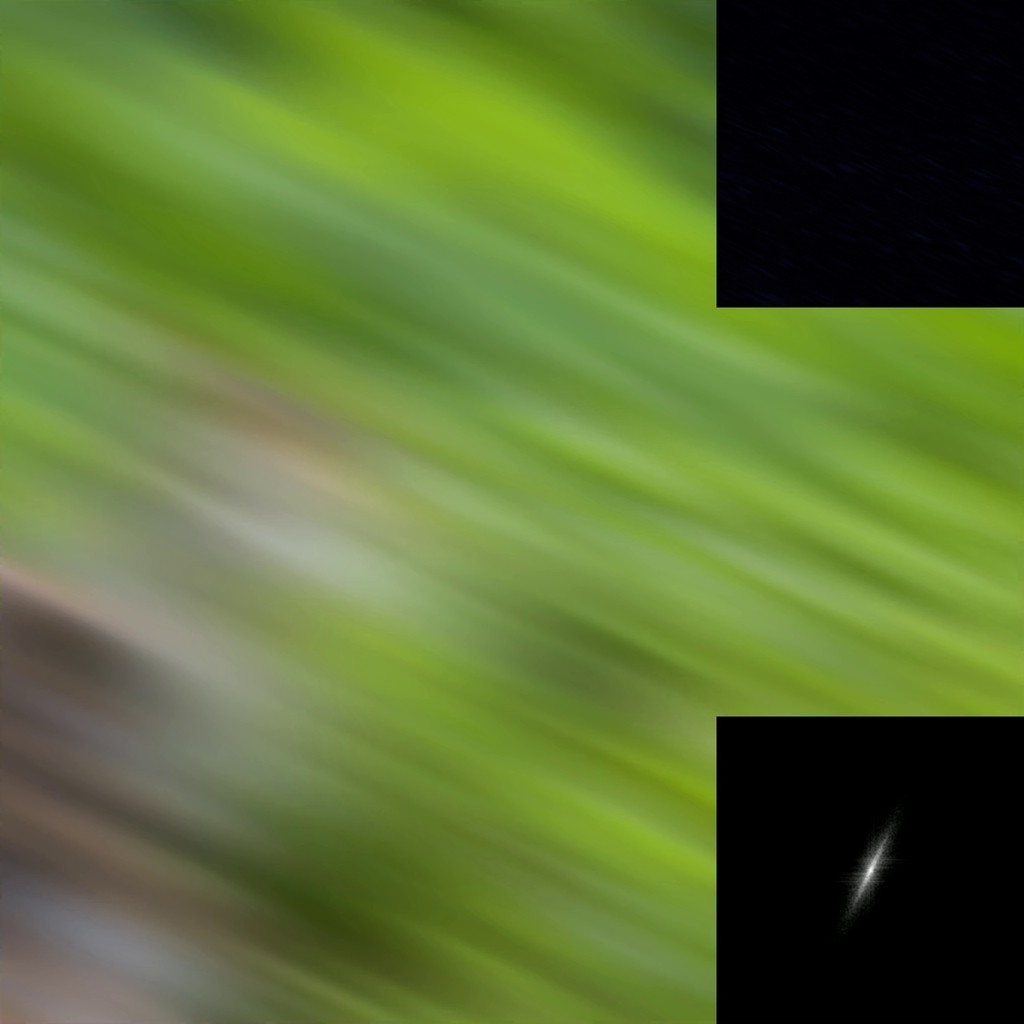} \\
            \includegraphics[width=0.155\textwidth, height=0.155\textwidth]{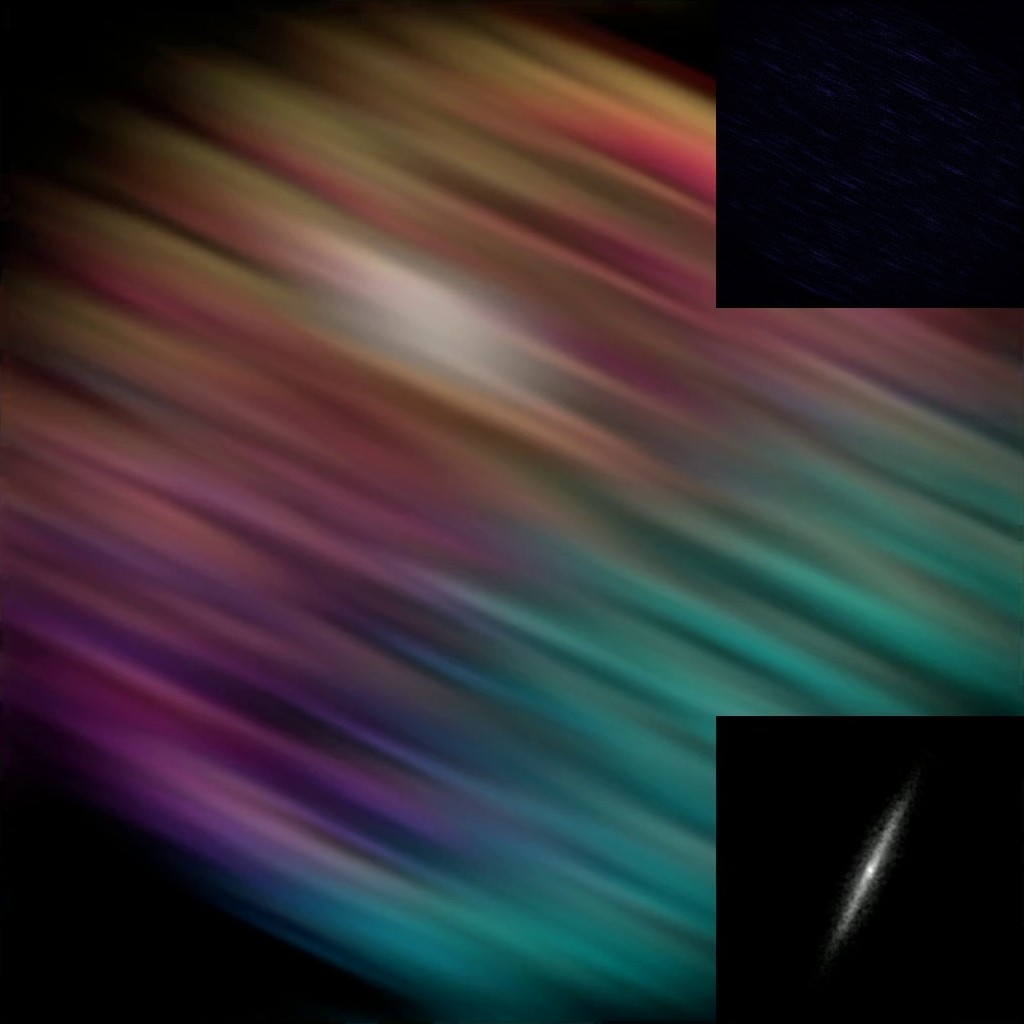}} & 
        \centeredtab{
            \includegraphics[width=0.155\textwidth, height=0.155\textwidth]{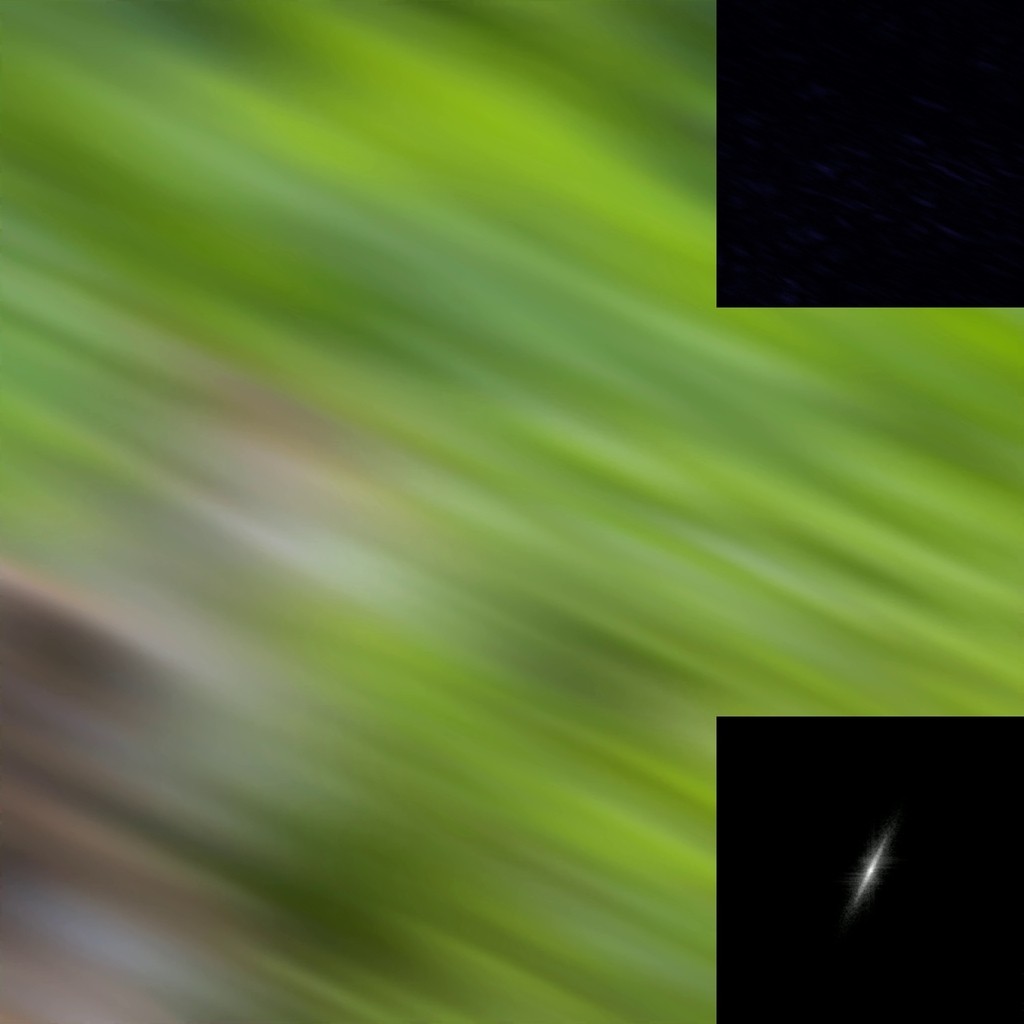} \\
            \includegraphics[width=0.155\textwidth, height=0.155\textwidth]{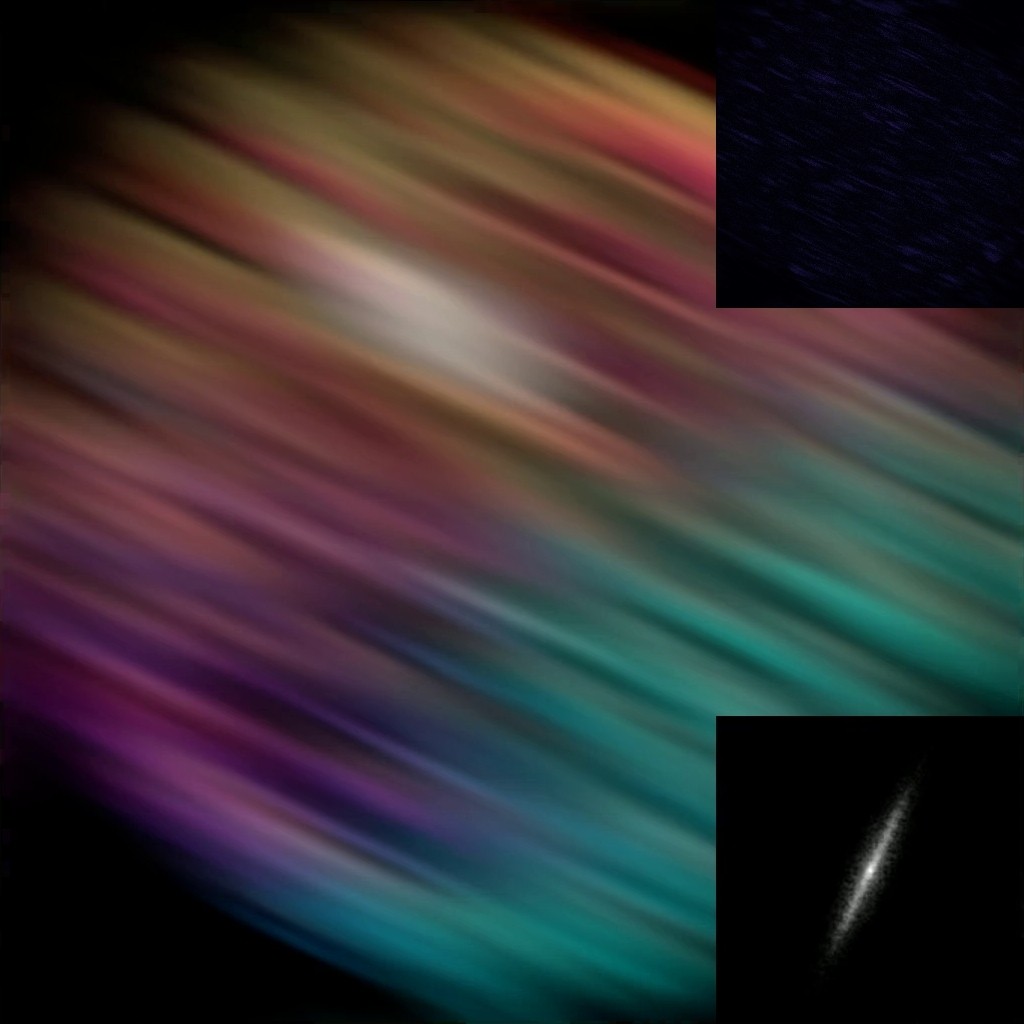}} & 
        \centeredtab{
            \includegraphics[width=0.155\textwidth, height=0.155\textwidth]{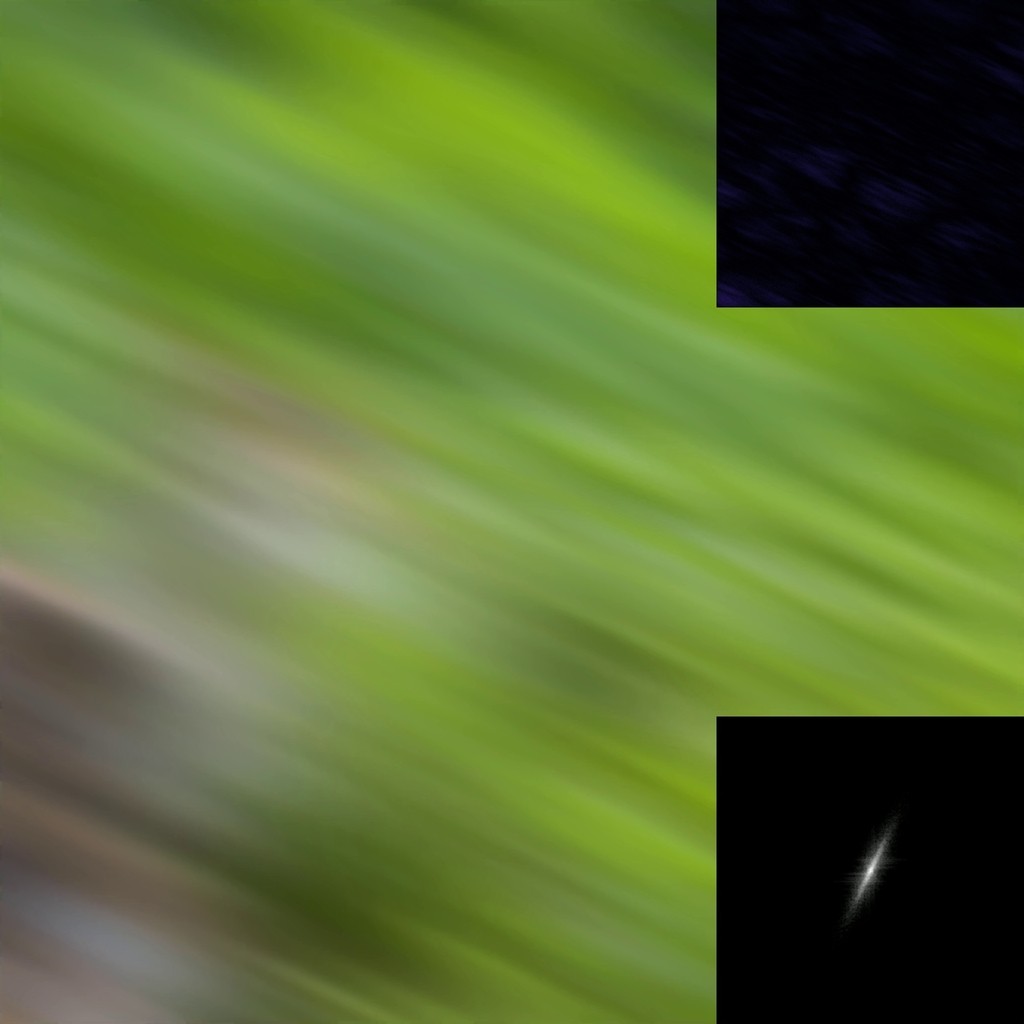} \\
            \includegraphics[width=0.155\textwidth, height=0.155\textwidth]{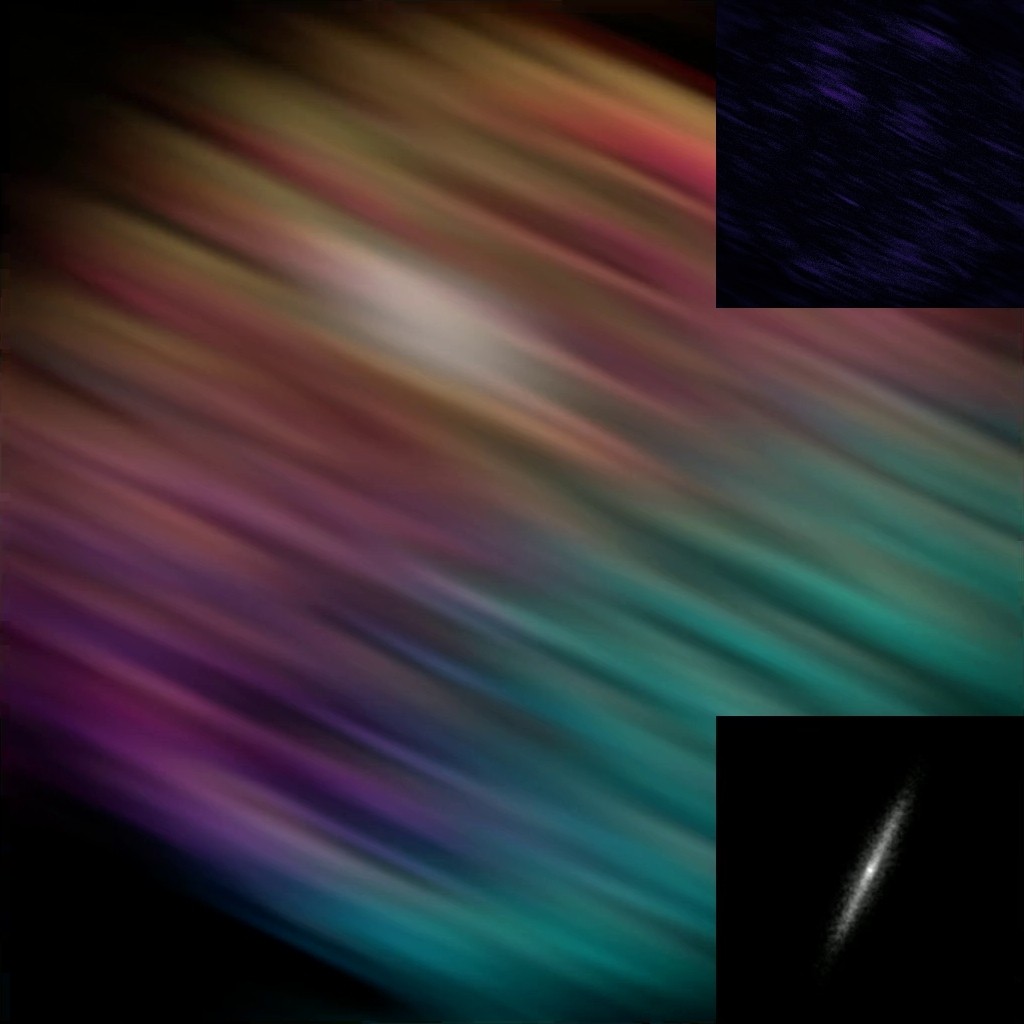}} & 
        \centeredtab{
            \includegraphics[width=0.155\textwidth, height=0.155\textwidth]{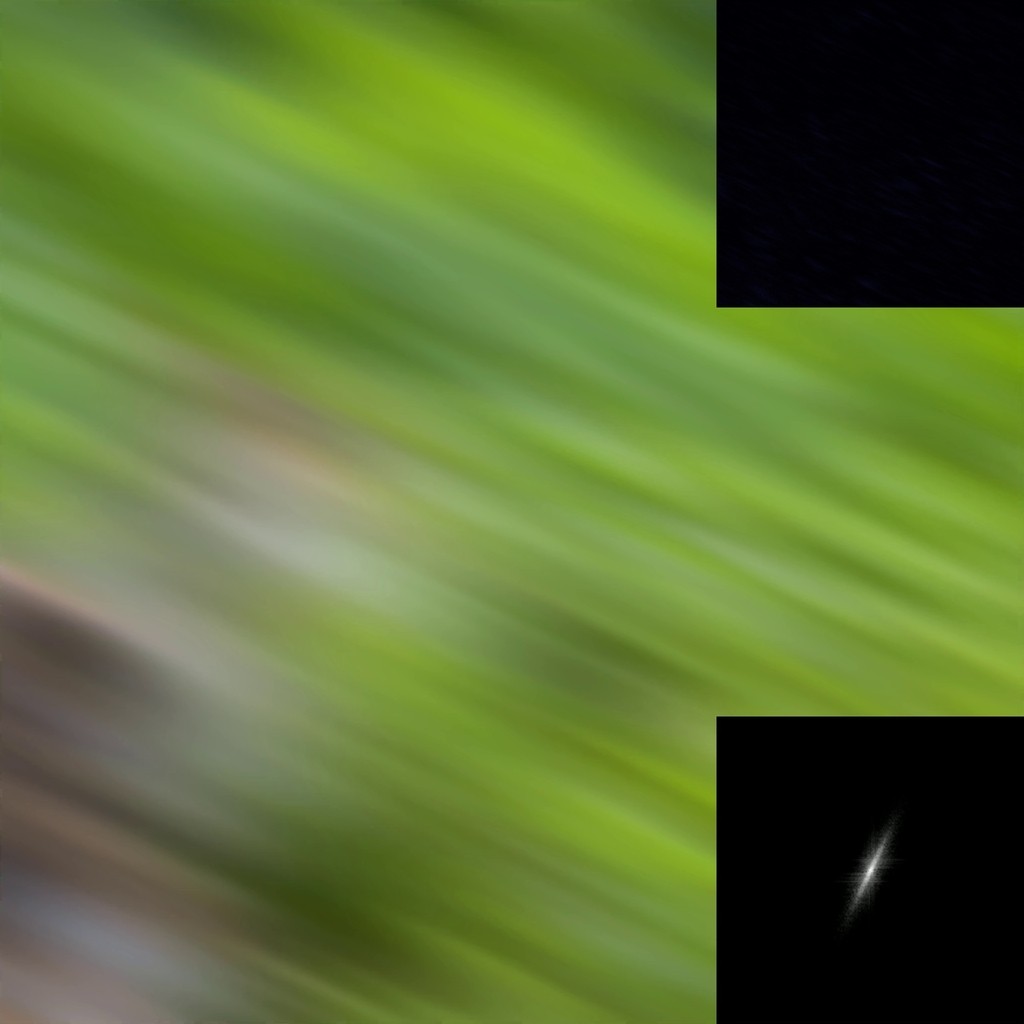} \\
            \includegraphics[width=0.155\textwidth, height=0.155\textwidth]{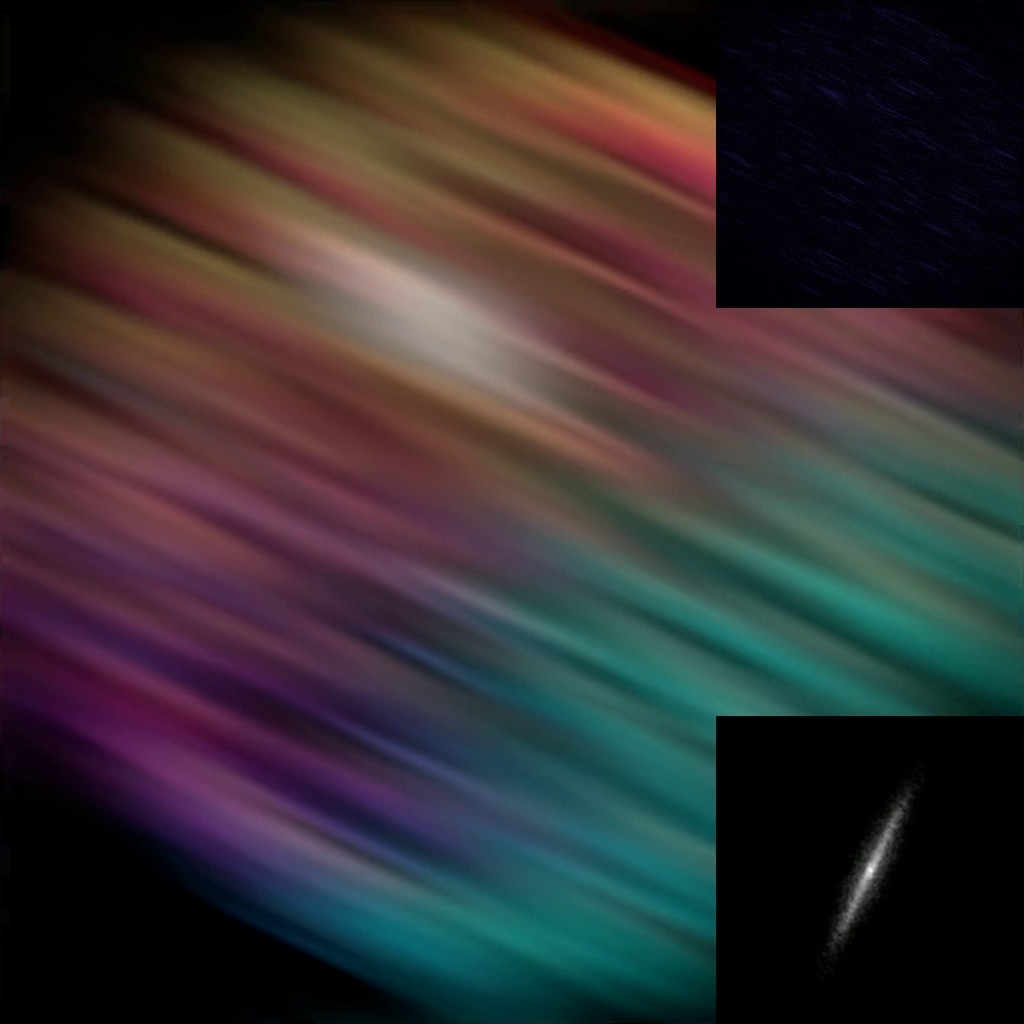}} & 
        \centeredtab{
            \includegraphics[width=0.155\textwidth, height=0.155\textwidth]{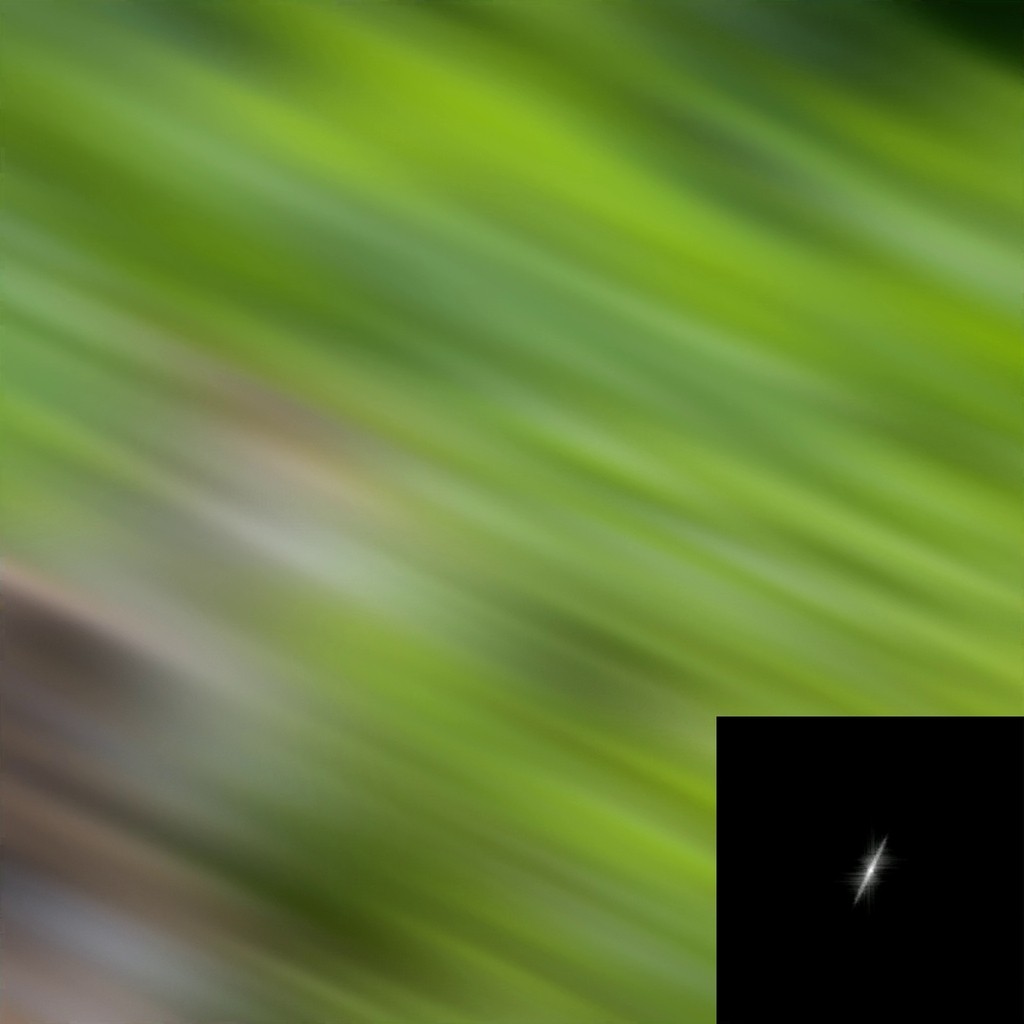} \\
            \includegraphics[width=0.155\textwidth, height=0.155\textwidth]{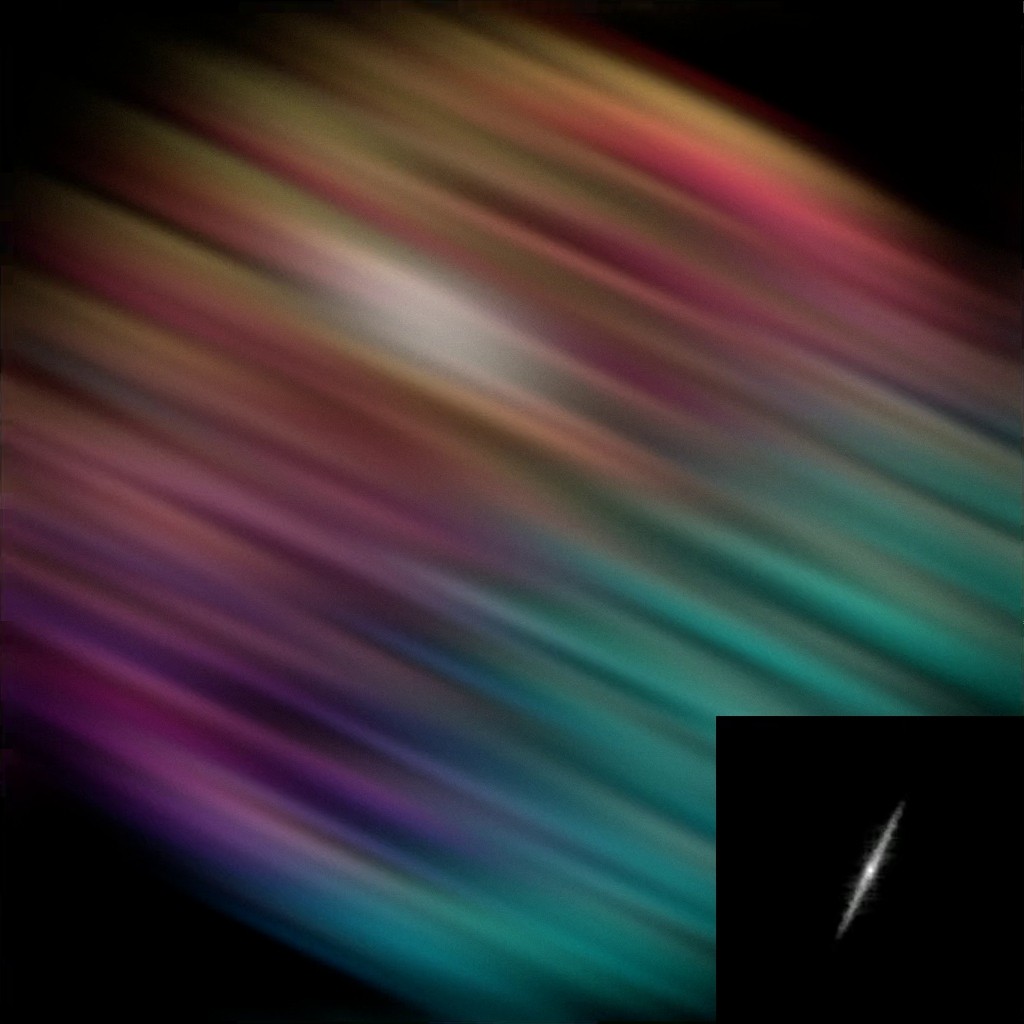}}\\
        \cmidrule(lr){3-8}
        \centeredtab{\rot{Box}} & &
        \centeredtab{
            \includegraphics[width=0.155\textwidth, height=0.155\textwidth]{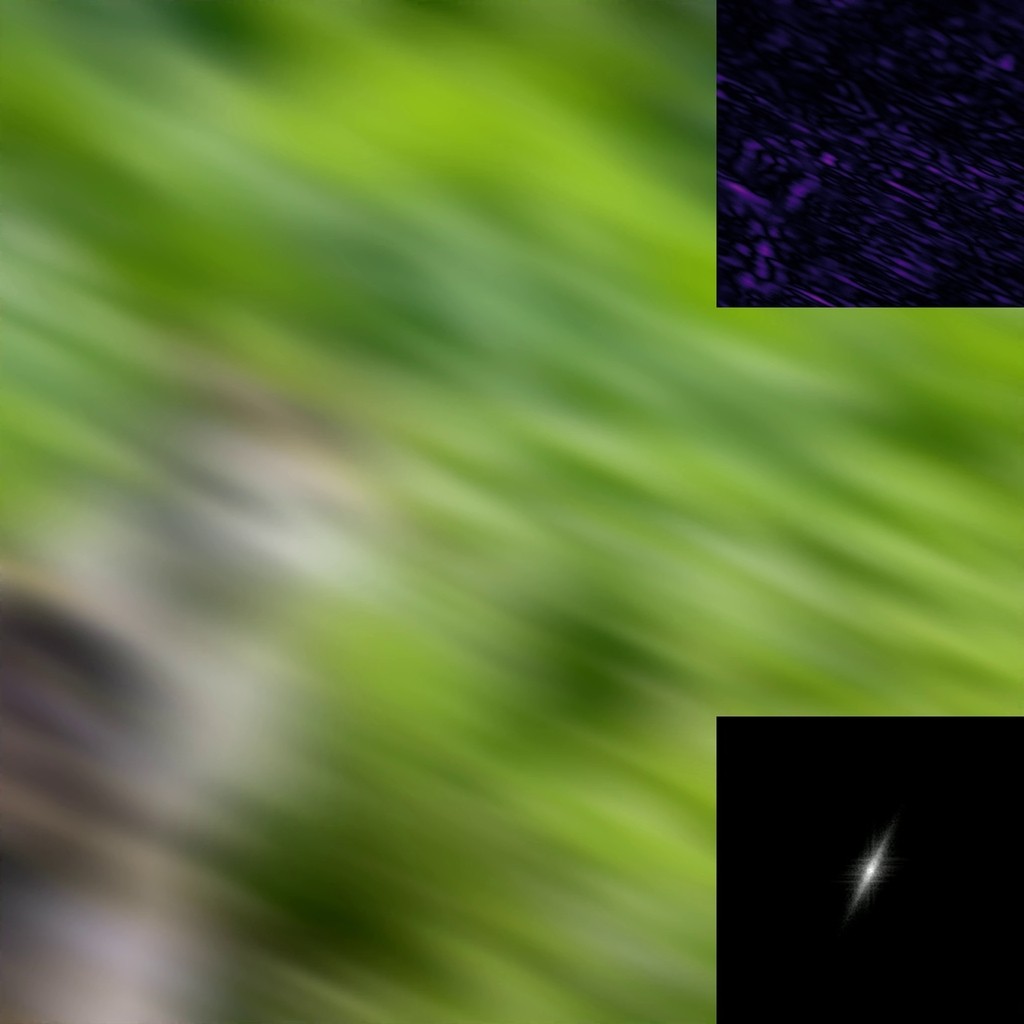} \\
            \includegraphics[width=0.155\textwidth, height=0.155\textwidth]{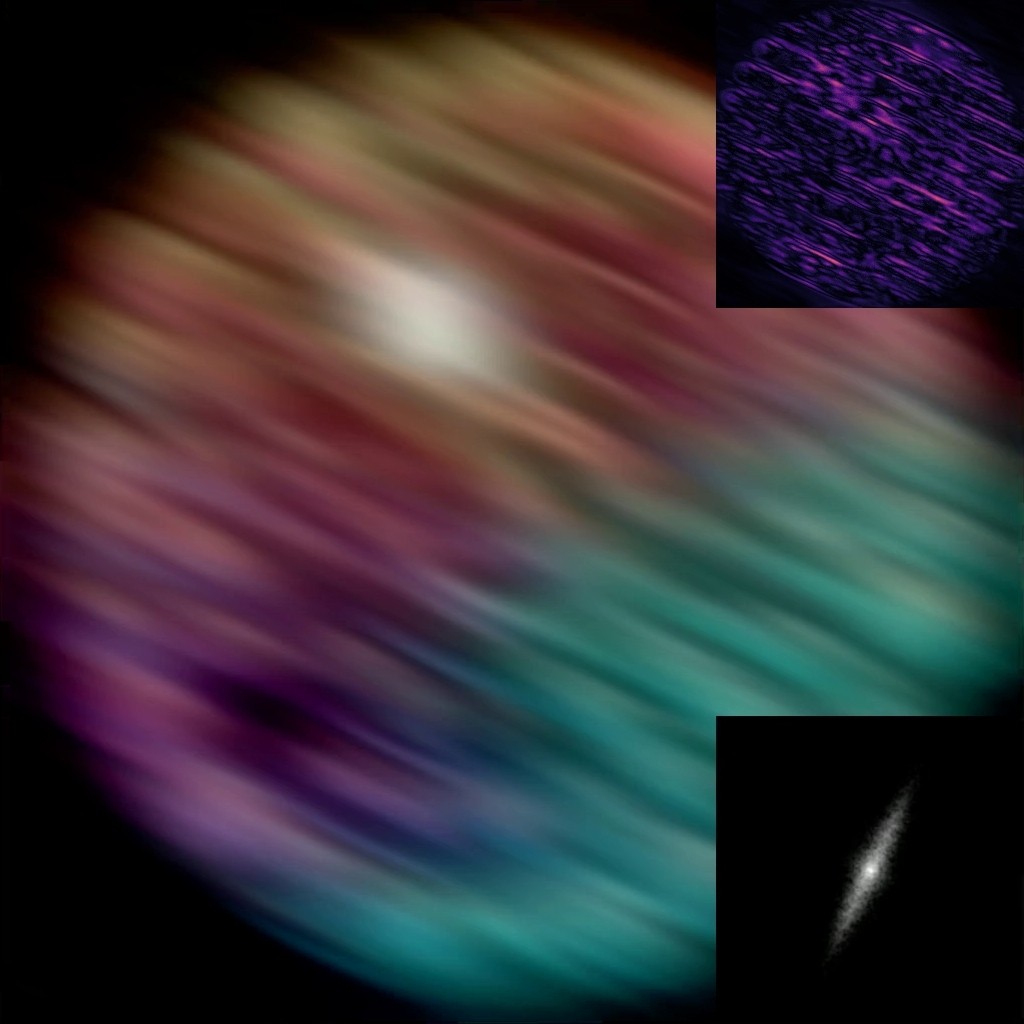}} & 
        \centeredtab{
            \includegraphics[width=0.155\textwidth, height=0.155\textwidth]{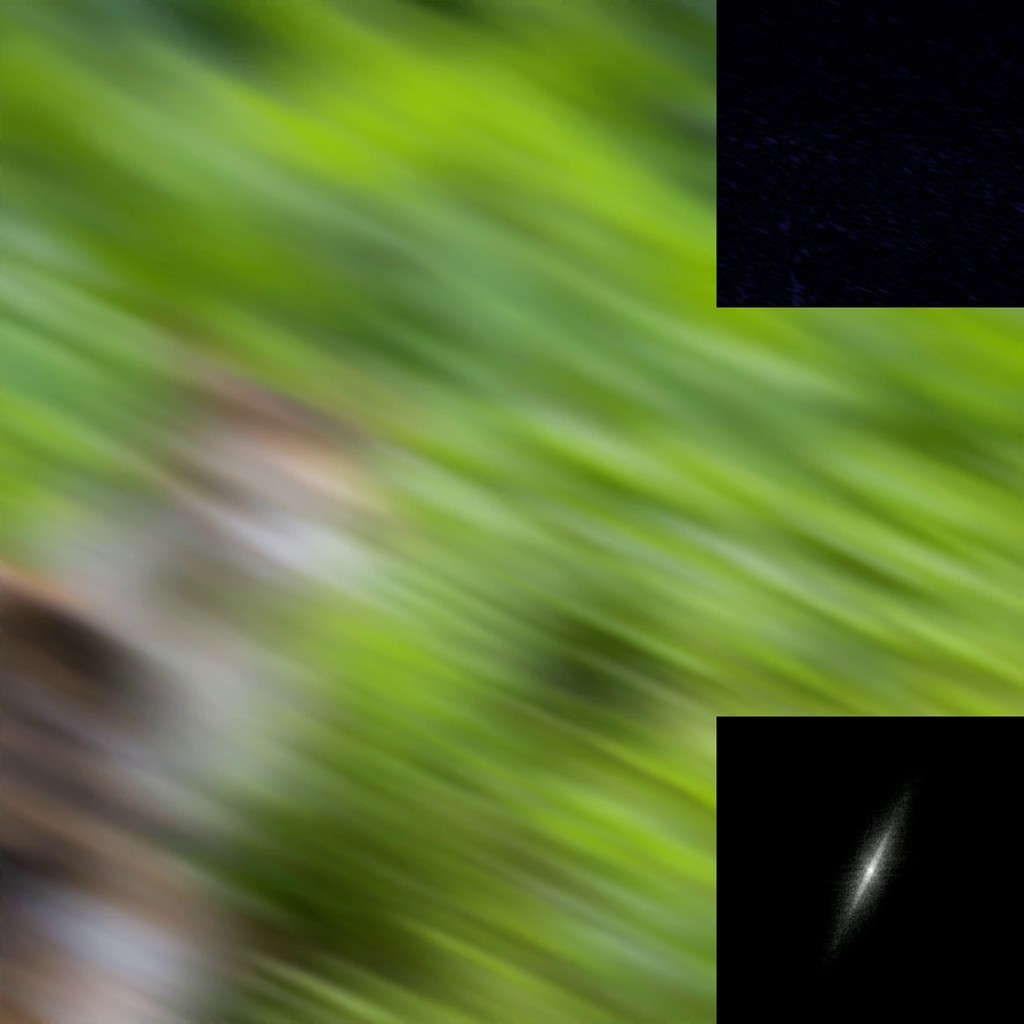} \\
            \includegraphics[width=0.155\textwidth, height=0.155\textwidth]{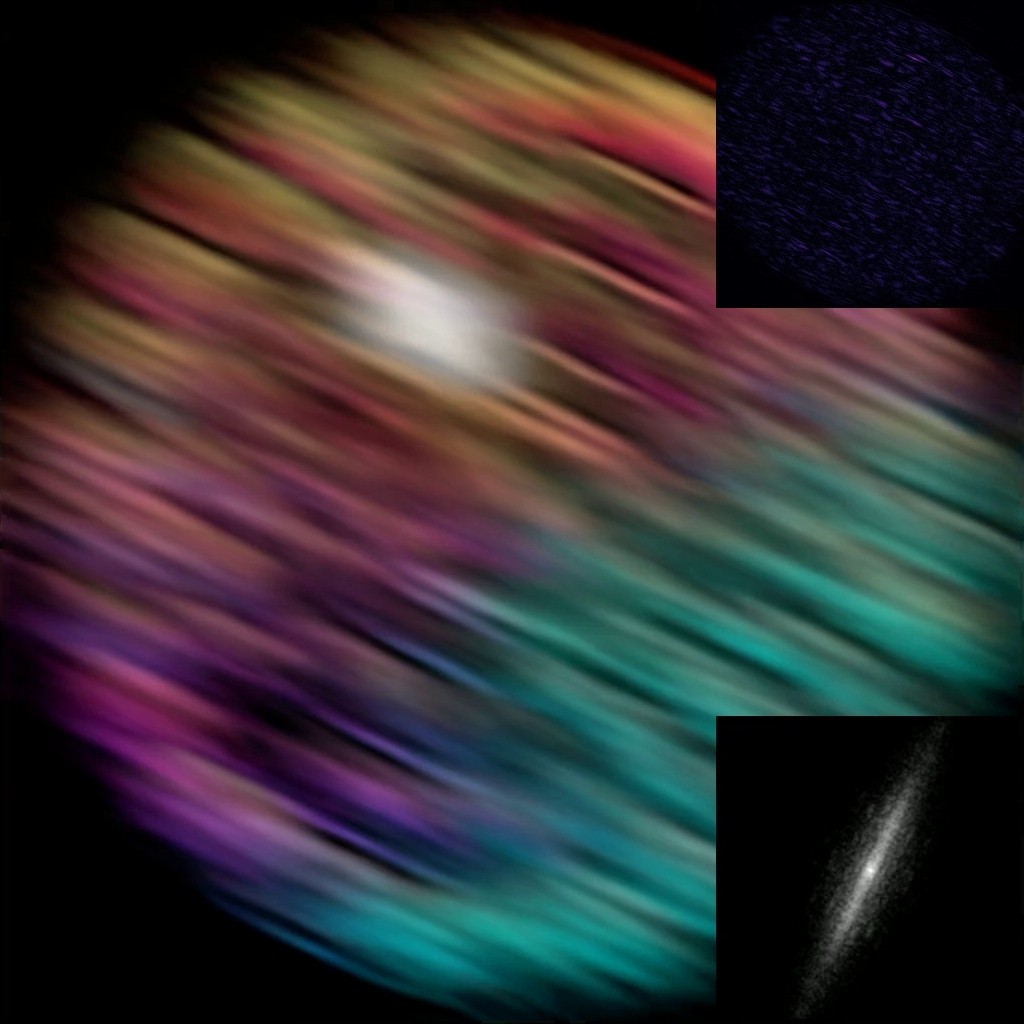}} & 
        \centeredtab{
            \includegraphics[width=0.155\textwidth, height=0.155\textwidth]{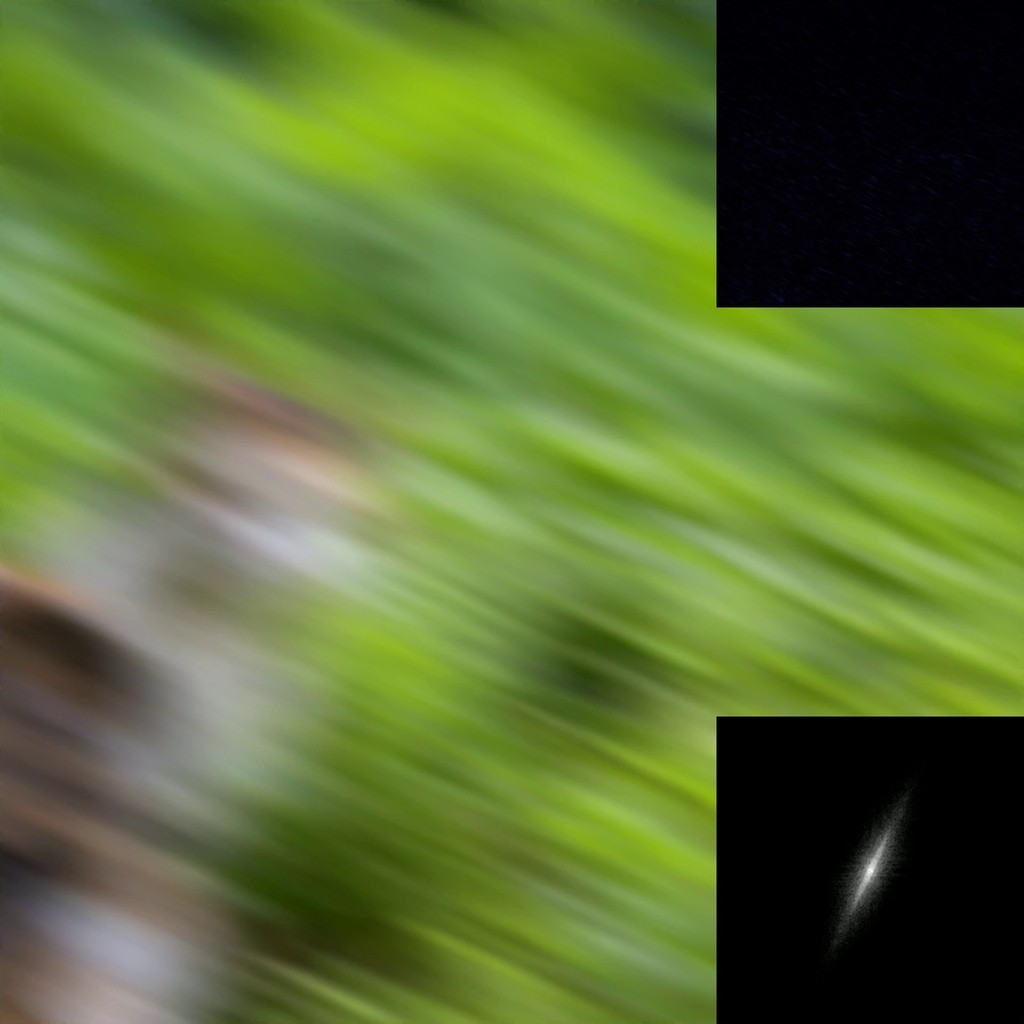} \\
            \includegraphics[width=0.155\textwidth, height=0.155\textwidth]{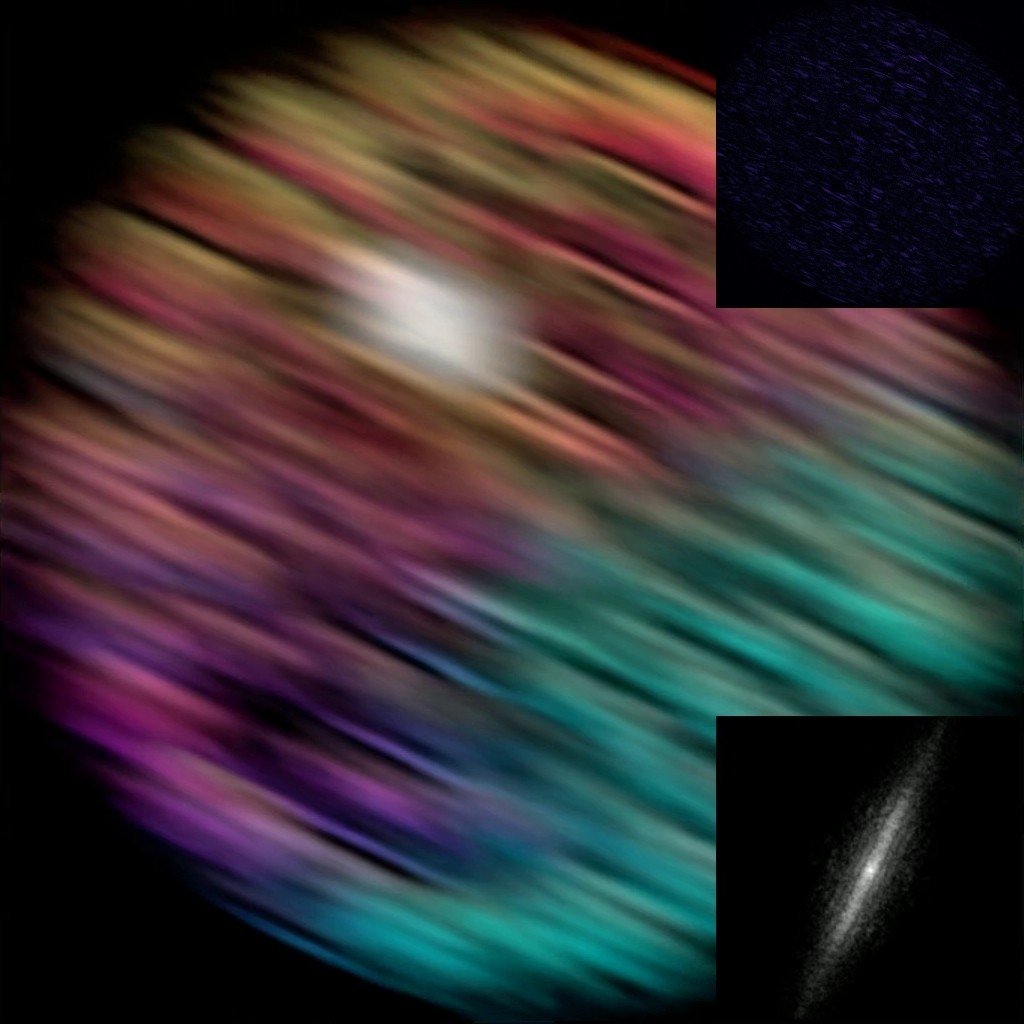}} & 
        \centeredtab{
            \includegraphics[width=0.155\textwidth, height=0.155\textwidth]{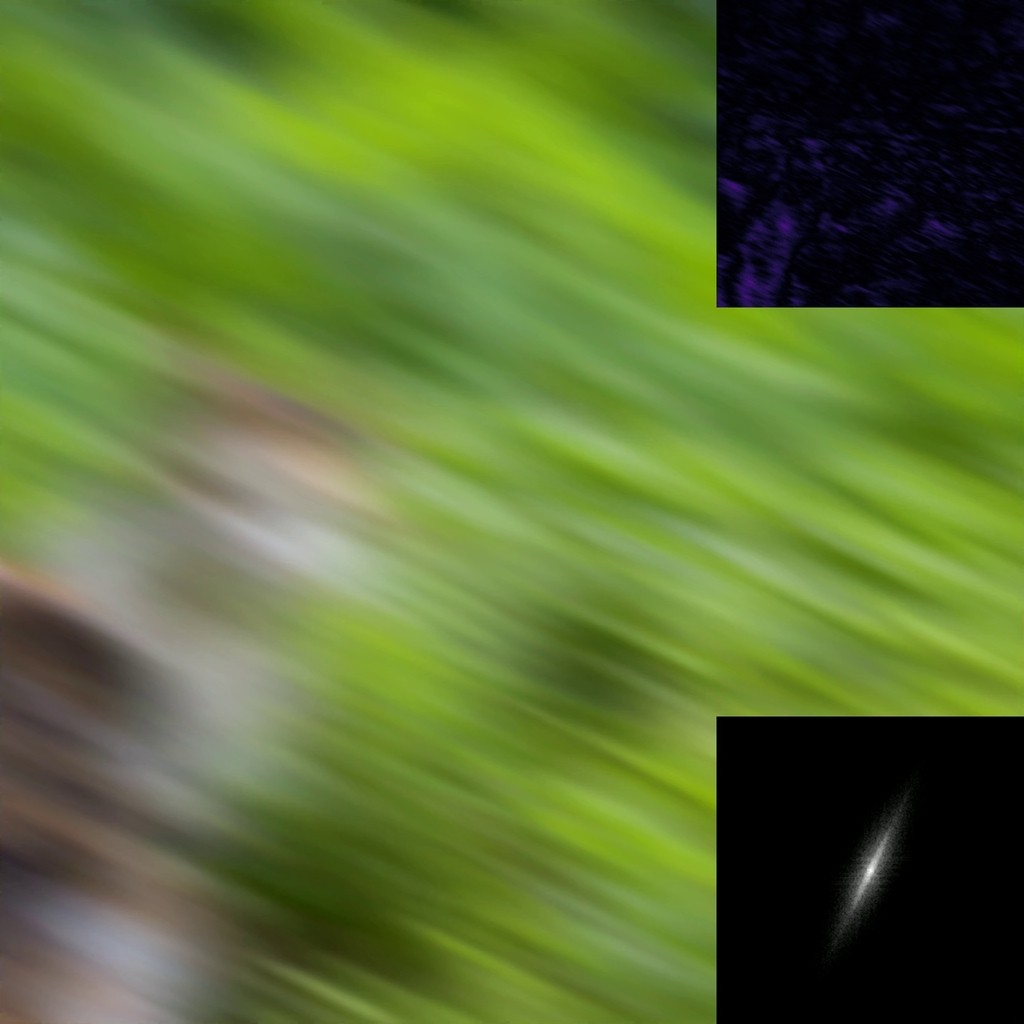} \\
            \includegraphics[width=0.155\textwidth, height=0.155\textwidth]{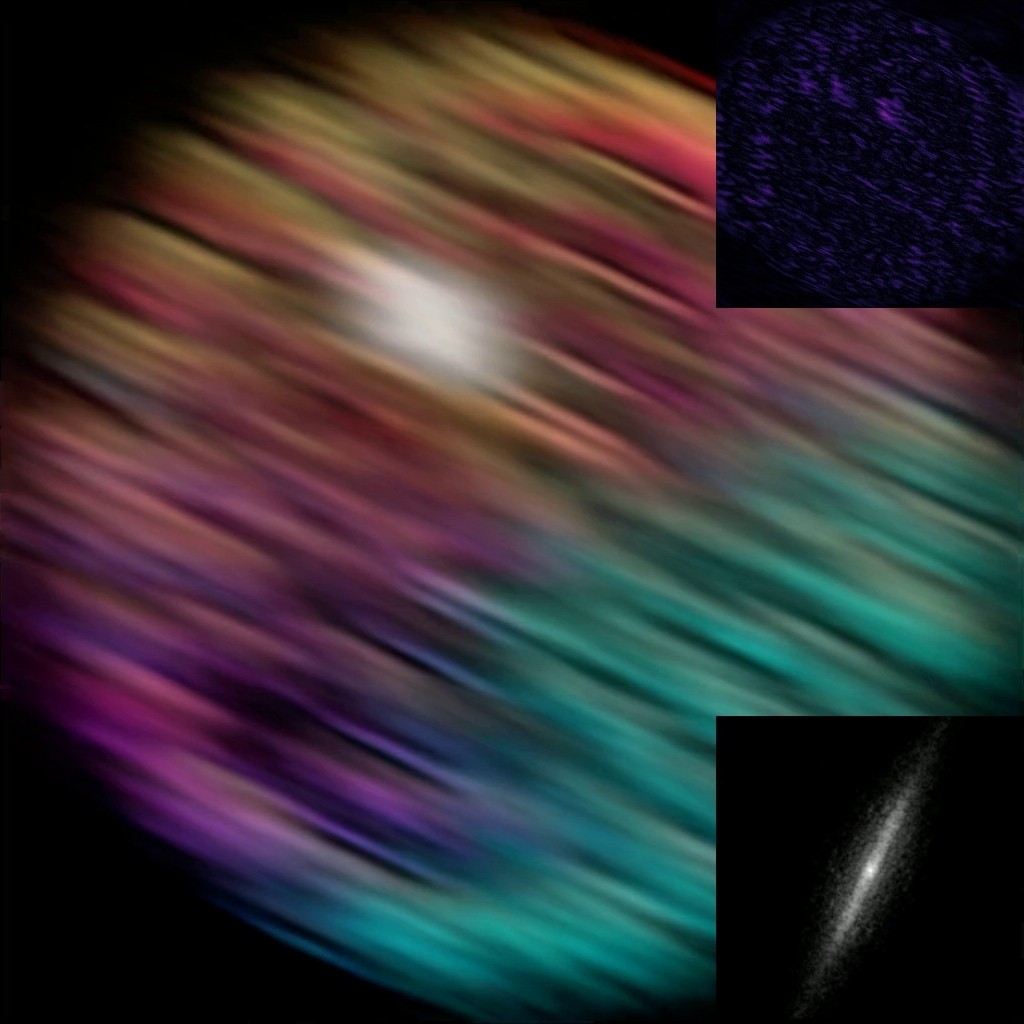}} & 
        \centeredtab{
            \includegraphics[width=0.155\textwidth, height=0.155\textwidth]{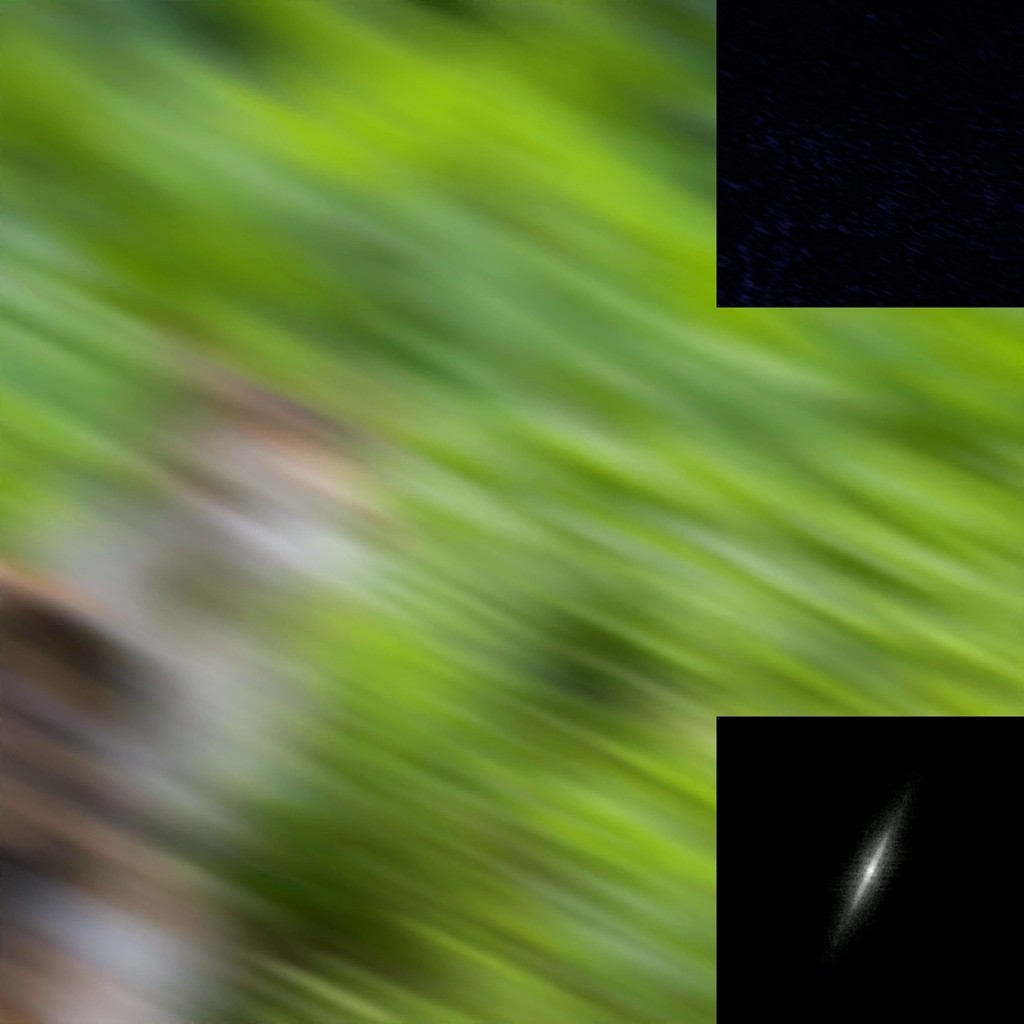} \\
            \includegraphics[width=0.155\textwidth, height=0.155\textwidth]{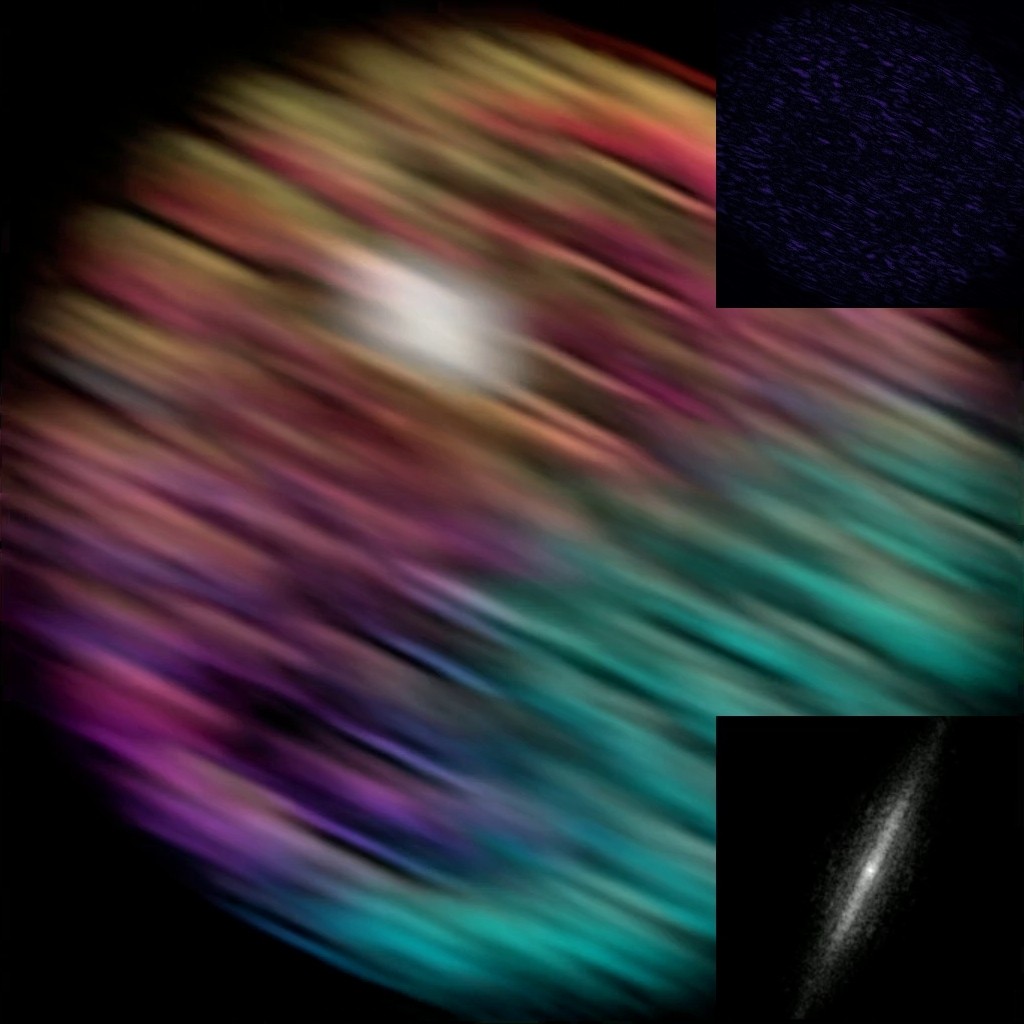}} & 
        \centeredtab{
            \includegraphics[width=0.155\textwidth, height=0.155\textwidth]{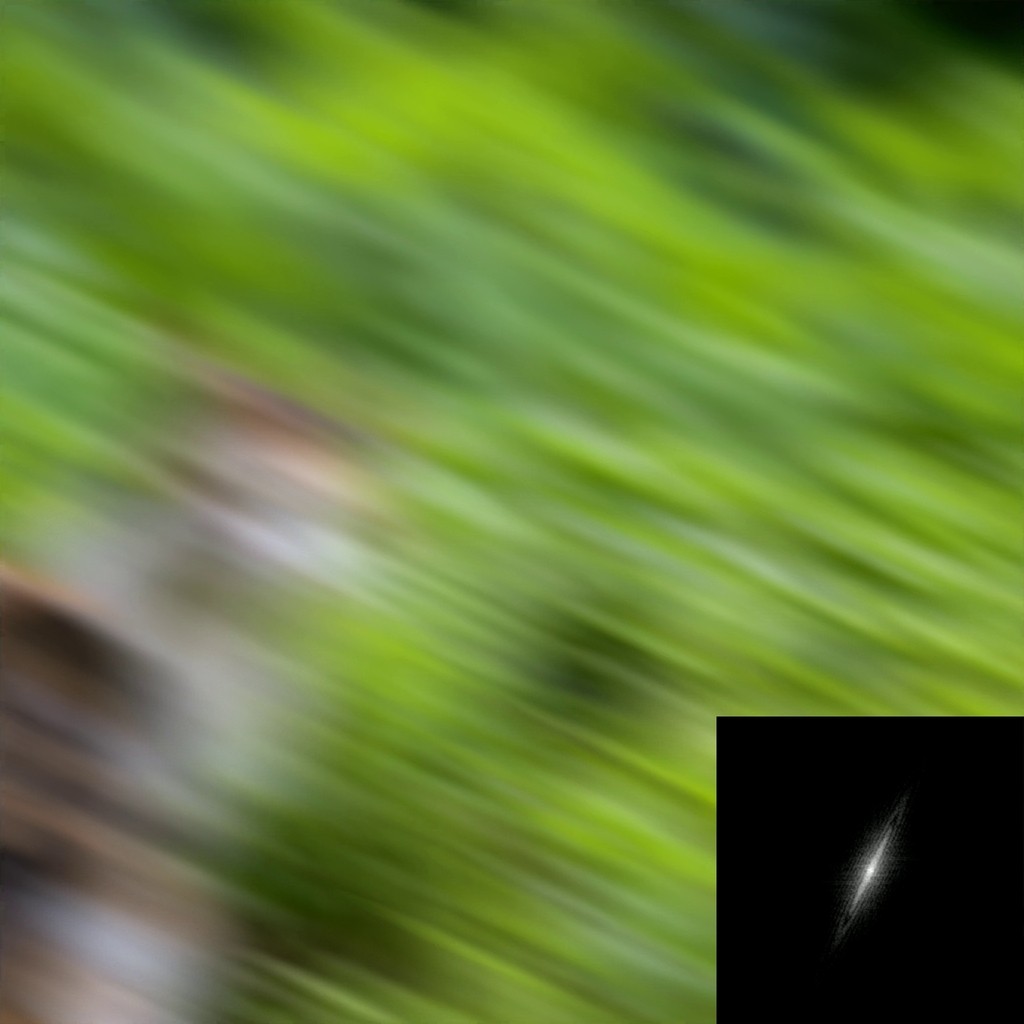} \\
            \includegraphics[width=0.155\textwidth, height=0.155\textwidth]{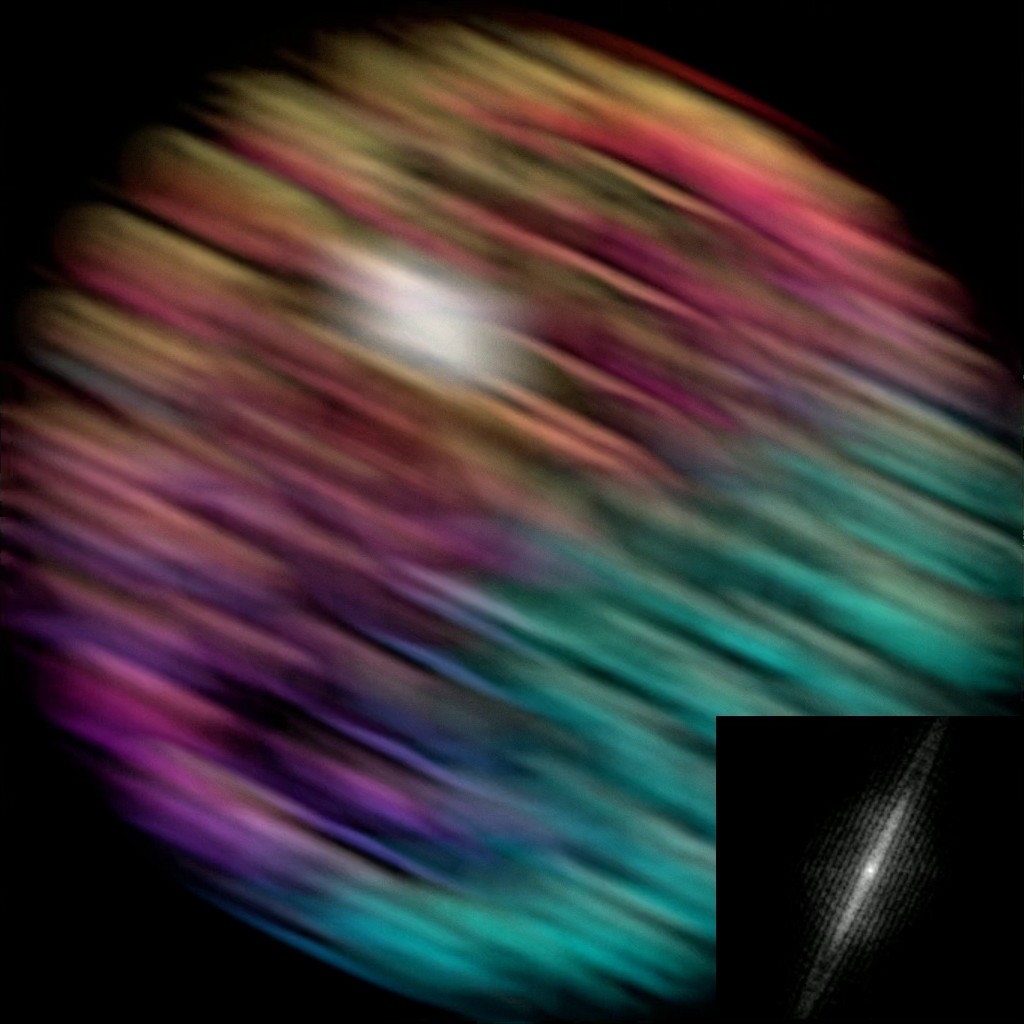}}\\
            \cmidrule(lr){3-8}
            \centeredtab{\rot{Lanczos}} & &
        \centeredtab{
            \includegraphics[width=0.155\textwidth, height=0.155\textwidth]{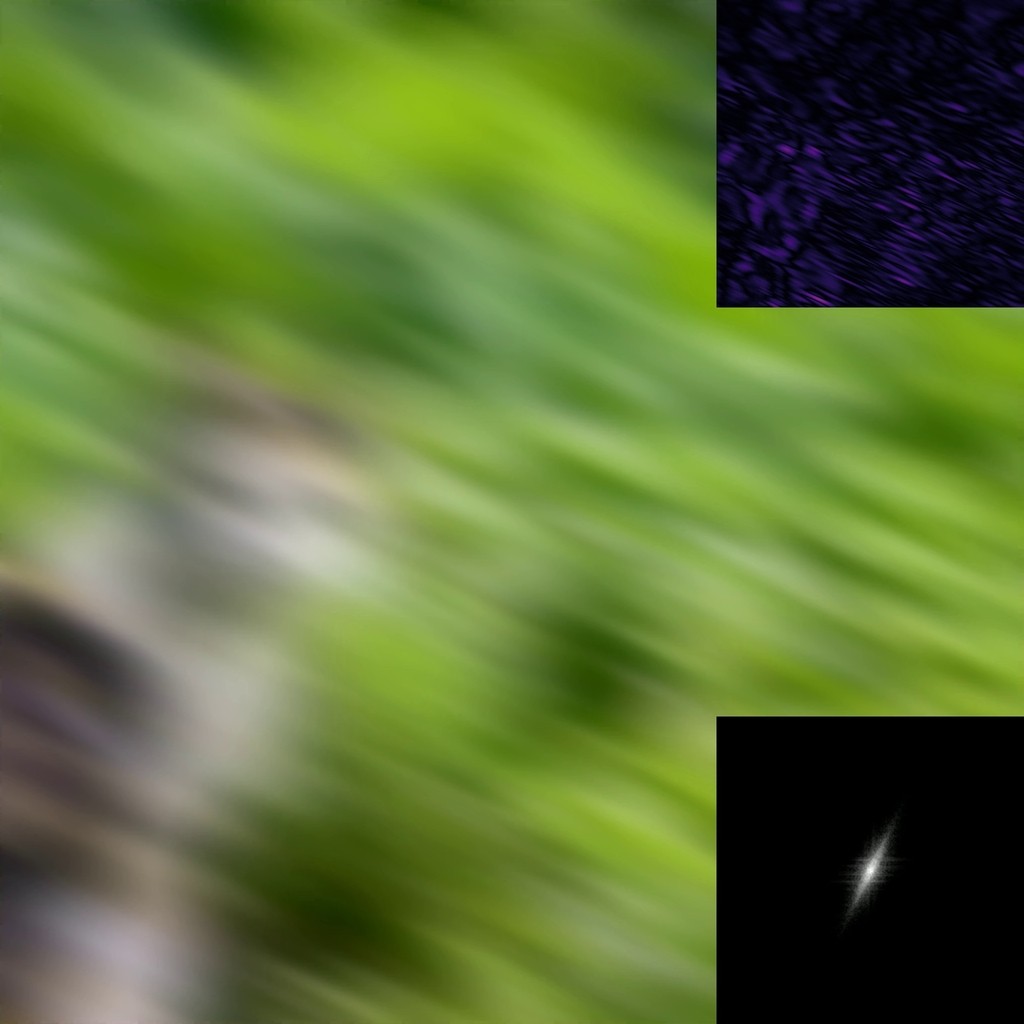} \\
            \includegraphics[width=0.155\textwidth, height=0.155\textwidth]{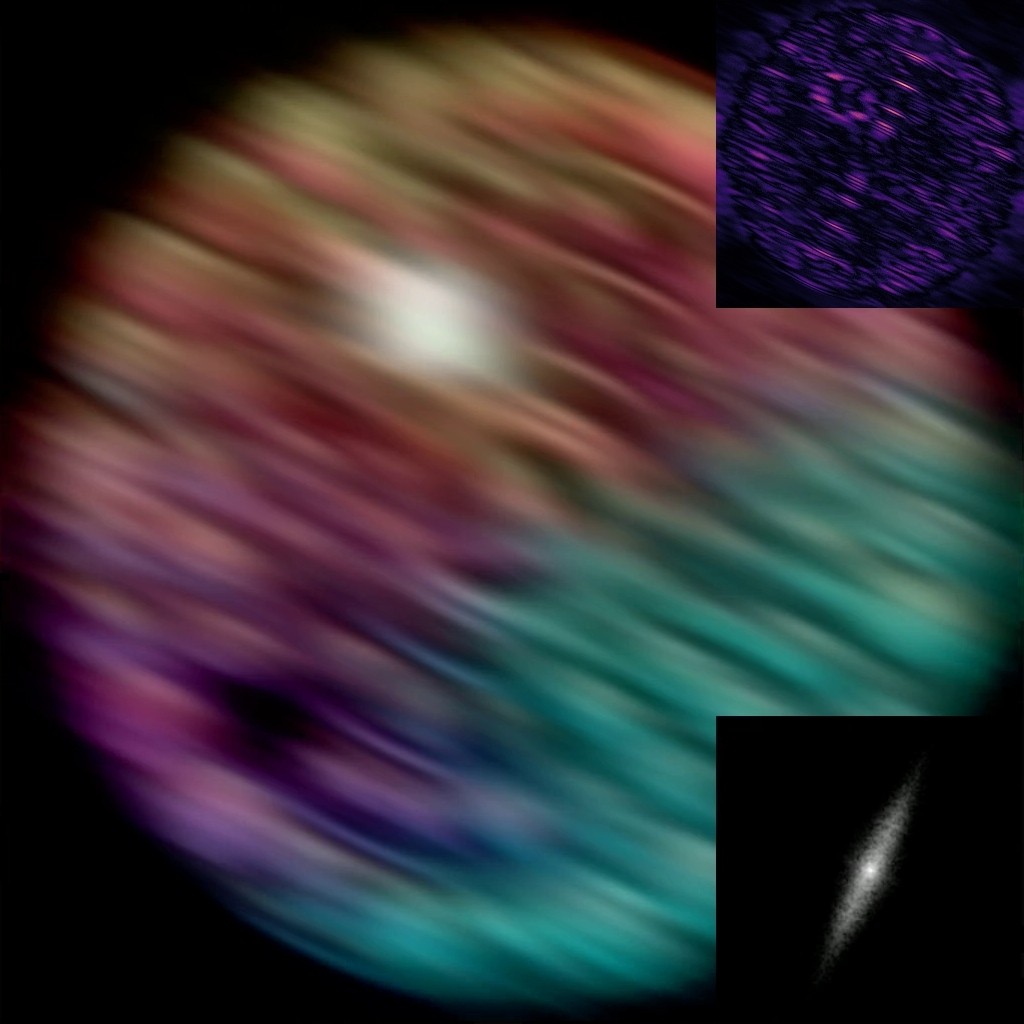}} & 
        \centeredtab{
            \includegraphics[width=0.155\textwidth, height=0.155\textwidth]{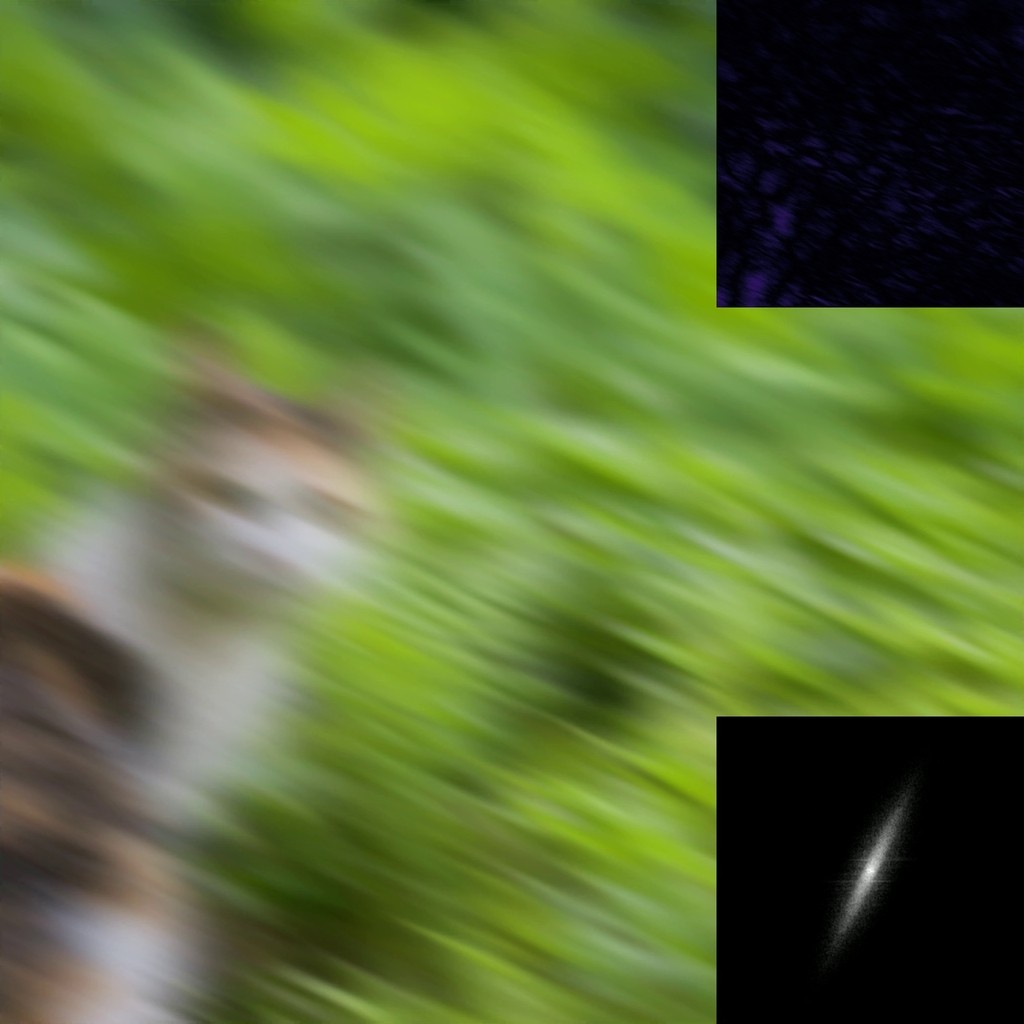} \\
            \includegraphics[width=0.155\textwidth, height=0.155\textwidth]{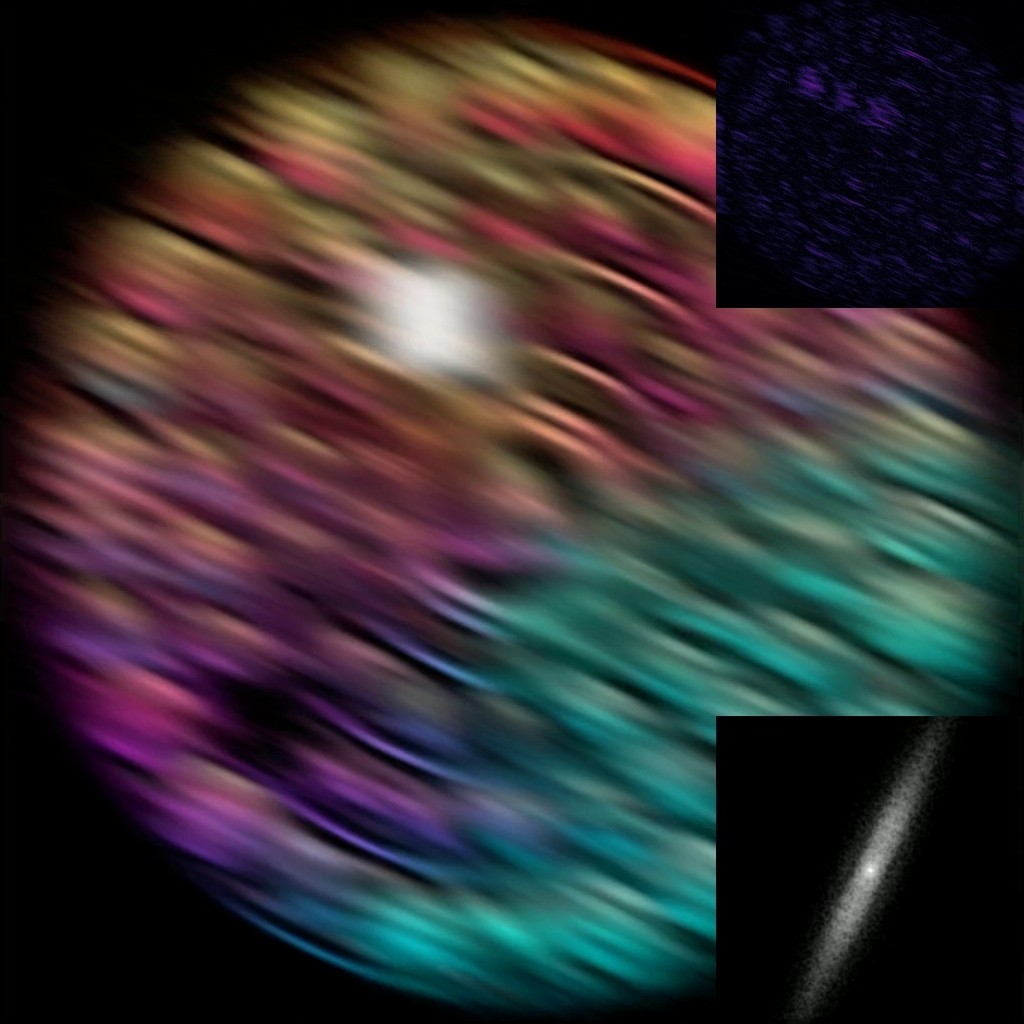}} & 
        \centeredtab{
            \includegraphics[width=0.155\textwidth, height=0.155\textwidth]{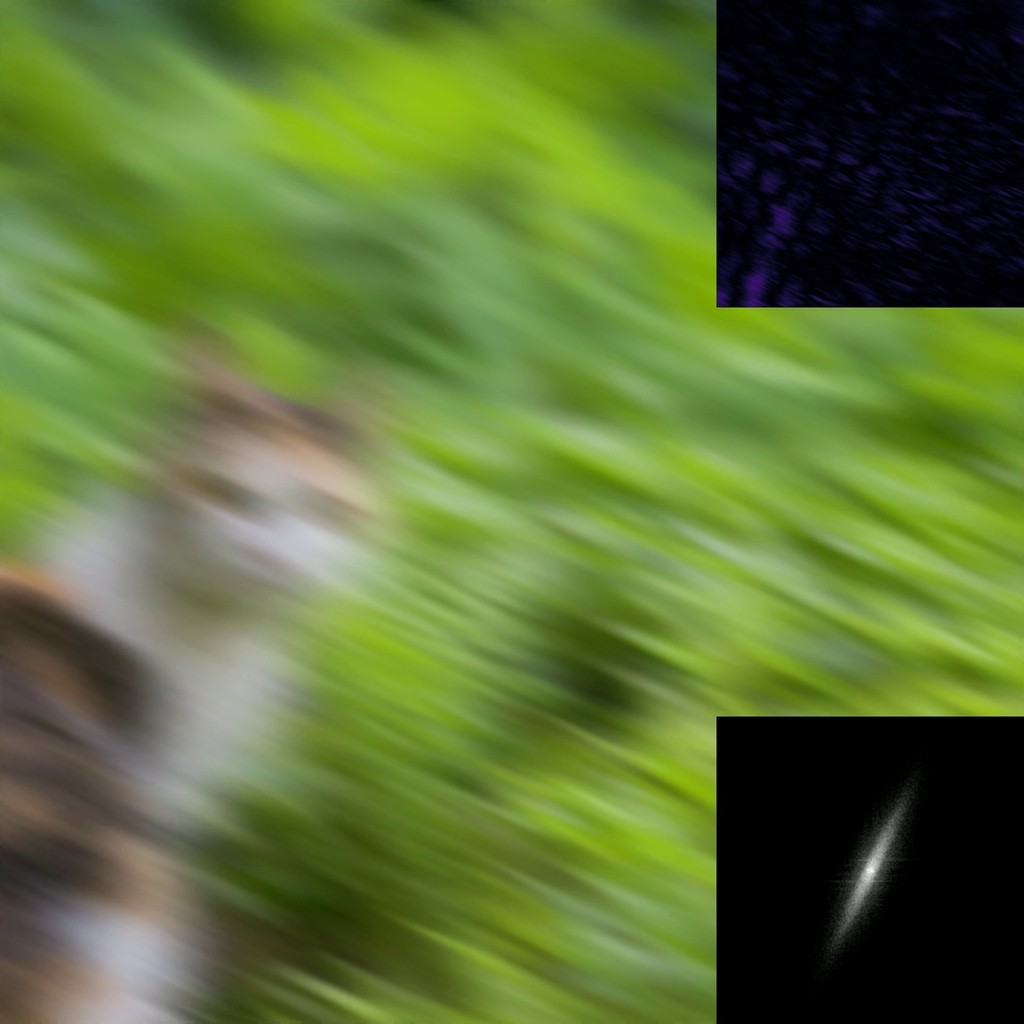} \\
            \includegraphics[width=0.155\textwidth, height=0.155\textwidth]{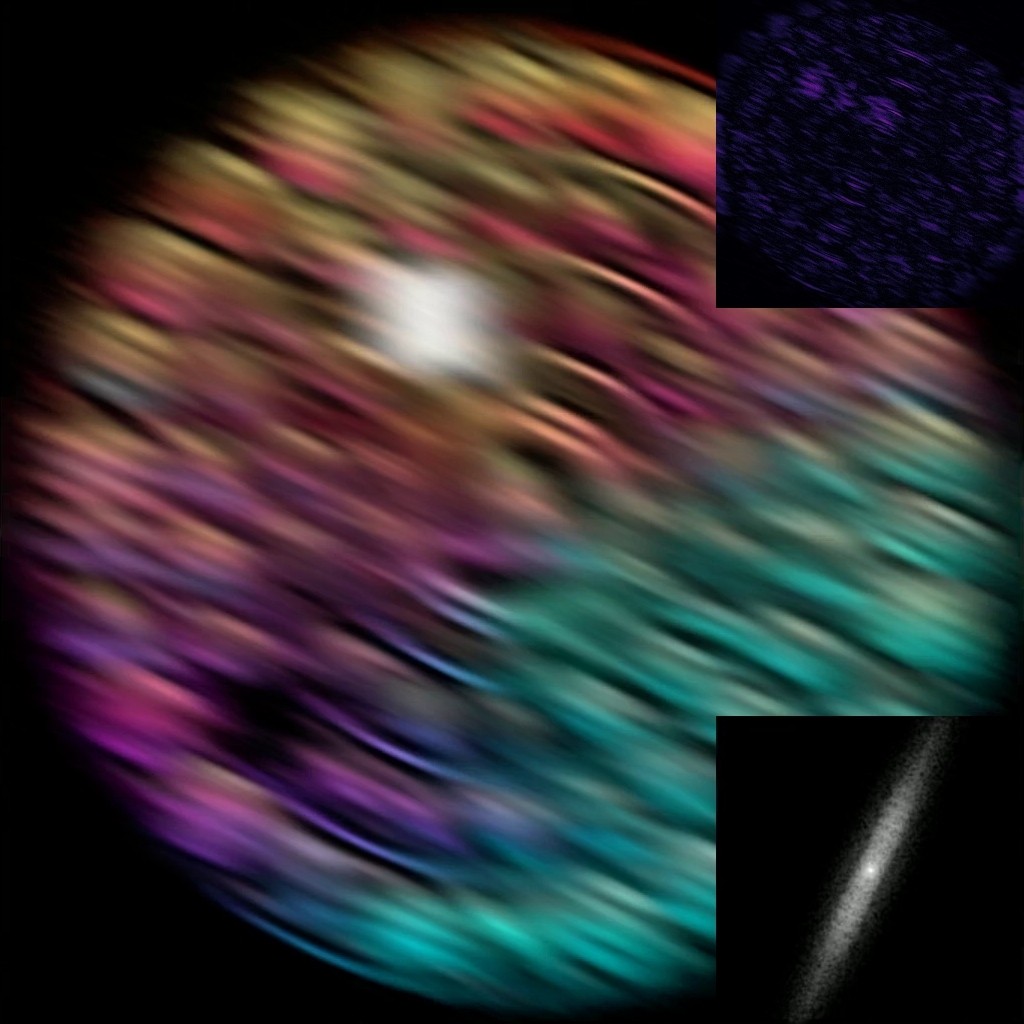}} & 
        \centeredtab{
            \includegraphics[width=0.155\textwidth, height=0.155\textwidth]{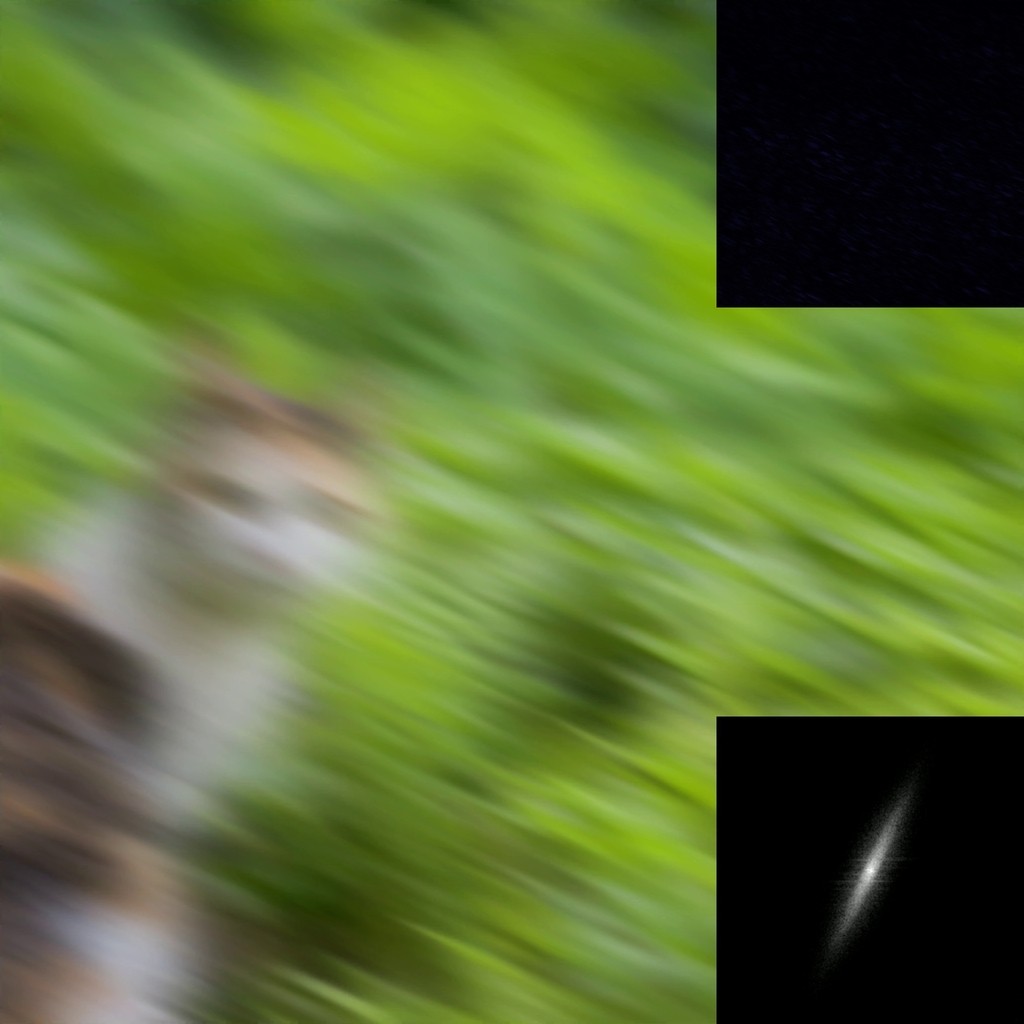} \\
            \includegraphics[width=0.155\textwidth, height=0.155\textwidth]{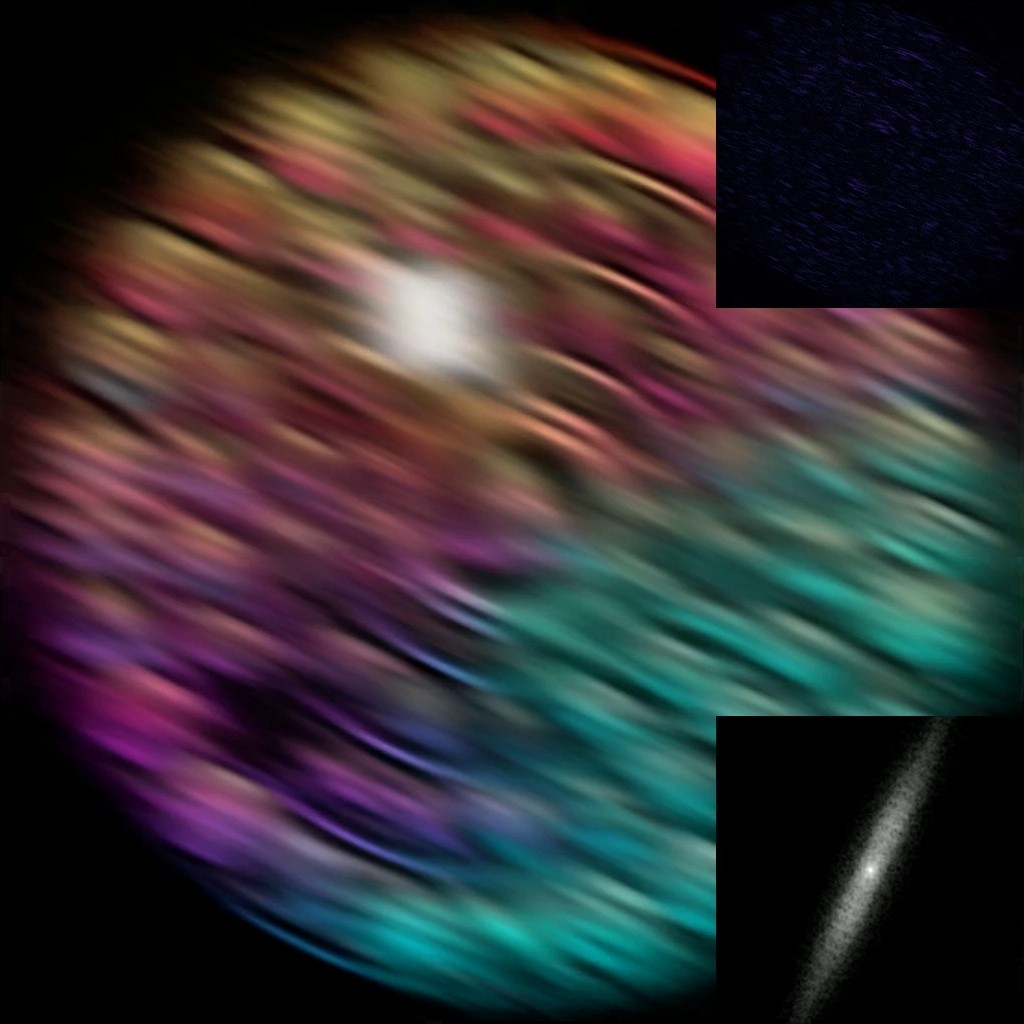}} & 
        \centeredtab{
            \includegraphics[width=0.155\textwidth, height=0.155\textwidth]{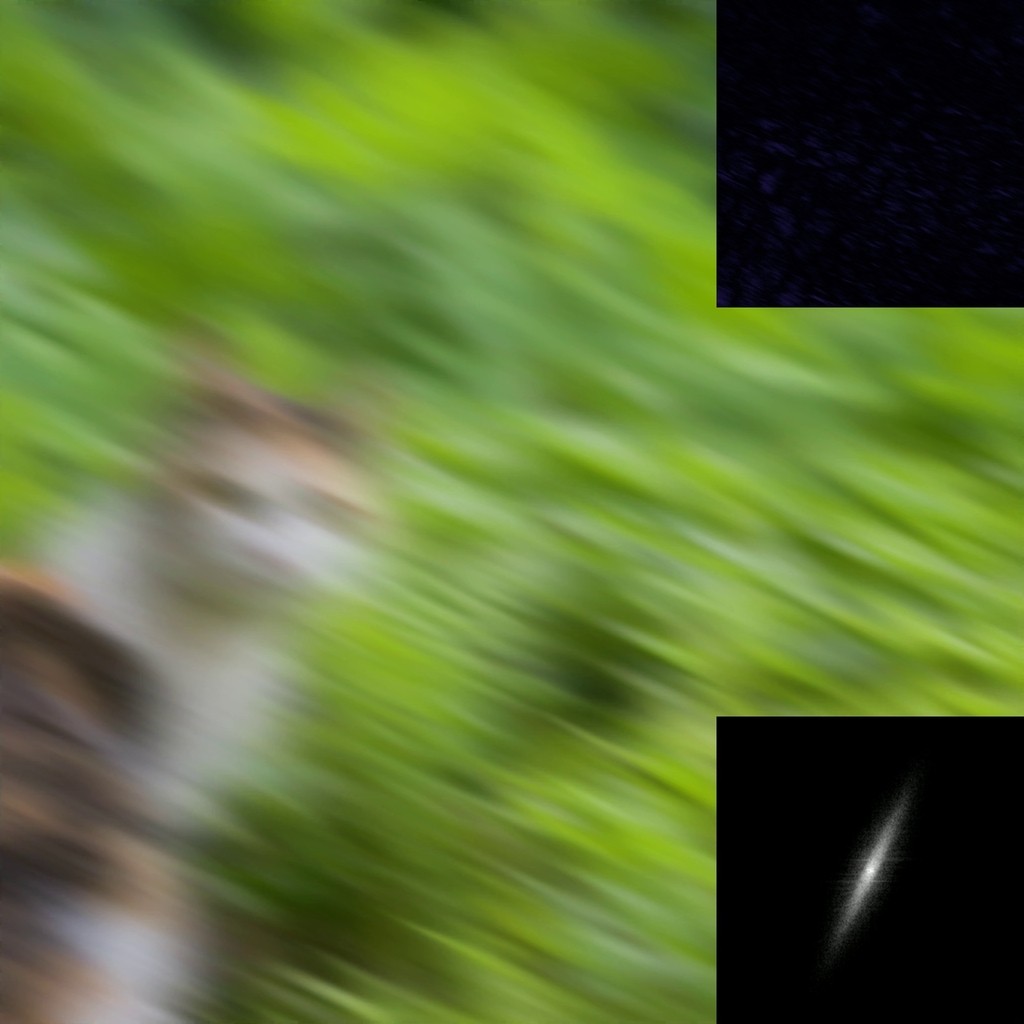} \\
            \includegraphics[width=0.155\textwidth, height=0.155\textwidth]{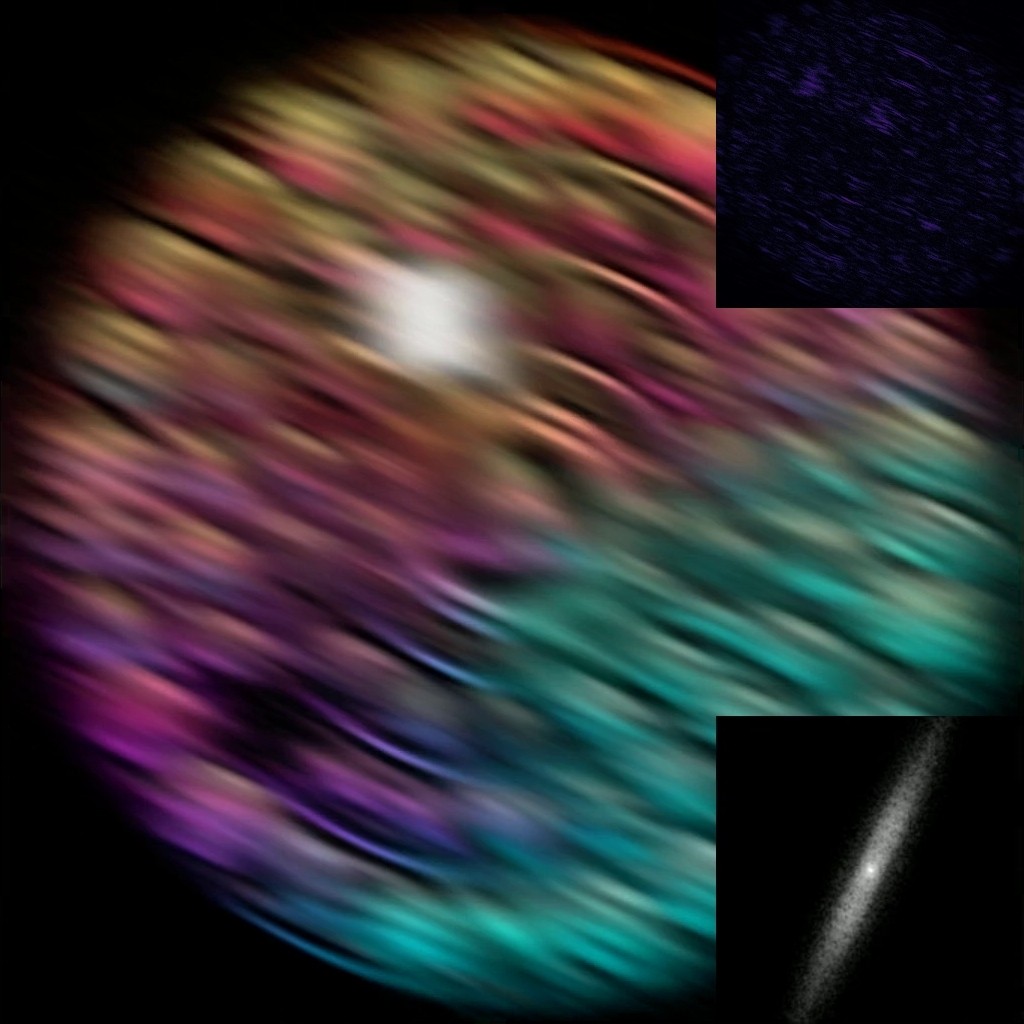}} & 
        \centeredtab{
            \includegraphics[width=0.155\textwidth, height=0.155\textwidth]{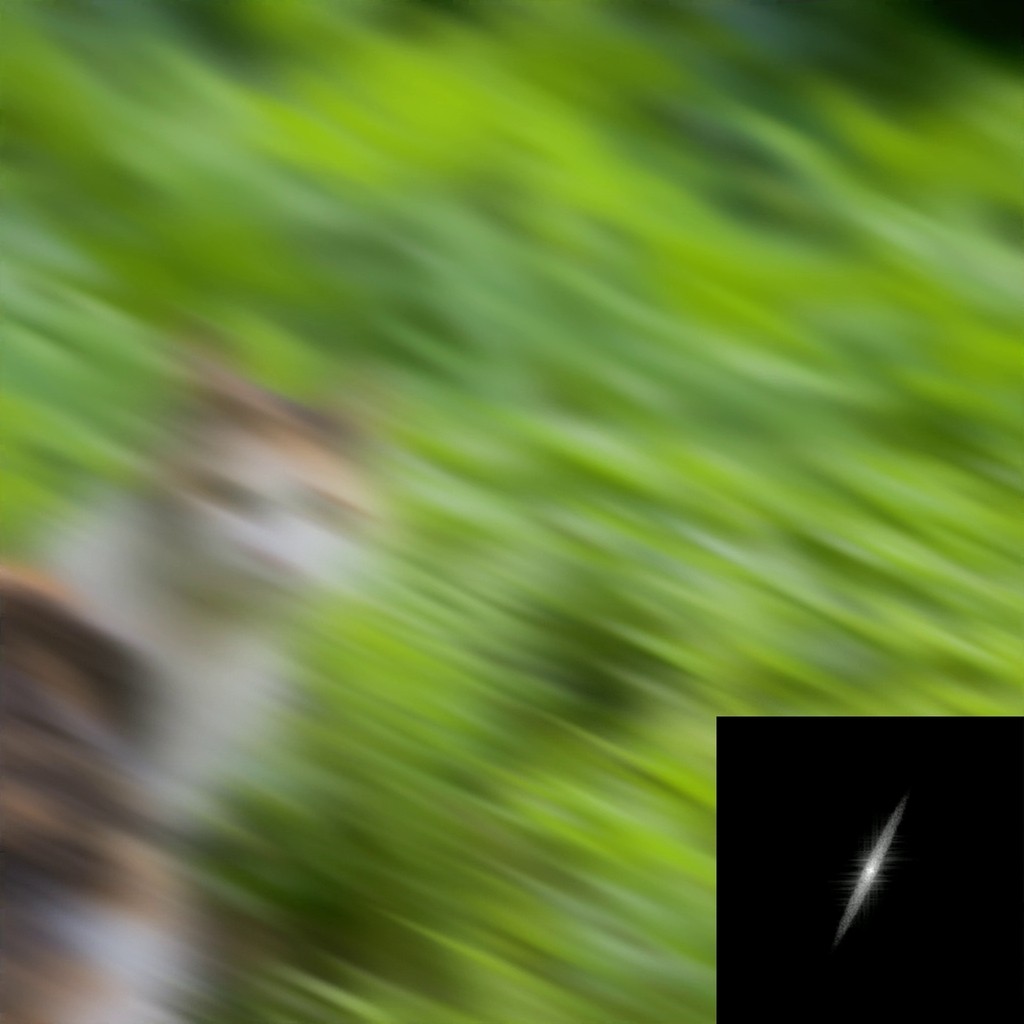} \\
            \includegraphics[width=0.155\textwidth, height=0.155\textwidth]{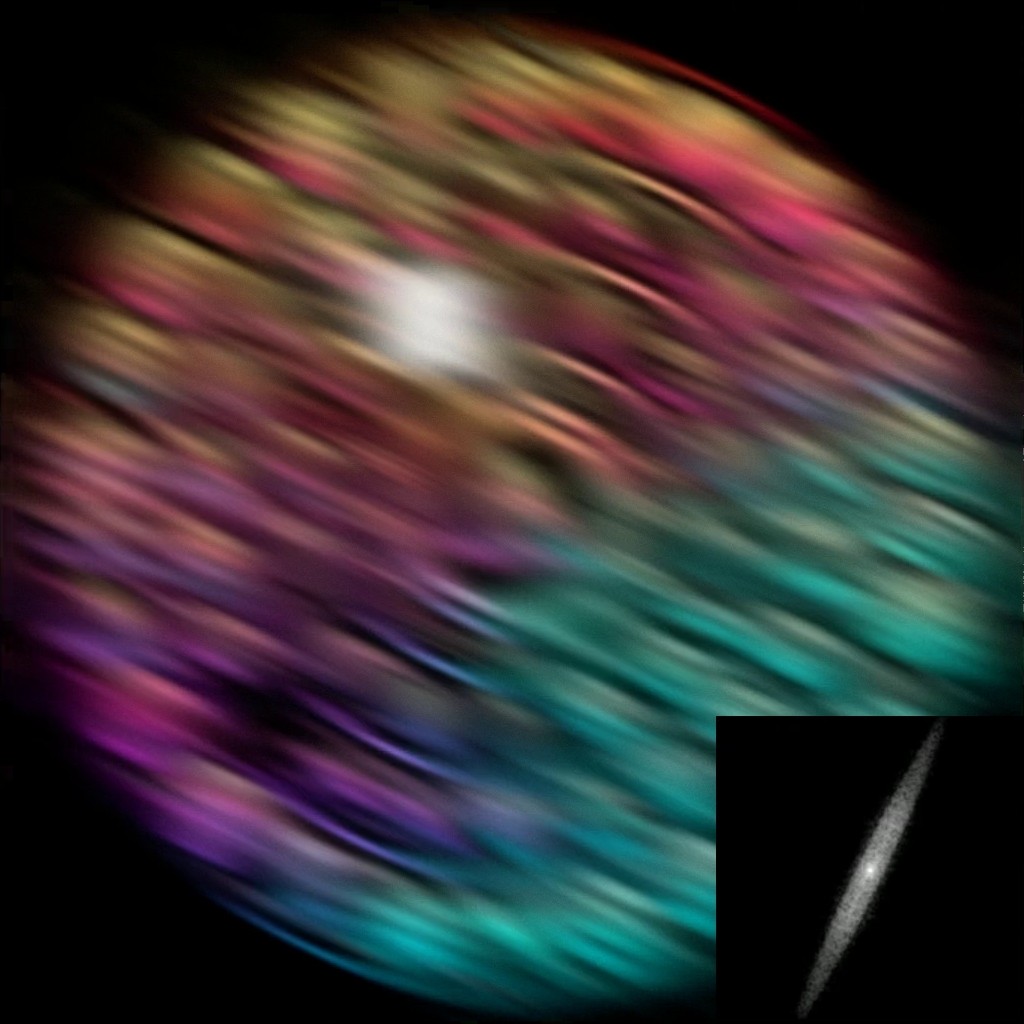}}\\
    \end{tabular}
    \caption{
    \label{fig:supp_results_images1}
        \captiontitle{Filter generalization on Images.}
        Comparisons between different training/testing pairs and recalibrated Neural Gaussian Scale-Space Fields~\cite{mujkanovic2024neural} (NGSSF) for image filtering with different kernels.
        Insets at the bottom-right of each image represent the frequency spectrum, 
        and insets at the top-right represent the absolute mean error map.
        Our model can generalize to unseen filters at testing time with less error compared to NGSSF.
        Images from Adobe FiveK; \textcopyright{} original photographers/Adobe.
    }
\end{figure*}

\begin{figure*}
    \centering
    \setlength{\tabcolsep}{1pt}
    \renewcommand{\arraystretch}{0.6}
    \fboxsep=-0.1pt
    \begin{tabular}{cccccccc}
        & & & \multicolumn{4}{c}{Ours} \\
        \cmidrule(lr){4-7} 
        & & NGSSF & Trained w/ Gaussian & Trained w/ Box & Trained w/ Lanczos & Trained with all & GT \\
        \centeredtab{\rot{Gaussian}} & &
        \centeredtab{
            \includegraphics[width=0.155\textwidth, height=0.155\textwidth]{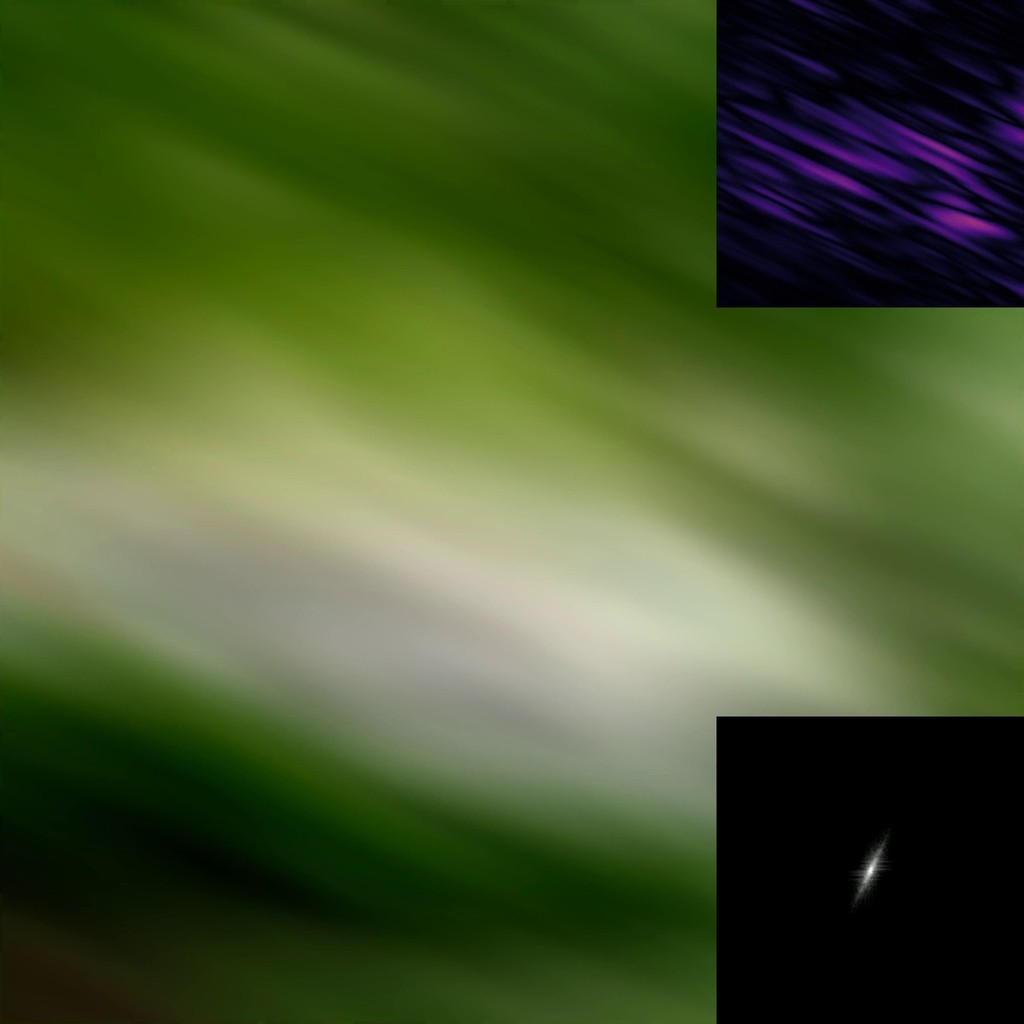} \\
            } & 
        \centeredtab{
            \includegraphics[width=0.155\textwidth, height=0.155\textwidth]{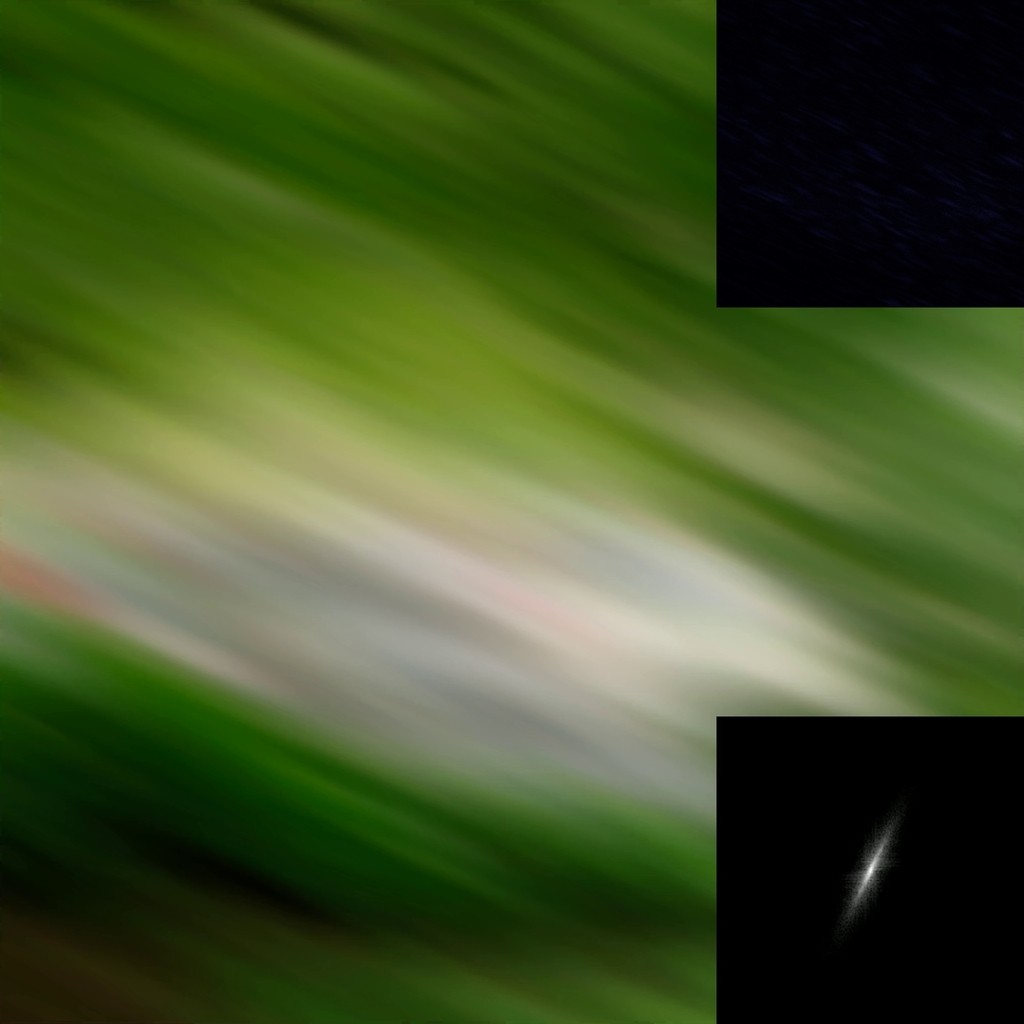} \\
            } & 
        \centeredtab{
            \includegraphics[width=0.155\textwidth, height=0.155\textwidth]{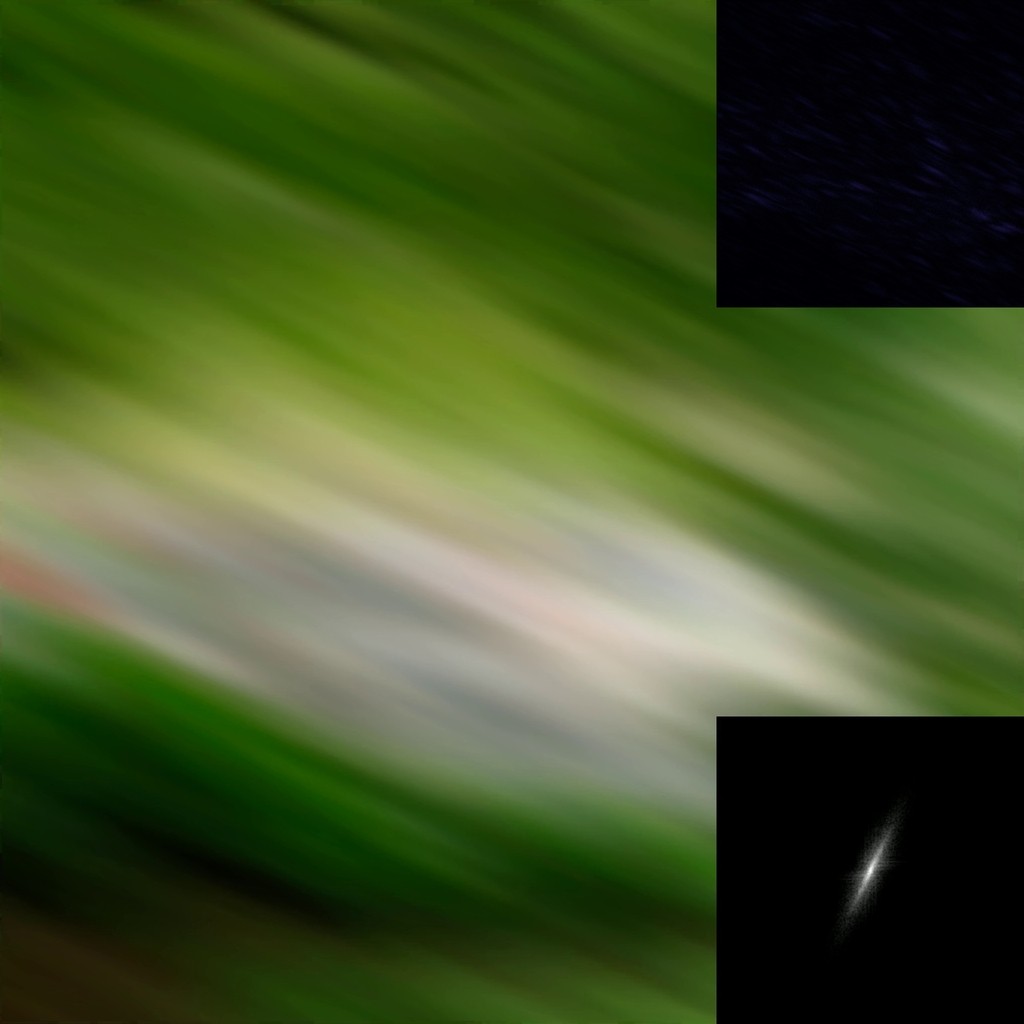} \\
            } & 
        \centeredtab{
            \includegraphics[width=0.155\textwidth, height=0.155\textwidth]{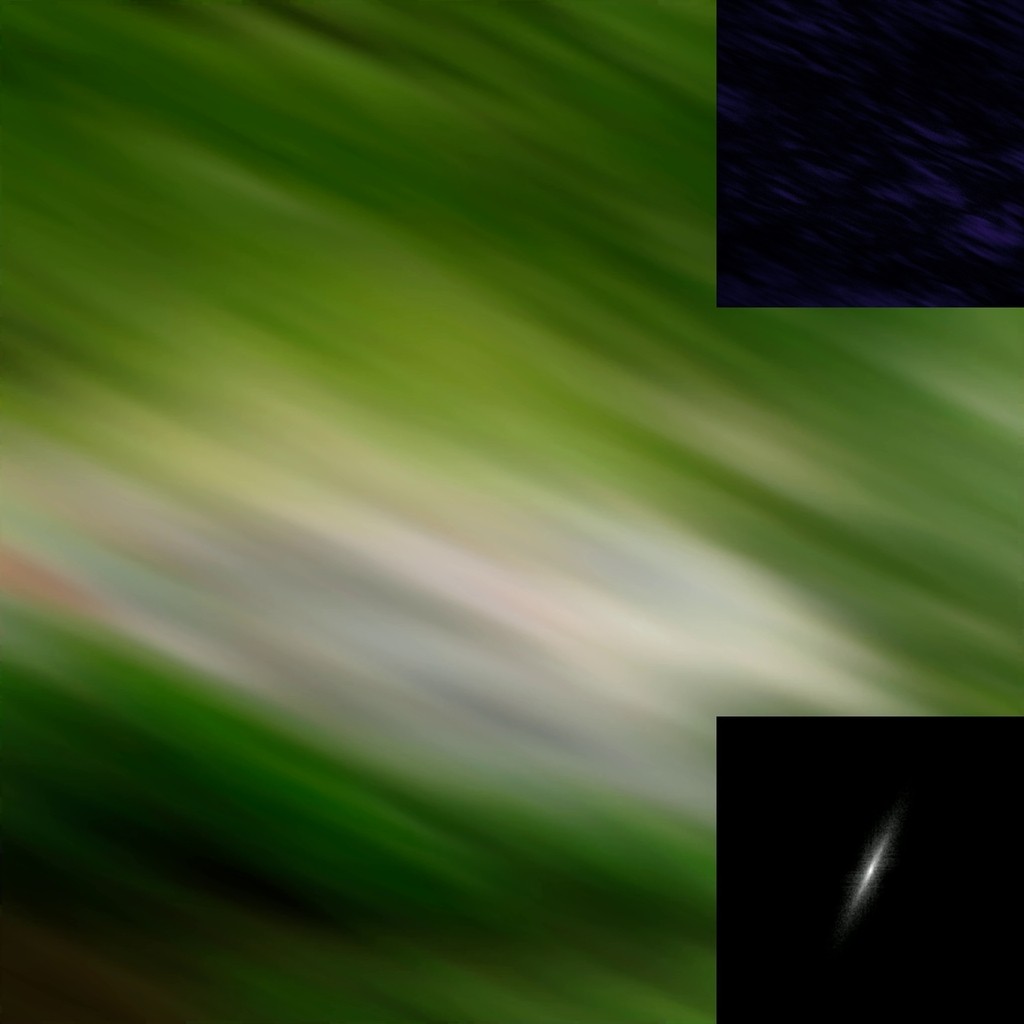} \\
            } & 
        \centeredtab{
            \includegraphics[width=0.155\textwidth, height=0.155\textwidth]{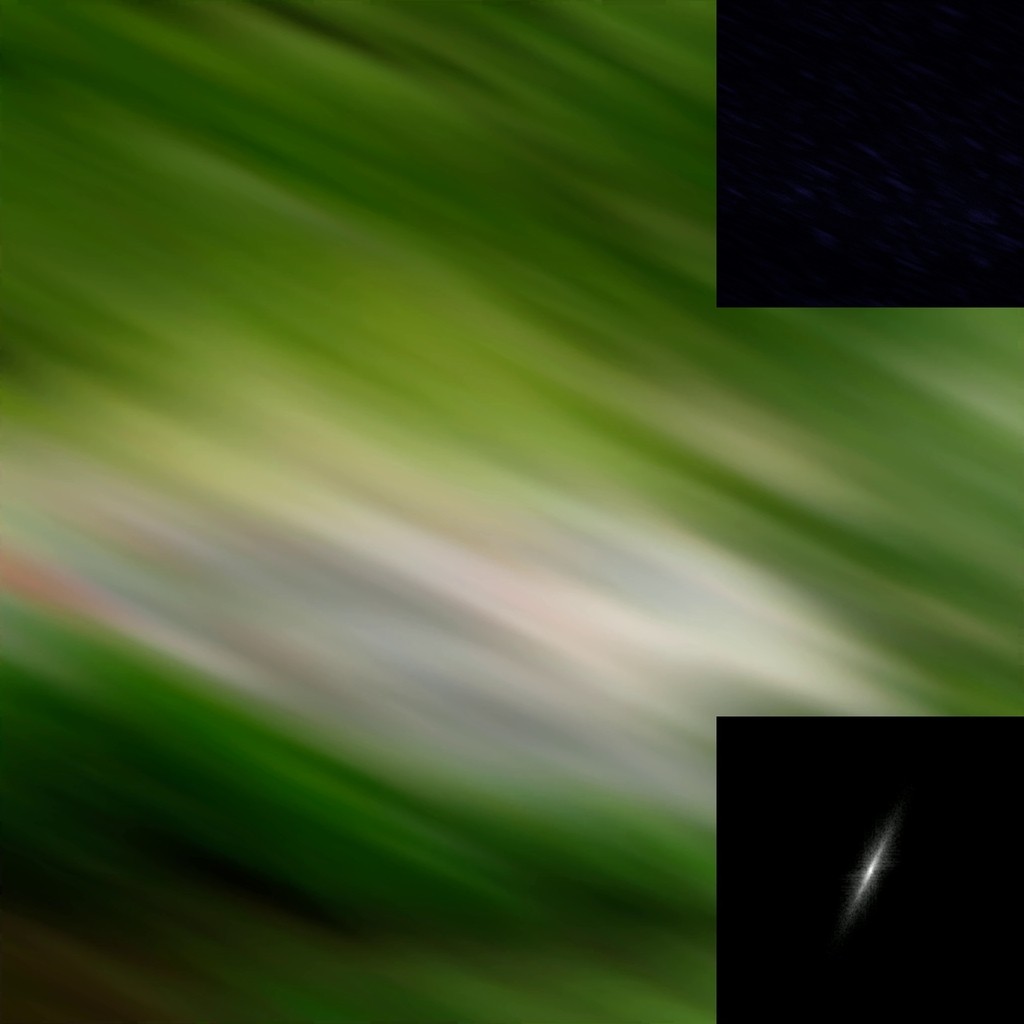} \\
            } & 
        \centeredtab{
            \includegraphics[width=0.155\textwidth, height=0.155\textwidth]{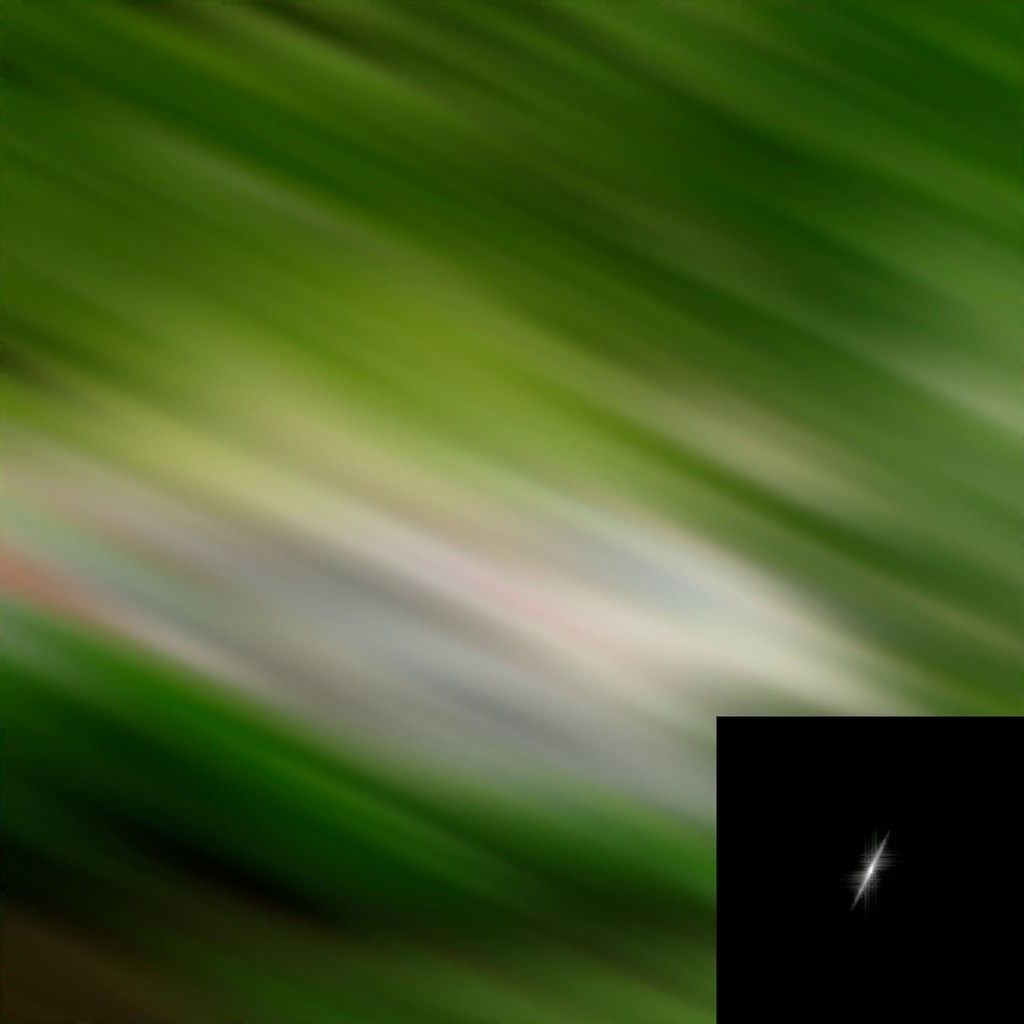} \\
            }\\
        \cmidrule(lr){3-8}
        \centeredtab{\rot{Box}} & &
        \centeredtab{
            \includegraphics[width=0.155\textwidth, height=0.155\textwidth]{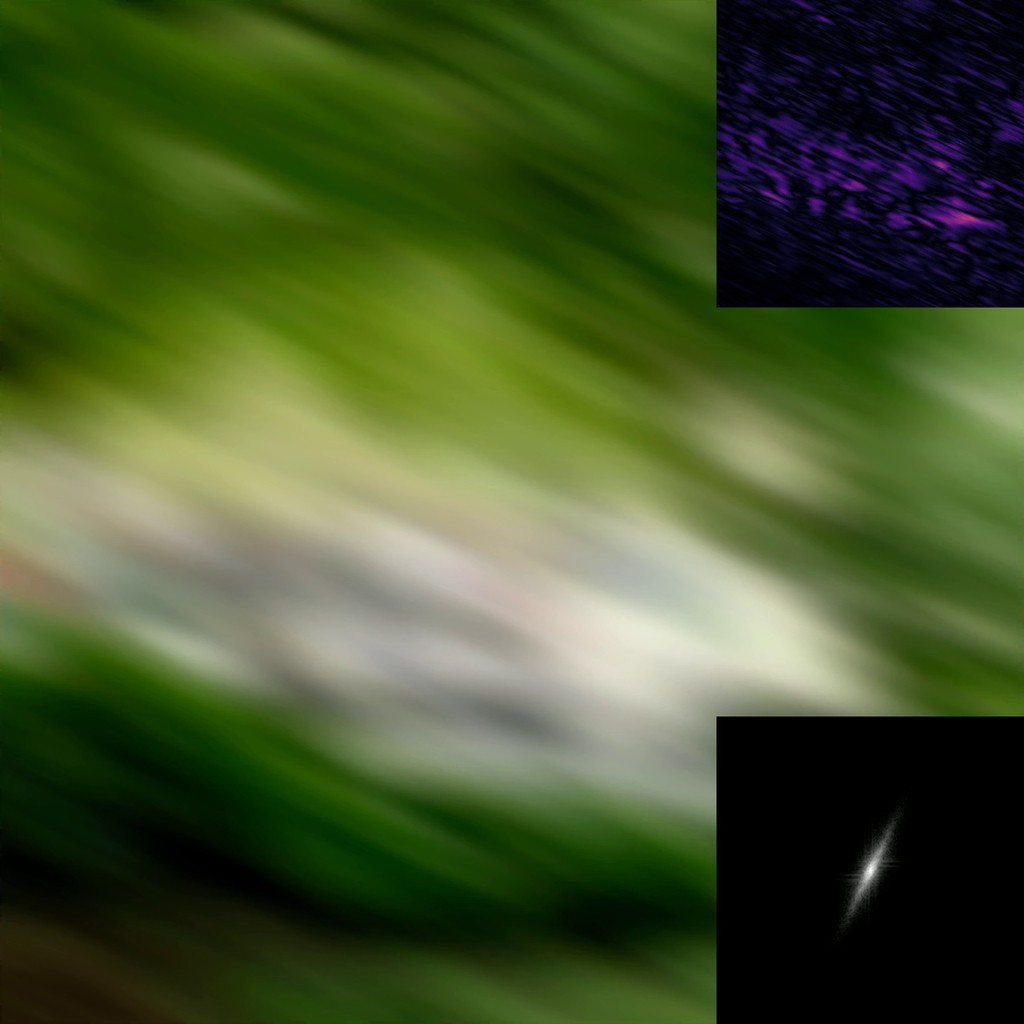} \\
            } & 
        \centeredtab{
            \includegraphics[width=0.155\textwidth, height=0.155\textwidth]{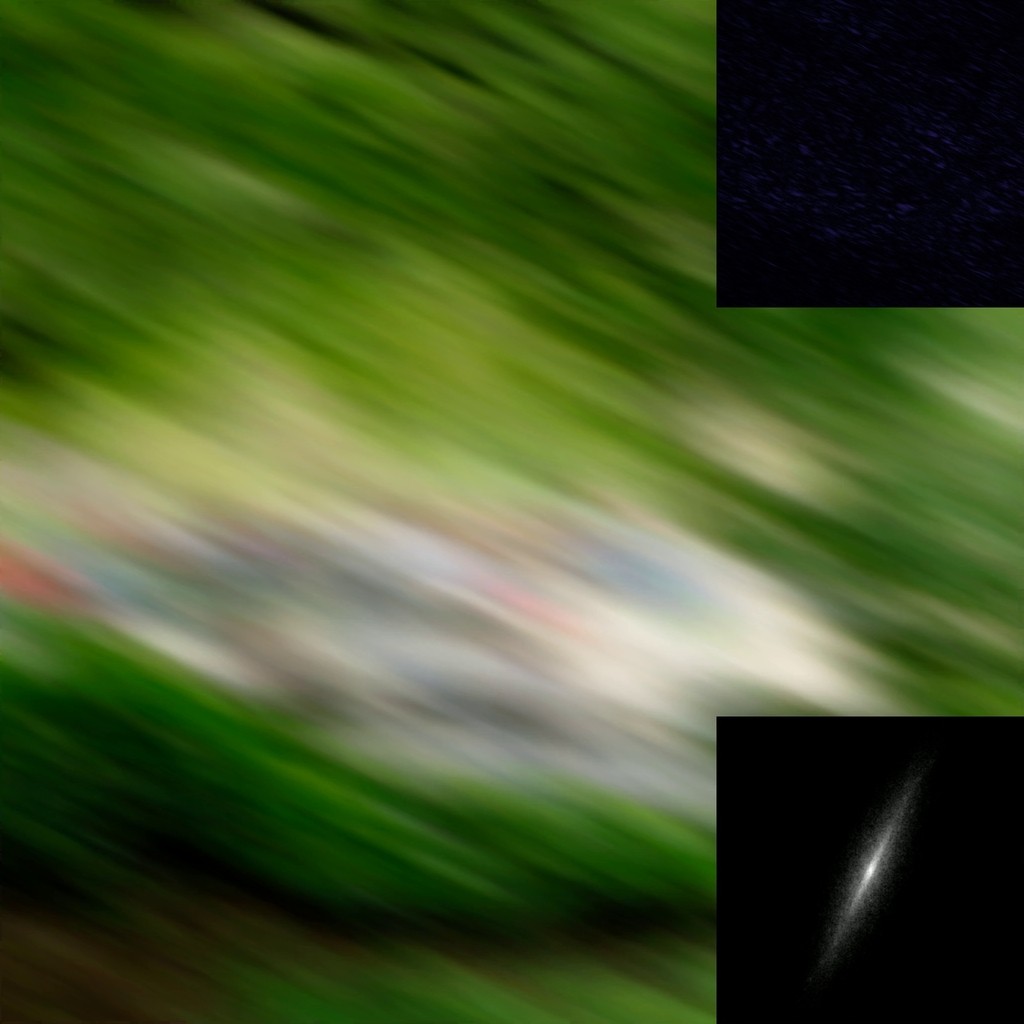} \\
            } & 
        \centeredtab{
            \includegraphics[width=0.155\textwidth, height=0.155\textwidth]{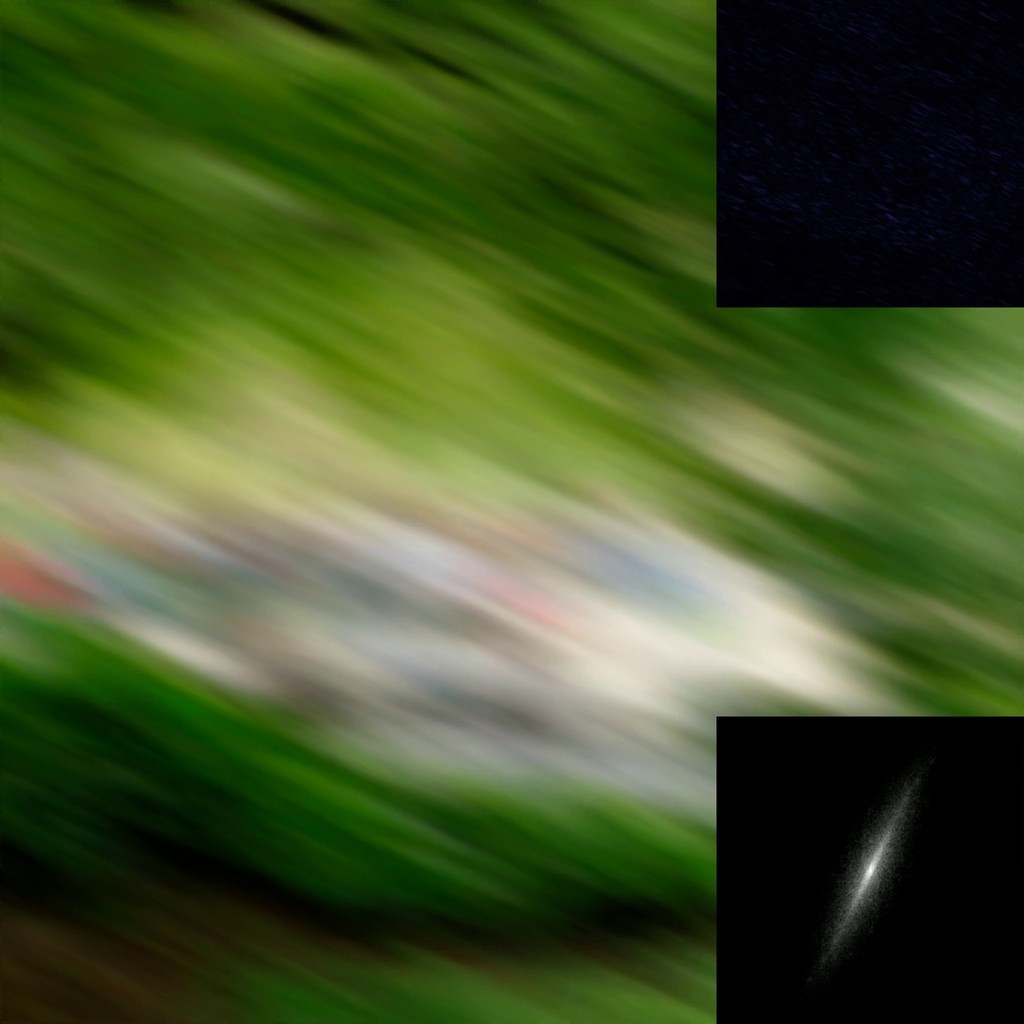} \\
            } & 
        \centeredtab{
            \includegraphics[width=0.155\textwidth, height=0.155\textwidth]{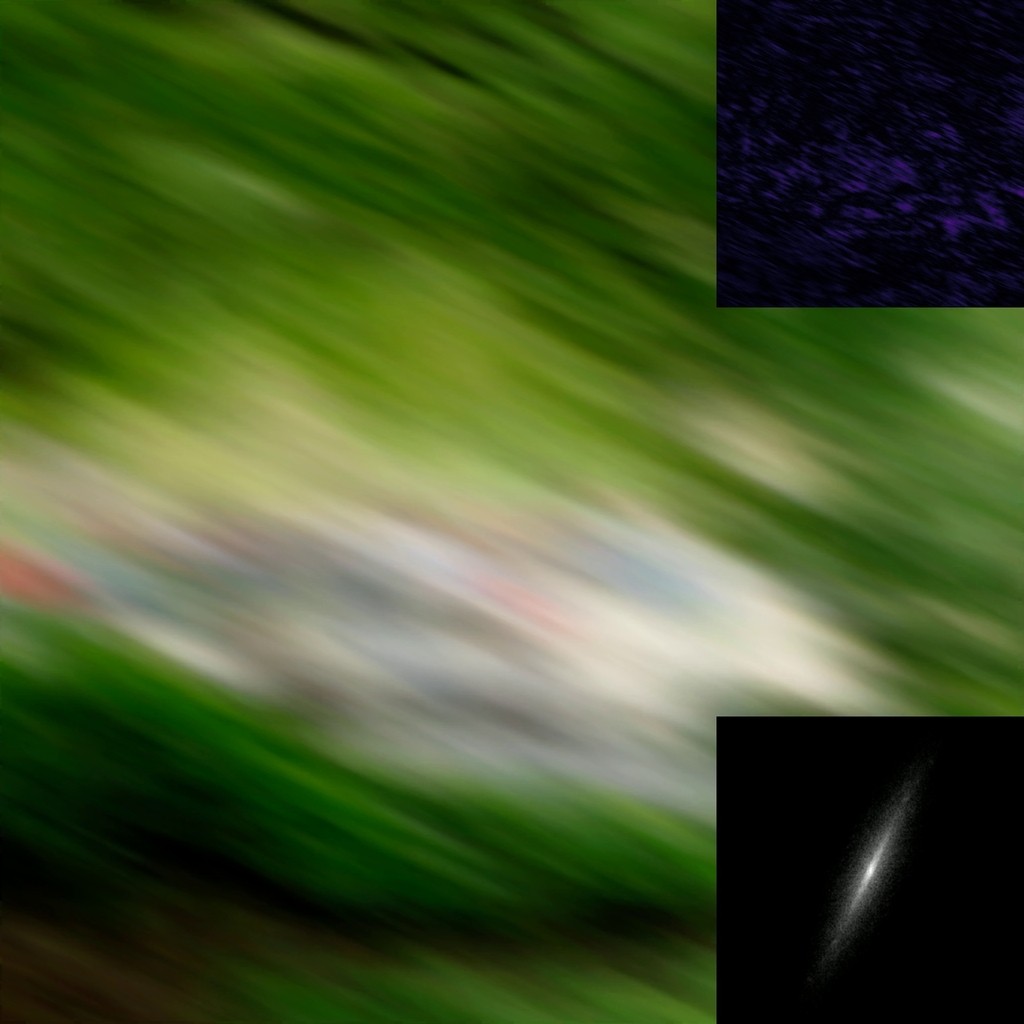} \\
            } & 
        \centeredtab{
            \includegraphics[width=0.155\textwidth, height=0.155\textwidth]{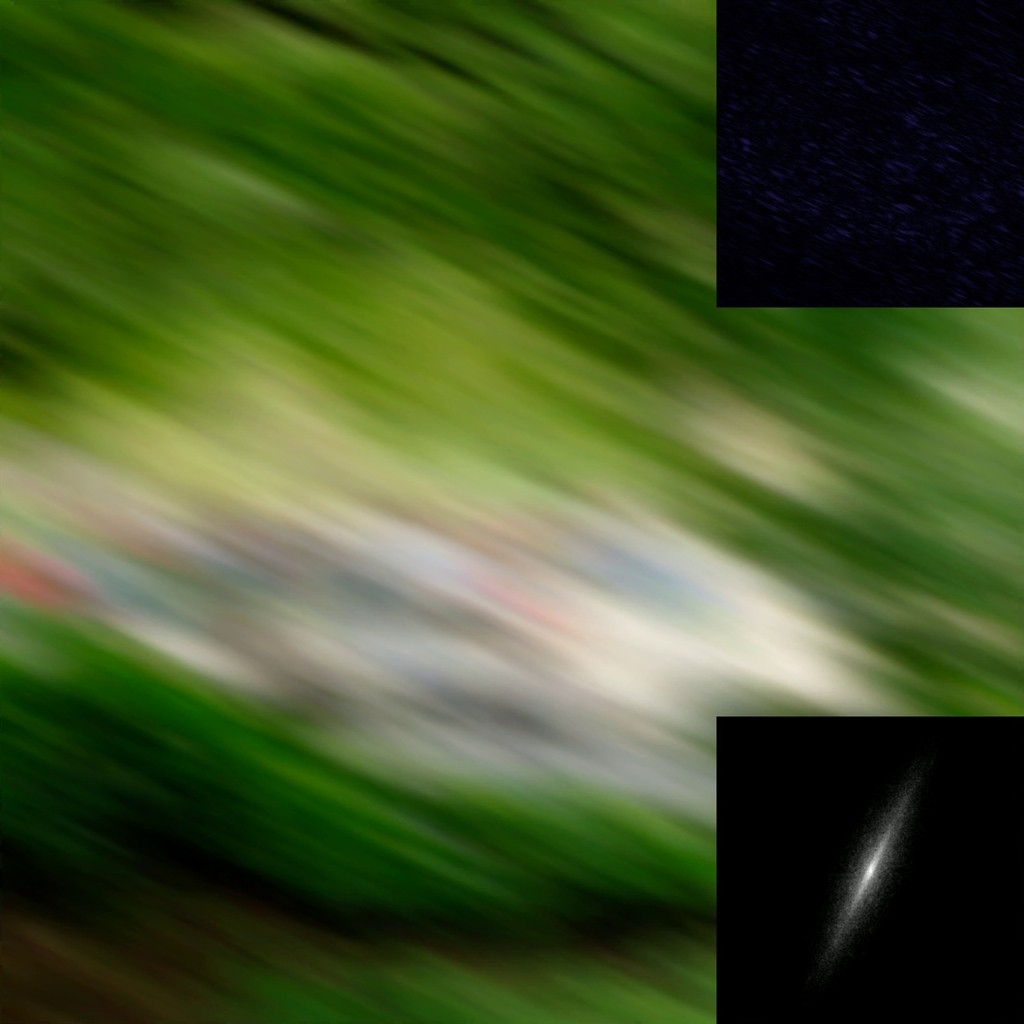} \\
            } & 
        \centeredtab{
            \includegraphics[width=0.155\textwidth, height=0.155\textwidth]{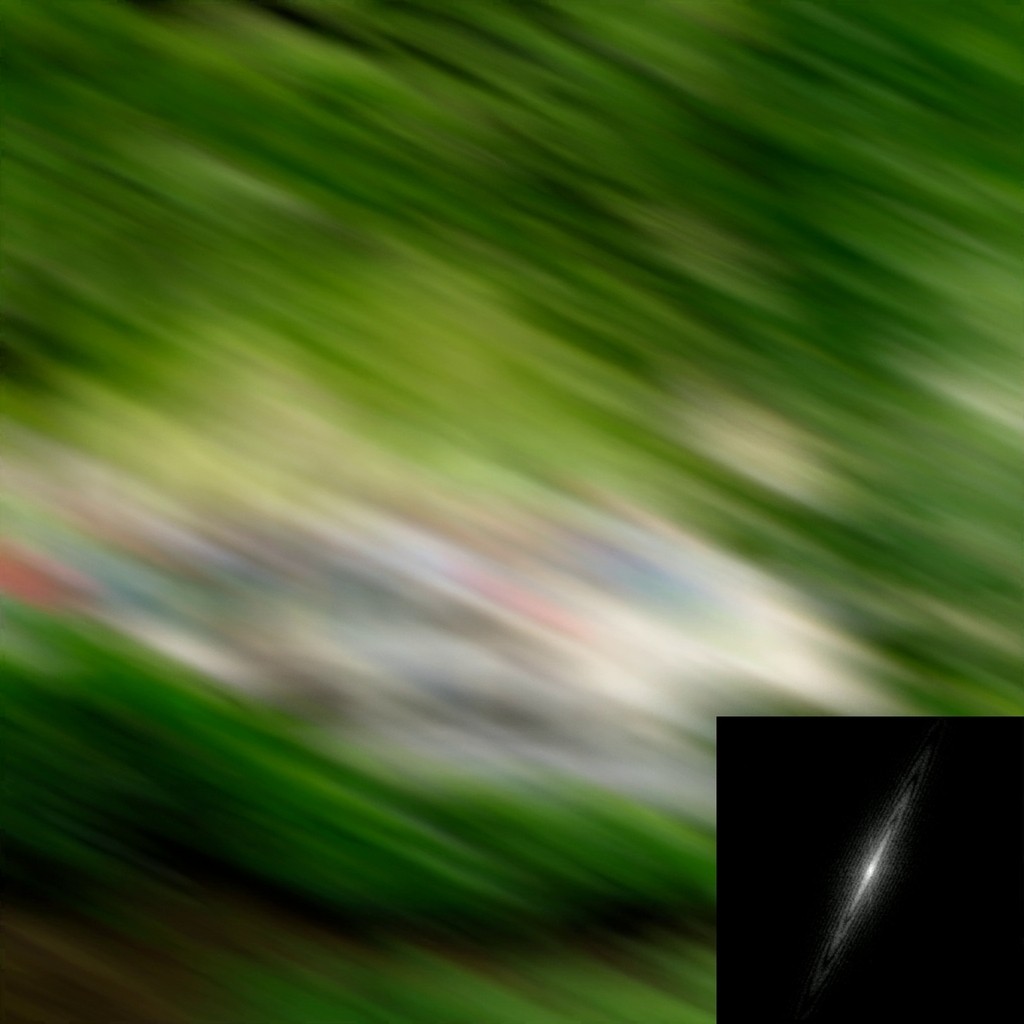} \\
            }\\
            \cmidrule(lr){3-8}
            \centeredtab{\rot{Lanczos}} & &
        \centeredtab{
            \includegraphics[width=0.155\textwidth, height=0.155\textwidth]{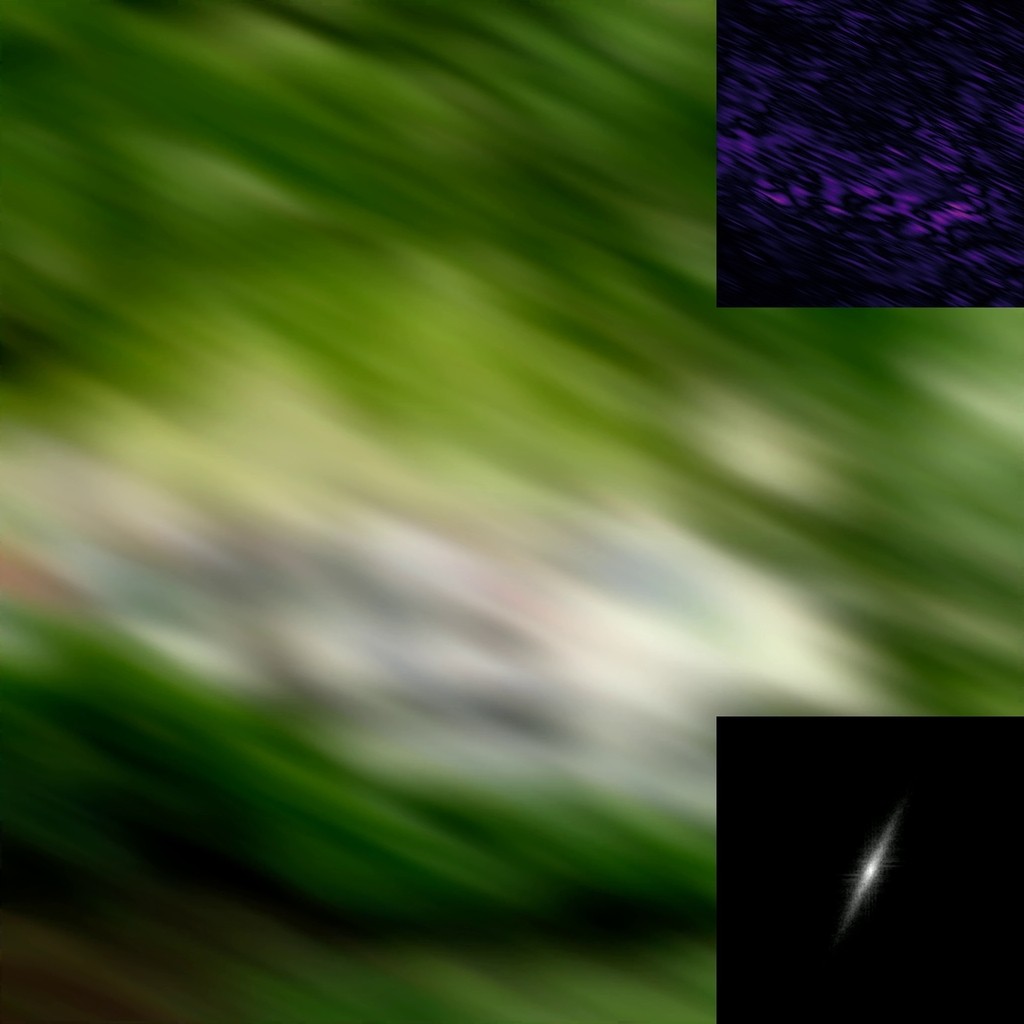} \\
            } & 
        \centeredtab{
            \includegraphics[width=0.155\textwidth, height=0.155\textwidth]{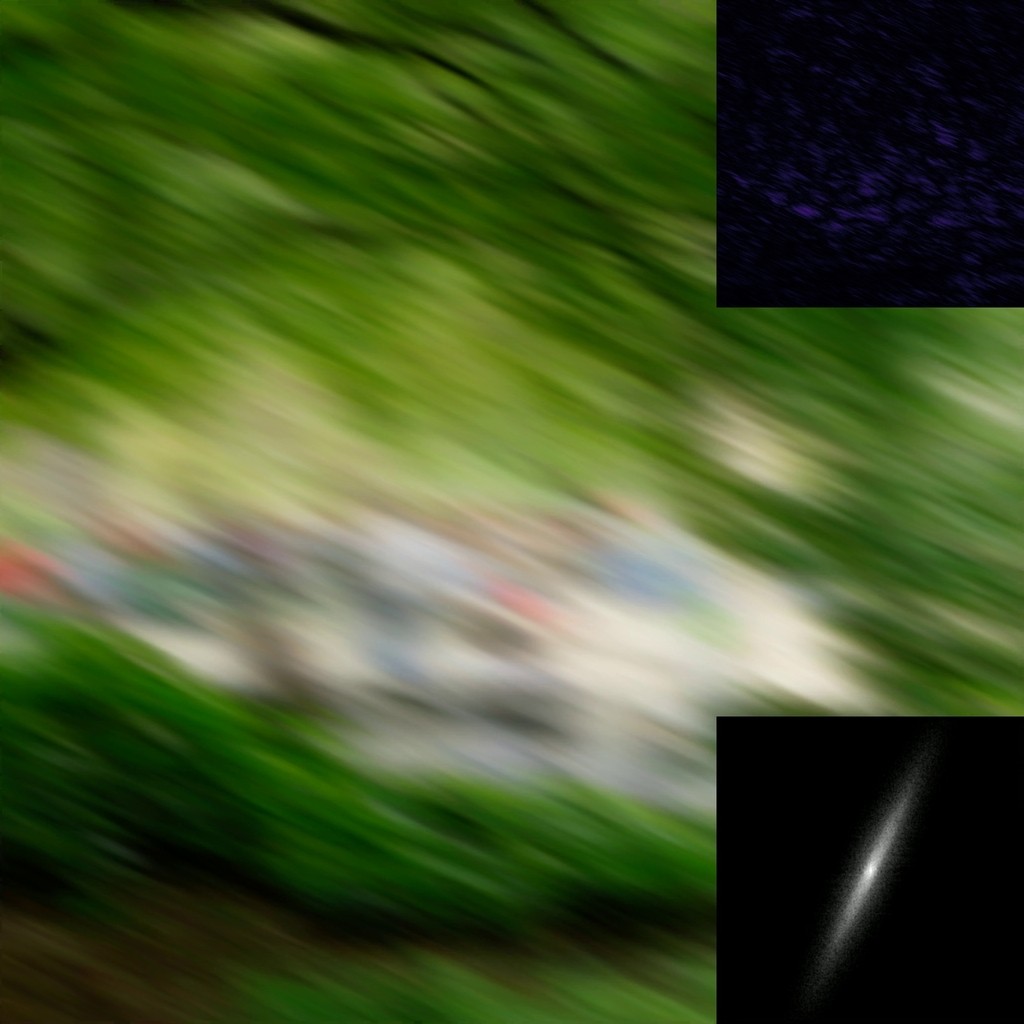} \\
            } & 
        \centeredtab{
            \includegraphics[width=0.155\textwidth, height=0.155\textwidth]{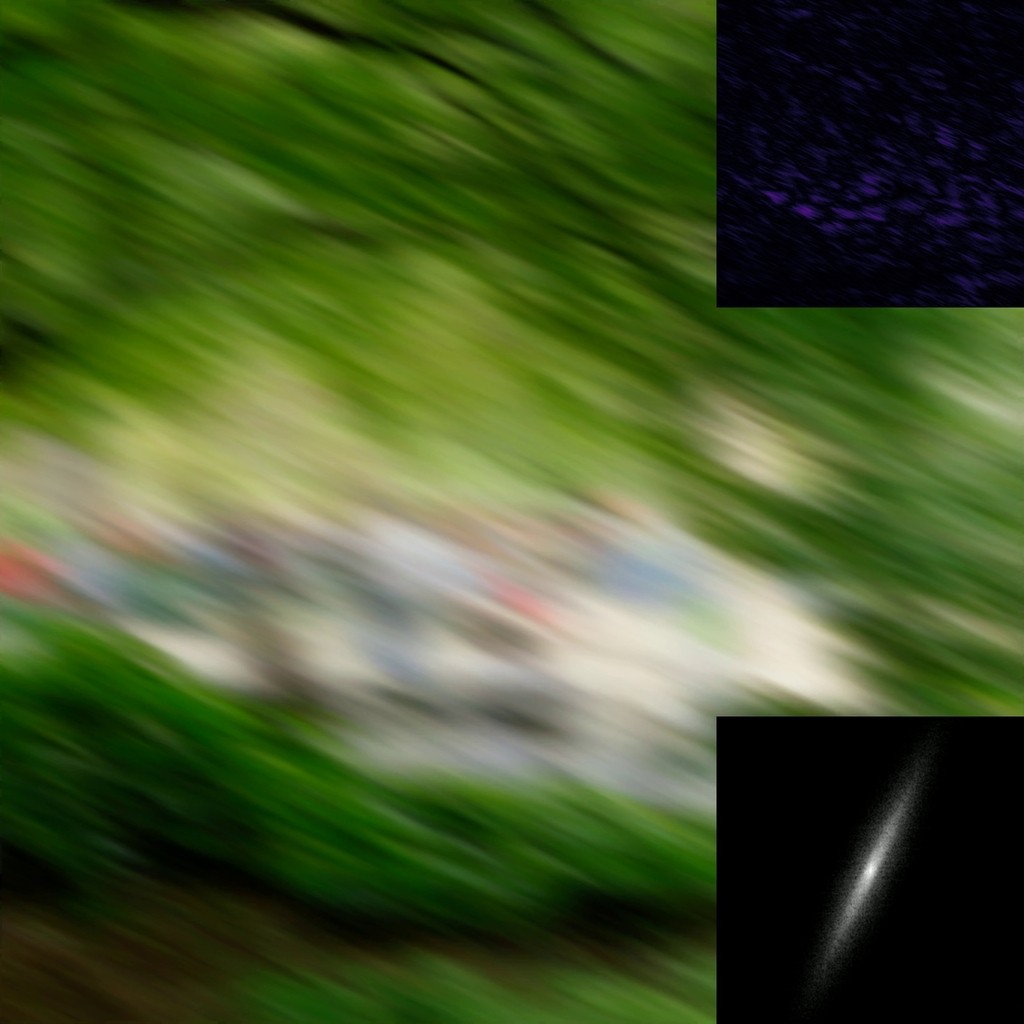} \\
            } & 
        \centeredtab{
            \includegraphics[width=0.155\textwidth, height=0.155\textwidth]{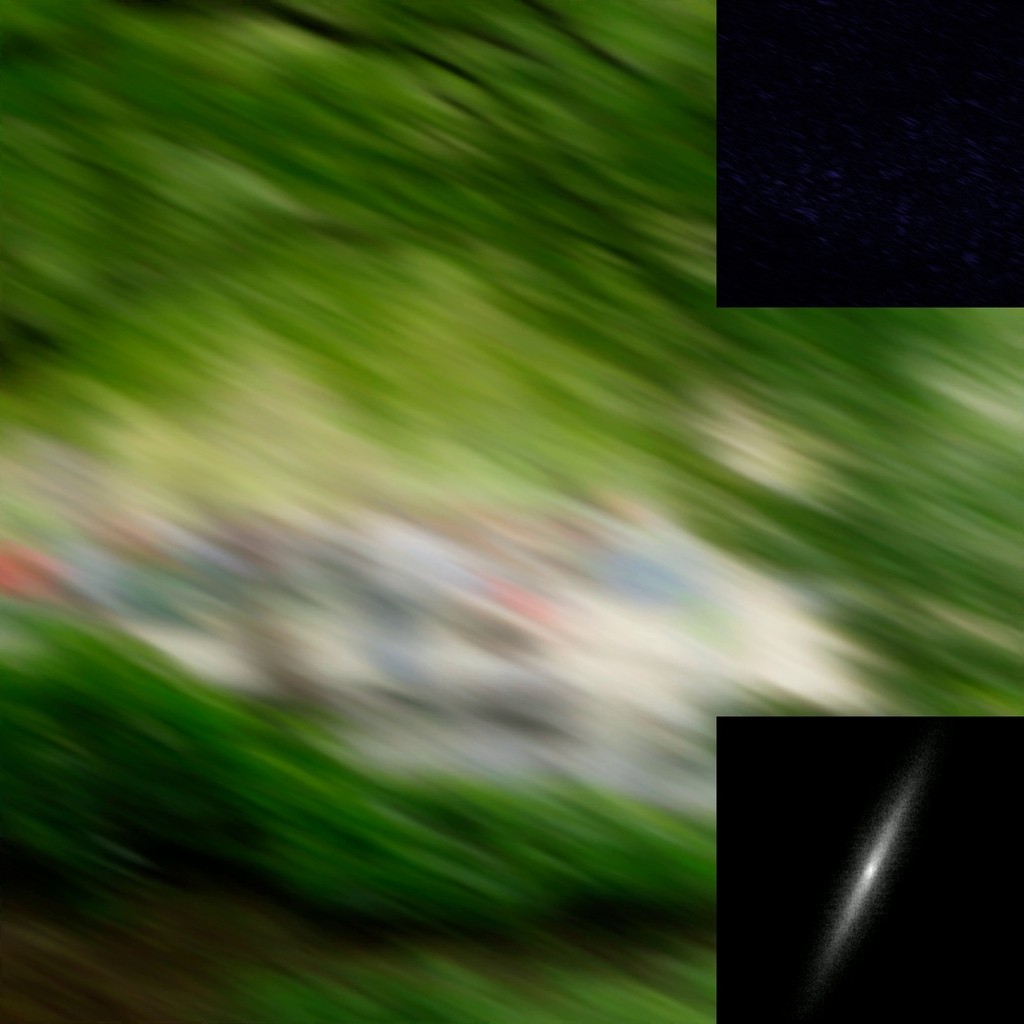} \\
            } & 
        \centeredtab{
            \includegraphics[width=0.155\textwidth, height=0.155\textwidth]{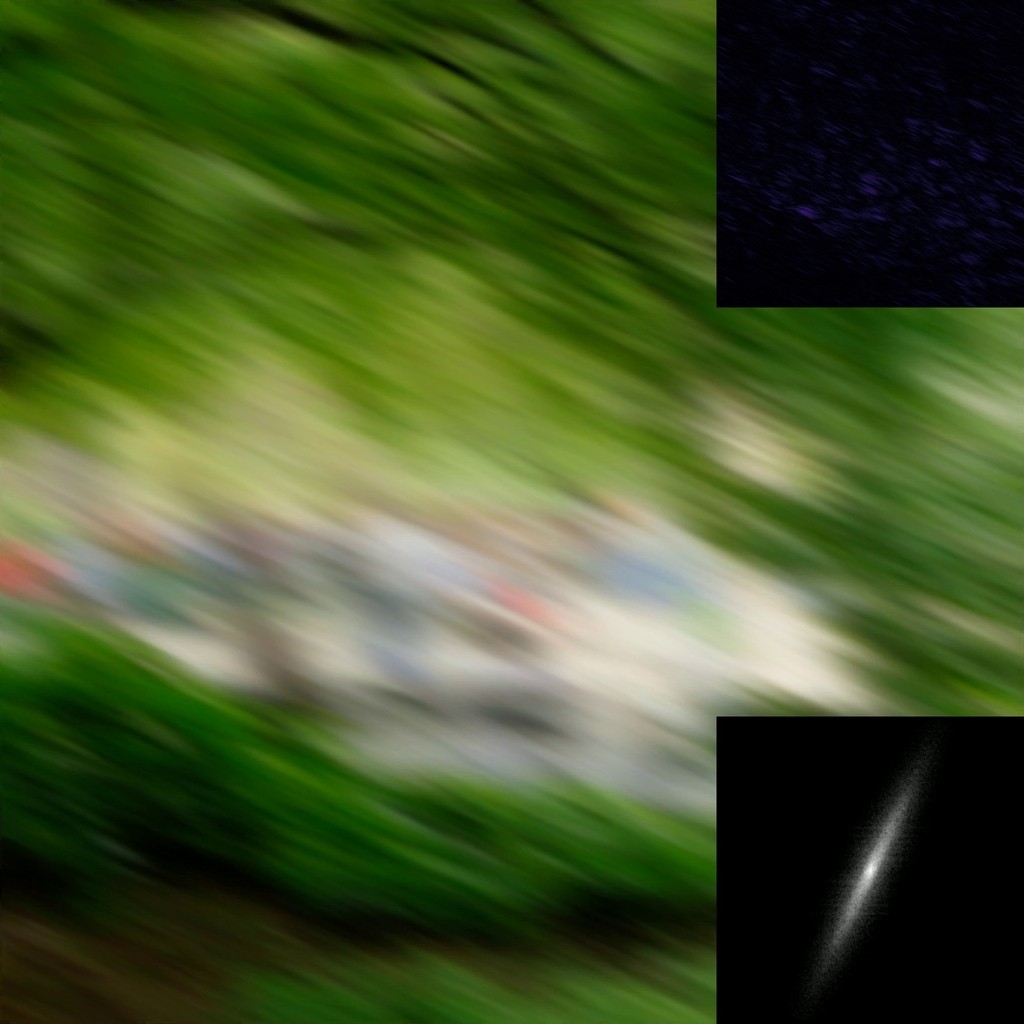} \\
            } & 
        \centeredtab{
            \includegraphics[width=0.155\textwidth, height=0.155\textwidth]{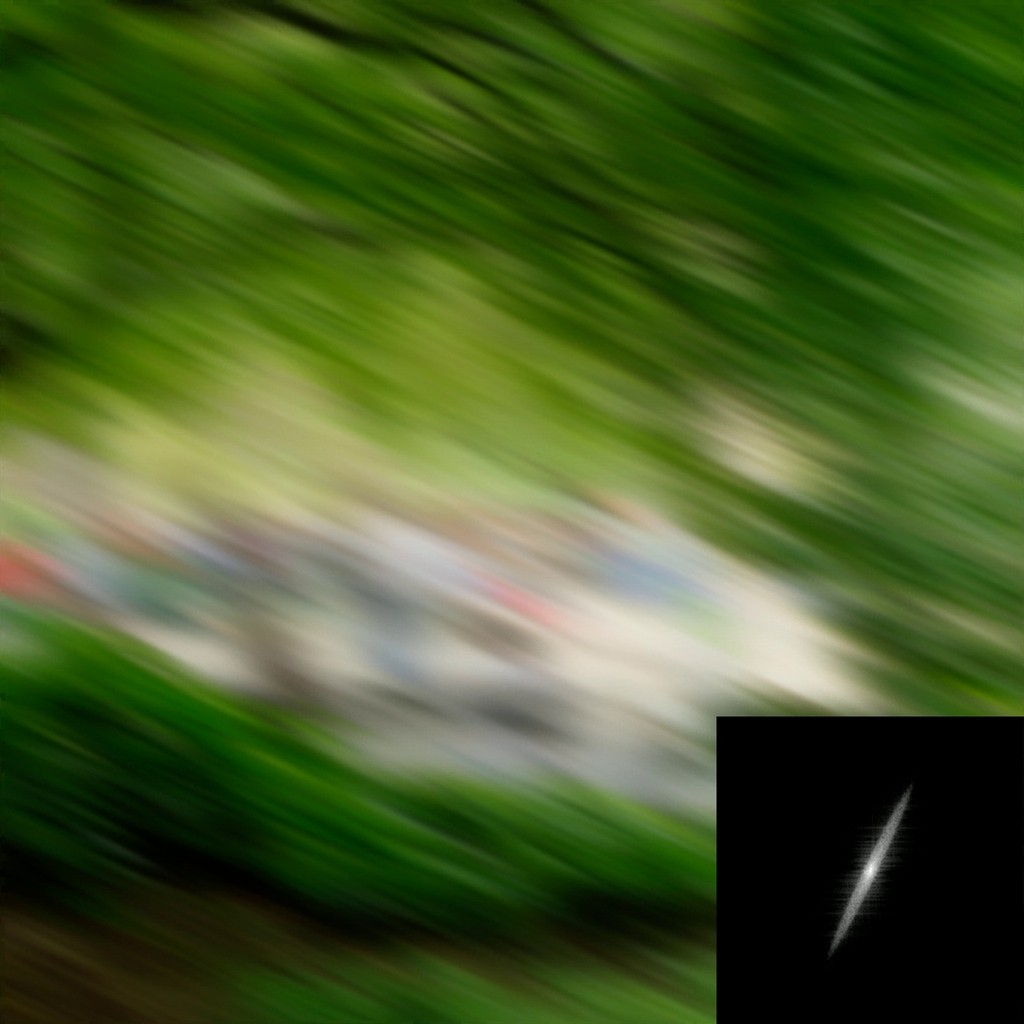} \\
            }
            \\
        & &  
        \multicolumn{2}{c}{
            \begin{tabular}{ccc}
                \centeredtab{0.0} & 
                \centeredtab{\fbox{\includegraphics[width=0.155\textwidth]{images/colorscale_magma.jpg}}} & 
                \centeredtab{0.25}
            \end{tabular}} \\
    \end{tabular}
    \caption{\label{fig:supp_results_images2}
        \captiontitle{Additional filter generalization comparisons.}
        Images from Adobe FiveK; \textcopyright{} original photographers/Adobe.
    }
\end{figure*}

\end{document}